\pdfoutput=1
\documentclass[rmp,twocolumn,amsmath,amssymb,seceqn]{revtex4}

\usepackage{graphicx}
\usepackage{graphics}
\usepackage{epsfig}
\usepackage{dcolumn}
\usepackage{bm}
\usepackage{makeidx}
\usepackage{subfigure}

\renewcommand{\Im}{\mathrm{Im}}

\makeatletter
\@addtoreset{equation}{section}
\makeatother

\newcommand{\rr}{\bm{r}}

\newcommand{\kk}{{\bm{k}}}

\newcommand{\bfig}{\begin{figure}}
\newcommand{\efig}{\end{figure}}
\newcommand{\incl}{\includegraphics}
\newcommand{\etal}{{\it et al}}
\newcommand{\be}{\begin{equation}}
\newcommand{\ee}{\end{equation}}

\begin{document}

\title{Anomalous Hall effect}

\author{Naoto Nagaosa}
\affiliation{Department of Applied Physics, University of Tokyo, Tokyo 113-8656, Japan}
\affiliation{Cross Correlated Research Materials Group (CMRG), ASI, RIKEN, Wako 351-0198, Saitama, Japan}

\author{Jairo Sinova}
\affiliation{Department of Physics, Texas A{\&}M University, College Station, Texas 77843-4242, USA}
\affiliation{Institute of Physics ASCR, Cukrovarnick\'a 10, 162 53 Praha 6, Czech
Republic }

\author{Shigeki Onoda}
\affiliation{Condensed Matter Theory Laboratory, ASI, RIKEN, Wako 351-0198, Saitama, Japan}

\author{A. H. MacDonald}
\affiliation{Department of Physics, University of Texas at Austin, Austin, Texas 78712-1081, USA}

\author{N. P. Ong}
\affiliation{Department of Physics, Princeton University, Princeton, New Jersey 08544, USA}

\date{\today}

\begin{abstract}

{We present a review of experimental and theoretical studies of the anomalous Hall effect (AHE),
focusing on recent developments 
that have provided a more complete framework for understanding this 
subtle phenomenon and have, in many instances, replaced 
controversy by clarity.  
Synergy between experimental and theoretical work, both playing a crucial role,
has been at the heart of these advances. 
On the theoretical front, the adoption of Berry-phase concepts has
established a link between the AHE and the topological nature of the Hall currents 
which originate from spin-orbit coupling. 
On the experimental front, new experimental studies of the AHE in transition metals, 
transition-metal oxides, spinels, pyrochlores, and  metallic dilute magnetic semiconductors,
have more clearly established systematic trends.  These two developments
in concert with first-principles electronic structure calculations, 
strongly favor the dominance of an intrinsic Berry-phase-related AHE mechanism in 
metallic ferromagnets with moderate conductivity.  
The intrinsic AHE can be expressed in terms of Berry-phase curvatures and it is therefore an intrinsic quantum mechanical property of a
perfect cyrstal.
An extrinsic mechanism, skew scattering from disorder, tends to dominate the 
AHE in highly conductive ferromagnets.  
We review the full modern semiclassical treatment of the AHE  
which incorporates an anomalous contribution to wavepacket group velocity 
due to momentum-space Berry curvatures and correctly combines
the roles of intrinsic and extrinsic (skew scattering and side-jump) scattering-related mechanisms.  
In addition, we review more rigorous quantum-mechanical treatments based on the Kubo and Keldysh formalisms,
taking into account multiband effects, and demonstrate the equivalence of all three linear response theories in the metallic regime.   
Building on results from recent experiment and theory, we propose a tentative global view of the AHE which 
summarizes the roles played by intrinsic and extrinsic contributions in the disorder-strength {\em vs.} temperature plane.
Finally we discuss outstanding issues and avenues for future investigation. 
}
\end{abstract}

\maketitle
\tableofcontents


\section{Introduction}\label{sec:intro}

	\subsection{A brief history of the AHE and new perspectives}\label{sec:brief}

The anomalous Hall effect has deep roots in the history of
electricity and magnetism.  In 1879 Edwin H. Hall~\cite{Hall:1879_a} made the
momentous discovery that, when a current-carrying conductor is placed in a magnetic field, the Lorentz force ``presses'' its electrons against one side of the conductor.  One year later, he reported that his ``pressing electricity'' effect was
ten times larger in ferromagnetic iron~\cite{Hall:1881_a} than in non-magnetic conductors.  
Both discoveries were remarkable, given how 
little was known at the time about how charge moves through conductors.  The first discovery provided a simple, elegant tool to measure  carrier concentration 
more accurately in non-magnetic conductors, and played a midwife's role in easing the birth of semiconductor physics and solid-state electronics in the late 1940's.  For this role, the Hall effect was frequently called the queen of solid-state transport experiments.  

The stronger effect that Hall discovered in ferromagnetic conductors came to be known as the anomalous Hall effect (AHE).   The AHE has been an enigmatic problem that has resisted theoretical and experimental assault for almost a century.  The main reason seems to be that, at its core, the AHE problem involves concepts based on topology and geometry that have been formulated only in recent times.  The early investigators grappled with notions that would not become clear and well defined until much later, such as the
concept of Berry-phase \cite{Berry:1984_a}.  What is now viewed as Berry phase curvature,  later dubbed ``anomalous velocity" by Luttinger, arose naturally in the first microscopic theory of the AHE by Karplus and Luttinger~\cite{Karplus:1954_a}.  However, because  understanding of these concepts, not to mention the odd intrinsic dissipationless Hall current they seemed to imply, would not be achieved for another 40 years, the AHE problem was quickly mired in a controversy of unusual endurance.  Moreover, the AHE seems to be a rare example of a pure, charge-transport problem whose elucidation has not -- to date -- benefited from the application of complementary spectroscopic and thermodynamic probes.

\bfig[h]
\incl[width=7cm]{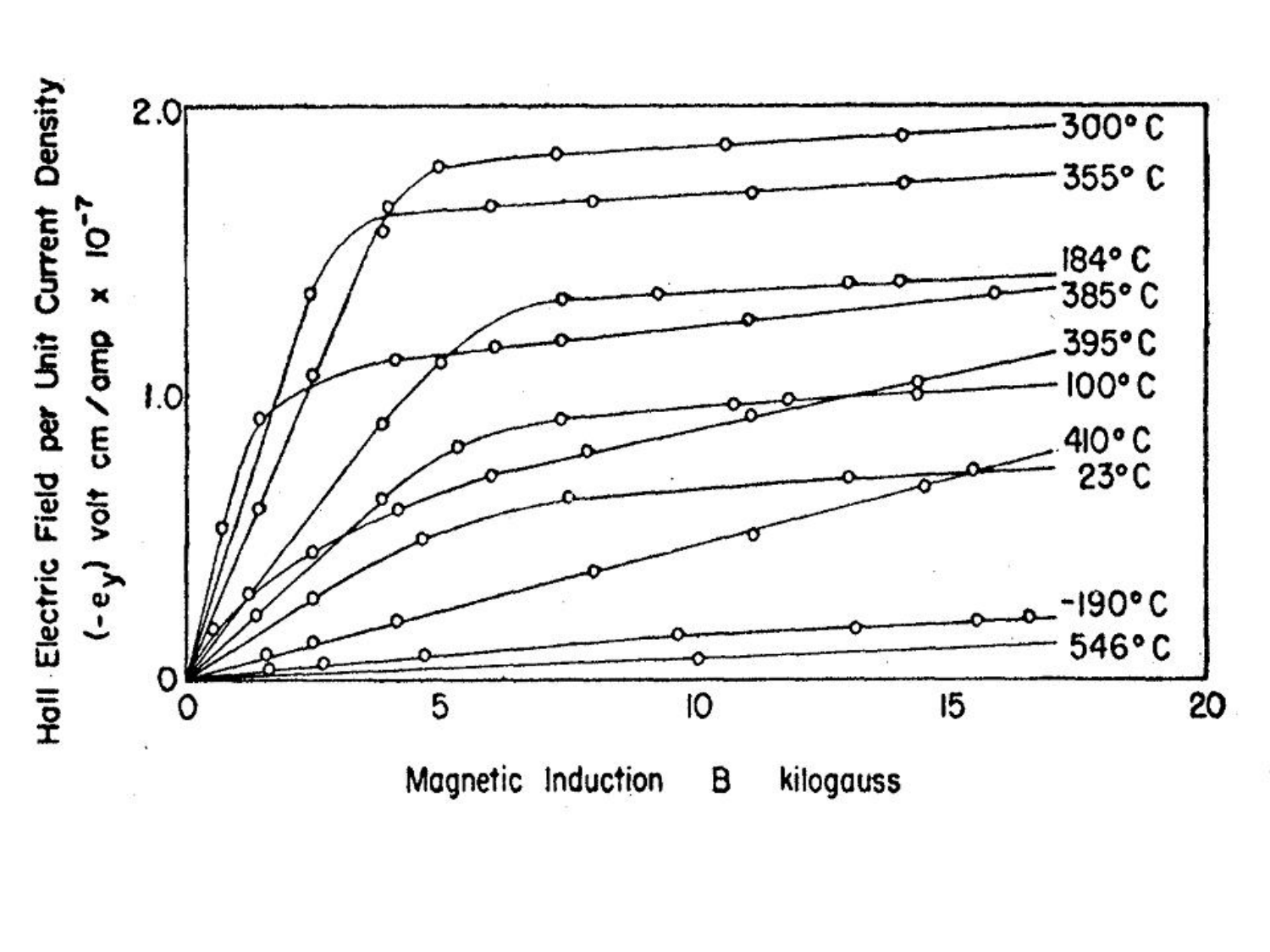}
\caption{\label{fig:Pugh}
The Hall effect in Ni [data from A. W. Smith, Phys. Rev. {\bf 30}, 1 (1910)]. [From Ref.~\onlinecite{Pugh:1953_a}.]
}
\efig
Very early on, experimental investigators learned that
the dependence of the Hall resistivity $\rho_{xy}$ on applied perpendicular field $H_z$
is qualitatively different in 
ferromagnetic and non-magnetic conductors.
In the latter, $\rho_{xy}$ increases linearly with $H_z$, as expected
from the Lorentz force.  In ferromagnets, however, $\rho_{xy}$
initially increases steeply in weak $H_z$, but saturates at a large value that is nearly $H_z$-independent (Fig. \ref{fig:Pugh}). 
Kundt noted that, in Fe, Co, and Ni, the saturation 
value is roughly proportional to the magnetization $M_z$~\cite{Kundt:1893_a} and has a weak
anisotropy when the field ($\hat{z}$) direction is rotated with respect to the cyrstal, corresponding to the weak magnetic anisotropy of Fe, Co, and Ni ~\cite{Webster:1925_a}.
Shortly thereafter, experiments by Pugh and coworkers~\cite{Pugh:1930_a,Pugh:1932_a} established 
that an empirical relation between $\rho_{xy}$, $H_z$, and $M_z$,
\begin{equation}
  \rho_{xy}=R_0H_z+R_s M_z,
  \label{eq:exp:fe:pugh}
\end{equation}
applies to many materials over a broad range of external magnetic fields.
The second term represents the Hall effect 
contribution due to the spontaneous magnetization.  This  AHE is the subject of this review.  
Unlike $R_0$, which was already understood to depend mainly on the density of carriers, $R_s$ was found to 
depend subtly on a variety of material specific
parameters and, in particular, on the longitudinal resistivity $\rho_{xx} = \rho$.

In 1954, Karplus and Luttinger (KL)~\cite{Karplus:1954_a}
proposed a theory for the AHE that, in hindsight, provided 
a crucial step in unraveling the AHE problem. KL showed that when an 
external electric field is applied to a solid, electrons acquire an additional contribution
to their group velocity.  KL's {\em anomalous velocity} was perpendicular to the
electric field and therefore could contribute to Hall effects.  In the case of ferromagnetic conductors,
the sum  of the {\em anomalous velocity} over all occupied band states
can be non-zero, implying a contribution to the Hall conductivity 
$\sigma_{xy}$.  Because this contribution depends only on the band structure and is largely independent of scattering, it has 
recently been referred to as the  
the {\em intrinsic} contribution to the AHE. When the conductivity tensor is inverted, the 
intrinsic AHE yields a contribution to $\rho_{xy} \approx \sigma_{xy}/\sigma_{xx}^2$ 
and therefore it is proportional to $\rho^2$.
The anomalous velocity is dependent only on the perfect crystal Hamiltonian and 
can be related to changes in the phase of Bloch state wavepackets when 
an electric field causes them to evolve in crystal momentum space \cite{Chang:1996_a,Sundaram:1999_a,Xiao:2009_a,Bohm:2003_a}.
As mentioned, the KL theory anticipated by several decades the modern interest 
in Berry phase and Berry curvature 
effects, particularly in momentum-space.  

\bfig[h]
\incl[width=7cm]{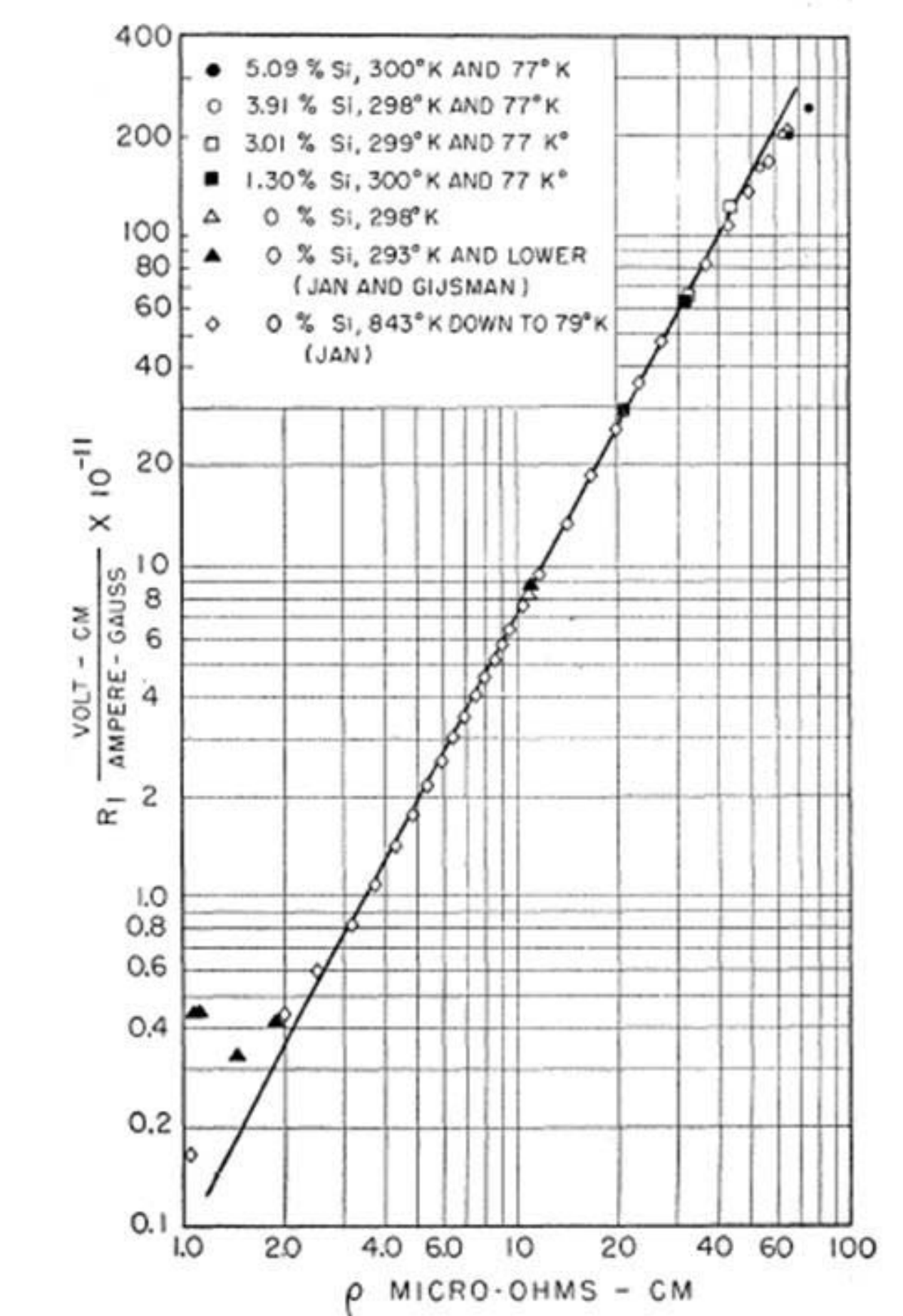}
\caption{\label{fig:Kooi}
Extraordinary Hall constant as a function of resistivity. The shown fit has the relation
$R_s\sim\rho^{1.9}$. [From Ref.~\onlinecite{Kooi:1954_a}.] 
}
\efig

Early experiments to measure the relationship between $\rho_{xy}$ and $\rho$ generally assumed to be of the power law 
form, {\it i.e.}, $\rho_{xy}\sim\rho^{\beta}$, mostly
involved plotting $\rho_{xy}$ (or $R_s$) vs. $\rho$, measured in 
a single sample over a broad interval of $T$ (typically 77 to 300 K).
As we explain below, competing theories in metals suggested either that $\beta=1$ or $\beta =2$.
A compiled set of results was published by Kooi~\cite{Kooi:1954_a} (Fig. \ref{fig:Kooi}).  
The subsequent consensus was that such plots do not settle
the debate.  At finite $T$, the carriers are strongly scattered 
by phonons and spin waves.  These inelastic processes -- 
difficult to treat microscopically even today -- lie far outside
the purview of the early theories.  Smit suggested that, in 
the skew-scattering theory (see below), phonon scattering increases the
value $\beta$ from 1 to values
approaching 2.  This was also found by other investigators. 
A lengthy calculation by Lyo~\cite{Lyo:1973_a} showed 
that skew-scattering 
at $T\gg \Theta_D$ (the Debye temperature) 
leads to the relationship $\rho_{xy}\sim (\rho^2+a \rho)$, with $a$ a constant.  
In an early theory by Kondo considering skew scattering from spin excitations \cite{Kondo:1962_a}, 
it may be seen that $\rho_{xy}$ also varies as $\rho^2$ at finite $T$.  

The proper test of the scaling relation in comparison with present theories involves measuring $\rho_{xy}$ and $\rho$ in a set of
samples at 4 K or lower (where impurity scattering dominates). By adjusting
the impurity concentration $n_i$, one may hope to change both quantities sufficiently to determine accurately the exponent $\beta$ 
and use this identification to tease out the underlying physics.

\begin{figure}[h] 
\fbox{\includegraphics[width=0.9\columnwidth]{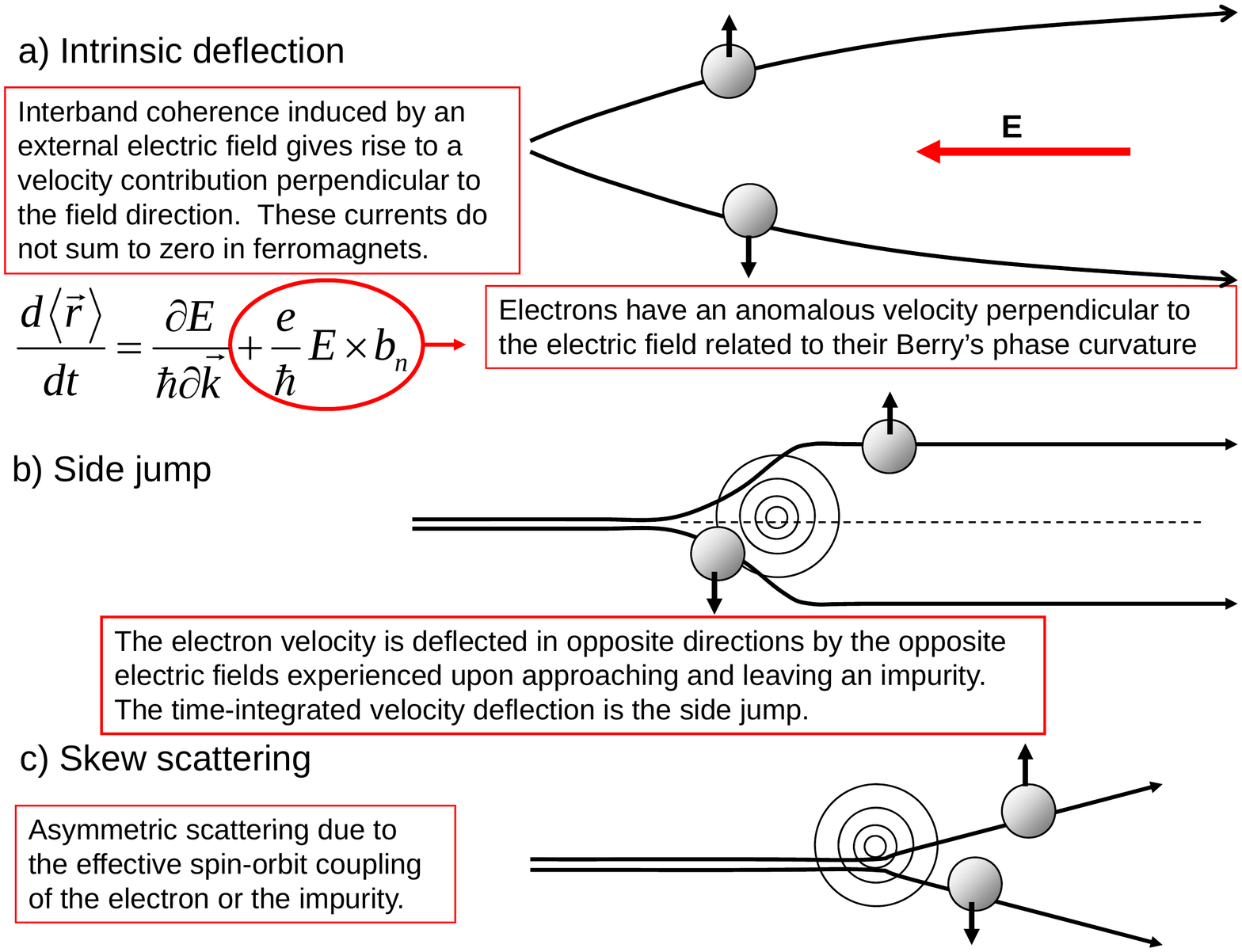}}
\caption{Illustration of the three main mechanisms that can give rise to an AHE.  In any real material all of these mechanisms act
to influence electron motion.}
\label{mechanisms}
\end{figure}

The main criticism of the KL theory centered on the complete absence of scattering from disorder
in the derived Hall response contribution.
The semi-classical AHE theories by Smit and Berger
focused instead on the influence of disorder scattering 
in imperfect crystals.
Smit argued that the main source of the AHE currents was 
asymmetric ({\em skew}) scattering from impurities caused by the spin-orbit
interaction (SOI)~\cite{Smit:1955_a,Smit:1958_a}.  
This AHE picture predicted that $R_s\sim \rho_{xx}$ ($\beta$ = 1).  Berger, on the other hand, 
argued that the main source of the AHE current was the {\em side-jump} experienced
by quasiparticles upon scattering from spin-orbit coupled impurities.
The side-jump mechanism could (confusingly) be viewed 
as a consequence of a KL anomalous velocity mechanism 
acting while a quasiparticle was under the influence of the electric field due to an impurity.  
The side-jump AHE current was viewed as the product of the side-jump per scattering event
and the scattering rate~\cite{Berger:1970_a}.  
One puzzling aspect of this semiclassical theory was that all dependence on the impurity density and strength seemingly dropped out.
As a result, it predicted $R_s\sim \rho_{xx}^2$ with an 
exponent $\beta$ identical to that of the KL mechanism.  
The side-jump mechanism therefore yielded a contribution to the Hall conductivity which was 
seemingly independent of the density or strength of scatterers.
In the decade 1970-80, a lively AHE debate was waged largely between the
proponents of these two extrinsic theories. 
The three main mechanisms considered in this early history are shown schematically in Fig.~\ref{mechanisms}.

Some of the confusion in experimental studies stemmed from 
a hazy distinction between the KL mechanism and the side-jump mechanism, a
poor understanding of how the effects competed at a microscopic level,
and a lack of systematic experimental studies in a diverse set of materials.

One aspect of the confusion may be illustrated by contrasting the case of a high-purity 
mono-domain ferromagnet, which produces a spontaneous 
AHE current proportional
to $M_z$, with  the case of a material containing magnetic impurities 
(e.g. Mn) embedded in a non-magnetic 
host such as Cu (the dilute Kondo system).  In a field $H$, the latter 
also displays an AHE current proportional to the induced $M = \chi H$, with $\chi$ the susceptibility
~\cite{Fert:1972_a}.  
However, in \emph{zero} $H$, time-reversal invariance (TRI)
is spontaneously broken in the former, 
but not in the latter. 
Throughout the period 1960-1989, 
the two Hall effects were often regarded as a common phenomenon that
should be understood microscopically on the same terms. 
It now seems clear that this view impeded progress. 

By the mid-1980s, interest in the AHE problem had  
waned significantly. The large body of Hall data garnered from experiments on 
dilute Kondo systems in the previous two decades showed that  
$\rho_{xy}\sim \rho$ and therefore appeared to favor the skew-scattering mechanism.
The points of controversy remained unsettled, however, and the topic was still mired in confusion. 

Since the 1980's, the quantum Hall effect 
in two-dimensional (2D) electron systems in semiconductor heterostructures has  become a major field
of research in physics~\cite{Prange:1987_a}.
The accurate quantization of the Hall conductance 
is the hallmark of this phenomenon.  Both the integer~\cite{Thouless:1982_a} and fractional 
quantum Hall effects can be explained in terms of the topological properties of the electronic wavefunctions.  For the case of electrons in a two-dimensional crystal, it has been found that the Hall conductance
is connected to the topological integer (Chern number) defined for
the Bloch wavefunction over the first Brillouin zone~\cite{Thouless:1982_a}.
This way of thinking about the quantum Hall effect began to have a deep impact on the AHE  problem starting around 1998.
Theoretical interest in the Berry phase and in its relation to transport phenomena,   
coupled with many developments in the growth of novel complex magnetic systems with strong spin-orbit coupling (notably the manganites, pyrochlores and
spinels) led to a strong resurgence of interest in the AHE and eventually to  deeper understanding.

Since 2003 many systematics studies, both theoretical and experimental, have led to a better understanding of 
the AHE in the metallic regime, and to the 
recognition of new unexplored regimes that  present challenges to future researchers.
As it is often the case in condensed matter physics, attempts to understand this complex and fascinating phenomenon
have motivated researchers to 
couple fundamental and sophisticated mathematical concepts to real-world materials
issues. The aim of this review is to survey recent 
experimental progress in the field, 
and to present the theories in a systematic fashion.
Researchers are now able to understand the 
links between different views on the AHE previously thought to be in conflict.  Despite the progress in recent years, understanding is still incomplete.  
We highlight some intriguing questions that remain and speculate on the most promising avenues for future exploration. 
In this review we focus, in particular, on reports
that have contributed significantly to the modern view of the AHE. For previous reviews, the reader may consult Pugh~\cite{Pugh:1953_a} and 
Hurd~\cite{Hurd:1972_a}.  For more recent short overviews focused on the topological aspects of the AHE,  we point the reader to the 
 reviews by Nagaosa~\cite{Nagaosa:2006_a}, and by Sinova \etal.\cite{Sinova:2004_b}. A review
 of the modern semiclassical treament of AHE was recently written by Sinitsyn \cite{Sinitsyn:2008_a}. 
 The present review has been informed by ideas explained in the earlier works.
 Readers who are not familair with Berry phase concepts may find it useful to consult the elementary review by Ong and Lee~\cite{Ong:2005_a} and the popular commentary by MacDonald and Niu \cite{MacDonald:2004_a}.

Some of the recent advances in the understanding of the AHE that will be covered in this review are:

\begin{enumerate}

\item
 When $\sigma_{xy}^{AH}$ is independent of $\sigma_{xx}$, the AHE can often be
 understood in terms of the geometric concepts of Berry phase and Berry curvature in momentum 
 space. This AHE mechanism is responsible for the intrinsic AHE. In this regime, the anomalous Hall current can be thought of as the unquantized 
 version of the quantum Hall effect. In 2D systems the intrinsic AHE is quantized in units of 
$e^2/h$ at temperature $T=0$ when the Fermi level lies between Bloch state bands.  
 
\item
  Three broad regimes have been identified when surveying a large body of experimental data for diverse materials:
(i) A high conductivity regime ($\sigma_{xx}> 10^6$ $(\Omega{\rm cm})^{-1}$) in which a linear contribution to $\sigma_{xy}^{AH}\sim \sigma_{xx}$
due to skew scattering dominates $\sigma_{xy}^{AH}$. In this regime  the normal Hall conductivity contribution can 
be significant and even dominate $\sigma_{xy}$, 
(ii) An intrinsic or scattering-independent regime  in which 
$\sigma_{xy}^{AH}$ is roughly independent of $\sigma_{xx}$ ($10^4$ $(\Omega{\rm cm})^{-1} < \sigma_{xx} <10^6$ $(\Omega{\rm cm})^{-1}$),
(iii) A bad-metal regime ($\sigma_{xx} < 10^4$ $(\Omega{\rm cm})^{-1}$) in which $\sigma_{xy}^{AH}$ decreases with decreasing $\sigma_{xx}$ at a rate faster than linear.

\item
The relevance of the intrinsic mechanisms can be studied in-depth in magnetic materials with strong spin-orbit coupling,
such as oxides and diluted magnetic semiconductors (DMS).
In these systems a systematic non-trivial comparison between the observed 
properties of systems with well controlled materials properties and 
theoretical model calculations can be achieved.

\item
The role of band (anti)-crossings near the Fermi energy has 
been identified using first-principles Berry curvature 
calculations as a mechanism which can lead to a large intrinsic AHE.

\item
 Semiclassical treatment by a generalized Boltzmann equation 
   taking into account the Berry curvature and coherent inter-band mixing effects 
   due to band structure and disorder has been formulated.  This 
   theory provides a clearer
   physical picture of the AHE than early theories by identifying correctly all the semiclassically defined
   mechanisms. This generalized semiclassical picture has been verified by comparison with
  controlled microscopic linear response treatments for identical models. 

\item
The relevance of non-coplanar spin structures with associated 
spin chirality and real-space Berry curvature to the AHE has been established both theoretically 
and experimentally in several materials.

\item
Theoretical frameworks based 
on the Kubo formalism and the Keldysh formalism have been developed which are capable of
treating transport phenomena in systems with multiple bands.
\end{enumerate}
 
The review is aimed at experimentalists and theorists interested in the AHE. We have structured the review as follows. In the remainder of this section, we provide the minimal theoretical background necessary to understand the different AHE mechanisms.  In particular we explain the
scattering-independent Berry phase mechanism which is more important for the AHE than for any other commonly measured transport coefficient. 
 In Sec. \ref{sec:experiments} we review recent experimental results on a broad range of materials, and compare them with relevant calculations where available. 
 In Sec. \ref{sec:theory}  we discuss AHE theory from an historical perspective,
  explaining links between different ideas which are not always recognized, and discussing the 
physics behind some of the past confusion.  The section may be skipped by readers who do not wish to be burdened by history.  
 Section \ref{sec:linear_transport}  discusses the present understanding of the metallic theory based on a careful comparison of the different linear response theories which are now finally consistent. In Sec. \ref{sec:summary} we present a summary and outlook.

	\subsection{Parsing the AHE:}\label{mechanisms-intro}

The anomalous Hall effect is at its core a quantum phenomena
which originates from quantum coherent  band mixing effects by both the external electric field and the disorder 
potential. Like other coherent interference transport phenomena (e.g. weak localization),
 it cannot be satisfactorily explained using traditional semiclassical Boltzmann transport theory.
Therefore, when parsing the different contributions to the AHE,  they can be defined semiclassically only in a carefully elaborated theory.

In this section we identify three distinct contributions which sum up to 
yield the full AHE: {\em intrinsic, skew scattering, and side-jump contributions}.
We choose this nomenclature to reflect the modern literature without breaking completely from the established AHE lexicon (see Sec.~\ref{sec:theory}). 
{\em However, unlike previous classifications, we base this parsing of the AHE on experimental and microscopic transport
theory considerations, rather than on the identification of one particular effect which could contribute to the AHE. } 
The link to semiclassically defined processes is established  
after developing a fully generalized Boltzmann transport theory which takes  inter-band coherence effects into
account  and is fully equivalent to 
microscopic theories (Sec.~\ref{sec:theory:Boltzmann}). 
In fact, much of the theoretical effort of the past few
years has been expended in understanding  this link between  semiclassical and microscopic theory
which has escaped cohesion for a long time.
 
A very natural classification of contributions to the AHE, which is guided by experiment and by microscopic theory of 
metals, is  to separate them according to their dependence on the Bloch state transport lifetime $\tau$. 
In the theory, disorder is treated perturbatively and 
higher order terms  vary with a higher power of the quasiparticle scattering rate $\tau^{-1}$.
 As we will discuss, it is relatively easy to identify contributions to the 
anomalous Hall conductivity, $\sigma_{xy}^{AH}$, which vary as $\tau^{1}$ and as $\tau^{0}$.  
In experiment a similar  separation can sometimes be achieved  by plotting $\sigma_{xy}$ vs.
 the  longitudinal conductivity $\sigma_{xx} \propto \tau$, when $\tau$ is 
varied by  altering disorder or varying temperature.  More commonly (and 
equivalently) the  Hall resistivity is separated into contributions proportional to $\rho_{xx}$  and 
 $\rho_{xx}^2$.  
 
This partitioning seemingly gives only two contributions to $\sigma_{xy}^{AH}$, one $\sim \tau$ and the other
$\sim \tau^0$. The first contribution we {\em define} as the {\em skew-scattering} contribution, $\sigma_{xy}^{AH-skew}$.
Note that in this parsing of AHE contributions it is the dependence on $\tau$ (or $\sigma_{xx}$) which defines it, not a particular mechanism linked to a microscopic or
semiclassical theory.
 The second contribution 
proportional to $\tau^0$  (or independent of $\sigma_{xx}$) we further separate into two parts: {\em intrinsic} and {\em side-jump}.
Although these two contributions cannot be separated experimentally by dc measurements, they {\em can} be separated experimentally (as well as theoretically)
by defining the intrinsic contribution, $\sigma_{xy}^{AH-int}$,
as the extrapolation of the ac-interband Hall conductivity to zero frequency in the limit of $\tau \rightarrow \infty$,
with $1/\tau\rightarrow 0$ faster than $\omega\rightarrow 0$. This then leaves a unique definition for the third and last contribution,
 termed side-jump, as $\sigma_{xy}^{AH-sj}\equiv \sigma_{xy}^{AH}-\sigma_{xy}^{AH-skew}-\sigma_{xy}^{AH-int}$. 

We examine these three contributions below ( still at an introductory level). It is important to note that the above definitions have 
not relied on identifications of semiclassical processes such as side-jump scattering \cite{Berger:1970_a} or skew-scattering from asymmetric 
contributions to the semiclassical scattering rates \cite{Smit:1955_a} identified in earlier theories. Not surprisingly, the contributions 
defined above contain these semiclassical processes. However, it is now understood (see Sec.~\ref{sec:linear_transport}), that other contributions 
are present in the fully generalized semiclassical theory which were not  precisely identified previously and which are necessary to
be fully consitent with microscopic theories.

The ideas explained briefly in this section are substantiated in Sec.~\ref{sec:experiments} 
by analyses of tendencies in the AHE data of several different material classes, 
and in Sec.~\ref{sec:theory} and Sec.~\ref{sec:linear_transport} by an extensive technical discussion of AHE theory.
We assume throughout that the ferromagnetic materials of interest 
are accurately described by a Stoner-like mean-field band theory.  In applications
to real materials we imagine that the band theory is based on spin-density-functional
theory \cite{Jones:1989_a} with a local-spin-density or similar approximation
for the exchange-correlation energy functional. 

\subsubsection{Intrinsic contribution to $\sigma_{xy}^{AH}$}\label{int-intro}
Among the three contributions,  
the easiest to evaluate accurately is the intrinsic contribution.  We have defined the intrinsic 
contribution microscopically as the {\em dc} limit of the 
interband conductivity, a quantity which is not zero in ferromagnets when SOI are included. There is however a 
direct link to  semiclassical theory   in which
the induced interband coherence is captured by a momentum-space
Berry-phase related contribution to the anomalous velocity. We show this equivalence below.  

This contribution to the AHE was first derived by KL \cite{Karplus:1954_a} but
its topological nature was not fully appreciated until recently \cite{MOnoda:2002_a,Jungwirth:2002_a}. 
The work of Jungwirth {\em et al.} \cite{Jungwirth:2002_a} 
was motivated by the experimental importance of the AHE
in ferromagnetic semiconductors and also by the thorough earlier analysis of the relationship
between momentum space Berry phases and anomalous velocities in semiclassical transport theory 
by Niu \etal. \cite{Chang:1996_a,Sundaram:1999_a}.  
The frequency-dependent inter-band Hall conductivity, which 
reduces to the intrinsic anomalous Hall conductivity in the 
{\em dc} limit, had been evaluated earlier for a number of materials 
by Mainkar {\em et al.} \cite{Mainkar:1996_a} and Guo and Ebert \cite{Guo:1995_a} but the topological
connection was not recognized.

The intrinsic contribution to the conductivity is 
dependent only on the band structure of the perfect crystal, hence
its name.  It can be calculated directly from the simple Kubo formula for the Hall conductivity 
for an ideal lattice, given the eigenstates $|n,\bm{k}\rangle$ and eigenvalues $\varepsilon_n(\bm{k})$
of a Bloch Hamiltonian $H$:   
\begin{eqnarray}
\sigma_{ij}^{AH-int}&=&{e^2\hbar }\sum_{n \ne n'}\int\frac{d\bm{k}}{(2\pi)^3}
[f(\varepsilon_n(\bm{k}))-f(\varepsilon_{n'}(\bm{k}))]
  \nonumber\\
  &&\times\Im \frac{\langle n,\bm{k}|v_i(\bm{k})|n',\bm{k}\rangle \langle 
n',\bm{k}|v_j(\bm{k})|n,\bm{k}\rangle}{(\varepsilon_n(\bm{k})-\varepsilon_{n'}(\bm{k}))^2}.
 \nonumber \\ 
\label{eq:Kubo}
\end{eqnarray}
In Eq.~(\ref{eq:Kubo}) $H$ is the $\bm{k}$-dependent Hamiltonian for the periodic 
part of the Bloch functions and the velocity operator is defined by
\begin{equation}
  \bm{v}(\bm{k}) =\frac{1}{i\hbar}\left[\bm{r},H(\bm{k})\right]
  =\frac{1}{\hbar} \bm{\nabla}_k H(\bm{k}).
  \label{eq:v}
\end{equation}
Note the restriction $n \ne n'$ in Eq.~(\ref{eq:Kubo}).

What makes this contribution quite unique is that, like the quantum Hall effect in a crystal, it is directly linked to the topological properties of the Bloch states.
(See Sec.~\ref{sec:top}.)  Specifically it is proportional to the integration 
over the Fermi sea of the Berry's curvature of each occupied band, or equivalently \cite{Haldane:2004_a,Wang:2007_a} to 
the integral of Berry phases over cuts of Fermi surface segments.  This 
result can be derived by noting that 
\begin{equation}
  \langle n,\bm{k}|\bm{\nabla}_k|n',\bm{k}\rangle=
\frac{\langle n,\bm{k}|\bm{\nabla}_k H(\bm{k})|n',\bm{k}\rangle}
{\varepsilon_{n'}(\bm{k})-\varepsilon_n(\bm{k})}.
  \label{eq:Feynman}
\end{equation}
Using this expression, Eq.~(\ref{eq:Kubo}) reduces to
\begin{equation}
  \sigma_{ij}^{AH-int} = -\epsilon_{ij\ell}\frac{e^2}{\hbar}
  \sum_n\int\!\frac{d\bm{k}}{(2\pi)^d}f(\varepsilon_n(\bm{k}))\ b_n^\ell(\bm{k}),
  \label{eq:TKNN}
\end{equation}
where  $\epsilon_{ij\ell}$ is the anti-symmetric tensor, $\bm{a}_n(\bm{k})$ is 
the Berry-phase connection $ \bm{a}_n(\bm{k}) = i \langle n,{\bm{k}}| \bm{\nabla}_k | n,{\bm{k}}\rangle$,
and $\bm{b}_n(\bm{k})$ the Berry-phase curvature 
\begin{equation}
  \bm{b}_n(\bm{k}) = \bm{\nabla}_k\times \bm{a}_n(\bm{k})
  \label{eq:b}
\end{equation}
corresponding to the states $\{|n,\bm{k}\rangle\}$.

This same linear response contribution to the AHE conductivity can be obtained from the 
semiclassical theory of wave-packets dynamics~\cite{Chang:1996_a,Sundaram:1999_a,Marder:1999_a}.
It can be shown that the wavepacket group velocity has an additional contribution in the 
presence of an electric field:
$\dot{\rr}_c={\partial E_n(\kk)}/{\hbar \partial \kk}-(\bm{E}/\hbar)\times \bm{b}_n(\kk)$.
(See Sec.~\ref{sec:theory:Boltzmann}.)  The intrinsic Hall conductivity formula, Eq.~(\ref{eq:TKNN}), 
is obtained simply by summing the second (anomalous) term 
over all occupied states. 

One of the motivations for identifying
the intrinsic contribution $\sigma_{xy}^{AH-int}$ is that it can be evaluated accurately
even for relatively complex materials using first-principles electronic structure theory techniques.
In many materials which have strongly spin-orbit coupled bands,
the intrinsic contribution seems to dominates the AHE.

\subsubsection{Skew scattering  contribution to $\sigma_{xy}^{AH}$}\label{skew-intro}

The skew scattering contribution to the AHE can be sharply defined; it is simply 
the contribution which is proportional to the Bloch state transport lifetime. It will  therefore  tend to dominate in nearly perfect crystals.  
It is  the only contribution to the AHE which appears within the confines of 
traditional Boltzmann transport theory in which interband coherence effects 
are completely neglected. Skew scattering is due to chiral features 
which appear in the disorder scattering of spin-orbit coupled ferromagnets.  
This mechanism was first identified by Smit \cite{Smit:1955_a,Smit:1958_a}. 

Treatments of semi-classical Boltzmann transport theory found in textbooks 
often appeal to the principle of detailed balance which states that the transition probability 
$W_{n \to m}$ from $n$ to $m$ is identical to the 
transition probability in the opposite direction ($W_{m \to n}$). 
Although these two transition probabilities are identical in a Fermi's golden-rule approximation, since
$
W_{n \to n'}= { ({ 2 \pi} / {\hbar})} |\langle n |V| n'\rangle|^2 
\delta(E_n - E_{n'}),
$
where $V$ is the perturbation inducing the transition,
detailed balance in this microscopic sense  is not generic.
In the presence of spin-orbit coupling, 
either in the Hamiltonian of the perfect crystal or in the disorder 
Hamiltonian, a transition which is right-handed with respect to the 
magnetization direction has a different transition probability than
the corresponding left-handed transition.     
When the transition rates are evaluated perturbatively, asymmetric 
chiral contributions appear first at third order. (See Sec. \ref{sec:theory:Boltzmann}).
In simple models the asymmetric chiral contribution to the
transition probability is often assummed to have the form (see Sec.~\ref{skew-detail}):
\begin{equation}
W^A_{\bm{k} \bm{k}'} = - \tau_A^{-1} \bm{k} \times \bm{k}'
\cdot \bm{M}_s.
\end{equation}
When this asymmetry is inserted into the Boltzmann equation it 
leads to a current proportional to the longitudinal current driven by $\bm{E}$
and perpendicular to both $\bm{E}$ and $\bm{M}_s$.
When this mechanism dominates, both the Hall conductivity $\sigma_H$ and 
the conductivity $\sigma$ are proportional to the transport lifetime $\tau$ 
and the Hall resistivity $\rho_H^{skew} = \sigma_H^{skew}\rho^2$ is therefore 
proportional to the longitudinal resistivity $\rho$.  
 
There are several specific mechanisms for skew scattering (see Sec.\ref{extrinsic} and Sec. \ref{sec:theory:Boltzmann}).  
Evaluation of the skew scattering contribution to the Hall 
conductivity or resistivity requires simply that the conventional linearized Boltzmann equation be solved using a 
collision term with accurate transition probabilities, since these will generically include a chiral contribution.  
In practice our ability to accurately estimate the skew scattering contribution to the AHE of a real material
is limited only by typically imperfect characterization of its disorder.    
We emphasize that skew scattering contributions to $\sigma_H$ are present 
not only because of spin-orbit coupling in the disorder Hamiltonian, but also because of 
spin-orbit coupling in the perfect crystal Hamiltonain combined with purely scalar disorder.
Either source of skew-scattering could dominate $\sigma_{xy}^{AH-skew}$ depending on the host material and also on the type of
 impurities.

We end this subsection with a small note directed to the reader who is more versed in the latest development of the full semiclassical
theory of the AHE and in its comparison to the microscopic theory (see Sec. \ref{sec:theory:Boltzmann} and \ref{kubo_semi_compar}).
 We have been careful above not to define the skew-scattering contribution to the AHE
as the sum of {\it all} the contributions arising from the asymmetric scattering rate present in the collision term of the Boltzmann transport equation. 
We know from microscopic theory that
this asymmetry also makes an AHE contribution or order $\tau^{0}$.
There exists a contribution from this asymmetry which is actually present in the microscopic theory treatment associated with the so called ladder diagram
corrections to the conductivity, and therefore of order $\tau^0$. In our experimentally practical parsing of AHE contributions 
we do not associate this contribution with skew-scattering but 
place it under the umbrella of side-jump scattering even though it does not physically originate from any side-step type of scattering.

\subsubsection{Side-jump contribution to $\sigma_{xy}^{AH}$}\label{sj-intro}

Given the sharp defintions we have provided for the intrinsic and skew scattering contributions 
to the AHE conductivity, the equation
\begin{equation} 
\sigma_{xy}^{AH} = \sigma_{xy}^{AH-int} + \sigma_{xy}^{AH-skew} + \sigma_{xy}^{AH-sj} 
\end{equation} 
defines the side-jump contribution as the difference between the full Hall conductivity and 
the two-simpler contributions.  In using the term side-jump for the remaining contribution,
we are appealing to the historically established taxonomy outlined in the previous section.
Establishing this connection mathematically 
has been the most controversial aspects of AHE theory,
and the one which has taken the longest to clarify from
a theory point of view.  Although this classification of Hall conductivity 
contributions is often useful (see below), it is not generically true 
that the only correction to the intrinsic and skew contributions can
be physically identified with the side-jump process defined as in the earlier studies of the AHE \cite{Berger:1964_a}.

The basic semiclassical argument for a side-jump contribution 
can be stated straight-forwardly: when considering the scattering of a Gaussian wavepacket from a
spherical impurity with SOI ( $H_{SO} = (1/2 m^2 c^2) (r^{-1} \partial V/ \partial r) S_z L_z$), a wavepacket with incident
wave-vector $\bm{k}$ will suffer a displacement transverse to $\bm{k}$ equal to ${ 1\over 6} k \hbar^2/m^2 c^2$. This type of contribution was 
first noticed, but discarded, by Smit \cite{Smit:1958_a} and reintroduced by Berger \cite{Berger:1964_a} who argued 
that it was the key contribution to the AHE.  This kind of mechanism clearly lies outside the bounds of 
traditional Boltzmann transport theory in which only the probabilities of transitions between 
Bloch states appears, and not microscopic details of the scattering processes.
This contribution to the conductivity ends up being independent of $\tau$ and therefore contributes
to the AHE at the same order as the intrinsic contribution in an expansion in powers of scattering rate. 
The separation between intrinsic and side-jump contributions, which cannot be distinquished by their dependence on $\tau$, has been perhaps the
most argued aspect of AHE theory since they cannot be distinquished by their 
dependence on scattering rate (see Sec.~\ref{sj-detail}). 

As explained clearly in a recent review by Sinitsyn \cite{Sinitsyn:2008_a}, 
side-jump and intrinsic contributions have quite different dependences on more specific system parameters,
particularly in systems with complex band structures.  Some of the initial controversy which 
surrounded side jump theories was associated with  physical meaning
ascribed to quantities which were plainly gauge dependent, like the Berry's connection which  in early theories
is typically identified as the definition of the side-step upon scattering.
Studies of simple models, for example models of semiconductor conduction bands, also gave
results in which the side-jump contribution seemed to be the same size but opposite in sign 
compared to the intrinsic contribution \cite{Nozieres:1973_a}.  We now understand \cite{Sinitsyn:2007_a} that these 
cancellations are unlikely, except in models with a very simple band structure, {\em e.g.} one with a constant Berry's curvature.
It is only through comparison between fully microscopic linear response theory calculations, based on 
equivalently valid microscopic formalisms such as Keldysh (non-equilibrium Grenn's function)
 or  Kubo formalisms, and the systematically developed semi-classical theory 
that the specific contribution due to the side-jump mechanism can be separately identified with 
confidence (see Sec. ~\ref{sec:theory:Boltzmann}). 

Having said this, all the calculations comparing the intrinsic and side-jump contibutions to the AHE from a microscopic point of view
have been performed for very simple models not immediately linked to real materials. A practical approch
which is followed at present for materials in which $\sigma^{AH}$ seems to be independent of $\sigma_{xx}$, is to first calculate the
intrisic contribution to the AHE.  If this explains the observation (and it appears that it usually does), then it is deemed that the 
intrinsic mechanism dominates.  
If not, we can take some comfort from understanding on the basis of 
simple model results, that there can be other contributions to $\sigma^{AH}$
which are also independent of $\sigma_{xx}$ and can for the most part be identified 
with the side jump mechanism.  Unfortunately it seems extremelly challenging, if not impossible, to develop a 
predictive theory for these contributions, partly because they require many higher
order terms in the perturbation theory that be summed, but more fundamentally because they
depend sensitively on details of the disorder in a particular material which are 
normally unknown.

\section{Experimental and theoretical studies on specific materials} \label{sec:experiments}

    \subsection{Transition-metals}\label{sec:tm} 

\subsubsection{Early experiments}
Four decades after the discovery of the AHE, 
an empirical relation between magnetization and Hall resistivity was proposed independently by
A. W. Smith and by E. M. 
Pugh~\cite{Smith,Pugh:1930_a,Pugh:1932_a} (see Sec. \ref{sec:brief}). 
Pugh investigated the AHE in Fe, Ni and Co and the alloys 
Co-Ni and Ni-Cu in magnetic fields up to 17 kG over large intervals in $T$
(10-800 K in the case of Ni), and found that the Hall
resistivity $\rho_{H}$ is comprised of 2 terms, {\it viz.}  
\begin{equation}
  \rho_{H}=R_0H+R_1 M(T,H),
  \label{eq:Pugh}
\end{equation}
where $M(T,H)$ is the magnetization averaged over the sample.
Pugh defined $R_0$ and $R_1$ as the ordinary and extra-ordinary
Hall coefficients, respectively.  The latter $R_1=R_s$
is now called the anomalous Hall coefficient (as in Eq.~\ref{eq:exp:fe:pugh})).

On dividing Eq. (\ref{eq:Pugh}) by $\rho^2$, we see that it
just expresses the additivity of the Hall currents: 
the total Hall conductivity
$\sigma^{tot}_{xy}$ equals $\sigma^{NH}_{xy}+\sigma^{AH}_{xy}$, where
$\sigma^{NH}_{xy}$ is the ordinary
Hall conductivity and $\sigma^{AH}_{xy}$ is the AHE conductivity.
A second implication of Eq. (\ref{eq:Pugh}) emerges when we consider 
the role of domains. 
The anomalous Hall coefficient in Eq.(2.1) is proportional to the AHE in a single domain.
As 
$H\rightarrow 0$, proliferation of domains rapidly 
reduces $\bm{M}(T,H)$
to zero (we ignore pinning).  Cancellations of $\sigma_{xy}^{AH}$
between domains result in a zero net Hall current.  Hence the observed
AHE term mimics the field profile of $M(T,H)$, as
implied by Pugh's term $R_1 M$.  The
role of $\bm{H}$ is simply to align the AHE currents by
rotating the domains into alignment.  The Lorentz force
term $\sigma_{xy}^{NH}$ is a ``background'' current with no bearing
on the AHE problem.  

The most interesting implication of Eq. (\ref{eq:Pugh})
is that, in the absence of $\bm{H}$, a single domain 
engenders a spontaneous Hall current transverse to both
$\bm{M}$ and $\bm{E}$.  
Understanding the origin of 
this spontaneous off-diagonal current has been a 
fundamental problem of charge transport
in solids for the past 60 years.
The AHE is also called the spontaneous Hall effect and
the extraordinary Hall effect in the older literature.

\subsubsection{Recent experiments}
The resurgence of interest in the AHE motivated by 
the Berry-phase approach (Sec. \ref{int-intro}) has
led to many new Hall experiments on 3$d$ transition
metals and their oxides.  
Both the recent and the older literature on 
Fe and Fe$_3$O$_4$ are reviewed in this section.
An important finding of these studies is the 
emergence of three distinct regimes roughly delimited by the conductivity $\sigma_{xx}$ and
characterized by the dependence of $\sigma_{xy}^{AH}$ on $\sigma_{xx}$.
The three regimes are:  
\begin{itemize}
\item[i)]
A high conductivity regime  for $\sigma_{xx}\gtrsim 10^{6}$ $ (\Omega\mathrm{cm})^{-1}$
in which $\sigma_{xy}^{AH-skew}\sim\sigma_{xx}^1$  dominates   $\sigma_{xy}^{AH}$, 
\item[ii)] A good metal regime for $\sigma_{xx}^{AH} \sim 10^{4}-10^6$ $(\Omega\mathrm{cm})^{-1}$ in which $\sigma_{xy}\sim \sigma_{xx}^0$,
\item[iii)] A bad metal/hopping regime for $\sigma_{xx} < 10^{4}$ $(\Omega\mathrm{cm})^{-1}$ in which 
$\sigma_{xy}^{AH}\sim \sigma_{xx}^{1.6-1.8} $.
\end{itemize}
We discuss each of these regimes below.
\\

\noindent{\it High conductivity regime --} The Hall conductivity in the high-purity regime, $\sigma_{xx}>0.5\times10^6$ $(\Omega\mathrm{cm})^{-1}$, is dominated by the skew scattering contribution $\sigma_{xy}^{skew}$.
The high-purity regime is one of the least studied experimentally. This regime is very challenging to investigate experimentally because
the field $H$ required for saturating $M$ also yields a very
large ordinary Hall effect (OHE) and $R_0$ tends to be of the order of $R_s$~\cite{Schad:1998_a}.  
In the limit $\omega_c\tau\gg 1$,  the OHE conductivity $\sigma_{xy}^{NH}$ may be nonlinear in $H$ 
($\omega_c$ is the cyclotron frequency). Although $\sigma_{xy}^{skew}$ increases as $\tau$, 
the OHE term $\sigma_{xy}^{NH}$ increases as $\tau^2$ and therefore   
the latter ultimately dominates, and the AHE 
current may be unresolvable.  
Even though the anomalous Hall current can not always be cleanly separated from the normal Lorentz-force Hall effect in the high conductivity regime, the total Hall current invariably increases with $\sigma_{xx}$ in a way which provides compelling evidence for a skew-scattering contribution.

\begin{figure}
\begin{center}
\includegraphics[width=0.9\columnwidth]{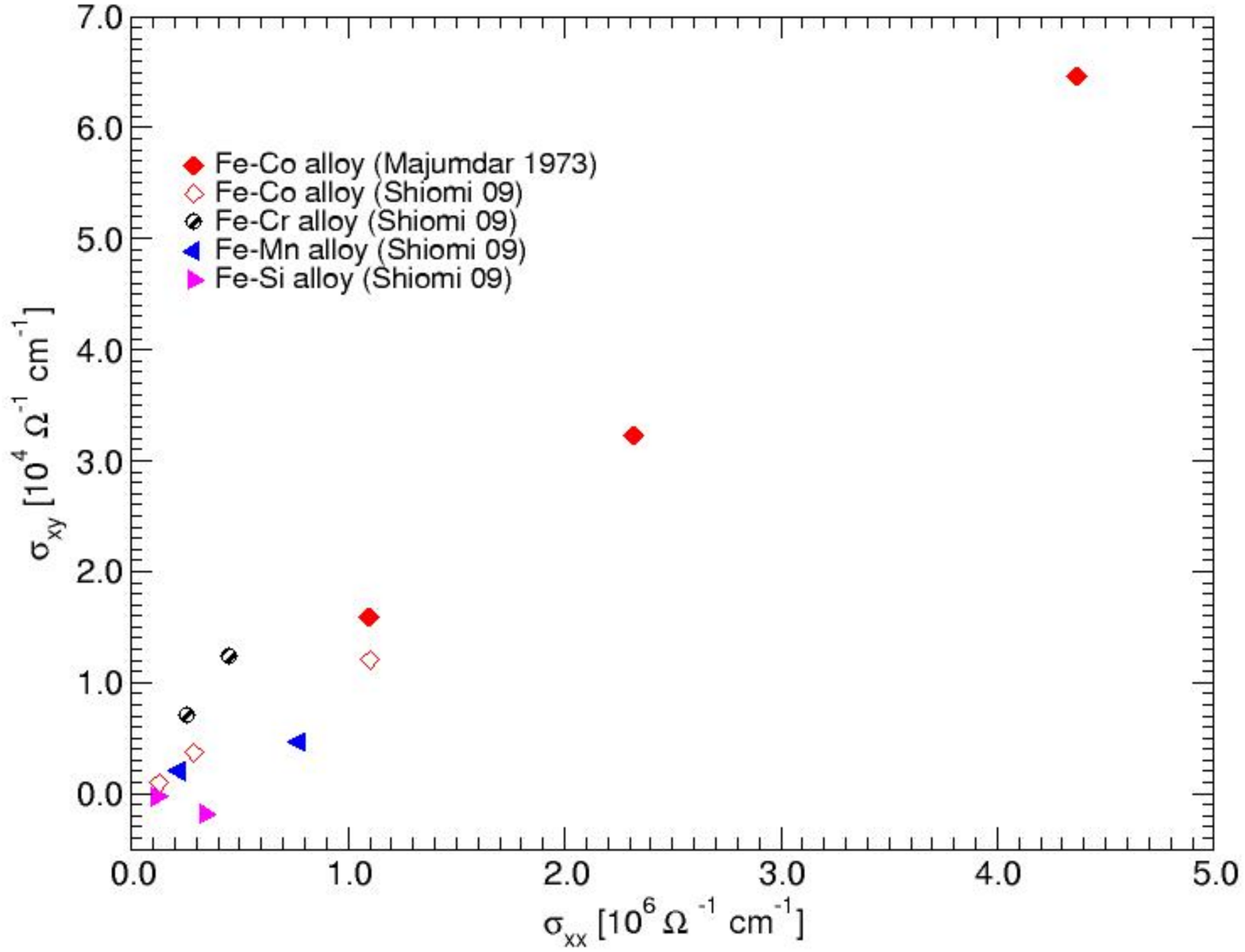}
\includegraphics[width=\columnwidth]{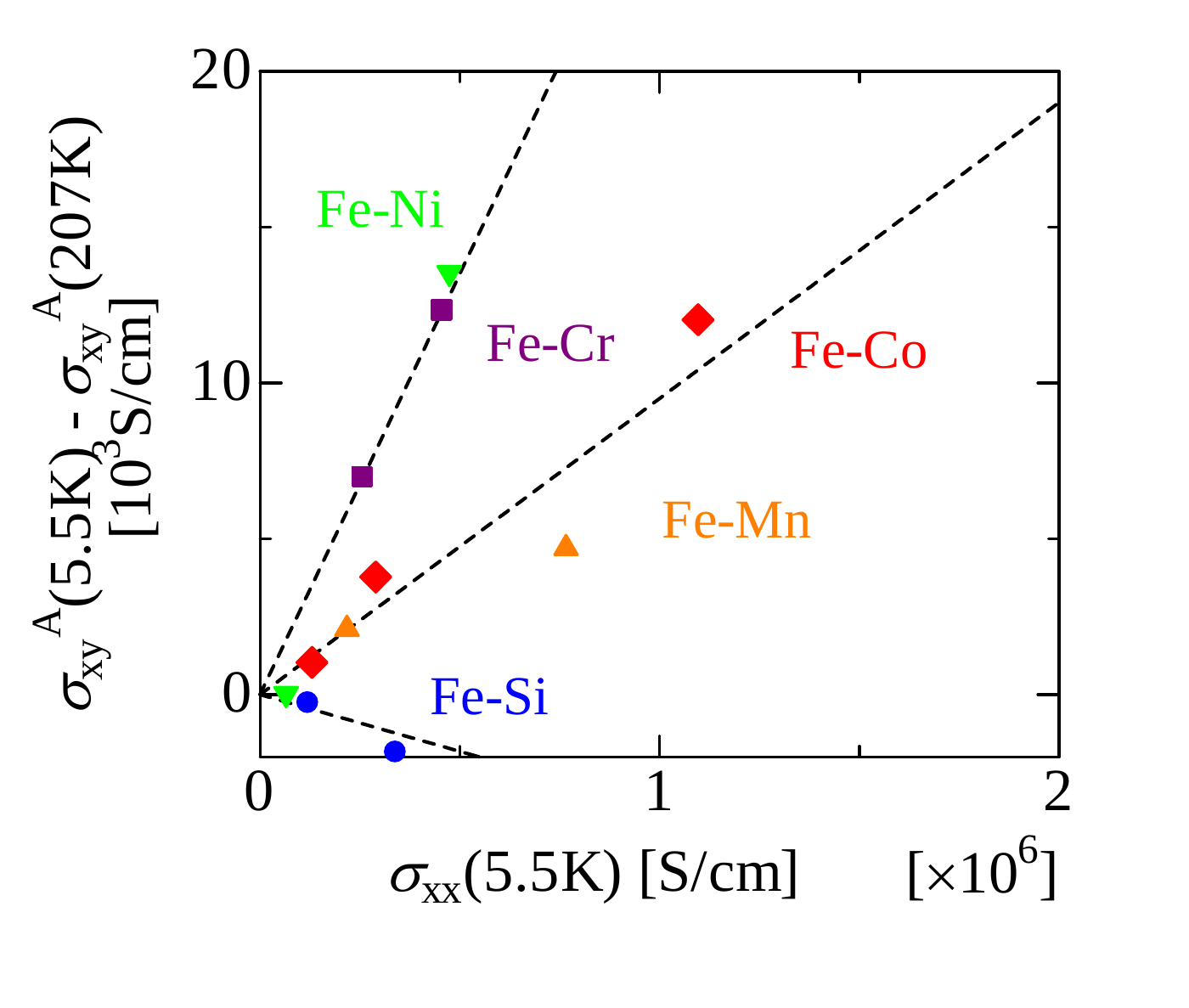}
\end{center}
\caption{$\sigma_{xy}^{AH}$ for pure  Fe film  doped with Cr, Co, Mn and Si vs. $\sigma_{xx}$ at low temperatures (
$T=4.2$ K and $T=5.5$ K ) (a). In many of the alloys, particularly in the Co doped system, 
 the linear scaling in the higher conductivy sector, $\sigma_{xx}>10^6$  $(\Omega\mathrm{cm})^{-1}$,
 implies that  skew scattering dominates $\sigma_{xy}^{AH}$. After Ref. \onlinecite{Majumdar:1973_a} and Ref. \onlinecite{Shiomi:2009_a}.
 The data from \cite{Shiomi:2009_a}, shown also at a larger scale in (b), is obtained by 
 substracting the high temperature contribuiton to $\sigma_{xy}$. In the data shown the ordinary Hall contribution has been identified and substracted.
 [Panel (b) From Ref. \onlinecite{Shiomi:2009_a}.]
}
\label{fig:Majumdar}
\end{figure}

In spite of these challenges several studies have managed to convincingly separate the competing 
contributions and have identifed a dominant linear relation between $\sigma_{xy}^{AH}$ and $\sigma_{xx}$  for 
$\sigma_{xx}\gtrsim 10^6$  $(\Omega\mathrm{cm})^{-1}$ \cite{Majumdar:1973_a,Shiomi:2009_a}. In an early study
Majumdar \etal. \cite{Majumdar:1973_a} grew highly pure Fe doped with  Co. 
The resulting $\sigma_{xy}^{AH}$, obtained from  Kohler plots extrapolation to zero field, 
 show a clear  dependence of $\sigma_{xy}^{AH}\sim\sigma_{xx}$ (Fig.~\ref{fig:Majumdar} a).
In a more recent study, a similar finding (linear dependence of $\sigma_{xy}^{AH}\sim \sigma_{xx}$)
was observed by Shiomi \etal. \cite{Shiomi:2009_a} in Fe doped with Co, Mn, Cr, and Si. 
In these studies the high temperature contribution to 
$\sigma_{xy}^{AH}$ (presumed to be intrinsic plus side jump) was substracted from $\sigma_{xy}$ and a linear dependence of the 
resulting $\sigma_{xy}^{AH}$ is observed (Fig.~\ref{fig:Majumdar} a and b).
In this recent study the conductivity is intentionally reduced by impurity doping to find the linear region and reliably exclude the Lorernz
contribution. The results of these authors show, in particular, that the slope of $\sigma_{xx}$ vs. $\sigma_{xy}^{AH}$ depends on the species of the impurities
as it is expected in the regime dominated by skew scattering.
It is reassuring to note that the 
skewness parameters ($ S_{skew} = \sigma^{AH}/\sigma_{xx}$) implied by the 
older and the more recent experiments
are consistent, in spite of differences in the conductivity ranges studied.  
$S_{skew}$ is independent of $\sigma_{xx}$ \cite{Majumdar:1973_a,Shiomi:2009_a}
 as it should be when the skew scattering mechanism dominates. 
 Further experiments in this regime are desirable to fully investiage the different dependence on doping, temperature,
and impurity type. Also, new approaches to reliably disentangle  the AHE and OHE currents will be needed to faicilitate such studies.

\begin{figure}
\begin{center}
\includegraphics[width=\columnwidth]{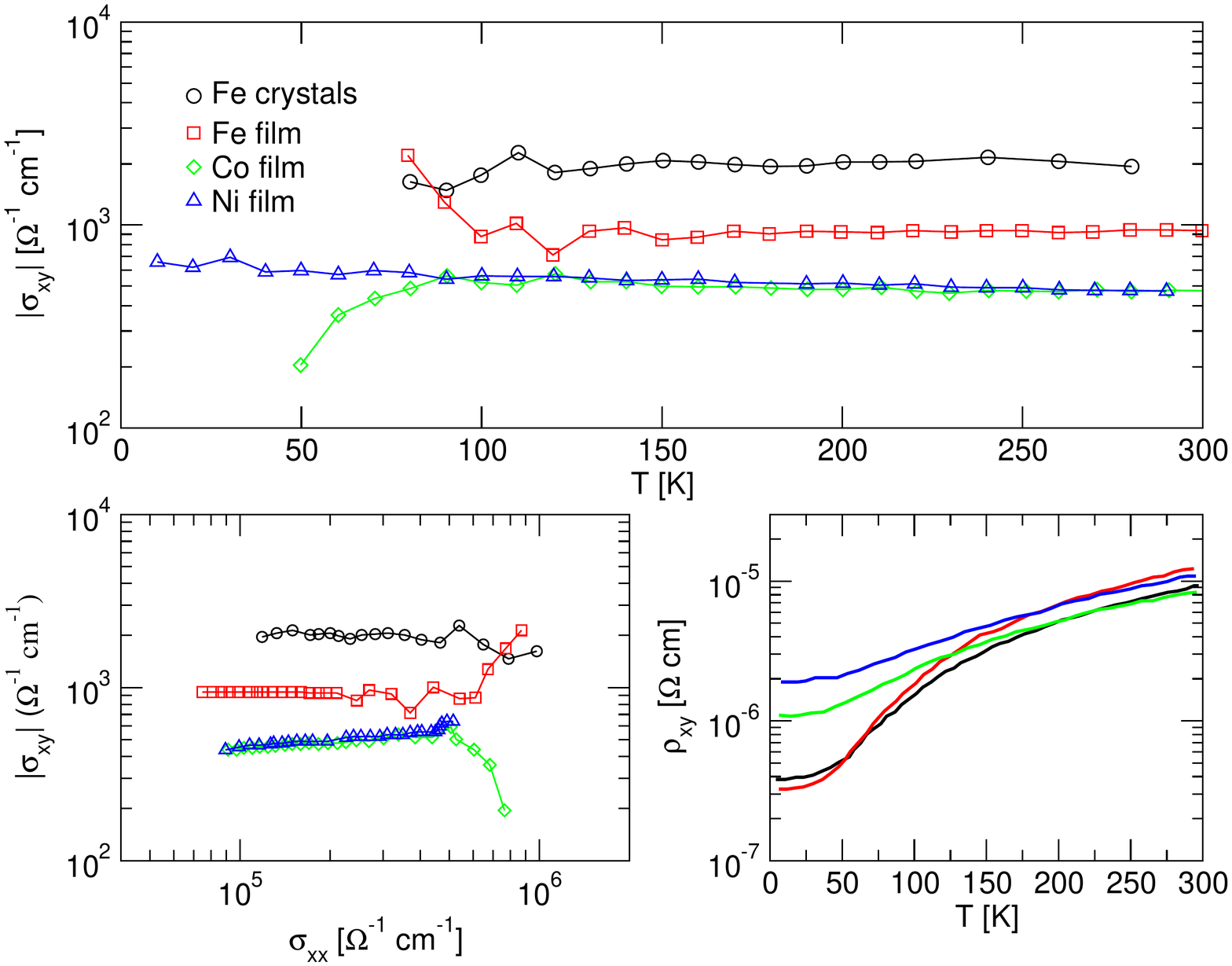}
\end{center}
\caption{Measurements of the Hall conductivity and resistivity
in single-crystal Fe and in thin foils of Fe, Co, and Ni.
The top and lower right panels show the $T$ dependence of $\sigma_{xy}$ 
and $\rho_{xx}$, respectively.  The lower left panel
plots $|\sigma_{xy}|$ against $\sigma_{xx}$.
After Ref. \onlinecite{Miyasato:2007_a}. 
}
\label{fig:miyasato:2007_a:1}
\end{figure}

\noindent {\it Good metal regime --} Experiments re-examining the AHE in 
Fe, Ni, and Co have been performed by Miyasato \textit{et al.}~\cite{Miyasato:2007_a}. These experiments
indicate a regime of  $\sigma_{xy}$ versus $\sigma_{xx}$ in which $\sigma_{xy}$
is insensitive to $\sigma_{xx}$ in the range $\sigma_{xx}\sim 10^4$-$10^6$ $(\Omega\mathrm{cm})^{-1}$
(see Fig.~\ref{fig:miyasato:2007_a:1}). 
This suggests that the scattering independent mechanisms (intrinsic and side-jump) dominates in this regime. 
However, in comparing this phenomenology to the discussion of AHE mechanisms in Sec.~\ref{mechanisms-intro},
one must  keep in mind that the temperature has been varied in the Hall data on  Fe, Ni, and Co in order to 
change the resistivity, even though it is restricted to the range well below $T_c$ (Fig.~\ref{fig:miyasato:2007_a:1} upper panel).
In the mechanisms discussed in Sec. \ref{mechanisms-intro} only elastic scattering was taken into account.
Earlier tests of the $\rho^2$ dependence of $\rho_{xy}$ carried by varying $T$ 
were treated as suspect in the early AHE period (see Fig. \ref{fig:Kooi}) because the role of inelastic 
scattering was not fully understood.  The effect of 
inelastic scattering from phonons and spin waves remains open in  AHE theory and is not addressed in this review.

\begin{figure}
\begin{center}
\includegraphics[width=\columnwidth]{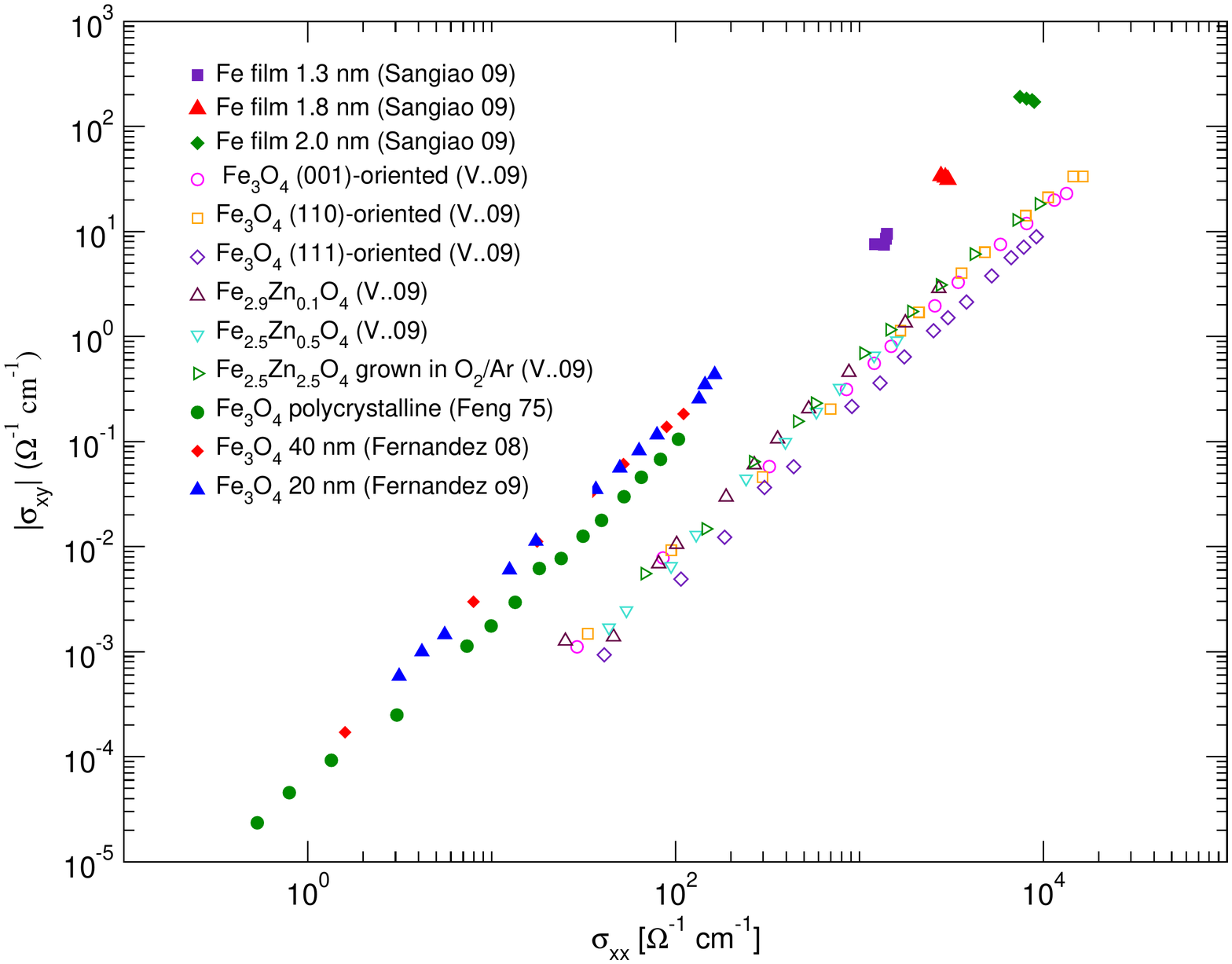}
\end{center}
\caption{Combined plot of the AHE conductivity $|\sigma_{xy}|$ versus the conductivity $\sigma_{xx}$ in epitaxial films of Fe grown on MgO with thickness $t$ = 2.5, 2.0, 1.8 and 1.3 nm  ~\cite{Sangiao:2009_a}, in polycrystalline Fe$_{3-x}$Zn$_x$O$_4$ \cite{Feng:1975_a},
in thin-film Fe$_{3-x}$Zn$_x$O$_4$ between 90 and 350 K \cite{Venkateshvaran:2008_a} and above the Verwey transition~\cite{Fernandez-Pacheco:2008_a}.}
\label{fig:trans_phase_diagram}
\end{figure}

\begin{figure}
\begin{center}
\includegraphics[width=7.0cm]{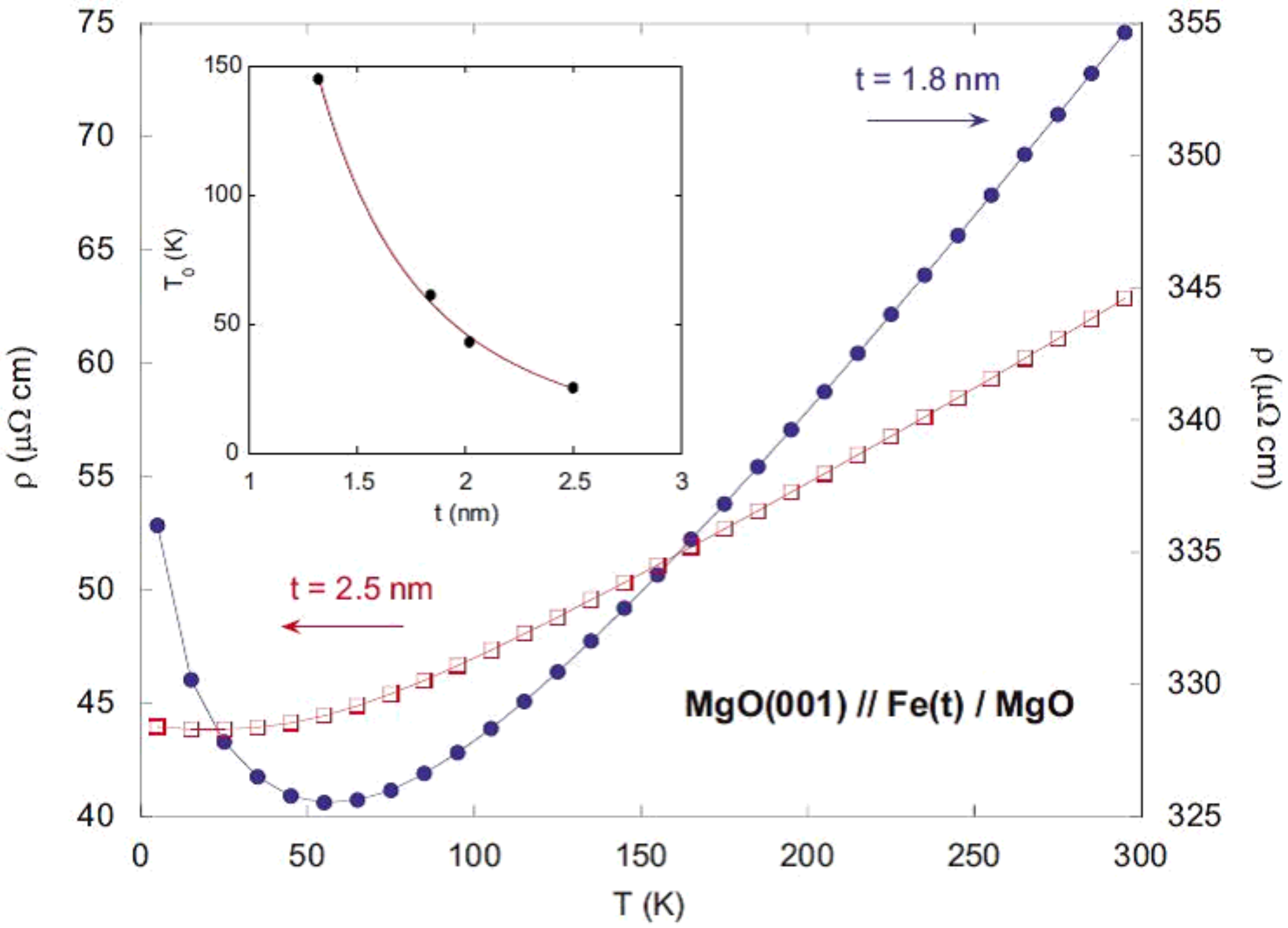}
\end{center}
\caption{The $T$ dependence of $\rho$ in epitaxial thin-films MgO(001)/Fe(t)/MgO with $t=1.8$ nm and 2.5 nm. The inset shows how $T_0$, the temperature of the resistivity minimum, varies with $t$.[From Ref.~\onlinecite{Sangiao:2009_a}.]}
\label{fig:sangiao:2009_a:1}
\end{figure}

\noindent{\em Bad metal/hopping regime--} Several groups have measured 
$\sigma_{xy}$ in Fe and Fe$_3$O$_4$ thin-film
ferromagnets~\cite{Miyasato:2007_a,Fernandez-Pacheco:2008_a,Venkateshvaran:2008_a,Sangiao:2009_a,Feng:1975_a}.
(See Fig. \ref{fig:trans_phase_diagram}.)
Sangiao \etal. \cite{Sangiao:2009_a} studied epitaxial 
thin-films of Fe deposited by pulsed-laser deposition (PLD) on single-crystal MgO (001) substrates at pressures $< 5\times 10^{-9}$ Torr.  
To vary $\sigma_{xx}$ over a broad range, they varied the film thickness $t$ from 1 to 10 nm. The $\rho$ vs. $T$ profile for the film with $t$ = 1.8 nm 
displays a resistance minimum near 50 K, below which $\rho$ shows an upturn which has been ascribed to localization or
electron interaction effects (Fig.~\ref{fig:sangiao:2009_a:1}).  
The magnetization $M$ is
nominally unchanged from the bulk value (except possibly
in the 1.3 nm film).  AHE experiments were carried out from 2--300 K on these films and displayed as $\sigma_{xy}$ vs. $\sigma_{xx}$ plots
together with previously published results
(Fig. \ref{fig:trans_phase_diagram}).  In the plot, 
the AHE data from films with $t\le$ 2 nm fall in the weakly localized regime.
The combined plot shows that Sangio \etal.'s data are collinear (on a logarithmic scale) with those measured on $1\ \mu$m-thick films by 
Miyasato \etal. \cite{Miyasato:2007_a}.  For the three samples
with $t$ = 1.3--2 nm, the inferred exponent in the dirty regime is on average $\sim 1.66$.  
A concern is that the data from the 1.3 nm film was obtained by subtracting a
$\ln T$ term from $\rho$  (the subtraction procedure was not described).
How localization affects the scaling plot is an open issue at present.
In Sec. \ref{sec:exp:loc}, we discuss recent 
AHE measurements in disordered polycrystalline Fe 
films with $t<$ 10 nm by Mitra \etal.~\cite{Mitra:2007_a}. Recent 
progress in understanding weak-localization corrections to the AHE
is also reviewed there.

\begin{figure}
\begin{center}
\includegraphics[width=7.0cm]{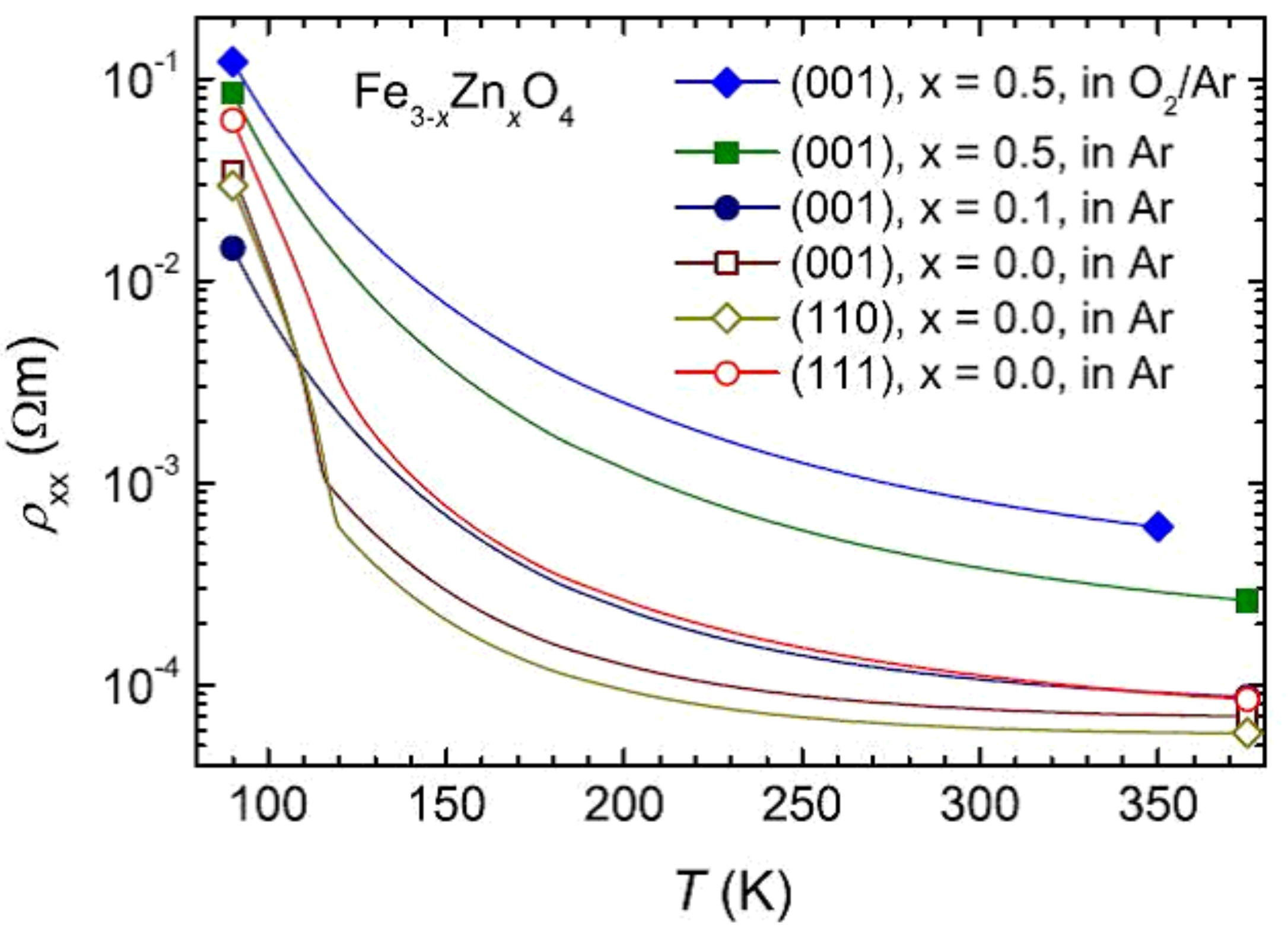}
\end{center}
\caption{Longitudinal resistivity $\rho_{xx}$ vs. 
$T$ for epitaxial Fe$_{3-x}$Zn$_x$O$_4$ films.
The (001), (110) and (111) oriented films were grown on
MgO(001), MgO(110) and Al$_2$O$_3$ substrates.
[From Ref. \onlinecite{Venkateshvaran:2008_a}.]
}
\label{fig:venkateshvaran:2008_a:1}
\end{figure}

In magnetite, Fe$_3$O$_4$,  scaling of 
$\sigma_{xy}\sim\sigma_{xx}^\alpha$ with $\alpha\sim 1.6-1.8$
was already apparent in early experiments on polycrystalline samples
~\cite{Feng:1975_a}.  Recently, two groups have re-investigated the
AHE in epitaxial thin films (data included in Fig.~\ref{fig:trans_phase_diagram}).
Fernandez-Pacheco \etal. \cite{Fernandez-Pacheco:2008_a}
measured a series of thin-film samples of Fe$_3$O$_4$ grown
by PLD on MgO (001) substrates in ultra-high vacuum,
whereas Venkateshvaran \etal. \cite{Venkateshvaran:2008_a}
studied both pure Fe$_3$O$_4$ and Zn-doped magnetite
Fe$_{3-x}$Zn$_x$O$_4$ deposited on MgO and Al$_2$O$_3$
substrates grown by laser molecular-beam epitaxy under 
pure Ar or Ar/O mixture.  In both studies,
$\rho$ increases monotonically by a factor of $\sim$10 as $T$ decreases 
from 300 K to the Verwey transition temperature $T_V$ = 120 K.  
Below $T_V$, $\rho$ further increases by a factor of 10 to 100.
The results for $\rho$ vs. $T$ from Venkateshvaran \etal. \cite{Venkateshvaran:2008_a} is shown in Fig.~\ref{fig:venkateshvaran:2008_a:1}.  

The large values of $\rho$ and its insulating trend imply 
that magnetite falls in the strongly localized regime, 
in contrast to thin-film Fe which lies partly in the weak-localization (or incoherent) regime. 

Both groups find good scaling fits extending over
several decades of $\sigma_{xx}$ with
$\alpha\sim 1.6-1.8$ when varying $T$.  Fernandez-Pacheco \etal.~\cite{Fernandez-Pacheco:2008_a} 
plot $\sigma_{xy}$ vs. $\sigma_{xx}$ in the range 150$<T<$300 K
for several thicknesses $t$ and infer an exponent 
$\alpha = 1.6$.  Venkateshvaran \etal. \cite{Venkateshvaran:2008_a} plot $\sigma_{xy}$
vs. $\sigma_{xx}$ from 90 to 350 K and obtain power-law fits
with $\alpha$ = 1.69, in both pure and Zn-doped magnetite
(data shown in Fig.~\ref{fig:trans_phase_diagram}).
Significantly, 
the 2 groups find that $\alpha$ is unchanged below $T_V$.
There is presently no theory in the poorly
conducting regime which predicts the observed scaling
($\sigma_{xx}< 10^{-1}$ $(\Omega\mathrm{cm})^{-1}$).

\subsubsection{Comparison to theories}
Detailed first-principles calculations of the intrinsic contribution 
to the AHE conductivity have been performed for
bcc Fe~\cite{Yao:2004_a,Wang:2006_a}, fcc Ni, and hpc Co~\cite{Wang:2007_a}.
In Fe and Co, the values of $\sigma_{xy}$ inferred from 
the Berry curvature $\Omega^z(\bm{k})$ are $7.5\times10^2$ and 
$4.8\times10^2$ $(\Omega\mathrm{cm})^{-1}$, 
respectively, in reasonable agreement with experiment. 
In Ni, however, the calculated value $-2.2\times10^3$ $(\Omega\mathrm{cm})^{-1}$ is only $30\%$ of the experimental value.

\begin{figure}
\begin{center}
\includegraphics[width=7.0cm]{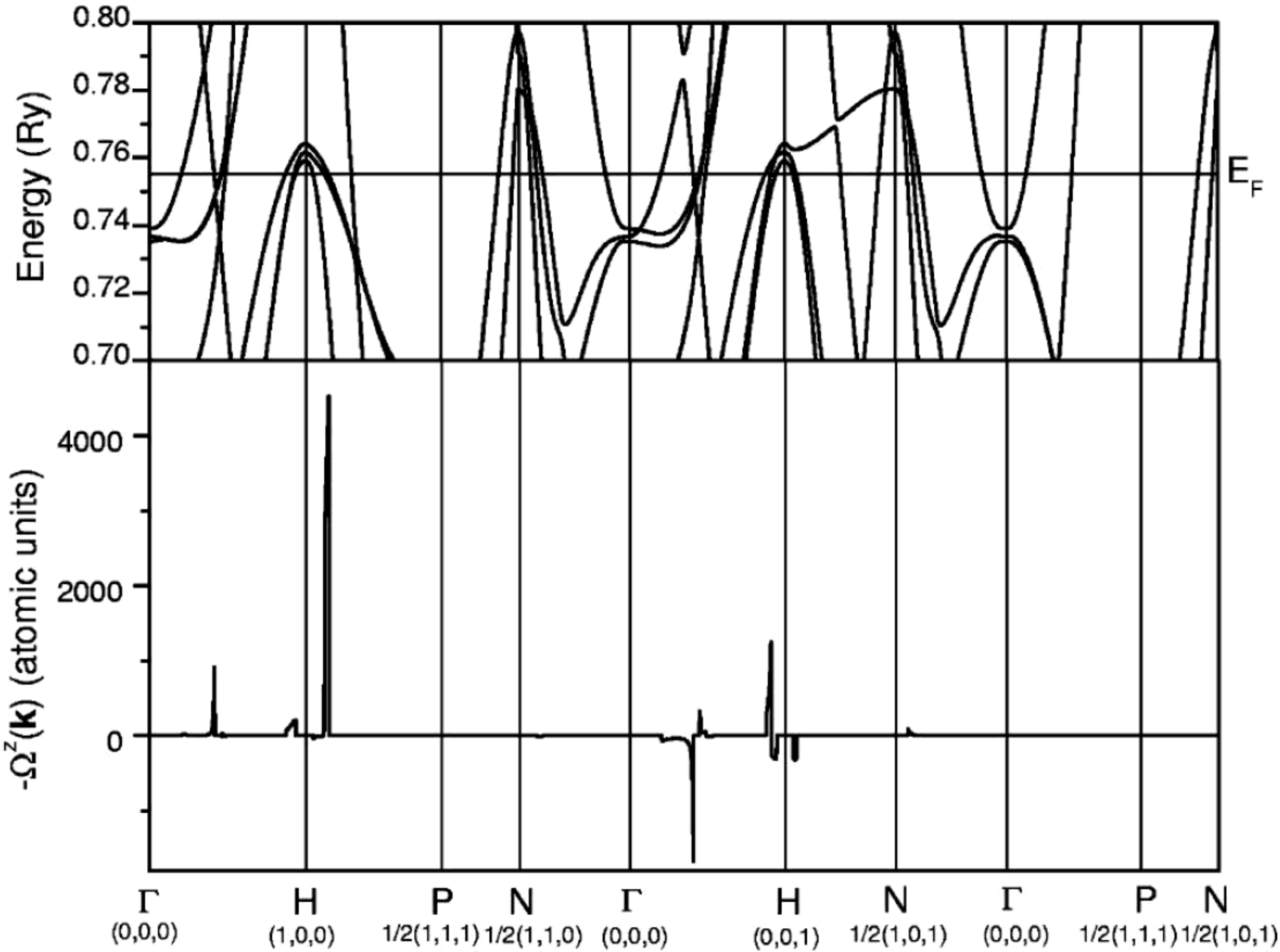}
\end{center}
\caption{First-principles calculation of the band dispersions and the Berry-phase curvature summed over  occupied bands. [From Ref.~\onlinecite{Yao:2004_a}.]}
\label{fig:Yao:2004_a:2}
\end{figure}

\begin{figure}
\begin{center}
\includegraphics[width=7.0cm]{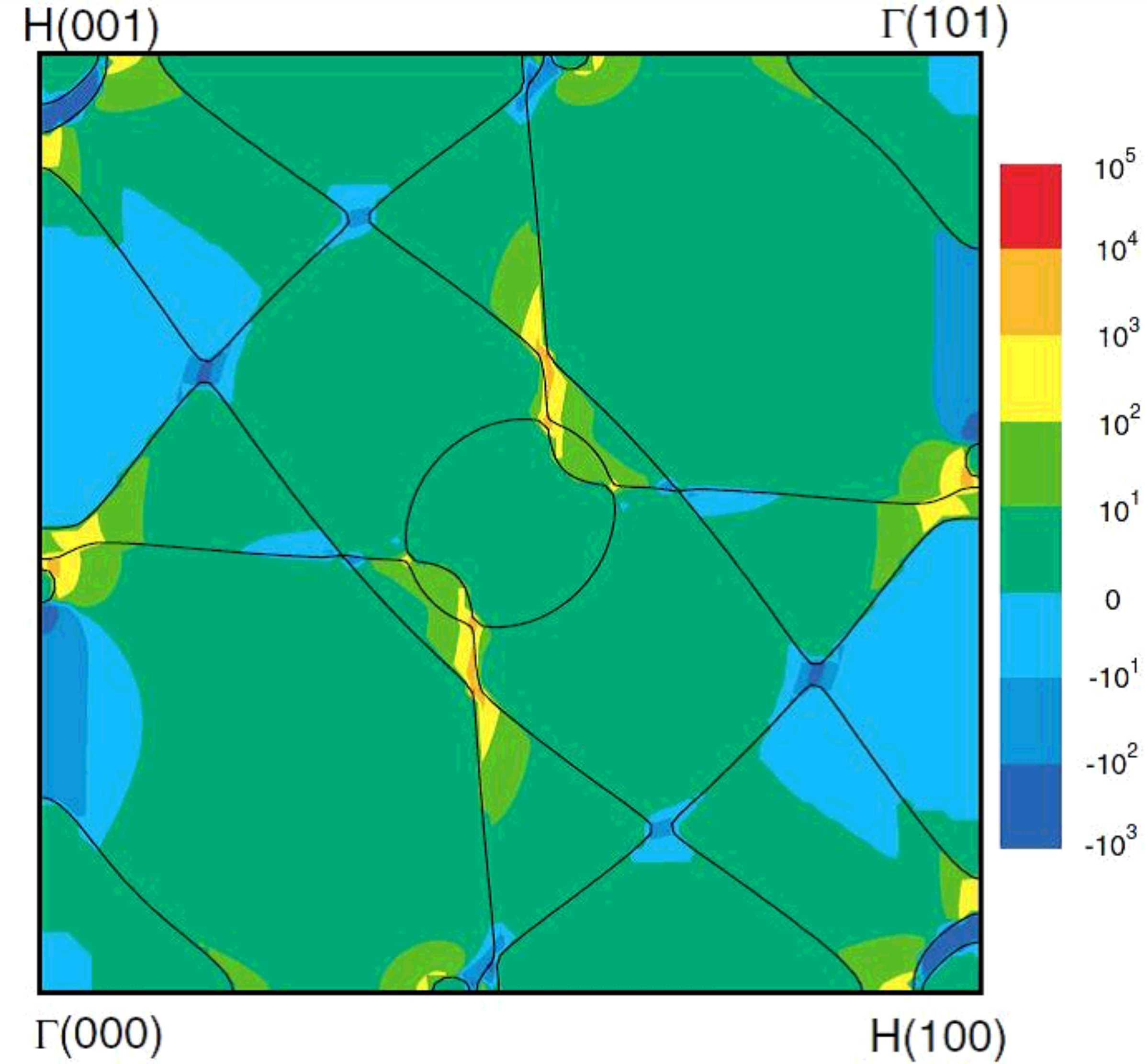}
\end{center}
\caption{First-principles calculation of the FS in the (010) plane (solid lines) and the Berry curvature in atomic units (color map).  
[From Ref.~\onlinecite{Yao:2004_a}.]}
\label{fig:Yao:2004_a:3}
\end{figure}

These calculations uncovered the crucial role played by 
avoided-crossings of band dispersions near the Fermi energy $\varepsilon_F$.  The Berry curvature $b_z$ is always strongly enhanced near avoided crossings,  opposite direction for the upper and lower bands.  A large contribution to $\sigma_{xy}^{int}$ when the crossing is at the Fermi energy so that only one of the two bands is occupied, {\it e.g.} near the point H in Fig.~\ref{fig:Yao:2004_a:2}. A map showing the contributions of different regions of the FS to $b^z(\bm{k})$ is shown in Fig.~\ref{fig:Yao:2004_a:3}. The SOI can lift an accidental degeneracy at certain wavevectors $\bm{k}$.  These points act as a magnetic monopole for 
the Berry curvature in $\bm{k}$-space~\cite{Fang:2003_a}. 
In the parameter space of  spin-orbit coupling, 
$\sigma_{xy}$ is nonperturbative in nature.  The effect of
these ``parity anomalies''~\cite{Jackiw:1984_a} on
the Hall conductivity was first 
discussed by Haldane~\cite{Haldane:1988_a}.  
A different conclusion on the role of  topological enhancement
in the intrinsic AHE was reported for a tight-binding calculation
with the 2 orbitals $d_{zx}$ and $d_{yz}$ on a square lattice~\cite{Kontani:2007_a}.

Motivated by the enhancement at the crossing points discussed above, Onoda \etal. \cite{SOnoda:2006_b,SOnoda:2008_a} 
proposed a minimal model that focuses on the topological 
and resonantly enhanced nature of the intrinsic AHE arising from
the sharp peak in $b^z(\bm{k})$ near  avoided crossings. 
The minimal model is essentially a 2D Rashba model~\cite{Bychkov:1984_a} with an exchange field which breaks symmetries and accounts for the magnetic order and a random impurity potential to account for disorder. 
  The model is discussed in Sec.~\ref{sec:theory:Rashba}.
In the clean limit, $\sigma_{xy}$ is dominated by the extrinsic skew-scattering contribution 
$\sigma_{xy}^{\mathrm{skew}}$, which almost masks the intrinsic contribution $\sigma_{xy}^{\mathrm{int}}$. Since $\sigma_{xy}^{\mathrm{skew}}\propto\tau$, it
is suppressed by increased impurity scattering, whereas
$\sigma_{xy}^{\mathrm{int}}$ -- an interband
effect -- is unaffected.  
In the moderately dirty regime where the quasiparticle damping is larger than the energy splitting at 
the avoided crossing (typically, the SOI energy) but less than the bandwidth, 
$\sigma_{xy}^{\mathrm{int}}$ dominates $\sigma_{xy}^{\mathrm{skew}}$.
As a result, one expects a crossover from the extrinsic to  the intrinsic regime.
When skew scattering is due to a spin-dependent scattering potential instead of spin-orbit coupling 
in the Bloch states, the skew to intrinsic crossover could be 
controlled by a different condition.
In this minimal model, 
a well-defined plateau is not well-reproduced in the intrinsic regime 
unless a weak impurity potential is assumed~\cite{SOnoda:2008_a}.
This crossover may be seen when the skew-scattering term shares the same sign as the intrinsic one~\cite{Kovalev:2009_a}.

With further increase in the scattering strength, 
spectral broadening leads to the scaling relationship 
$\sigma_{xy}\propto\sigma_{xx}^{1.6}$, as discussed
above~\cite{SOnoda:2006_a,SOnoda:2008_a,Kovalev:2009_a}. 
In the strong-disorder regime, $\sigma_{xx}$ is no longer linear
in the scattering lifetime $\tau$.  A different scaling, $\sigma_{xy}\propto\sigma_{xx}^2$, attributed to broadening of the 
electronic spectrum in the intrinsic regime, has been proposed by Kontani \etal. ~\cite{Kontani:2007_a}.

As discussed above, 
there is some experimental evidence that this scaling prevails not only in the
dirty metallic regime, but also deep into the hopping regime.  
Sangio \etal.~\cite{Sangiao:2009_a}, for e.g., obtained 
the exponent $\alpha \simeq$1.7 in epitaxial thin-film Fe in the dirty regime.  
In manganite, scaling seems to hold, 
with the same nominal value of $\alpha$,
even below the Verwey transition where charge transport 
is deep in the hopping regime~\cite{Fernandez-Pacheco:2008_a,
Venkateshvaran:2008_a}.
These regimes are well beyond the purview of either the minimal model, which 
considers only elastic scattering, or the 
theoretical approximations used to model its properties~\cite{SOnoda:2006_a,SOnoda:2008_a}. 

Nonetheless, the experimental reports have uncovered a
robust scaling relationship with $\alpha$ near 1.6, which 
extends over a remarkably large range of $\sigma_{xx}$.  The origin 
of this scaling is an open issue at present.

	\subsection{Complex oxide ferromagnets}\label{sec:exp:oxcide}

\subsubsection{First-principles calculations and experiments on SrRuO$_3$} 

\begin{figure}
\incl[width=6.5cm]{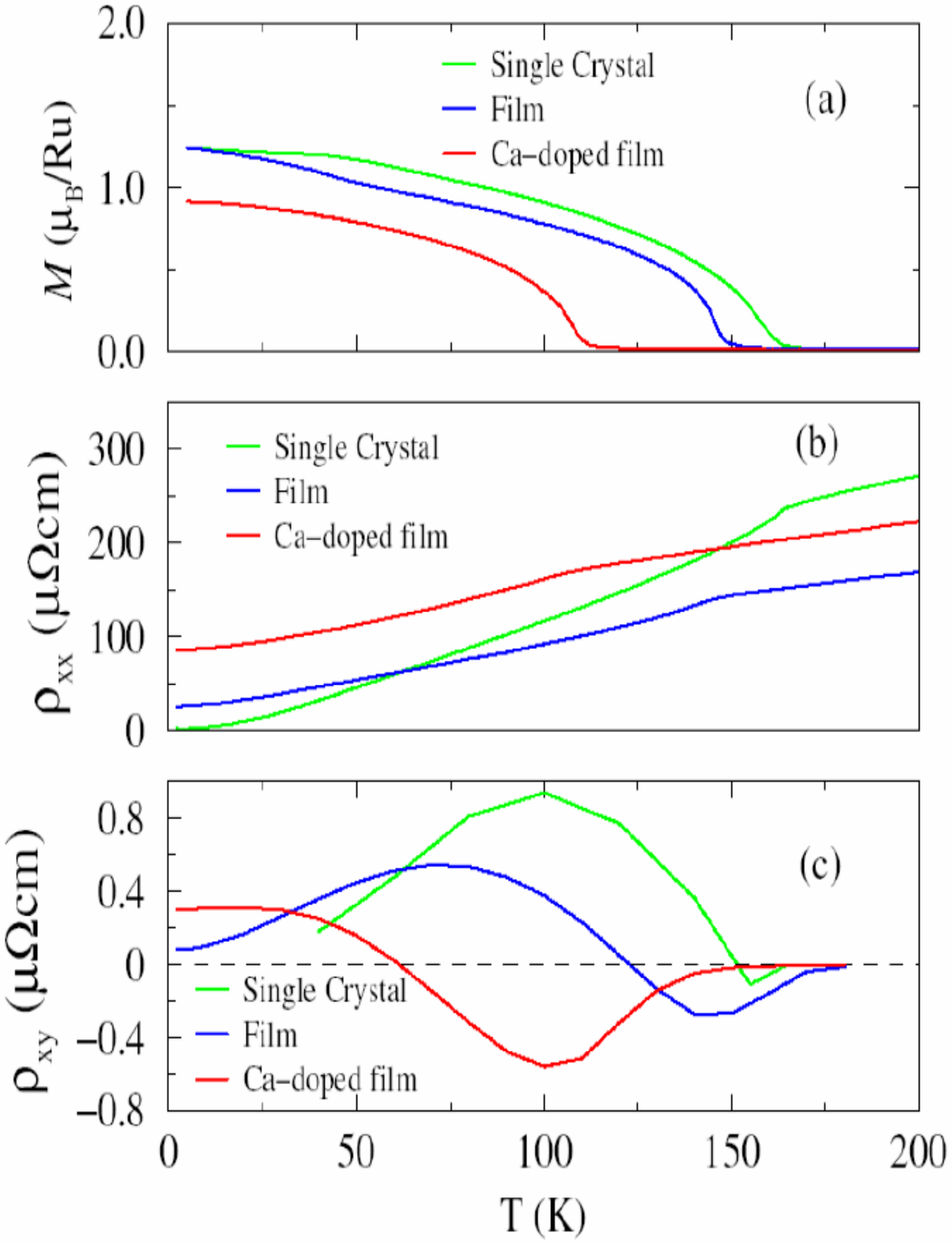}
\caption{\label{fig:SRO}
 Anomalous Hall effect in SrRuO$_3$.  
(A) The magnetization $M$, (B) longitudinal resistivity $\rho_{xx}$,
and (C) transverse resistivity $\rho_{xy}$ as functions of the temperature
$T$  for the single crystal and thin-film SrRuO$_3$,
as well as for  Ca-doped Sr$_{0.8}$Ca$_{0.2}$RuO$_3$ thin film. 
$\mu_B$ is the Bohr magneton. [From Ref. \onlinecite{Fang:2003_a}.]
}
\end{figure}

\begin{figure}
\incl[width=6.5cm]{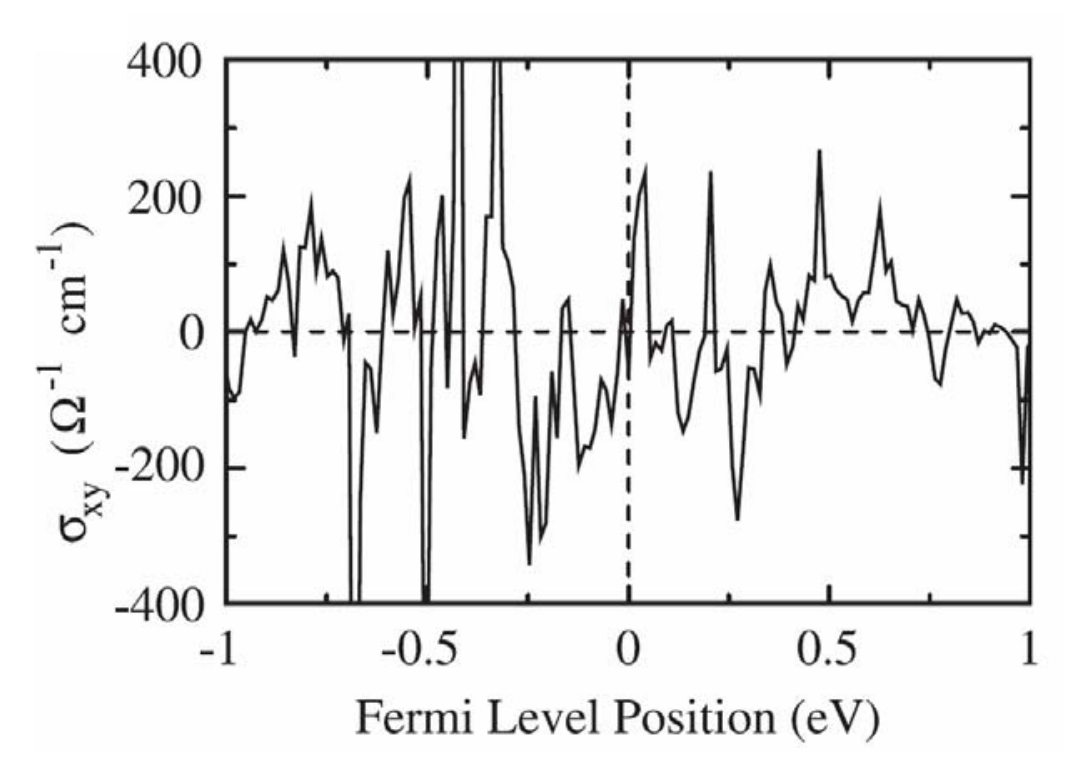}
\caption{\label{fig:Chaos}
The calculated transverse conductivity 
$\sigma_{xy}$ as a function of the chemical potential $\mu$
for SrRuO$_3$. The chaotic behavior is the fingerprint of 
the Berry curvature distribution illustrated in Fig.~\ref{fig:Monopole}. [From Ref. \onlinecite{Fang:2003_a}.]
}
\end{figure}

The perovskite oxide SrRuO$_3$ is an itinerant ferromagnet with a critical temperature $T_c$ of 165 K.  
The electrons occupying the 4$d$ $t_{2g}$ orbitals in $Ru^{4+}$ have a
SOI energy much larger than that for 
3$d$ electrons. Early transport investigations of this 
material were reported in Refs. \onlinecite{Allen:1996_a}
and \onlinecite{Izumi:1997_a}.  The latter authors also reported
results on thin-film SrTiO$_3$.  Recently, the Berry-phase
theory has been applied to account for its AHE 
\cite{Fang:2003_a,Mathieu:2004_a,Mathieu:2004_b},
which is strongly $T$ dependent
(Fig.\ref{fig:SRO}c). Neither the KL theory nor
the skew-scattering theory seemed adequate for
explaining the $T$ dependence of the inferred AHE conductivity $\sigma_{xy}$~\cite{Fang:2003_a}.

\begin{figure}
\vskip 1.0 cm
\incl[width=5.5cm]{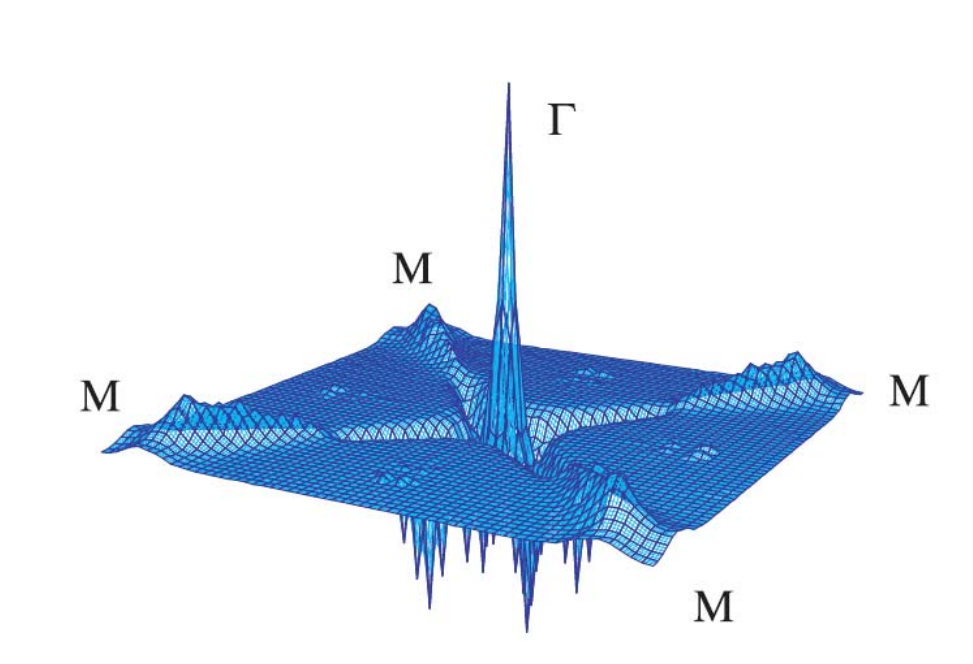}
\caption{\label{fig:Monopole}
The Berry curvature $b_z(\bm{k})$ for a band as a function of
$\bm{k}_\perp = (k_x,k_y)$ with the fixed $k_z=0$. [From Ref. \onlinecite{Fang:2003_a}.]}
\end{figure}

\bfig
\incl[width=6cm]{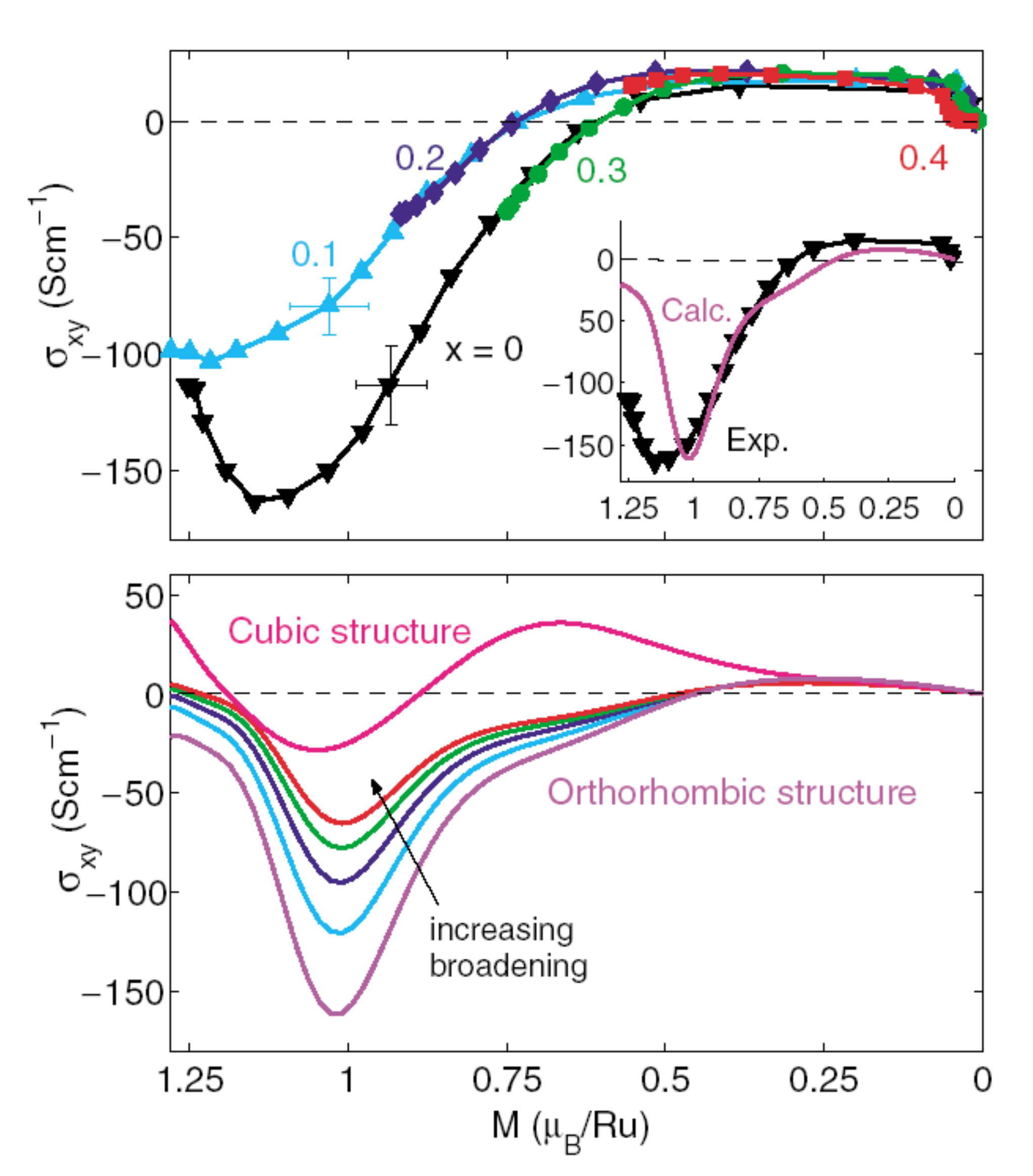}
\caption{\label{fig:Mathieu}
(Upper panel) Combined plots of the Hall conductivity $\sigma_{xy}$ vs. 
magnetization $M$ in 5 samples of the 
ruthenate $\rm Sr_{1-x}Ca_xRuO_3$ ($0\le x\le 0.4$). The
inset compares data  $\sigma_{xy}$ at $x=0$ (triangles) 
with  calculated values (solid curve).  
(Lower panel) First-principles calculations of $\sigma_{xy}$
vs. $M$ for cubic and orthorhombic structures.
The effect of broadening on the curves is 
shown for the orthorhombic case. 
[From Ref.~\onlinecite{Mathieu:2004_b}.]
}
\efig  

The experimental results motivated a detailed first-principles, band-structure calculation 
that fully incorporated the SOI.   The AHE conductivity $\sigma_{xy}$ was calculated directly using the Kubo 
formula Eq.~(\ref{eq:Kubo})~\cite{Fang:2003_a}. 
To handle numerical instabilities which arise near certain critical points, a fictitious energy 
broadening $\delta$ = 70 meV was introduced in the energy denominator.  Fig.~\ref{fig:Chaos} shows the 
dependence of $\sigma_{xy}(\mu)$ on the chemical potential $\mu$. 
In sharp contrast to the diagonal conductivity $\sigma_{xx}$, $\sigma_{xy}(\mu)$ 
fluctuates strongly, displaying sharp peaks and numerous
changes in sign. The fluctuations may be understood 
if we map the momentum dependence of the
Berry curvature $b_z(\bm{k})$ in the occupied band.  
For example, Fig.~\ref{fig:Monopole} displays $b_z(\bm{k})$ plotted as a function of $\bm{k}_\perp = (k_x, k_y)$, with $k_z$ fixed at 0.  The 
prominent peak at $\bm{k}_\perp = \bm{0}$
corresponds to the avoided crossing of the
energy band dispersions, which are split by the 
SOI.  As discussed in Sec. \ref{mechanisms-intro}, variation of the exchange 
splitting caused by a change in the spontaneous magnetization $M$ strongly affects $\sigma_{xy}$ in a nontrivial way.

From the first-principles calculations, one may estimate
the temperature dependence of $\sigma_{xy}$ by assuming that it is due to the temperature-dependence of the Bloch state 
exchange splitting and that this splitting is proportional to the temperature-dependent magnetization.  The new insight
is that the $T$-dependence of $\sigma_{xy}(T)$ simply reflects the $M$ dependence of $\sigma_{xy}$: at a
finite temperature $T'$, the magnitude and sign of $\sigma_{xy}$ may be deduced by using the value of 
$M(T')$ in the zero-$T$ curve.
This proposal was tested against the results on both the pure material and the Ca-doped material 
Sr$_{1-x}$Ca$_x$RuO$_3$.  In the latter, Ca doping suppresses both $T_c$ and $M$ systematically~\cite{Mathieu:2004_a,Mathieu:2004_b}. 
As shown in Fig.~\ref{fig:Mathieu} (upper panel),
the measured values of $M$ and $\sigma_{xy}$, obtained from
5 samples with Ca content 0.4$\le x\le$ 0, fall 
on 2 continuous curves.  In the inset, the curve for the
pure sample ($x=0$) is compared with the calculation.
The lower panel of Fig.~\ref{fig:Mathieu} compares
calculated curves of $\sigma_{xy}$ for cubic and
orthorhombic lattice structures.  The sensitivity of $\sigma_{xy}$ to the lattice symmetry reflects the 
dominant contribution of the avoided crossing near $\varepsilon_F$. The sensitivity to broadening is
shown for the orthorhombic case.

Kats \etal.~\cite{Kats:2004_a} also have studied the magnetic field dependence of $\rho_{xy}$ in an epitaxial 
film of SrRuO$_3$.  They have observed sign-changes in $\rho_{xy}$ near  a magnetic field $B=3$ T at $T$ = 130 K and near $B=8$ T at 134 K.  
This seems to be qualitative consistent with the Berry-phase scenario. On the other hand the authors  suggest that the 
intrinsic-dominated picture is likely incomplete (or incorrect) near Tc~\cite{Kats:2004_a}.

\subsubsection{Spin chirality mechanism of the AHE in manganites}
In the manganites, e.g. $\rm La_{1-x}Ca_xMnO_3$ (LCMO),
the three $t_{2g}$ electrons on each Mn ion form a 
core local moment of spin $S=\frac32$.  A large Hund
energy $J_H$ aligns the core spin $\bf S$ with the $s=\frac12$ spin of 
an itinerant electron that momentarily occupies the $e_g$ orbital.  
Because this Hund coupling leads to an
extraordinary magnetoresistance in weak $H$, 
the manganites are called colossal magnetoresistance (CMR) materials~\cite{Tokura:1999_a}.  The double
exchange theory summarized by  Eq.~(\ref{eq:DE}) (below) is widely
adopted to describe the onset of ferromagnetism in the
CMR manganites.

\begin{figure}
\incl[width=7cm]{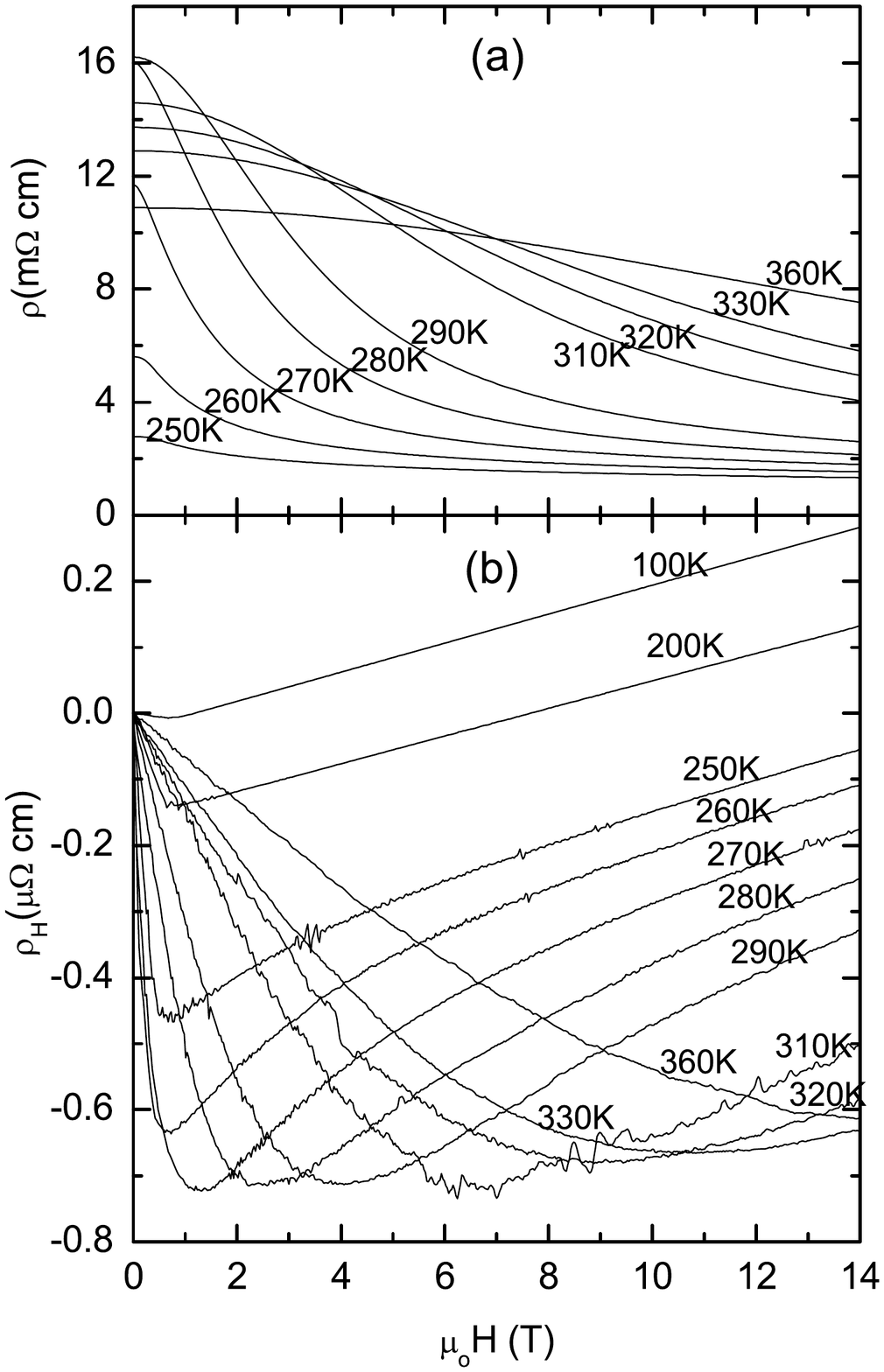}
\caption{\label{fig:Matl}
(a) The colossal magnetoresistance $\rho$ vs. $H$ in $\rm La_{1-x}Ca_xMnO_3$
($T_C$ = 265 K) at selected $T$. (b) The Hall resistivity $\rho_{xy}$
vs. $H$ at temperatures 100 to 360 K. Above $T_C$, $\rho_{xy}$ is
strongly influenced by the MR and the susceptibility $\chi$.
[From Ref. \onlinecite{Matl:1998_a}.]
}
\end{figure}

As $T$ decreases below the Curie temperature $T_C\simeq$ 270 K in LCMO, 
the resistivity $\rho$ falls rapidly from $\sim$15
m$\Omega$ cm to metallic values $<$2 m$\Omega$cm. 
CMR is observed over a significant interval of temperatures above and below $T_C$, where charge transport occurs by 
hopping of electrons between adjacent Mn ions (Fig. \ref{fig:Matl}a).  At each Mn site $i$, the Hund energy
tends to align the carrier spin $\bf s$ with the core spin $\bm{S}_i$.  
 
\begin{figure}[h]
\vskip 1.0cm
\incl[width=6.0cm]{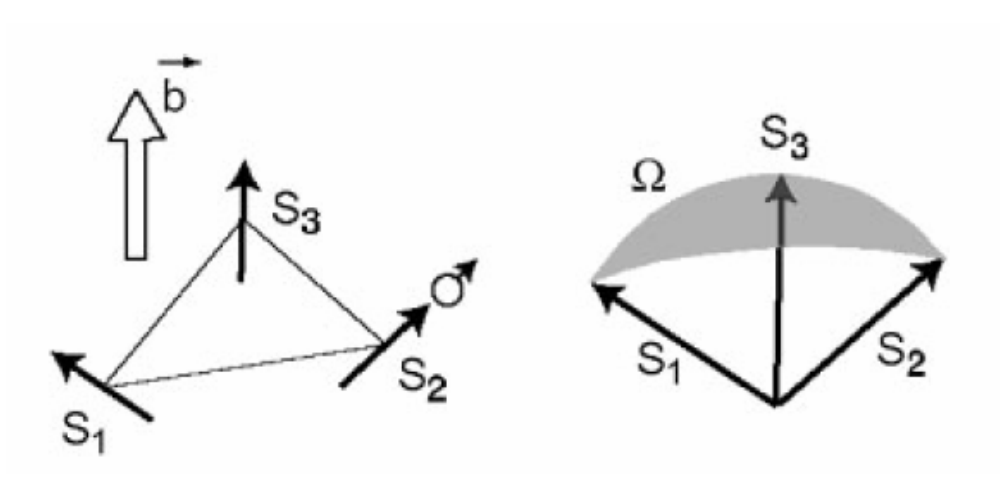}
\caption{\label{fig:spinchirality}
Schematic view of spin chirality. When circulates among three spins to which it is exchange-coupled,
it feels a fictitious magnetic field ${\vec b}$ with  flux given by 
the half of the solid angle $\Omega$ subtended by the three spins. [From Ref. \onlinecite{Lee:2006_a}.]
}
\end{figure}

Early theories of hopping conductivity~\cite{Holstein:1961_a} predicted
the existence of a Hall current produced by the
phase shift (Peierls factor) associated with the magnetic flux $\phi$ piercing the area defined by 3 non-collinear atoms.
However, the hopping Hall current is weak.
The observation of a large $\rho_{xy}$ in LCMO
that attains a broad maximum in modest $H$ (Fig.~\ref{fig:Matl}b) 
led Matl \etal.~\cite{Matl:1998_a} to propose that
the phase shift is geometric in origin, arising from the solid angle described by $\bm{s}$ as the electron visits each Mn site ($\bm{s} ||\bm{S}_i$ at each site $i$
as shown in Fig.~\ref{fig:spinchirality}).  
To obtain the large $\rho_{xy}$ seen, one 
requires $\bm{S}_i$ to define a finite solid angle $\Omega$.  
Since $\bm{S}_{i}$ gradually
aligns with $\bf H$ with increasing field, this effect should disappear along with $\langle\Omega\rangle$,
as observed in the experiment.  This appears to be the first application of a geometric-phase mechanism to account for an AHE experiment.

Subsequently, Ye \etal. \cite{Ye:1999_a} considered 
the Berry phase due to the thermal excitations of the Skyrmion (and anti-Skyrmion).
They argued that the SOI gives rise
to a coupling between the uniform magnetization $\bm{M}$
and the gauge field $\bm{b}$ by the term $\lambda \bm{M} \cdot \bm{b}$.
In the ferromagnetic state, the spontaneous uniform 
magnetization $\bm{M}$ leads to a finite and uniform
$\bm{b}$, which acts as a uniform magnetic field.
Lyanda-Geller \etal.~\cite{Lyanda-Geller:2001_a}
also considered the AHE due to the spin chirality fluctuation in the incoherent limit where the hopping is treated perturbatively. This approach, applicable to the high-$T$ limit, complements the theory of Ye \etal. \cite{Ye:1999_a}.

The Berry phase associated with 
non-coplanar spin configurations, the scalar spin chirality, was first considered in theories of high-temperature superconductors 
in the context of  the flux distribution generated by the 
complex order parameter of the resonating valence bond (RVB) correlation 
defined by $\chi_{ij}$, which acts as the transfer integral of the 
``spinon'' between the sites $i$ and $j$~\cite{Lee:2006_a}. 
The complex transfer integral also
appears in the double-exchange model specified by 
\begin{equation}
H = - \sum_{ij,\alpha} t_{ij}( c^\dagger_{i \alpha} c_{j \alpha} + h.c.)
   - J_H \sum_i \bm{S}_i \cdot c^\dagger_{i \alpha} 
\bm{\sigma}_{\alpha \beta} c_{i \beta},
\label{eq:DE}
\end{equation} 
where $J_H$ is the ferromagnetic Hund's coupling between the spin $\bm{\sigma}$ of the conduction electrons and the localized spins $\bm{S}_i$.

In the manganese oxides, $\bm{S}_i$ represents
the localized spin in $t_{2g}$-orbitals, while
$c^\dagger$ and $c$ are the operators for $e_g$-electrons. 
(The mean-field approximations of Hubbard-like theories
of magnetism, the localized spin may also be regarded as 
the molecular field created by the conduction electrons themselves, in which case $J_H$ is replaced by the on-site Coulomb interaction energy $U$.)
In the limit of large $J_H$, the conduction electron 
spin $\bm{s}$ is forced to align with $\bm{S}_i$ at each site.  The matrix element for hopping from $i \to j$ is then given by 
\begin{equation}
t^{\rm eff}_{ij} = t_{ij} \langle \chi_i | \chi_j \rangle 
= t_{ij} e^{i a_{ij}} \cos \biggl( {{\theta_{ij}} \over 2} \biggr),
\end{equation}
where $| \chi_i\rangle$ is the two-component spinor spin wave function with quantization axis $||\bm{S}_i$.
The phase factor $e^{ia_{ij}}$ acts like a Peierls phase and 
can be viewed as originating from a 
fictitious magnetic field which 
influences the orbital motion of the conduction electrons.

We next discuss how the Peierls phase leads to a
gauge field, i.e. flux, in the presence of non-coplanar spin configurations.
Let $\bm{S}_i$, $\bm{S}_j$, and 
$\bm{S}_k$ be the local spins at sites $i$, $j$, and $k$, respectively. 
The product of the three transfer integrals 
corresponding to the loop $i \to j \to k \to i$ is 
\begin{eqnarray}
&&\langle\bm{n}_i | \bm{n}_k\rangle\langle\bm{n}_k | \bm{n}_j\rangle\langle\bm{n}_j | \bm{n}_i\rangle 
\nonumber \\
&=& ( 1 + \bm{n}_i \cdot \bm{n}_j + \bm{n}_j \cdot \bm{n}_k + 
\bm{n}_k \cdot \bm{n}_j )
+ i \bm{n}_i \cdot ( \bm{n}_j \times  \bm{n}_k)  
\nonumber \\
&\propto& e^{i(a_{ij} + a_{jk} + a_{ki})} = e^{i \Omega/2}
\label{eq:3spin}
\end{eqnarray}
where $| \bm{n}_i\rangle$ is the two-component spinor wavefunction 
of the spin state polarized along $\bm{n}_i = \bm{S}_i/|\bm{S}_i|$.
Its imaginary part is proportional to 
$ \bm{S}_i \cdot ( \bm{S}_j \times \bm{S}_k)$,
which corresponds to 
the solid angle $\Omega$ subtended by the three spins on the unit sphere, and is
called the scalar spin chirality (Fig. \ref{fig:spinchirality}). 
The phase acquired by the electron's wave function
around the loop is $e^{i\Omega/2}$, which leads to the Aharonov-Bohm (AB) effect and, as a consequence, 
to a large Hall response.

In the continuum approximation, this phase factor is
given by the flux of the ``effective'' magnetic field
$\bm{b} \cdot d \bm{S} = \nabla \times \bm{a}
\cdot d \bm{S}$ where $d \bm{S}$ is the elemental directed surface area defined by the three sites. 
The  discussion implies that a large Hall
current requires the unit vector $\bm{n(x)= S(x)/|S(x)|}$ to fluctuate strongly
as a function of $\bm{x}$, the position coordinate in the sample.  An insightful way to quantify this fluctuation is to regard $\bm{n(x)}$ as a map from the
$x$-$y$ plane to the surface of the unit sphere
(we take a 2D sample for simplicity).  An important defect in a ferromagnet -- the Skyrmion~\cite{Sondhi:1993_a} -- occurs when $\bm{n(x)}$ points down
at a point $\bm{x}'$ in a region $\cal A$ of the $x$-$y$ plane, but gradually relaxes back to up at the boundary of $\cal A$.  The map of this spin texture wraps around the sphere once as $\bm{x}'$ roams over $\cal A$.  The number of Skyrmions in the sample is given 
by the topological index
\begin{equation}
N_s = \int_{\cal A} dx dy\;\; \bm{n} \cdot \biggl( 
{{\partial \bm{n}} \over {\partial x}} \times
{{\partial \bm{n}} \over {\partial y}} \biggr)
= \int_{\cal A} dx dy\; b_z,
\end{equation}
where the first integrand is the directed area of
the image on the unit sphere. $N_s$ counts the number of
times the map covers the sphere as ${\cal A}$ extends over the sample.
The gauge field $\bm{b}$ produces a Hall conductivity.
Ye \etal. \cite{Ye:1999_a} derived in the continuum approximation 
the coupling between the field $\bm{b}$ and the 
spontaneuos magnetization $\bm{M}$ through the SOI. 
The SOI coupling produces an excess of thermally excited
positive Skyrmions over negative ones.  This imbalance
leads to a net uniform ``magnetic field'' $\bm{b}$ (anti)parallel to $\bm{M}$, and the AHE.
In this scenario, $\rho_{xy}$ is predicted to attain a maximum slightly below $T_c$, before falling exponentially to zero as $T\to 0$.

The Hall effect in the hopping regime has been discussed 
by Holstein~\cite{Holstein:1961_a} in the context of impurity conduction in semiconductors. 
Since the energies $\varepsilon_j$ and $\varepsilon_k$ of adjacent impurity sites may differ significantly,  
charge conduction must proceed by phonon-assisted hopping. 
To obtain a Hall effect, we consider
3 non-collinear sites (labelled as $i$ = 1, 2 and 3). 
In a field $H$,
the magnetic flux $\phi$ piercing the area enclosed 
by the 3 sites plays the key role in
the Hall response.  According to Holstein,
the Hall current arises from interference between
the direct hopping path $1\to 2$ and the path
$1\to 3\to 2$ going via 3 as an intermediate step.  Taking into account the changes in 
the phonon number in each process, we have
\begin{eqnarray}
(1,N_\lambda,N_{\lambda'}) &\to & (1, N_\lambda \mp 1,N_{\lambda'} ) 
\to (2, N_\lambda \mp 1,N_{\lambda'} \mp 1 ), 
\nonumber \\
(1,N_\lambda,N_{\lambda'}) &\to & (3, N_\lambda \mp 1,N_{\lambda'} ) 
\to (2, N_\lambda \mp 1,N_{\lambda'} \mp 1 ), \nonumber
\\
\label{eq:TwoProc}
\end{eqnarray}
where $N_\lambda,N_{\lambda'}$ are the phonon 
numbers for the modes $\lambda, \lambda'$, respectively.

In a field $H$, the hopping matrix element 
from $\bm{R}_i$ to $\bm{R}_j$ includes the Peierls phase factor
$ \exp[ - i (e/c) \int_{\bm{R}_i}^{\bm{R}_j} 
d \bm{r}  \cdot \bm{A} ( \bm{r})] $.
When we consider the interference between the two processes in Eq.~(\ref{eq:TwoProc}), the
Peierls phase factors combine to produce the
phase shift $\exp(i2\pi\phi/\phi_0)$ where $\phi_0 = hc/e$ is the flux quantum.  By the Aharonov-Bohm effect, 
this leads to a Hall response.

As discussed, this idea was generalized for the
manganites by replacing the Peierls phase factor
with the Berry phase factor in Eq.~(\ref{eq:3spin})
~\cite{Lyanda-Geller:2001_a}.
The calculated Hall conductivity is 
\begin{equation}
\sigma_H = G(\{ \varepsilon \}) \cos ( \theta_{ij}/2 ) \cos ( \theta_{jk}/2 ) 
 \cos ( \theta_{ki}/2 ) \sin ( \Omega/2),
\end{equation}
where $\{ \varepsilon \}$ is the set of the energy levels 
$\varepsilon_a\;\;(a=i,j,k)$,
and $\theta_{ij}$ is the angle between  $\bm{n}_i$ and $\bm{n}_j$.  
When the average of $\sigma_H$ over all directions of $\bm{n}_a$ is taken, 
it vanishes even for finite spontaneous magnetization $m$.
To obtain a finite $\sigma_{xy}$, it is necessary to incorporate SOI. 

\begin{figure}
\incl[width=0.71\columnwidth]{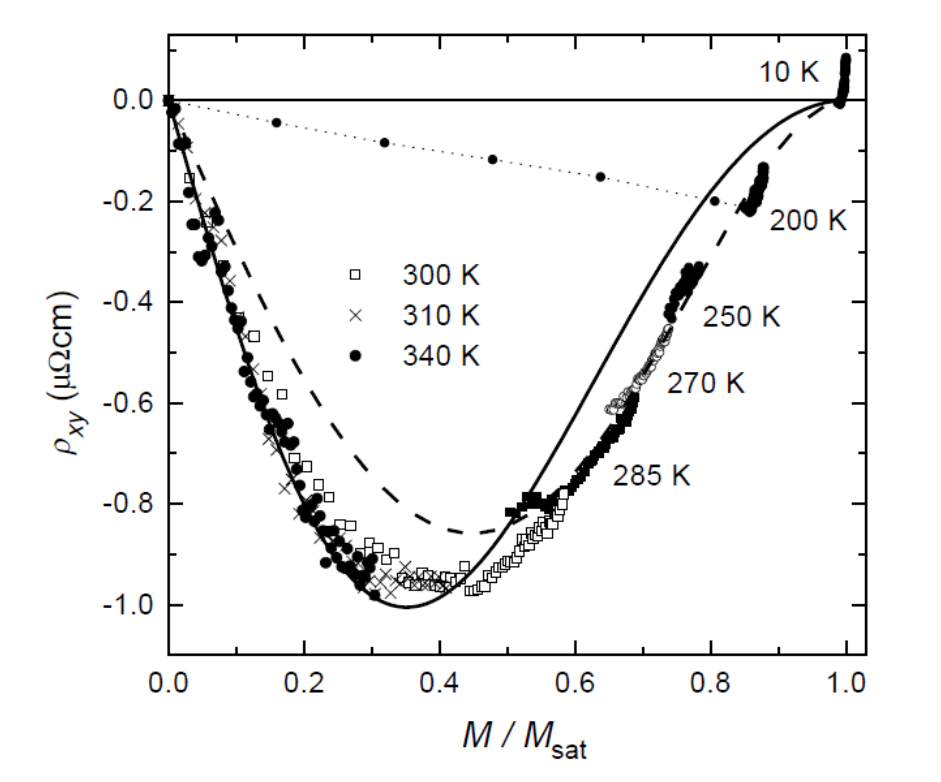}
\caption{\label{fig:MnScaling}
Comparison between experiment and the theoretical prediction 
Eq.(\ref{eq:MnScaling}). 
Scaling behavior between the Hall 
resistivity $\rho_{H}$ and the 
magnetization $M$ is shown. The solid line is a fit to Eq.(\ref{eq:MnScaling}); 
the dashed line is the numerator of Eq.(\ref{eq:MnScaling}) only. 
There are no fitting parameters except the overall scale. [From Ref.~\onlinecite{Chun:2000_a}.]
}
\end{figure}

Assuming the form of hopping integral with the SOI given
by
\begin{equation}
V_{jk} = V^{\rm orb}_{jk} ( 1 + i \bm{\sigma} \cdot \bm{g}_{jk} ),
\end{equation}
the Hall conductivity is proportional to the average of
\begin{equation}
[ \bm{g}_{jk} \cdot ( \bm{n}_j \times \bm{n}_k )] [ \bm{n}_1 \cdot 
\bm{n}_2 \times \bm{n}_3 ].
\end{equation}
Taking the average of $\bm{n}$'s with 
$m=M/M_{sat}$ where $M_{sat}$ is the saturated magnetization,
we finally obtain
\begin{equation}
\rho_{xy} = \rho_{xy}^0 
{ { m ( 1 - m^2)^2 } \over { ( 1 + m^2)^2} }.
\label{eq:MnScaling}
\end{equation}
This prediction has been tested by the experiment of  Chun \etal. \cite{Chun:2000_a}
shown in Fig.~\ref{fig:MnScaling}.
The scaling law for the anomalous $\rho_{xy}$ as a function of $|\bf M|$ 
obtained near $T_C$ is in good agreement with the experiment.   

Similar ideas have been used by Burkov and Balents \cite{Burkov:2003_a} to analyze the
variable range hopping region in 
(Ga,Mn)As.  The spin-chirality mechanism for 
the AHE has also been applied to 
CrO$_2$~\cite{Yanagihara:2002_a,Yanagihara:2007_a}, and 
the element Gd~\cite{Baily:2005_a}. In the former case, 
the comparison between $\rho_{xy}$ and the specific heat 
supports the claim that the critical properties of $\rho_{xy}$ are governed by the Skyrmion density.

The theories described above assume large Hund coupling. 
In the weak-Hund coupling limit, a perturbative treatment
in $J_H$ has been developed to relate the AHE 
conductivity to the scalar spin chirality \cite{Tatara:2002_a}. This theory has been applied 
to metallic spin-glass systems~\cite{Kawamura:2007_a}.

\subsubsection{Lanthanum cobaltite}
The subtleties and complications involved in analyzing the Hall 
conductivity of tunable ferromagnetic oxides are well illustrated by the cobaltites.
Samoilov \etal.~\cite{Samoilov:1998_a}, and Baily and Salamon~\cite{Baily:2003_a}
investigated the AHE in Ca-doped lanthanum cobaltite
$\rm La_{1-x}Ca_xCoO_3$, which displays a number of 
unusual magnetic and transport properties.
They found an unusually large AHE near $T_C$ as well
as at low $T$, and proposed the relevance of
spin-ordered clusters and orbital disorder scattering
to the AHE in the low-$T$ limit. Subsequently, a more detailed 
investigation of $\rm La_{1-x}Sr_xCoO_3$ was reported by Onose and Tokura~\cite{Onose:2006_a}.  
Figs.~\ref{fig:LSCO} a, b and c summarize the $T$ dependence of $M$, $\rho_{xy}$ and $\sigma_{xy}$, 
respectively in four crystals with 
$0.17\leq x \leq 0.30$.  The variation of $\rho_{xx}$ vs. $x$ suggests
that a metal-insulator transition occurs between 0.17 and 0.19.
Whereas the samples with $x\ge 0.2$ have a metallic $\rho_{xx}$-$T$ profile,
the sample with $x$ = 0.17 is non-metallic (hopping conduction).  
Moreover, it displays a very large MR at low $T$ and large hysteresis
in curves of $M$ vs. $H$, features that are consistent with a ferromagnetic cluster-glass state.  

When the Hall conductivity is plotted vs. $M$ (Fig.~\ref{fig:LSCO} d),
$\sigma_{xy}$ shows a linear dependence on $M$ for the most metallic
sample ($x$ = 0.30). 
However, for $x$ = 0.17 and 0.20,
there is a pronounced downturn suggestive of the appearance of
a different Hall term that is electron-like in sign.  This is
most apparent in the trend of the curves of $\rho_{xy}$ vs. $T$ in Fig.~\ref{fig:LSCO} b. 
Onose and Tokura~\cite{Onose:2006_a} 
propose that the negative term may arise from 
hopping of carriers between local moments
which define a chirality that is finite,
as discussed above.

\bfig
\incl[width=8.5cm]{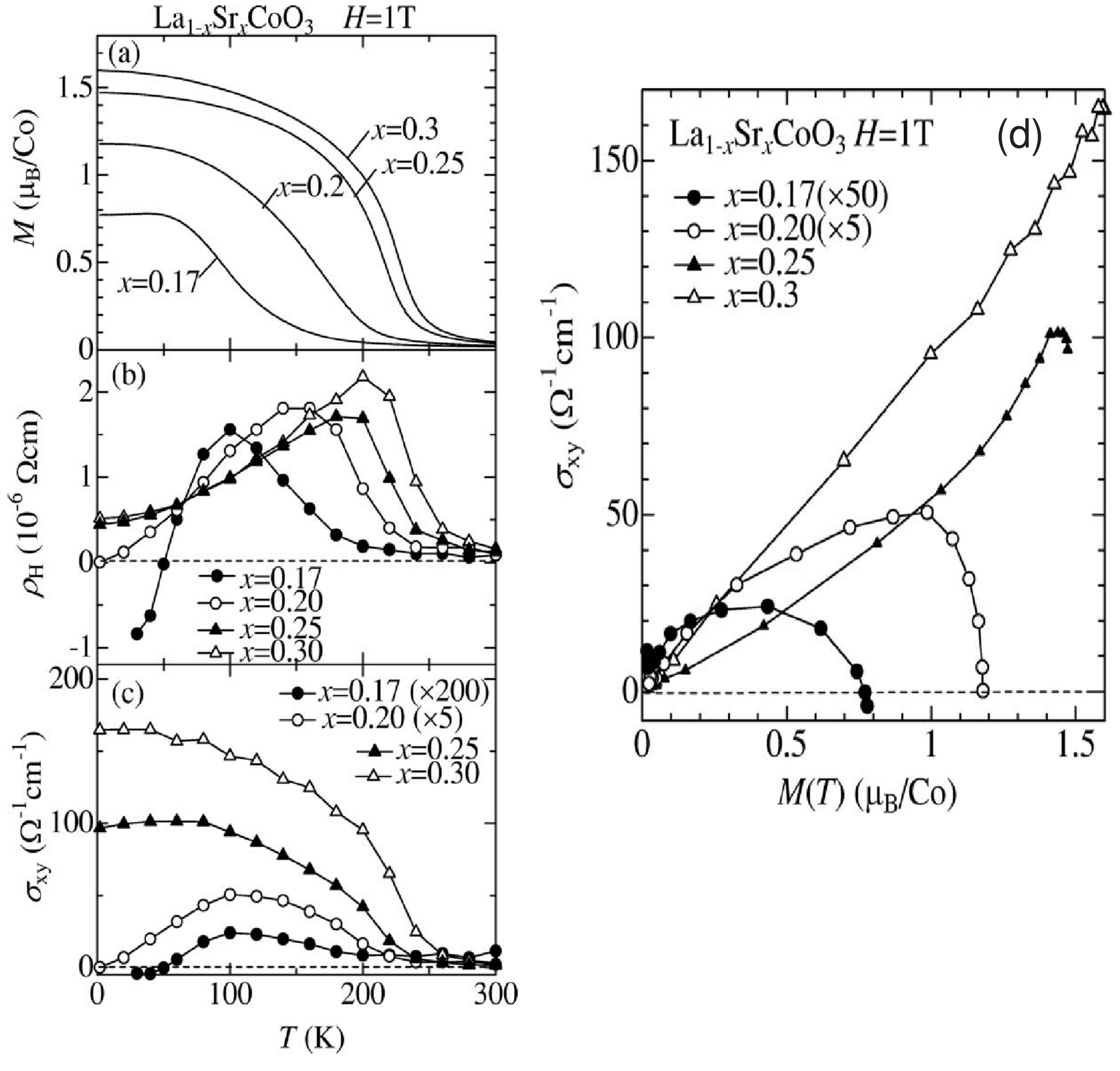}
\caption{\label{fig:LSCO}
Temperature dependence of the magnetization $M$ (a),
Hall resistivity $\rho_{xy}$ (b), Hall conductivity $\sigma_{xy}$ (c),
in four crystals of $\rm La_{1-x}Sr_xCoO_3$ ($0.17\leq x\leq 0.30$)
(all measured in a field $H$ = 1 T).  
In Panels a, b and c, the data for $x$ = 0.17 and 0.20 were
multiplied by a factor of 200 and 50, respectively.
(d) The Hall conductivity $\sigma_{xy}$ at 1 T
in the four crystals plotted against $M$.  Results for 
$x$ = 0.17 and 0.20 were multiplied by factors 50 
and 5, respectively.
[From Ref. \onlinecite{Onose:2006_a}.]
}
\efig

\subsubsection{Spin chirality mechanism in pyrochlore ferromagnets}
 
In the examples discussed in the previous subsection, the spin-chirality mechanism 
leads to a large AHE at finite temperatures.  An interesting question is whether  or not
there exist ferromagnets in which the spin chirality is finite in the ground state.

Ohgushi \etal. \cite{Ohgushi:2000_a} considered the ground state of the non-coplanar spin configuration in 
the Kagome lattice, which may be obtained as a
projection of the pyrochlore lattice onto the plane normal to (1,1,1) axis.
Considering the double exchange model Eq.~(\ref{eq:DE}), they obtained the band structure 
of the conduction electrons and the Berry phase distribution. Quite similar to the 
Haldane model~\cite{Haldane:1988_a} or the model discussed in Eq.~(\ref{eq:spmodel}),
the Chern number of each band becomes nonzero, and a quantized Hall effect
results when the chemical potential is in the energy gap.

Turning to real materials, the pyrochlore ferromagnet 
Nd$_2$Mo$_2$O$_7$ (NMO) provides a test-bed for exploring these issues.  Its lattice structure consists
 of two interpenetrating sublattices comprised of tetrahedrons of 
Nd and Mo atoms, respectively (the sublattices are shifted along the $c$-axis) 
\cite{Taguchi:2001_a,Yoshii:2000_a}. 
While the exchange between spins on either sublattice is ferromagnetic, the exchange coupling 
$J_{df}$ between spins of the conducting $d$-electrons of Mo and 
localized $f$-electron spins on Nd is antiferromagnetic.

\begin{figure}[h]
\incl[width=8.5cm]{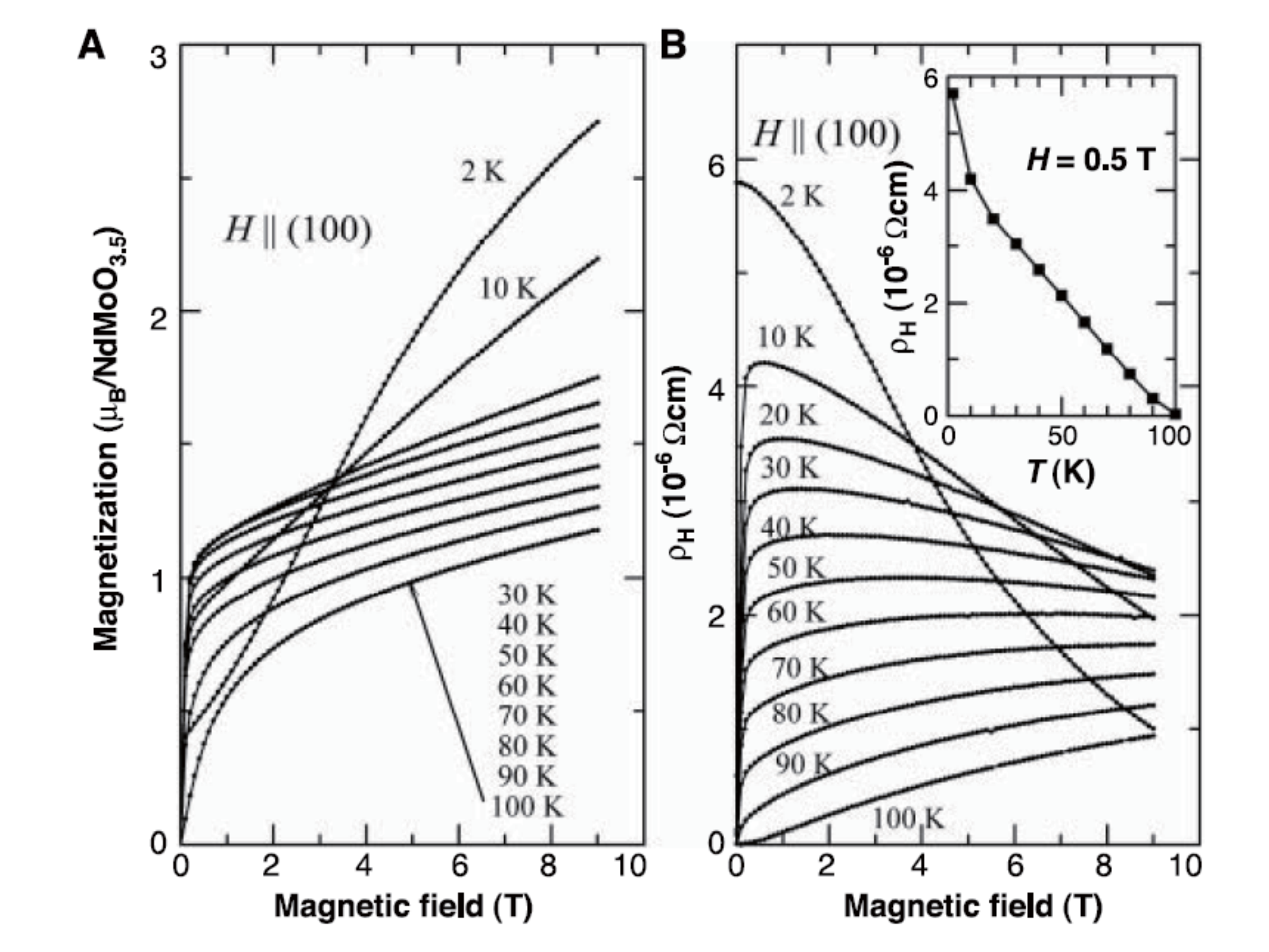}
\caption{
Anomalous Hall effect in
Nd$_2$Mo$_2$O$_7$. 
Magnetic field dependence of (A) the magnetization 
and (B) the  transverse resistivity 
($\rho_{xy}$) for different temperatures.  [From Ref. \onlinecite{Taguchi:2001_a}.]
}
\label{fig:NMO}
\end{figure}

The dependences of the anomalous Hall resistivity
$\rho_{xy}$ on $H$ at selected temperatures are shown in Fig. \ref{fig:NMO}.
The spins of Nd begin to align antiparallel 
to those of Mo below the crossover temperature $T^*\cong 40$ K.
Each Nd spin is subject to a strong easy-axis anisotropy 
along the line from a vertex of the Nd tetrahedron to its center.  The resulting noncoplanar spin configuration
induces a transverse component of the Mo spins. 
Spin chirality is expected to be produced by the 
coupling $J_{df}$, which leads to the AHE of $d$-electrons.
An analysis of the neutron scattering 
experiment has determined the magnetic structure \cite{Taguchi:2001_a}. 
The tilt angle of the Nd spins is close to that expected from the 
strong limit of the spin anisotropy, and the exchange coupling $J_{fd}$ 
is estimated as $J_{fd} \sim 5$ K. This leads to a
tilt angle of the Mo spins of $\sim$5$^{\rm o}$. 
From these estimates, a calculation of the anomalous Hall 
conductivity in a tight-binding Hamiltonian of triply degenerate $t_{2g}$ 
bands leads to $\sigma_H \sim 20 \  (\Omega {\rm cm})^{-1}$, consistent with the value measured at low $T$. 
In a strong $H$, this tilt angle is expected to be
reduced along with $\rho_{xy}$.  This is in agreement with 
the traces displayed in Fig.~\ref{fig:NMO}.

The $T$ dependence of the Hall conductivity
has also been analyzed in the spin-chirality scenario
by incorporating spin fluctuations \cite{SOnoda:2003_a}. 
The result is that frustration of the Ising 
Nd spins leads to large fluctuations, which accounts for the large $\rho$ observed. 
The recent observation of a sign-change in $\sigma_{xy}$ in a field $\bm H$ applied in the $[1,1,1]$ 
direction \cite{Taguchi:2003_a} is consistent with
the sign-change of the spin chirality. 

In the system Gd$_2$Mo$_2$O$_7$ (GMO),
in which Gd$^{3+}$ ($d^7$) has no spin anisotropy,
the low-$T$ AHE is an order-of-magnitude smaller than that in NMO.  
This is consistent with the spin-chirality scenario~\cite{Taguchi:2004_a}. 
The effect of the spin chirality mechanism on the
finite-frequency conductivity $\sigma_H(\omega)$ has been
investigated~\cite{Kezsmarki:2005_a}. 

In another work, Yasui \etal. \cite{Yasui:2006_a,Yasui:2007_a} 
performed neutron scattering
experiments over a large region in the $(H,T)$-plane with $\bm{H}$ along the $[0,\bar{1},1]$ and $[0,0,1]$ directions.
By fitting the magnetization $M_{\rm Nd}(H,T)$ of Nd, the 
magnetic specific heat $C_{\rm mag}(H,T)$, and the magnetic
scattering intensity $I_{\rm mag}(Q,H,T)$, they estimated
$J_{dj} \cong 0.5$ K, which was considerably smaller than estimated previously \cite{Taguchi:2001_a}.
Furthermore, they calculated the thermal average of the spin chirality  
$\langle \bm{S}_i \cdot \bm{S}_j \times \bm{S}_k \rangle$ and compared its value with that inferred from the AHE
resistivity $\rho_{xy}$. They have emphasized that,
when a 3-Tesla field is applied in the $[0,0,1]$ direction (along this direction $H$ cancels the exchange field from the Mo spins), 
no appreciable reduction of $\rho_{xy}$ is observed.  
These recent conclusions have cast doubt on the 
spin-chirality scenario for NMO.

A further puzzling feature is that, 
with $\bm H$ applied in the
$[1,1,1]$ direction, one expects a discontinuous transition from the two-in, 
two-out structure (i.e. 2 of the Nd spins point
towards the tetrahedron center while 2 point
away)
to the three-in, one-out structure for the Nd spins.  
However, no Hall features that might be identified with this
cancellation have been observed down to very low $T$.
This seems to suggest that quantum fluctuations 
of the Nd spins may play an important role,
despite the large spin quantum number ($S=\frac32$).

Machida has discussed the possible relevance of spin chirality to the AHE in the pyrochlore 
Pr$_2$Ir$_2$O$_7$ ~\cite{Machida:2007_a,Machida:2007_b}. In this system, a novel ``Kondo effect'' is observed 
even though the Pr$^{3+}$ ions with $S=1$ are 
subject to a large magnetic anisotropy.   
The magnetic and transport properties of $R_2$Mo$_2$O$_7$
near the phase boundary between the spin glass 
Mott insulator and ferromagnetic metal by changing 
the rare earth ion $R$ has been studied~\cite{Katsufuji:2000_a}.

\subsubsection{Anatase and Rutile $\rm Ti_{1-x}Co_xO_{2-\delta}$}

In thin-film samples of the ferromagnetic semiconductor anatase 
$\rm Ti_{1-x}Co_xO_{2-\delta}$, Ueno \etal. \cite{Ueno:2008_a} have
reported  scaling between the 
AHE resistance and the magnetization $M$.
The AHE conductivity $\sigma^{AH}_{xy}$ scales with the
conductivity $\sigma_{xx}$ as $\sigma^{AH}_{xy} \propto \sigma_{xx}^{1.6}$
(Fig.~\ref{fig:Ueno}).  A similar scaling relation 
was observed in another polymorph rutile.
See also Ref.~\onlinecite{Ramaneti:2007_a} for related work on Co-doped TiO$_2$.

\bfig
\incl[width=6cm]{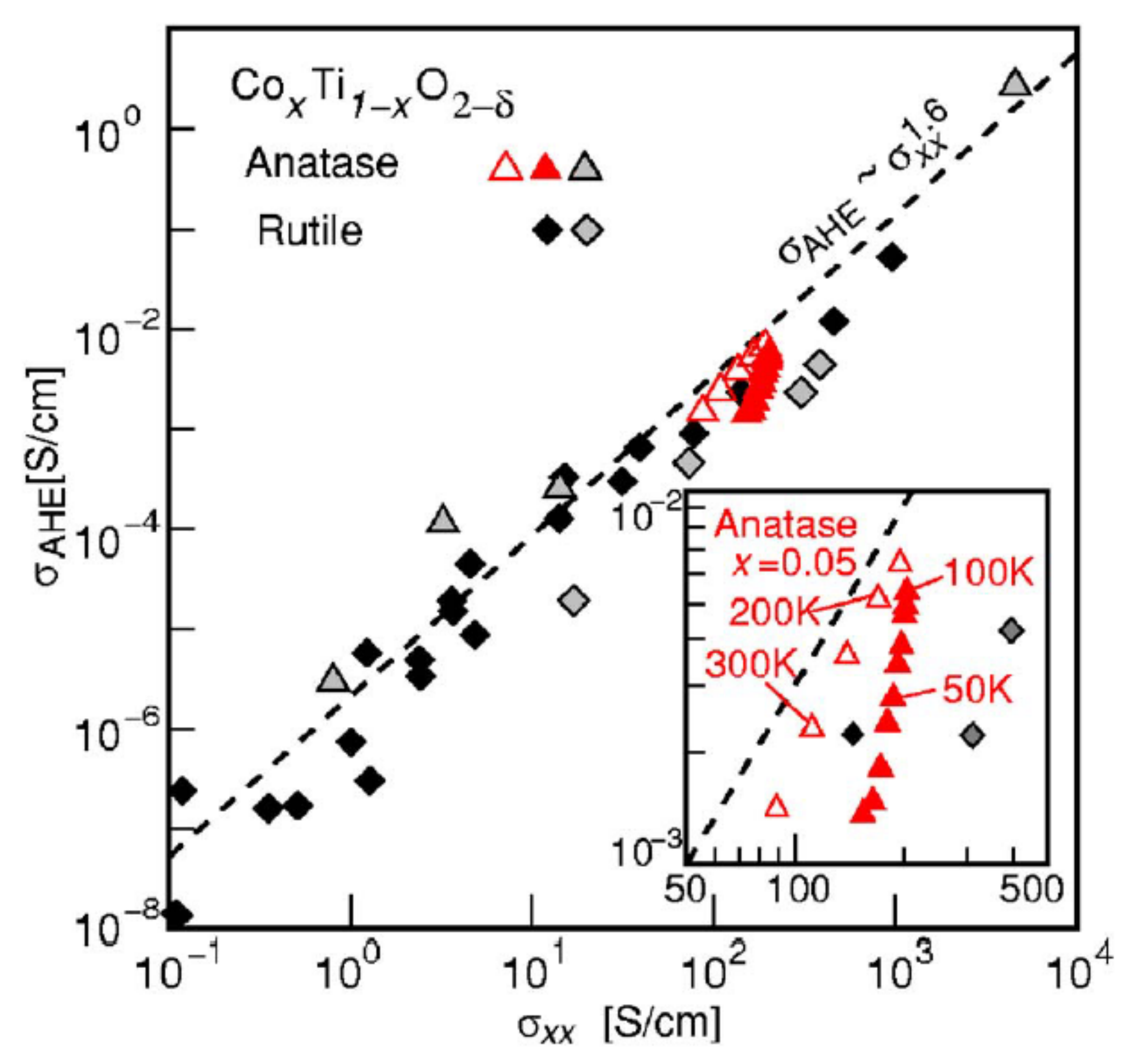}
\caption{ \label{fig:TiCoO2}
Plot of AHE conductivity $\sigma_{AHE}$ vs. conductivity
$\sigma$ for anatase $\rm Ti_{1-x}Co_xO_{2-\delta}$ (triangles)
and rutile $\rm Ti_{1-x}Co_xO_{2-\delta}$ (diamonds).  
Grey symbols are data taken by other groups.
The inset shows the expanded view of data 
for anatase with $x$= 0.05 (the open and closed triangles are 
for $T>$ 150 K and $T<$ 100 K, respectively.
[From Ref.~\onlinecite{Ueno:2007_a}.]}
\label{fig:Ueno}
\efig

	\subsection{Ferromagnetic semiconductors}\label{sec:exp:semi}

Ferromagnetic semiconductors combine
semiconductor tunability  and collective ferromagnetic properties
in a single material. The most widely studied ferromagnetic semiconductors 
are diluted magnetic semiconductors 
(DMS)  created by doping a host semiconductor with a transition metal 
which provides a localized large moment (formed by the d-electrons)
and by introducing  carriers which can mediate a ferromagnetic coupling between these local
moments. The most extensively studied  are the Mn based (III,Mn)V DMSs, 
in which substituting Mn for the cations in a (III,V) semiconductor  can dope the system with hole
carriers;  (Ga,Mn)As becomes ferromagnetic beyond a concentration of   1\%. 
 
 The simplicity of this basic but generally correct model hides within it a 
 cornucopia of physical and materials science effects  present in  these materials.
 Among the phenomena which have been studied are  metal-insulator
transitions,   carrier  mediated  ferromagnetism,   disorder  physics,
magneto-resistance effects, magneto-optical effects, coupled
magnetization dynamics, post-growth dependent properties, {\it etc}.
A more in-depth discussion of these  materials, both from 
the experimental and theoretical point of view, can be found in the  recent 
review by Jungwirth {\it et al} \cite{Jungwirth:2006_a}. 

The AHE  has been one of the most fundamental
characterization tools in DMSs,     
allowing, for example, direct electrical measurement of transition temperatures.
The reliability of electrical measurement of magnetic properties in these materials has been verified by comparison
 with  remnant magnetization measurements  using a  
SQUID magnetometer~\cite{Ohno:1992_a}. The relative 
simplicity of the effective band structure of the carriers  in metallic DMSs, has made them a 
playing ground to understand AHE of ferromagnetic systems with strong spin-orbit coupling.

\begin{figure}
\includegraphics[width=\columnwidth]{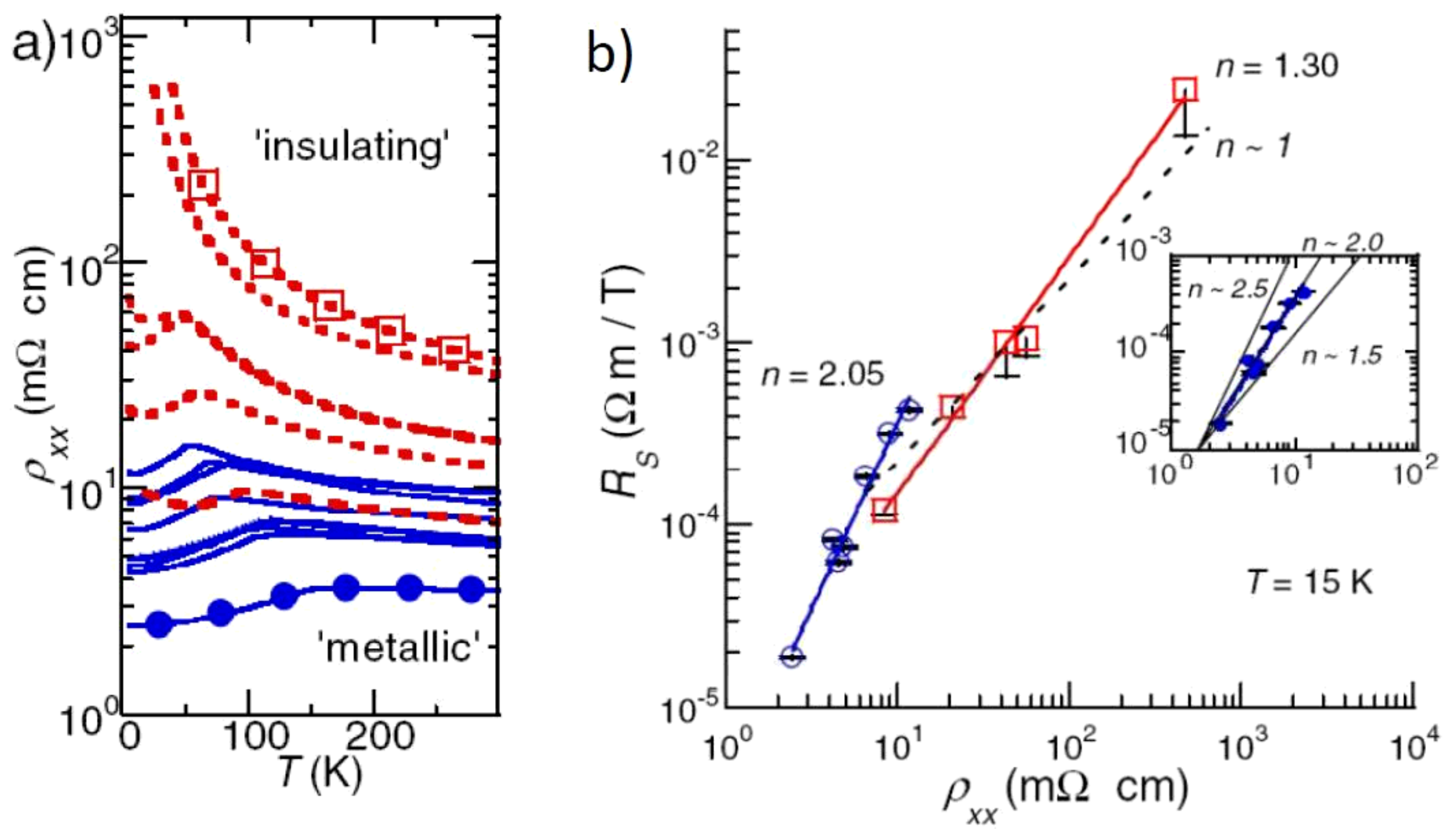}
\caption{(a) Ga$_{1-x}$Mn$_x$As samples that show insulating and metallic behavior defined by $\partial \rho_{xx}/\partial T$ near $T=0$.
(b) R$_s$ vs. $\rho_{xx}$ extrapolated from $\rho_{xy}(B)$ data to zero field and low temperatures. [From Ref. \onlinecite{Chun:2007_a}.]}
\label{ChunPRL2007fig}
\end{figure}

Experimentally, it has been established that the AHE in the archetypical DMS system  (Ga,Mn)As is in the metallic regime
 dominated by a scattering-independent mechanism, i.e. $\rho^{AH}_{xy}\propto \rho_{xx}^2$
 \cite{Edmonds:2002_a,Ruzmetov:2004_a,Chun:2007_a,Pu:2008_a}.
The studies of \onlinecite{Edmonds:2002_a} and \onlinecite{Chun:2007_a} have established this relationship in the non-insulating materials by extrapolating the 
low temperature $\rho_{xy}(B)$ to zero field and zero temperature. This is 
illustrated in Fig.~\ref{ChunPRL2007fig} where  metallic samples, which span a larger range than the ones
studied by \onlinecite{Edmonds:2002_a}, show a clear  $R_S\sim\rho_{xx}^2$ dependence. 

DMSs  grown requires  non-equilibrium (low temperature) conditions
and the as-grown (often insulating) materials and post-grown annealed metallic materials 
show typically different behavior in the AHE response. A similar extrapolating procedure performed on  insulating
(Ga,Mn)As seems to exhibit a somewhat linear dependence of $R_S$ on $\rho_{xx}$. On the other hand, 
considerable uncertainty is introduced by the
extrapolation to low temperatures because 
$\rho_{xx}$ diverges and the complicated 
magetoresistance of $\rho_{xx}$ is a priori not known in the low $T$ range. 

\begin{figure}
\includegraphics[width=0.9\columnwidth]{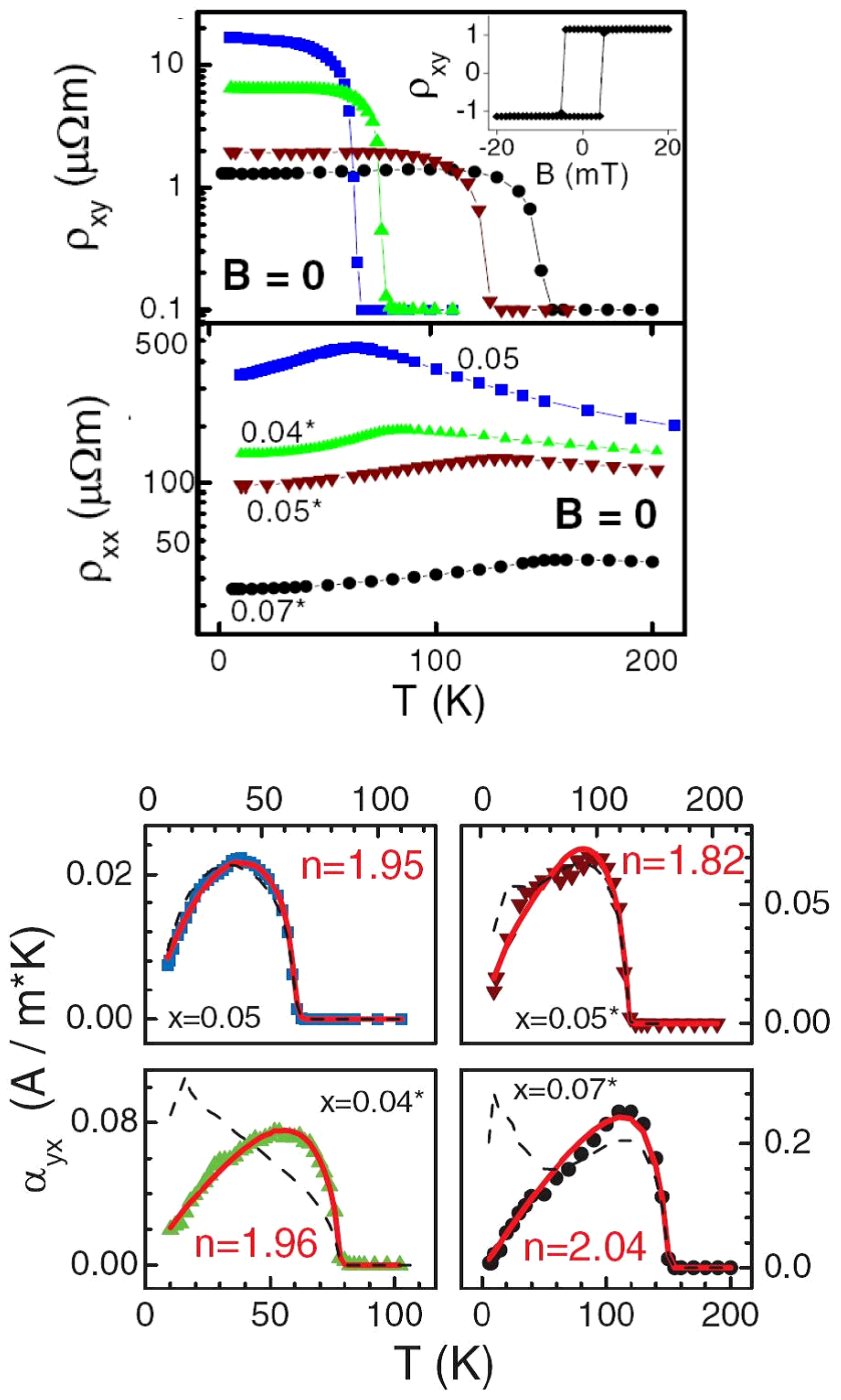}
\caption{(Top) Zero $B$ field $\rho_{xy}$ and $\rho_{xx}$ for four samples grown on InAs substrates 
(i.e. perpendicular to plane easy axis). Annealed samples, which produce perpendicular to plane easy axes, are marked by a $*$. The inset indicates the $B$ 
dependence of the 7\% sample at 10 K. (Bottom) Zero-field Nerst coefficient $\alpha_{xy}$ for the four samples.
The solid red curves indicate the best fit using Eq.~(\ref{thermoelectro}) and the dashed curves  the best fit 
setting n=1. [From Ref. \onlinecite{Pu:2008_a}.]}
\label{Pu_figure}
\end{figure}
A more recent study by Pu \etal.~\cite{Pu:2008_a}of (Ga,Mn)As grown on InAs, such that the tensile strain 
creates a perpendicular anisotropic ferromagnet, has established the dominance of 
the intrinsic mechanism in metallic (Ga,Mn)As samples beyond any doubt. 
Measuring the longitudinal thermo-electric transport coefficients ( $\rho_{xx}$,
$\rho_{xy}$, $\alpha_{xy}$, and $\alpha_{xx}$ where $\bm{J}=\sigma \bm{E}+\alpha(-\bm{\nabla} T)$), one 
can show that given the Mott relation $\alpha=\frac{\pi^2 k_B^2 T}{3e} (\frac{\partial \sigma}{\partial E})_{E_F}$
and the empirical relation $\rho_{xy}(B=0)=\lambda M_z \rho_{xx}^n$, the relation between the four separately measured 
transport coefficients is:
\begin{equation}
\alpha_{xy}=\frac{\rho_{xy}}{\rho_{xx}^2}\left(\frac{\pi^2 k_B^2 T}{3e }\frac{ \lambda'}{ \lambda}- (n-2)\alpha_{xx}\rho{xx}\right).
\label{thermoelectro}
\end{equation}
The fit to the $\frac{ \lambda'}{ \lambda}$ and $n$ parameters are shown in Fig.~\ref{Pu_figure}. Fixing $n=1$ does not produce any 
good fit to the data as indicated by the dashed lines in Fig.~\ref{Pu_figure}. These data excludes the possibility of a $n=1$ type 
contribution to the AHE in metallic (Ga,Mn)As and further verify the validity the Mott relation in these materials.

Having established that the main contribution  to the AHE in metallic (Ga,Mn)As is scattering-independent contributions, rather than
skew-scattering contributions, DMSs are an ideal system to test  our understanding of AHE. 
In the regime  where  the  largest  ferromagnetic critical 
temperatures  are   achieved  (  for   doping  levels  above   1.5\% ),
semi-phenomenological models  that are built  on Bloch states  for the
band  quasiparticles, rather than  localized basis  states appropriate
for  the  localized   regime \cite{Berciu:2001_a},  provide  the  natural
starting point  for a model  Hamiltonian which reproduces many  of the
observed  experimental effects \cite{Jungwirth:2006_a,Sinova:2005_a}.   Recognizing that  the  length scales
associated with  holes in the DMS  compounds are still  long enough, a
$\bm{k}\cdot\bm{p}$ envelope  function   description  of  the
semiconductor   valence  bands   is  appropriate.    To understand
the  AHE and magnetic anisotropy, it is
necessary to incorporate intrinsic  spin-orbit coupling in a realistic
way. 

A successful model for (Ga,Mn)As
is specified by the  effective Hamiltonian
\begin{equation}
{\cal H}={\cal H}_{KL}+J_{pd}\sum_{I} \bm{S}_I\cdot \hat{\bm{ s}}(\bm{ r}) \delta(\bm{r}-\bm{ R}_I)+{\cal H}_{dis},
\label{Heff}
\end{equation}
where ${\cal  H}_{KL}$ is the  six-band Kohn-Luttinger (KL)
$\bm{k}\cdot\bm{p}$  Hamiltonian \cite{Dietl:2001_b},
the second  term is  the short-range antiferromagnetic  
kinetic-exchange interaction
between local spin $\bm{S}_I$  at site $\bm{R}_I$ and the itinerant
hole spin (a finite range  can  be incorporated  in  more  realistic  models), and  ${\cal
H}_{dis}$  is   the  scalar  scattering   potential  representing  the
difference  between a  valence  band electron  on  a host  site and  a
valence  band  electron  on  a   Mn  site  and  the  screened  Coulomb
interaction of the itinerant electrons with the ionized impurities.

Several approximations can  be used to vastly simplify the  above model, 
namely, the virtual crystal approximation  (replacing the spatial dependence of
the  local  Mn   moments  by  a  constant  average)   and   mean  field
theory in which quantum and thermal fluctuations of the local moment spin-orientations are ignored \cite{Dietl:2001_b,Jungwirth:2006_a}.  
In the  metallic regime, disorder can be treated by a Born approximation or by more
sophisticated,    exact-diagonalization       or       Monte-Carlo
methods  \cite{Jungwirth:2002_c,Sinova:2002_a,Yang:2003_b,Schliemann:2002_a}.

\begin{figure}
\includegraphics[width=0.9\columnwidth]{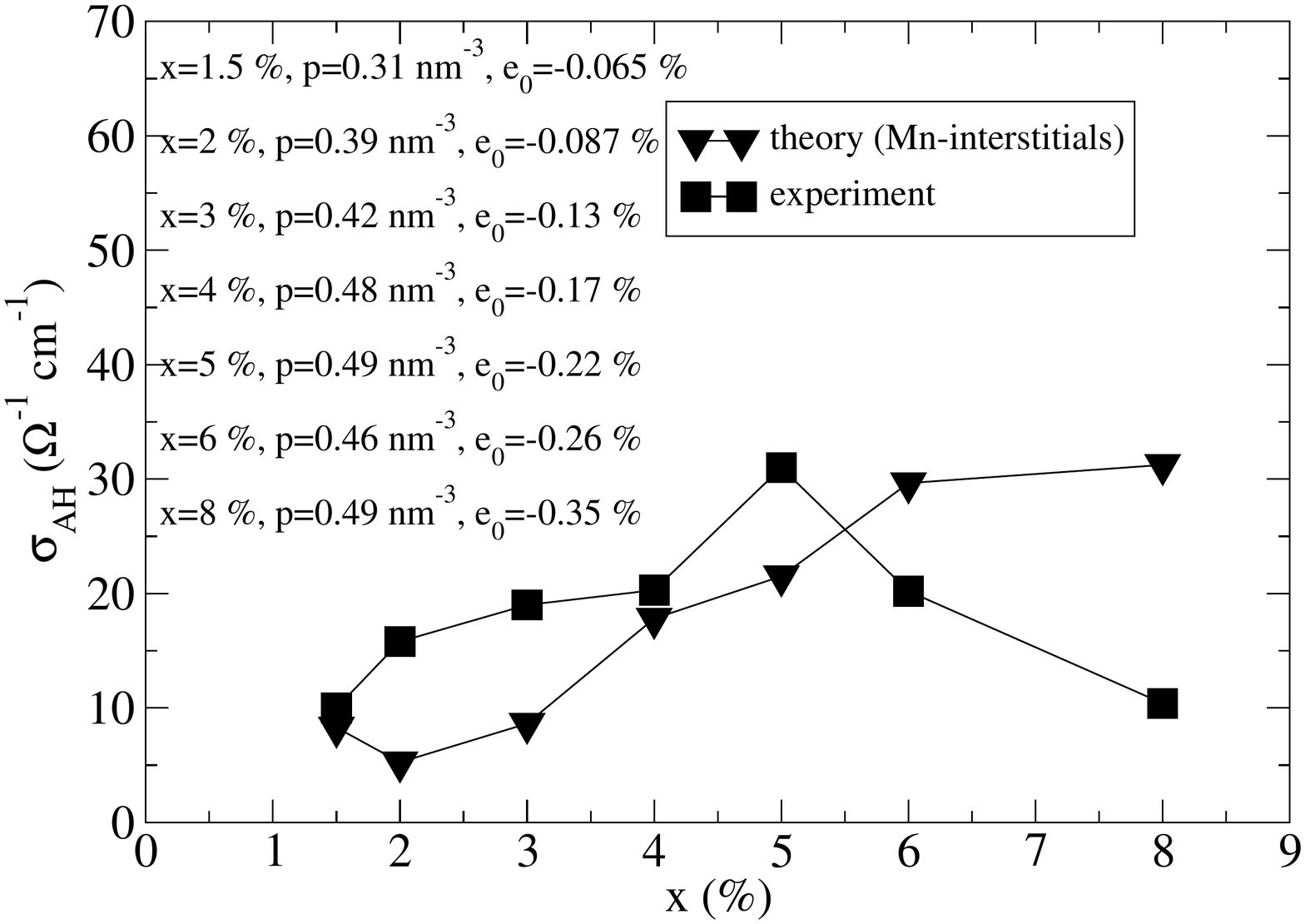}
\caption{Comparison between experimental and theoretical anomalous Hall
conductivities. After Ref. \onlinecite{Jungwirth:2003_b}.}
\label{AHE_teor_exp}
\end{figure}

Given the above simple model Hamiltonian the AHE can be computed if one 
assumes that the intrinsic
Berry's phase (or Karplus-Luttinger) contribution will most likely be dominan, because of the large SOI of the carriers and the experimental 
evidence showing the dominance of scattering-independent mechanisms.
For practical calculations it is useful to use, as in the  intrinsic AHE studies in the oxides ~\cite{Fang:2003_a,Mathieu:2004_a,Mathieu:2004_b}, 
the Kubo formalism given in Eq.~(\ref{eq:Kubo}) with disorder induced broadening, $\Gamma$, of the band-structure (but no side-jump contribution).
The broadening is achieved by
 substituting one of the $(\varepsilon_n(\bm{k})-\varepsilon_{n'}(\bm{k}))$ factors in the denominator
by $(\varepsilon_n(\bm{k})-\varepsilon_{n'}(\bm{k})+i\Gamma)$.
Applying  this theory to metallic  (III,Mn)V materials using  both the 4-band  and 6-band 
$\bm{k}\cdot\bm{p}$ description of the  valence band electronic 
structure one obtains results in   quantitative agreement 
with    experimental data in (Ga,Mn)As and (In,Mn)As DMS  \cite{Jungwirth:2002_a}.
In    a   follow    up   calculation Jungwirth \etal.~\cite{Jungwirth:2003_b},  
a  more  quantitative  comparison  of  the  theory with experiments was made in order  to account
for   finite  quasiparticle  lifetime   effects  in these strongly disorder systems.  
  The effective
lifetime   for  transitions  between   bands  $n$   and  $n^{\prime}$,
$\tau_{n,n^{\prime}}\equiv 1/\Gamma_{n,n'}$,  can be  calculated by
averaging  quasiparticle  scattering  rates  obtained from  Fermi's
golden rule including both screened Coulomb and exchange potentials of
randomly  distributed substitutional  Mn and  compensating  defects as
done      in      the      dc     Boltzman      transport      studies
\cite{Jungwirth:2002_c,Sinova:2002_a}.   A systematic comparison between  theoretical and experimental AHE data
is   shown  in  Fig.~\ref{AHE_teor_exp}  \cite{Jungwirth:2003_b}. 
The
results  are plotted  vs.  nominal Mn  concentration  $x$ while  other
parameters  of the  seven samples  studied  are listed  in the  figure
legend.   The measured  $\sigma_{AH}$ values  are indicated  by filled
squares; triangles are theoretical results obtained for  a  disordered  system  assuming  
Mn-interstitial compensationg  defects.
The valence  band  hole   eigenenergies  $\varepsilon_{n\bm{k}}$  and  eigenvectors
$|n\bm{k}\rangle$  are
obtained  by solving  the six-band  Kohn-Luttinger Hamiltonian  in the
presence  of  the   exchange  field,  $\bm{h}=N_{Mn}S  J_{pd}\hat{z}$
\cite{Jungwirth:2006_a}.  Here $N_{Mn}=4x/a_{DMS}^3$  is the Mn density
in the Mn$_x$Ga$_{1-x}$As epilayer  with a lattice constant $a_{DMS}$,
the  local  Mn  spin  $S=5/2$,  and  the  exchange  coupling  constant
$J_{pd}=55$ meV nm$^{-3}$.

    In general,  when disorder is
accounted for,  the theory  is in a  good agreement  with experimental
data over  the full  range of   Mn densities studied from  $x=1.5\%$ to
$x=8\%$.   The   effect   of   disorder,  especially   when   assuming
Mn-interstitial  compensation, is particularly  strong in  the $x=8\%$
sample   shifting  the  theoretical   $\sigma_{AH}$  much   closer  to
experiment,  compared   to  the  clean  limit   theory.  The  remaining
quantitative  discrepancies between  theory and  experiment  have been
attributed  to the resolution  in measuring  experimental hole  and Mn
densities \cite{Jungwirth:2003_b}.

We conclude this section by  mentioning the anomalous
Hall  effect in the  non-metallic or insulating/hopping regimes.
Experimental studies of (Ga,Mn)As  digital ferromagnetic heterostructures, which consist of submonolayers of MnAs separated by spacer layers of GaAs, have shown longitudinal and Hall resistances of the hopping conduction type, $R_{xx}\propto T^\alpha \exp[(T_0/T)^\beta]$, and have shown that the anomalous Hall resistivity is dominated by hopping with a sublinear dependence of $R_{AH}$ on $R_{xx}$~\cite{Allen:2004_a,Shen:2008_a}, similar to experimental observations on other materials. Studies in this regime are still not as systematic as  their metallic counterparts which clearly indicate 
a scaling power of 2. Experiments find a sublinear dependence of $R_{AH}\sim R_{xx}^\beta$, with $\beta\sim 0.2-1.0$ depending 
on the sample studies~\cite{Shen:2008_a}.
A theoretical understanding for this hopping regime 
remains  to be worked out still. In previous
theoretical calculations based on the hopping conduction with the Berry phase~\cite{Burkov:2003_a} showed the insulating behavior  $R_{AH}\to\infty$ as $R_{xx}\to\infty$ but failed to explain the scaling dependence of $R_{AH}$ on $R_{xx}$.

	\subsection{Other classes of materials}\label{sec:exp:other}

\subsubsection{Spinel CuCr$_2$Se$_4$}

\bfig
\incl[width=6cm]{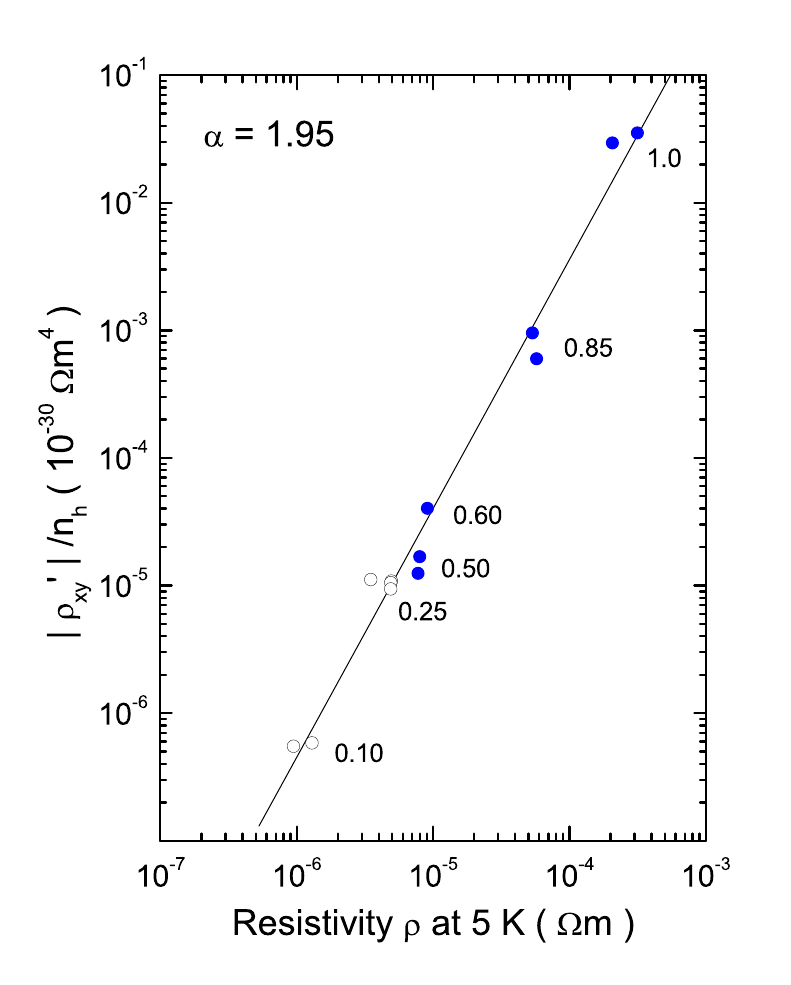}
\caption{\label{fig:spinelrxy}
Log-log plot of the quantity $|\rho_{xy}'|/n_h$ vs. $\rho$
in twelve crystals of Br-doped spinel $\rm  CuCr_2Se_{4-x}Br_x$
with $n_h$ the hole density ($\rho$ is measured at 5 K; $\rho_{xy}$ is measured at 2 and 5 K).
Samples in which $\rho_{yx}$ is electron-like (hole-like) are shown
as open (closed) circles. The straight-line fit implies
$|\rho_{xy}'|/n_h= A\rho^{\alpha}$ with $\alpha = 1.95\pm 0.08$
and $A$ = 2.24$\times 10^{-25}$ (SI units).
[From Ref. \onlinecite{Lee:2004_a}.]
}
\efig

As mentioned in earlier sections, a key prediction of the KL theory (and its modern generalization, based on the Berry-phase approach) 
is that the AHE conductivity 
$\sigma_{yx}^{AH}$ is independent of the carrier lifetime $\tau$
(dissipationless Hall current) in materials with moderate conductivity.
This implies that the anomalous Hall resistivity $\rho_{yx}^{AH}$ varies
as $\rho^2$.  Moreover, $\sigma^{AH}_{xy}$ tend to be proportional
to the observed magnetization $M_z$.  We write
\be
\sigma^{AH}_{xy} = S_HM_z,
\label{eq:SHM}
\ee
with $S_H$ a constant.

Previously, most tests were performed by
comparing $\rho_{yx}$ {\it vs.} $\rho$ measured on the same sample 
over an extended temperature range.  However, because the
skew-scattering model can also lead to the same prediction
$\rho_{yx}\sim\rho^2$ when inelastic scattering predominates,
tests at finite $T$ are inconclusive.  The proper test requires
a system in which $\rho$ at 4 K can be varied over a very large range
without degrading the exchange energy and magnetization $M$.

In the spinel $\rm CuCr_2Se_4$, the ferromagnetic state is stabilized
by 90$^{\rm o}$ superexchange between the local moments of adjacent
Cr ions.  The charge carriers play only a weak role in 
the superexchange energy.  The experimental
proof of this is that when the carrier density $n$ is varied 
by a factor of 20 (by substituting Se by Br), the 
Curie temperature $T_C$ decreases by only 100 K from 380 K.
Significantly, $M$ at 4 K changes negligibly.  The resistivity
$\rho$ at 4 K may be varied by a factor of 10$^3$ without 
weakening $M$.  Detailed Hall and resistivity measurements 
were carried out by Lee \etal.~\cite{Lee:2004_a} on twelve crystals
of $\rm  CuCr_2Se_{4-x}Br_x$.  They found that $\rho_{yx}$
measured at 5 K changes sign (negative to positive) when $x$  
exceeds 0.4.  At $x$ = 1, $\rho_{yx}$ attains very large
values ($\simeq 700\;\mu\Omega$cm at 5 K).

Lee \etal.~\cite{Lee:2004_a} showed that the magnitude $|\rho_{yx}|/n$ varies
as $\rho^2$ over 3 decades in $\rho$ (Fig.~\ref{fig:spinelrxy}), consistent with the prediction of  
the KL theory.

The AHE in the related materials $\rm CuCr_2S_4$,
$\rm Cu_x Zn_x Cr_2Se_4$ ($x=\frac12$) and $\rm Cu_3Te_4$ has been
investigated by Oda \etal.~\cite{Oda:2001_a}.

\subsubsection{Heusler Alloy}
The full Heusler alloy $\rm Co_2CrAl$ has the L2$_1$ lattice
structure and orders ferromagnetically below 333 K.  
Several groups~\cite{Galanakis:2002_a,Block:2004_a} have argued
that the conduction electrons are fully spin 
polarized (``half metal'').  The absence of minority
carrier spins is expected to simplify the analysis of
the Hall conductivity. Hence this system is potentially
an important system to test theories of the AHE.  

The AHE has been investigated on single crystals with 
stoichiometry $\rm Co_{2.06}Cr_{1.04}Al_{0.90}$
~\cite{Husmann:2006_a}.  Below the Curie temperature 
$T_C$ = 333 K, the magnetization $M$ increases rapidly,
eventually saturating to a low-$T$ value that corresponds 
to 1.65 $\mu_B$ (Bohr magneton) per formula unit.  
Husmann and Singh show that, below $\sim$310 K, 
$M(T)$ fits well to the form $[1-(T/T_C)^2]^{\frac12}$.  
Assuming that the ordinary coefficient $R_0$ is negligible,
they found that the Hall conductivity $\sigma_{xy}= \rho_{yx}/\rho^2$
is strictly linear in $M$ (expressed as 
$\sigma_{xy} = \sigma_H^1 M$) over the $T$ interval 36--278 K
(Fig. \ref{fig:Heusler}).  They interpret the linear 
variation as consistent with the intrinsic AHE theory.
The value of $\sigma_H^1$ = 0.383
G$/(4\pi\Omega$cm) inferred is similar to values derived from
measurements on the dilute Ni alloys, half Heuslers and silicides.

\bfig
\incl[width=7cm]{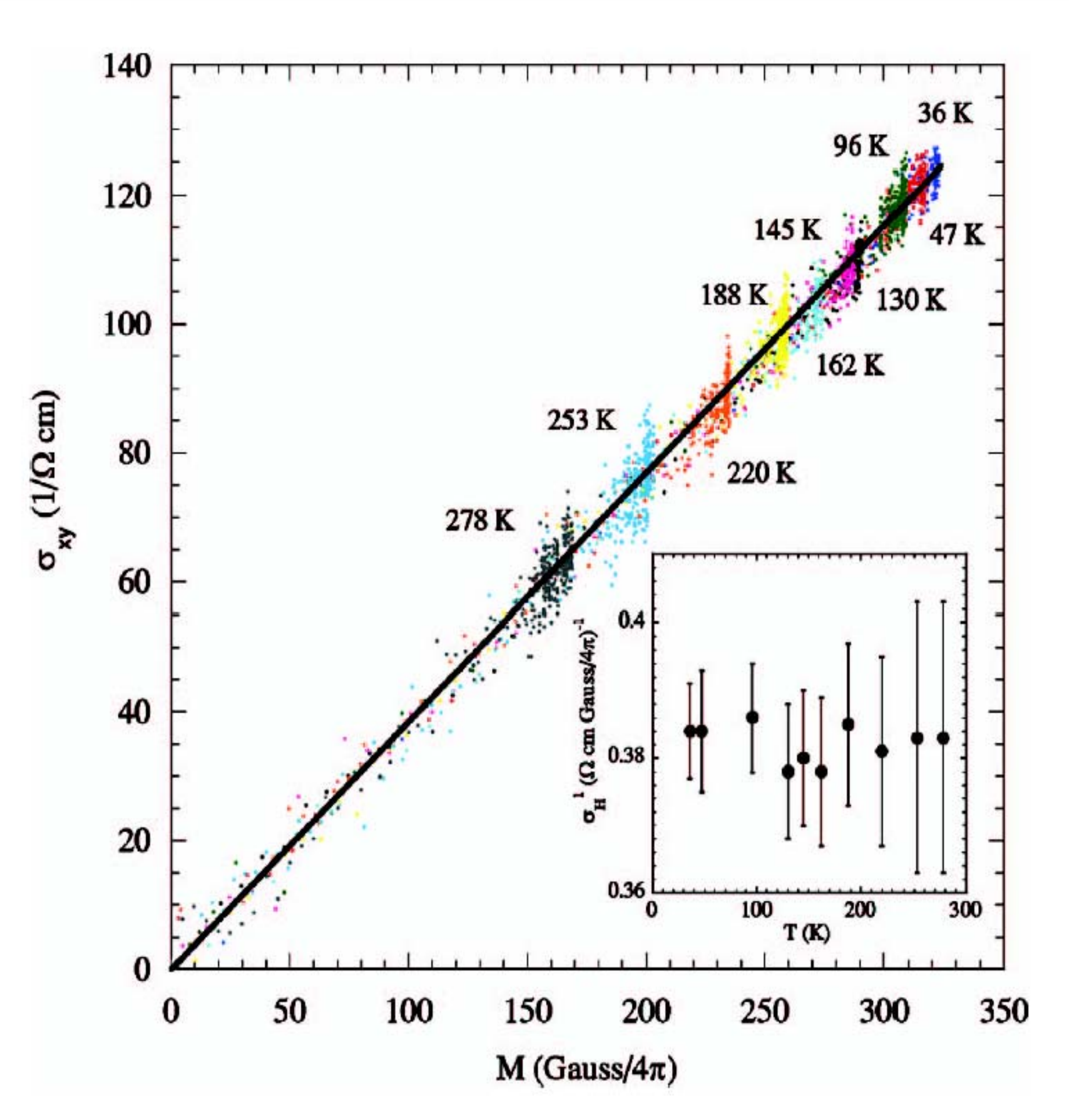}
\caption{\label{fig:Heusler}
Combined plots of $\sigma_{xy}$ in the Heusler 
alloy $\rm Co_2CrAl$ versus the 
magnetization $M$ at selected temperatures from 
36 to 278 K. The inset shows the inferred values of $\sigma_{xy}^{AH}\equiv \sigma_{xy}/M$ at each $T$.
[From Ref. \onlinecite{Husmann:2006_a}.]
}
\efig

\subsubsection{Fe$_{1-y}$Co$_y$Si}
The silicide $\rm FeSi$ is a non-magnetic Kondo insulator.  Doping
with Co leads to a metallic state with a low density $p$ of holes.  
Over a range of Co doping ($0.05<y<0.8$)
the ground state is a helical magnetic state with a peak Curie temperature $T_C\sim$ 50 K.  The magnetization corresponds 
to 1 $\mu_B$ (Bohr magneton) per Co ion.  The tunability allows investigation of transport in a magnetic system with low
$p$.  Manyala \etal.~\cite{Manyala:2004_a} observe that the Hall 
resistivity $\rho_H$ increases to $\sim$1.5 $\mu\Omega$cm (at 5 K)
at the doping $y$= 0.1. By plotting the observed Hall 
conductivity $\sigma_{xy}$ against 
$M$, they find $\sigma_{xy} = S_HM$ with $S_H\sim$ 0.22 G$/(4\pi\Omega$cm).
In contrast, in heavy fermion systems (which include FeSi)
$\sigma_{xy}\sim M^{-3}$.

\bfig
\incl[width=9cm]{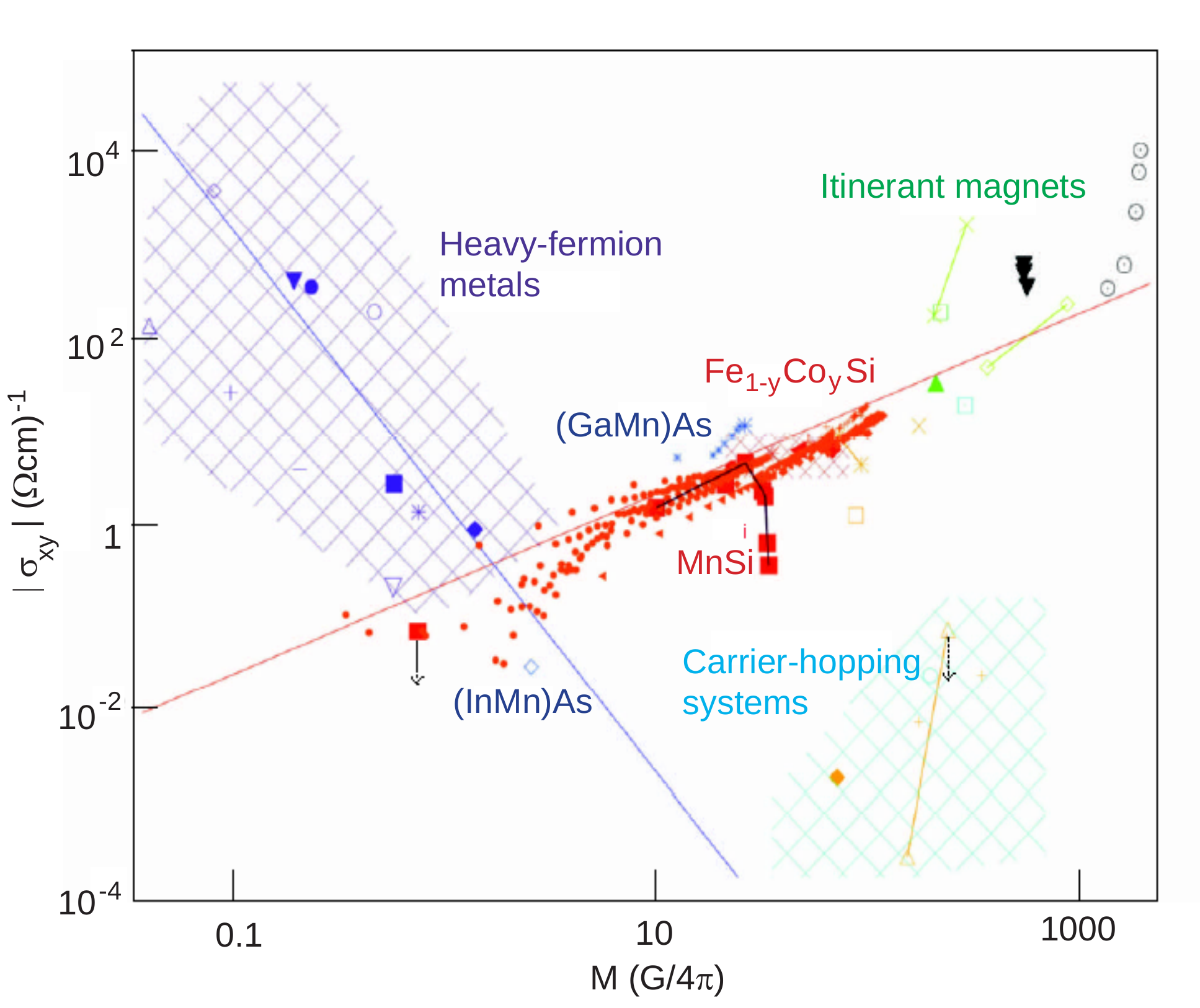}
\caption{\label{fig:Fisk}
Hall conductivity $\sigma_{xy}$ of ferromagnetic metals and heavy fermion
materials (collected at 1 kG and 5 K unless otherwise noted).  
Small solid circles represent Fe$_{1-y}$Co$_y$Si for $5<T<75$ K and
500 G$<H<$ 50 kG with $y$ = 0.1 
(filled circles), 0.15 (filled triangles),
$y$ = 0.2 (+), and 0.3 (filled diamonds).  MnSi data (5-35 K) are large, 
solid squares connected by black line.  Small asterisks are (GaMn)As data
for 5$<T<$ 120 K.  The rising line $\sigma_{xy}\sim M$
is consistent with the intrinsic AHE in itinerant ferromagnets while
the falling line $\sigma_{xy}\sim M^{-3}$ applies to heavy fermions.  
[From Ref. \onlinecite{Manyala:2004_a}.]
}
\efig

\subsubsection{MnSi}
MnSi grows in the non-centrosymmetric 
B20 lattice structure which lacks inversion symmetry.  Competition between the 
exchange energy $J$ and Dzyaloshinsky-Moriya term $D$ 
leads to a helical magnetic state with a long
pitch $\lambda$ ($\sim$180 \AA).  At ambient pressure, 
the helical state forms at the critical temperature 
$T_C$ = 30 K. Under moderate hydrostatic 
pressure $P$, $T_C$ decreases monotonically, reaching
zero at the critical pressure $P_c$ = 14 kbar.  Although
MnSi has been investigated for several decades,  interest has been revived recently by a neutron scattering experiment
which shows that, above $P_c$, MnSi displays an unusual magnetic phase in which the sharp magnetic Bragg spots at $P<P_c$ are 
replaced by a Bragg sphere~\cite{Pfleiderer:2004_a}.  
Non-Fermi liquid exponents in the
resistivity $\rho$ vs. $T$ are observed above $P_c$.

Among ferromagnets, MnSi at 4 K has a low
resistivity ($\rho\sim$ 2-5 $\mu\Omega$cm).  The unusually
long carrier mean-free-path $\ell$ implies that the ordinary
term $\sigma_{xy}^{NH}\sim\ell^2$ is greatly enhanced.
In addition, the small $\rho$  renders the total Hall voltage
difficult to resolve.  Both factors greatly complicate the 
task of separating $\sigma_{xy}^{NH}$ from
the AHE conductivity.  However, the long $\ell$ in MnSi
presents an opportunity to explore the AHE in the high-purity 
limit of ferromagnets. Using high-resolution 
measurements of the Hall 
resistivity $\rho_{yx}$ (Fig.~\ref{fig:MnSi_rxy} a),
Lee \etal.~\cite{Lee:2007_a}
recently accomplished this separation by exploiting
the large longitudinal magnetoresistance (MR) $\rho(H)$.  
From Eq.~(\ref{eq:Pugh}) and (\ref{eq:SHM}), we have $\rho_{yx}'(H) = S_H\rho(H)^2M(H)$, where
$\rho_{yx}'=\rho_{yx}^{AH}=\rho_{yx}-R_0B$.  At each $T$, the field profiles of
$\rho_{yx}'(H)$ and $M(H)$ are matched by adjusting the two 
$H$-independent parameters $S_H$ and $R_0$ (Fig.~\ref{fig:MnSi_rxy} b).
The inferred parameters $S_H$ and $R_0$ are found
to be $T$ independent below $T_C$ (Fig.~\ref{fig:MnSi_SH} a).

\bfig
\incl[width=6cm]{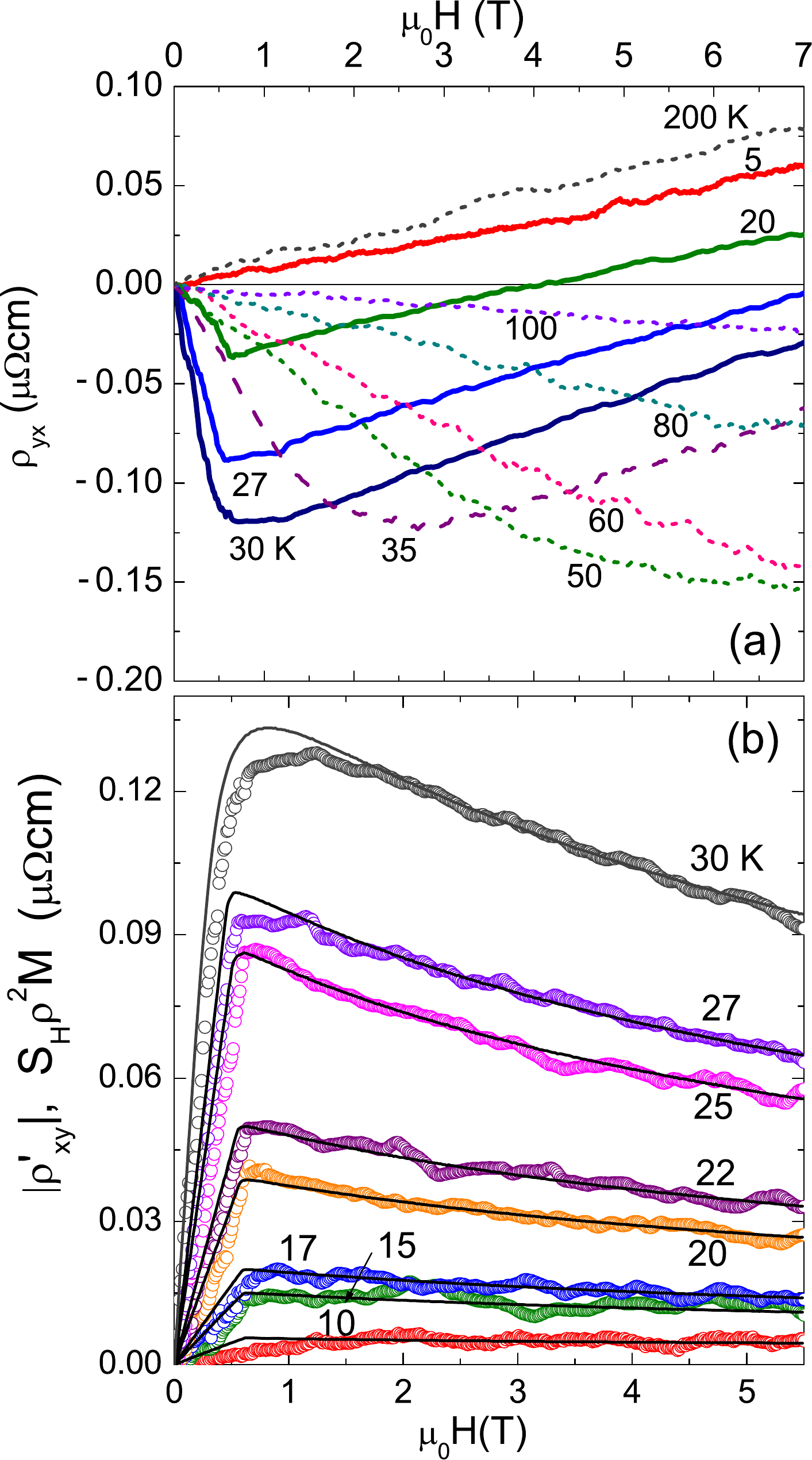}
\caption{\label{fig:MnSi_rxy}
(a) Hall resistivity $\rho_{yx}$ vs. $H$
in MnSi at selected $T$ from 5 to 200 K. At high $T$, $\rho_{yx}$
is linear in $H$, but gradually acquires an 
anomalous component
$\rho_{yx}' = \rho_{yx}- R_0B$ with a prominent ``knee''
feature below $T_C$ = 30 K.
(b) Matching of the field profiles of the anomalous Hall resistivity
$\rho_{yx}'$ to the profiles of $\rho^2M$, treating $S_H$ and $R_0$
as adjustable parameters.  Note the positive curvature of the 
high-field segments.
[From Ref. \onlinecite{Lee:2007_a}.]
}
\efig

\bfig
\incl[width=7cm]{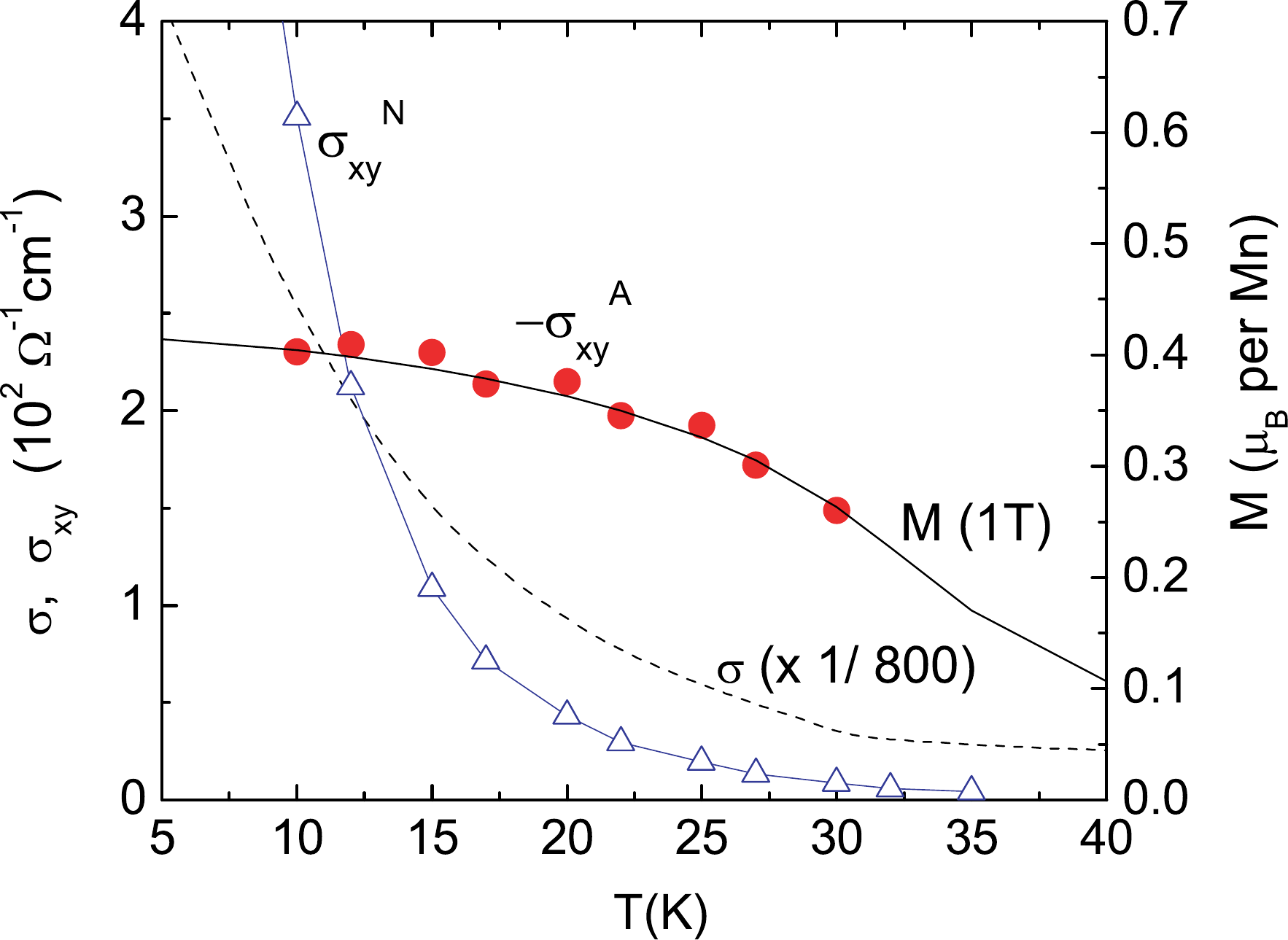}
\caption{\label{fig:MnSi_SH}
Comparison of the anomalous Hall conductivity
$\sigma_{xy}^A$ and the ordinary term $\sigma_{xy}^N$ measured
in a 1-Tesla field in MnSi. $\sigma_{xy}^A$, inferred from the measured $M$ (solid curve) and Eq. (\ref{eq:SHM}), is strictly independent
of $\ell$. Its $T$ dependence reflects that of $M(T)$. 
$\sigma_{xy}^N\sim\ell^2$ is calculated from $R_0$. The 
conductivity at zero $H$, $\sigma\sim \ell$, is shown as 
a dashed curve.
[From Ref. \onlinecite{Lee:2007_a}.]
}
\efig

The Hall effect of MnSi under 
hydrostatic pressure (5--11.4 kbar)
was measured recently~\cite{MLee:2008_a}.  
In addition to the AHE and OHE terms $\sigma^{AH}_{xy}$ 
and $\sigma^{NH}_{xy}$, Lee \etal.~observed a well-defined
Hall term $\sigma^C_{xy}$ with an unusual profile.  As shown in Fig.~\ref{fig:MnSi_chiral}
(note that in the figure $\sigma^{AH}_{xy}$ 
and $\sigma^{NH}_{xy}$ are labeled $\sigma^{A}_{xy}$ 
and $\sigma^{N}_{xy}$ respectively), the new term appears
abruptly at 0.1 T, rapidly rises to a large plateau value,
and then vanishes at 0.45 T (curves at 5, 7 and 10 K).
From the large magnitude of $\sigma^C_{xy}$, and
its restriction to the field interval in which the
cone angle is non-zero, the authors argue that it arises
from the coupling of the carrier spin to the
chiral spin textures in the helical magnetization, as
discussed in Sec.~\ref{sec:exp:oxcide}.  The authors note
that MnSi under pressure provides a very rare example in which the three Hall conductivities co-exist at the same $T$.

\bfig
\incl[width=7cm]{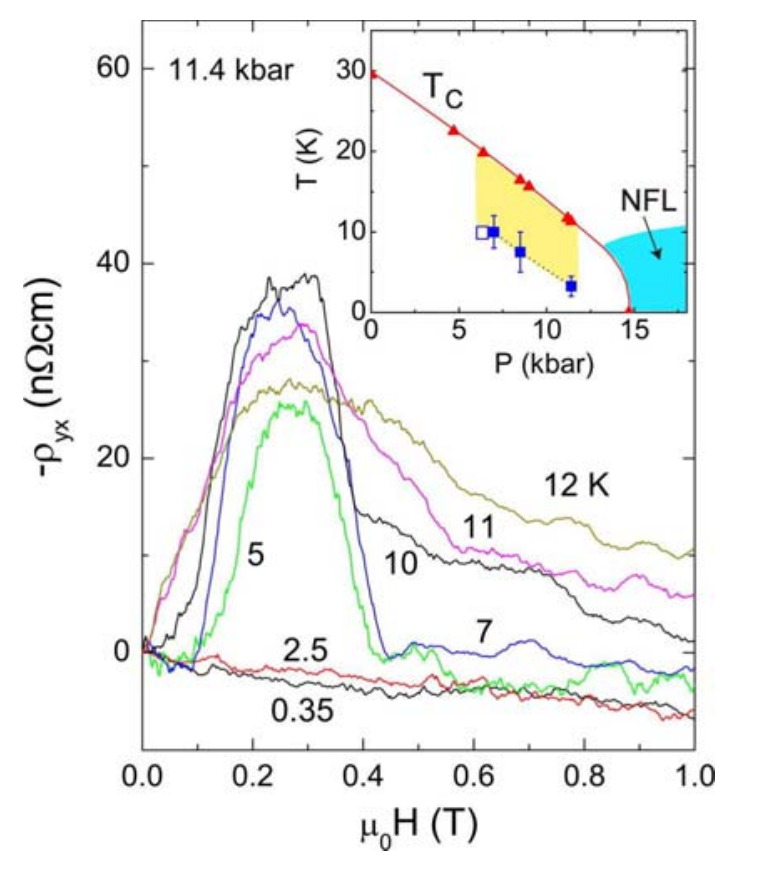}
\caption{\label{fig:MnSi_chiral}
(Main panel)  $-\rho_{yx}$ vs. $H$
in MnSi under hydrostatic pressure $P$ = 11.4 kbar
at several $T<T_C$, with $\bm{H}$ nominally along (111).  
The large Hall anomaly observed (electron-like in sign)
arises from a new chiral contribution $\sigma_{xy}^C$ to
the total Hall conductivity.  In the phase diagram (inset)
the shaded region is where $\sigma_{xy}^C$ is resolved. 
The non-Fermi liquid region is shaded blue.
[From Ref. \onlinecite{MLee:2008_a}.]
}
\efig

\subsubsection{Mn$_5$Ge$_3$}
The AHE in thin-film samples of Mn$_5$Ge$_3$ was investigated by Zeng \etal.~\cite{Zeng:2006_a}.  They express the AHE 
resistivity $\rho_{AH}$, which is strongly $T$ dependent
(Fig.~\ref{fig:Niu}a), as the sum of the skew-scattering term $a(M)\rho_{xx}$
and the intrinsic term $b(M)\rho_{xx}^2$, viz.
\be
\rho_{AH} = a(M)\rho_{xx} + b(M)\rho_{xx}^2.
\label{eq:rxyNiu}
\ee
The quantity $b(M)$ is the intrinsic
AHE conductivity $\sigma_{AH-int}$.

\bfig
\incl[width=7cm]{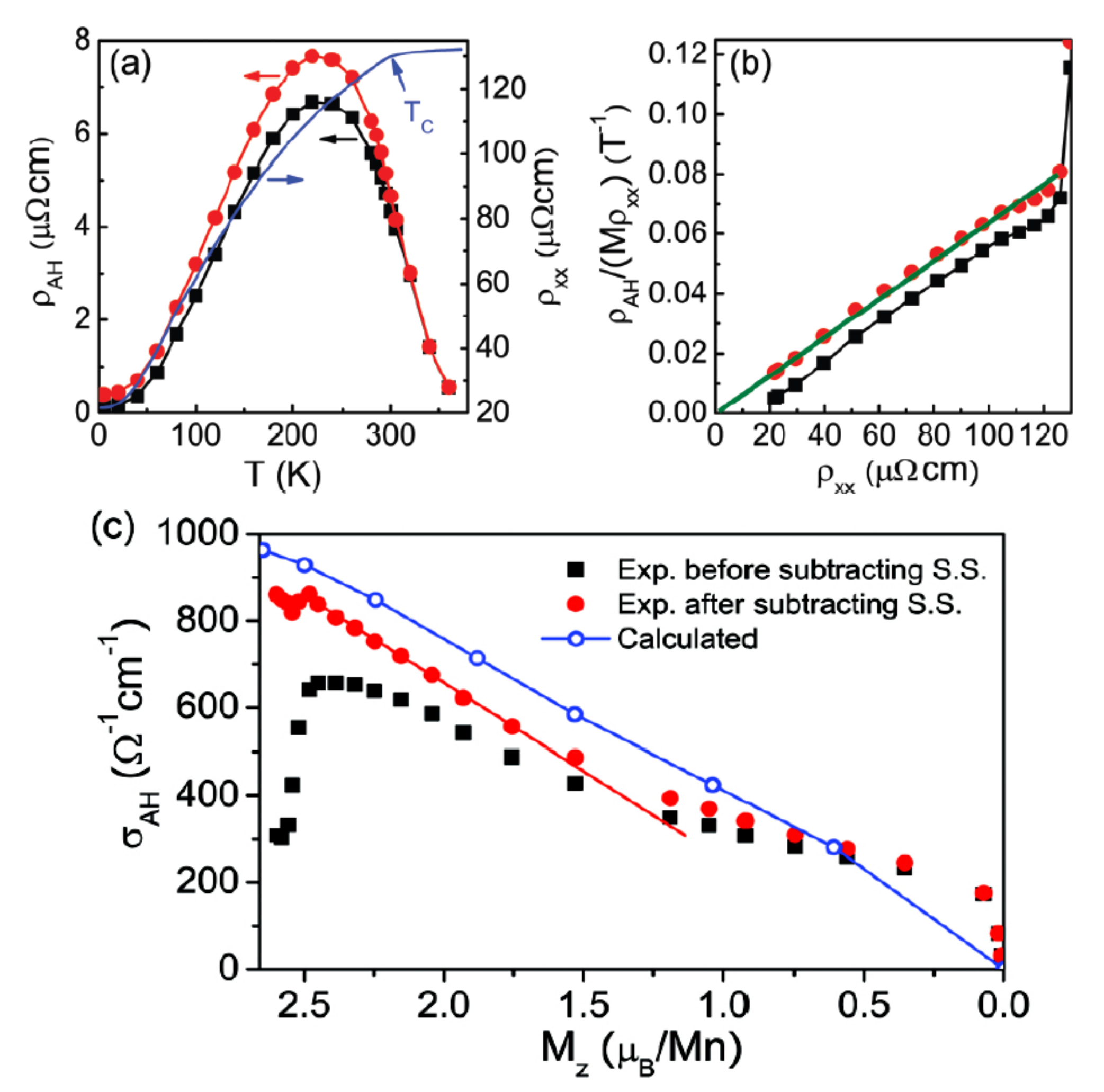}
\caption{\label{fig:Niu}
(a) The $T$ dependence of $\rho_{AH}$ in Mn$_5$ Ge$_3$ before (solid squares)
and after (solid circles) subtraction of the skew-scattering
contribution. The thin curve is the resistivity $\rho$. 
(b) Plots of $\rho_{AH}/(M\rho_{xx})$ vs. $\rho_{xx}$ before
(solid squares) and after (solid circles) subtraction of skew-scattering
term. (c) Comparison of calculated $\sigma_{AH-int}$ vs. $M_z$ (open
circles) with experimental values before (squares) and after
(solid cicles) subtraction of skew-scattering term.
[From Ref.~\onlinecite{Zeng:2006_a}.]
}
\efig

To separate the 2 terms, Zeng \etal.~plotted the quantity $\rho_{AH}/(M(T)\rho_{xx})$ 
against $\rho_{xx}$ with $T$
as a parameter.  For $T<0.8 T_C$, the plot falls on a straight line
with a small negative intercept (solid squares in Fig. \ref{fig:Niu}b).
The intercept yields the skew-scattering term $a(M)/M$
whereas the slope gives the intrinsic term $b(M)/M$.
From the constant slope, they derive their main conclusion that
$\sigma_{AH-int}$ varies linearly with $M$.  

To account for the linear-$M$ dependence, 
Zeng \etal.~\cite{Zeng:2006_a} identify the role of long-wavelength spin waves
which cause fluctuations in the local direction 
of $\bm{M}(x)$ (and hence of $\bm{\Omega}$).  They calculate the reduction in $\sigma_{AH-int}$ 
and show that it varies linearly with $M$
(Fig.~\ref{fig:Niu} c).

\subsubsection{Layered dichalcogenides}
The layered transition-metal dichalcogenides are comprised
of layers weakly bound by the van der Waals force.
Parkin and Friend~\cite{Parkin:1980_a}
have shown that a large number of interesting magnetic systems may be synthesized by intercalating 3$d$ magnetic ions between the layers.

The dichalcogenide $\rm Fe_xTaS_2$ typically displays properties
suggestive of a ferromagnetic cluster-glass state at low $T$
for a range of Fe content $x$.  However, at the composition
$x=\frac14$, the magnetic state is homogeneous.  
In single crystals of $\rm Fe_{\frac14}TaS_2$, the easy
axis of $\bm{M}$ is parallel to $\bf\hat{c}$ (normal to the 
$\rm TaS_2$ layers).  Morosan \etal.~\cite{Morosan:2007_a} observed that
the curves of $M$ vs. $H$ display very sharp switching 
at the coercive field at all $T<T_C$ (160 K).  In this system,
the large ordinary term $\sigma_{xy}^{NH}$ complicates the extraction
of the AHE term $\sigma_{xy}^{AH}$. Converting the $\rho_{yx}$-$H$ curves
to $\sigma_{xy}$-$H$ curves (Fig.~\ref{fig:FTSSxy}), Checkelsky
\etal.~\cite{Checkelsky:2008_a} infer that the jump magnitude 
$\Delta\sigma_{xy}$ equals $2\sigma_{xy}^{AH}$ by assuming that
Eq.~(\ref{eq:SHM}) is valid.  This method provides
a direct measurement of $\sigma_{xy}^{AH}$ without knowledge of $R_0$.
As shown in Fig.~\ref{fig:FTSSxy} b, both the inferred $\sigma_{xy}^{AH}$
and measured $M$ are nearly $T$-independent below 50 K, but
the former deviates sharply downwards above 50 K.  
Checkelsky \etal.~\cite{Checkelsky:2008_a} propose that the deviation represents a 
large, negative inelastic-scattering contribution $\sigma_{xy}^{in}$ 
that involves scattering from chiral textures of the spins
which increase rapidly as $T$ approaches $T_C^-$.
In support, they show that the curve of $\sigma_{xy}^{in}/M(T)$ vs. $T$
matches (within the resolution) that of $\Delta\rho(T)^2$ with $\Delta\rho(T)= \rho(T)-\rho(0)$.

\bfig
\incl[width=7cm]{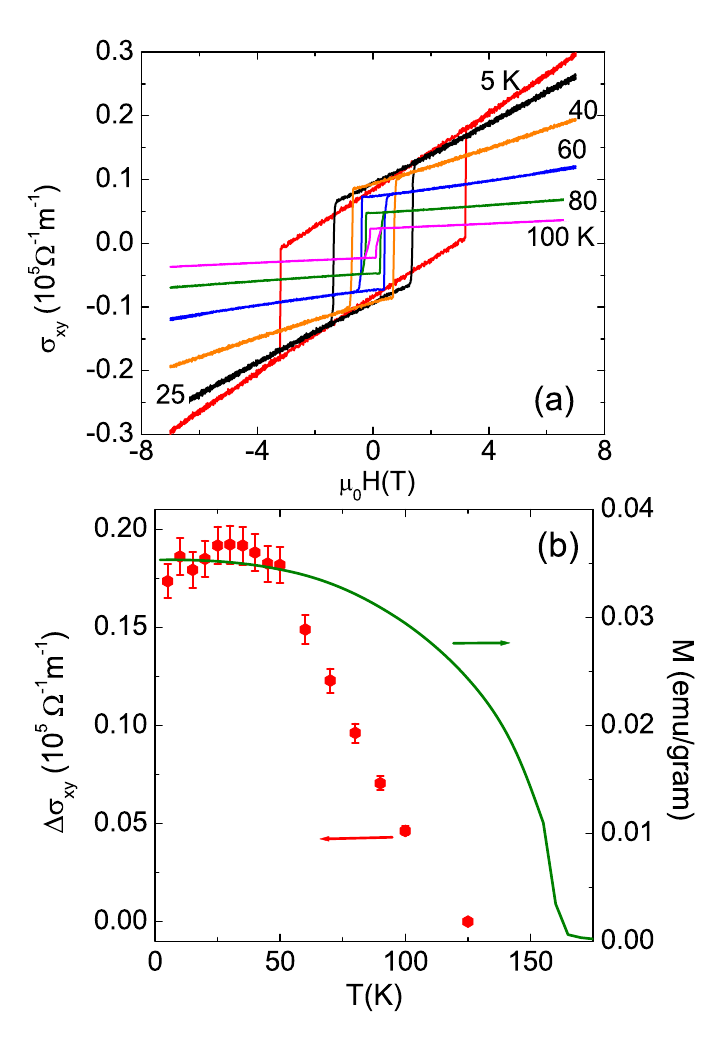}
\caption{\label{fig:FTSSxy}
(a) Hystersis loops of $\sigma_{xy}$ vs. $H$ in $\rm Fe_{\frac14}TaS_2$
calculated from the measured $\rho$ and $\rho_{yx}$ curves.
The linear portions correspond to $\sigma_{xy}^{NH}$ while the jump
magnitude $\Delta\sigma_{xy}$ equals 2$\sigma_{xy}^{AH}$.
(b) Comparison of $\Delta\sigma_{xy}$ with the magnetization
$M$ measured at 0.1 T.  Within the resolution, $\Delta\sigma_{xy}$  seems to be proportional to
$M$ below 50 K, but deviates sharply from $M$ at higher temperatures,
reflecting the growing dominance of a negative, inelastic-scattering 
term $\sigma_{xy}^{in}$.
[From Ref.~\onlinecite{Checkelsky:2008_a}.]
}
\efig

	\subsection{Localization and AHE}\label{sec:exp:loc}

\begin{figure}
\incl[width=0.7\columnwidth]{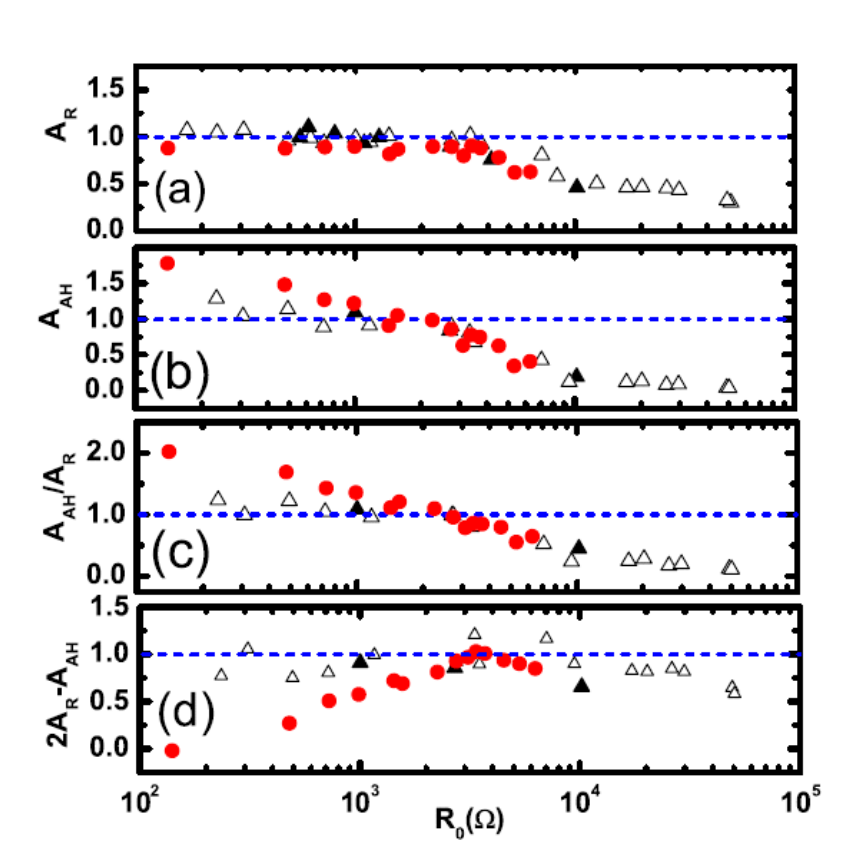}
\caption{\label{fig:WL}
The resistance ($R_0$)-temperature dependence of the coefficients $A_R$ and 
$A_{AH}$  defined in Eq. (\ref{eq:As}).
Different symbols correspond to different methods of sample preparation. [From Ref. \onlinecite{Mitra:2007_a}.]
}
\end{figure}

The role of  localization in the anomalous Hall effect is an 
important issue, and there have been several works on this subject.
( For a review of theoretical works, see Ref. \onlinecite{Woelfle:2006_a}.) 
The weak localization effect on the normal Hall effect due to the 
external magnetic field has been studied by Fukuyama \cite{Fukuyama:1980_a}, 
and the relation $\delta \sigma^{WL}_{xy}/\sigma_{xy} = 2\delta 
\sigma^{WL}_{xx}/\sigma_{xx}$ has been obtained where $\delta O^{WL}$ 
represents the correction of the physical quantity $O$ by the weak 
localization effect. This means that the Hall coefficient is not 
subject to the change due to the weak localization, {\it i.e.} $\delta \rho^{NH}_{xy}=0$. An early 
experiment by Bergmann and Ye \cite{Bergmann:1991_a} on the anomalous 
Hall effect in the ferromagnetic amorphous metals showed almost no 
temperature dependence of the anomalous Hall conductivity, 
while the diagonal conductivity shows the logarithmic temperature 
dependence. Langenfeld and Woelfle \cite{Langenfeld:1991_a,Woelfle:2006_a} studied 
theoretically the model
\begin{eqnarray}
H &=& \sum_{ \bm{k}, \sigma} \varepsilon_{\bm{k}} 
c^\dagger_{\bm{k}, \sigma} c_{\bm{k}, \sigma}
\nonumber \\
&+& \sum_i \sum_{ \bm{k},{\bf k}',\sigma}
e^{i( \bm{k}-\bm{k}') \cdot \bm{R}_i }
[ V + i( \bm{k} \times \bm{k}') \cdot \bm{J}_i ] 
c^\dagger_{ \bm{k}', \sigma'} c_{\bm{k}, \sigma}\nonumber\\
\end{eqnarray}
where $\bm{J}_i$ is proportional to the angular momentum 
of the impurity at $\bm{R}_i$, and the term containing it
describes the spin-orbit scattering. They discussed the logarithmic 
correction to the anomalous Hall conductivity in this model, and found 
that Coulomb anomaly terms vanished identically, and the weak localization 
correction was cut-off by the phase-breaking lifetime $\tau_{\phi}$ due 
to the skew scattering, explaining the experiment by Bergmann and Ye 
 \cite{Bergmann:1991_a}. 
Dugaev, Crepieux, and Bruno \cite{Dugaev:2002_a} found that for the
two-dimensional case the correction to the Hall conductivity 
was logarithmic in the ratio 
${\rm max}( \tau_{\rm tr}/\tau_{SO},\tau_{\rm tr}/\tau_{\phi} )$, with 
$\tau_{\rm tr}$ being the transport lifetime and $\tau_{SO}$ the lifetime 
due to the spin-orbit scattering. 
More recently, Mitra \etal.~\cite{Mitra:2007_a} first observed the 
logarithmic temperature dependence of the anomalous Hall conductivity 
in the ultrathin film of polycrystalline Fe of sheet resistance $R_{xx}$ 
less than $\sim 3$ k$\Omega$. 
They defined the quantities $\Delta^N(Q) = (1/ R_0 L_{00}) (\delta Q/Q)$
for the physical quantity $Q$ 
with respect to the reference temperature $T=T_0=5$ K by
$\delta Q = Q(T) - Q(T_0)$, $R_0 = R_{xx} (T_0)$, and
$L_{00} = e^2/ \pi \hbar$. 
Defining the coefficients $A_R$ and $A_H$ by 
\begin{eqnarray}
\Delta^N (\sigma_{xx}) &=& A_R \ln \biggl( {{ T_0} \over T } \biggr)
\nonumber \\
\Delta^N (\sigma_{xy}) &=& (2A_R - A_{AH})  \ln \biggl( {{ T_0} \over T } \biggr)
\label{eq:As}
\end{eqnarray}
Fig.~\ref{fig:WL} shows the $R_0$-dependence of 
the coefficients $A_R$ and $A_{AH}$ defined in the above
Eq.~(\ref{eq:As}). 

The change in the interpretation comes 
from the fact that the phase-breaking lifetime $\tau_{\phi}$ in the Fe 
film is mostly from scattering by the magnons and 
not from  skew scattering, which allows a temperature regime where 
max$(1/\tau_s, 1/\tau_{SO}, \omega_H)<< \tau_{\phi} << 1/\tau_{tr}$
( $\tau_s$: spin-flip scattering time, $\omega_H$: internal magnetic 
field in the ferromagnet ). In this region, they found that the weak 
localization effect can lead to coefficient of the logarithmic temperature 
dependence of $\sigma_{xy}$ proportional to the factor 
$\sigma_{xy}^{SSM}/( \sigma_{xy}^{SSM} +\sigma_{xy}^{SJM})$ 
where $\sigma_{xy}^{SSM} (\sigma_{xy}^{SJM})$ is the contribution 
from the skew scattering (side jump) mechanism. Assuming that that 
ratio $\sigma_{xy}^{SJM}/\sigma_{xy}^{SSM}$ decreases as the sheet 
resistance $R_0$ increases, they were able to explain the sample-dependence of the 
logarithmic correction to the anomalous Hall  conductivity~\cite{Mitra:2007_a}. 
 The absence of the logarithmic term in \cite{Bergmann:1991_a} is interpreted as  being due to
a large ratio of $\sigma_{xy}^{SJM}/\sigma_{xy}^{SSM}$ in their sample. It is interesting 
that the ratio $\sigma_{xy}^{SJM}/\sigma_{xy}^{SSM}$ can be estimated from the 
coefficient of the logarithmic term.

Up to now, we have discussed the weak localization effect. When 
the disorder strength increases, a metal-insulator transition will occur. 
For a normal metal under external magnetic field, the system belongs to the unitary class, and  in 2D all the states 
are localized for any disorder \cite{Lee:1985_a}. 
In the quantum Hall system, however, the extended states can survive at 
discrete energies at the center of the broadened density of states at 
each Laudau level. This extended state carries the quantum Hall current. 
Field theoretical formulation of this localization problem has been 
developed~\cite{Prange:1987_a}. In the presence of the 
external magnetic field, a Chern-Simons term appears in the 
non-linear sigma model whose coefficient is $\sigma_{xy}$. Therefore, 
the scaling variables are $\sigma_{xx}$ and $\sigma_{xy}$, {\it i.e.}, the 
two-parameter scaling theory should be applied instead of the single-parameter 
scaling. It has been discussed that the scaling trajectory in the 
$\sigma_{xy}-\sigma_{xx}$ plane has the fixed point at 
$(\sigma_{xy},\sigma_{xx})  = ((n + 1/2) (e^2/h), \sigma_0)$ 
where $\sigma_0$ is some finite value, and $\sigma_{xy}$ scales to the 
quantized value $n(e^2/h)$ when the initial value (given by the Boltzmann 
transport theory) lies in the range 
$(n-1/2)(e2/h) < \sigma_{xy}^{(0)} < ( n+1/2)(e^2/h)$. 
In contrast to this quantum Hall system, there is no external magnetic field 
or the Landau level formation in the anomalous Hall system, and it is not 
trivial that the same two-parameter scaling theory applies to this case.

Onoda and Nagaosa \cite{MOnoda:2003_a} studied this problem using the 
generalized Haldane model \cite{Haldane:1988_a} which shows the quantum Hall effect 
without the external magnetic field. 
They calculated the scaling function of the 
localization length in the finite-width stripe sample 
in terms of the iterative 
transfer matrix method by MacKinnon \cite{MacKinnon:1983_a}, and found that two-parameter scaling 
holds even without an external magnetic field or Landau level formation. 
For the experimental realization of this quantized anomalous Hall effect, 
the $|\sigma_{xy}^{(0)}|$ given by Boltzmann transport theory (without 
the quantum correction ) is  larger than $e^2/(2h)$, and the temperature is 
lower than $T_{Loc} \sim \varepsilon_F 
e^{ - c \sigma_{xx}^{(0)}/ \sigma_{xy}^{(0)} }$,
where $\varepsilon_F$ is the Fermi energy and $c$ is a constant of 
the order of unity. 
Therefore, the Hall angle $\sigma_{xy}^{(0)}/ \sigma_{xx}^{(0)}$ 
should not be so small, hopefully of the order of 0.1.
However, in the usual case, the Hall angle is at most 0.01, 
which makes the quantized anomalous Hall effect rather difficult to 
realize. 

\section{Theoretical aspects of the AHE and early theories}\label{sec:theory}
	In this section, we review recent theoretical developments as well as the early theoretical studies of the AHE.
	First, we give a pedagogical discussion on the difference between the normal Hall effect due to the Lorentz force and 
	the AHE (Sec.~\ref{sec:elemB}). In Sec.~\ref{sec:top}  we discuss the topological nature of the AHE.
	In this Sec.~\ref{sec:theory:early} we present a wide survey of the early theories from a modern viewpoint 
	in order to bring them into the context of the present linear transport theory formalisms now used as a framework for AHE theories.
    \subsection{Symmetry considerations and analogies between normal Hall effect and AHE}\label{sec:elemB}

Before describing these recent developments, we provide an elementary discussion that may facilitate 
understanding of the following sections.

The Hall effect is one of the fundamental transport phenomena in solid-state physics. 
Its occurrence is a direct consequence of broken time-reversal symmetry in the ferromagnetic state,  ${\cal T}$.
The charge current $\bm{J}$ is ${\cal T}$-odd,{\it i.e.}, it changes sign under 
time reversal. On the other hand, the electric field 
$\bm{E}$ is ${\cal T}$-even. Therefore, Ohm's law 
\begin{equation}
\bm{J} = \sigma \bm{E}
\label{eq:Ohm}
\end{equation}
relates two quantities with different ${\cal T}$-symmetries, which implies
that the conductivity $\sigma$ must be associated with 
diispative irreversible processes, and indeed we know that the Joule heating
$Q=\sigma \bm{E}^2/2$ always accompanies the conductivity in 
Eq. (\ref{eq:Ohm}).
This irreversibility appears only when we consider  macroscopic systems with continuous energy spectra.

Next, we consider the other aspect of the ${\cal T}$-symmetry, 
{\it i.e.}, the consequences of the ${\cal T}$-symmetry of 
the Hamiltonian which governs the microscopic dynamics of the system. 
This important issue has been formulated by Onsager, 
who showed that the response functions satisfy the following relation~\cite{Landau:1984_a} 
\begin{equation}
K_{\alpha \beta} (\omega;, r, r'; B) 
= \varepsilon_\alpha \varepsilon_\beta 
K_{\beta \alpha} (\omega;, r', r; -B),
\end{equation}
where $K_{\alpha \beta} (\omega;, r, r'; B)$
is the response of a physical quantity $\alpha$
at position $r$ to the stimulus conjugate to the quantity $\beta$
at position $r'$ with frequency $\omega$. Here
$\varepsilon_\alpha(\varepsilon_\beta) = \pm 1$ 
specifies the symmetry property of $\alpha$ ($\beta$) with respect to 
the ${\cal T}$-operation. $B$ suggests a magnetic field, but represents any time-reversal breaking field.  In the case of a ferromagnet $B$ can be associated with the magnetization $M$, the spontaneously generated time-reversal symmetry breaking field of a ferromagnet.
The conductivity tensor $\sigma_{ab}$ at a given frequency is proportional to the current-current response function.  We can therefore make the identification
$\alpha \to J_a$, $\beta \to J_b$, where $J_i$ ($i = x,y,z$) are the 
components of the current operator. Since, 
$\varepsilon_\alpha= \varepsilon_\beta=-1$,  we can conlcude that 
\begin{equation}
\sigma_{ab} (\omega;B) 
= \sigma_{ba} (\omega; -B).
\end{equation}
Hence, we conclude that $\sigma_{ab}$ is symmetric 
with respect to $a$ and $b$ in systems with  ${\cal T}$-symmetry.  
The anti-symmetric part $\sigma_{ab}(\omega) - \sigma_{ba}(\omega)$ 
is finite only if ${\cal T}$-symmetry is broken.

Now we turn back to  irreversibility for the general
form of the conductivity $\sigma_{ab}$, for which 
the dissipation is given by~\cite{Landau:1984_a} 
\begin{eqnarray}
Q &=& \sum_{ab} { 1 \over 4} ( \sigma_{ab}^* + \sigma_{ba} ) E_a E_b^*
\nonumber \\
&=&{ 1 \over 4}  \sum_{ab}[ {\rm Re}(\sigma_{ab} + \sigma_{ba}) {\rm R}e(E_a E_b^*)
\nonumber \\                    
    &+&{\rm  Im}(\sigma_{ab} - \sigma_{ba}) {\rm Im}(E_a E_b^*) ].
\end{eqnarray}
 The real part of the symmetric combination 
 $\sigma_{ab} + \sigma_{ba}$ and the imaginary part of 
 the antisymmetric combination $\sigma_{ab} - \sigma_{ba}$
 contribute to the dissipation, while the imaginary 
 part of $\sigma_{ab} + \sigma_{ba}$ and the 
 real part of $\sigma_{ab} - \sigma_{ba}$ represent the 
 dispersive (dissipationless) responses.
 Therefore, ${\rm Re}(\sigma_{ab} - \sigma_{ba})$,
 which corresponds to the Hall response, does not produce dissipation. 
 This means that the physical
 processes contributing to this quantity {\it can be} reversible, {\it i.e.}, dissipationless,
 even though they are not necessarily so. This point is directly related to the controversy between intrinsic and extrinsic mechanism for the AHE. 

The origin of the ordinary Hall effect is 
the Lorentz force due to 
the magnetic field $\bm{H}$:
\begin{equation}
\bm{F} = - { e \over c} \bm{v} \times \bm{H},
\end{equation}
which produces an acceleration of the electron
perpendicular to its velocity 
$\bm{v}$ and $\bm{H}$. For free electrons this leads to circular cyclotron motion with frequency $\omega_c = eH/(mc)$.  
The Lorentz force leads to charge 
accumulations of opposite signs on the two edges of the sample.
In the steady state, the Lorentz force is balanced by the
resultant transverse electric field, which is observed as 
the Hall voltage $V_H$.  In the standard Boltzmann-theory approach,
the equation for the electron distribution 
function $f( \bm{p}, \bm{x})$ is given by~\cite{Ziman:1967_a}: 
\begin{equation}
{ {\partial f} \over {\partial t}} + \bm{v} \cdot 
{{\partial f} \over {\partial \bm{x}} }
+ \bm{F} \cdot  {{\partial f} \over {\partial \bm{p}} }
= \left( { {\partial f} \over {\partial t}} \right)_{\rm coll.},
\end{equation}
where $\bm{p}$ and $\bm{x}$ are the momentum and real space 
coordinates, respectively. 
Putting $\bm{F} = -e ( \bm{E} + {\bm{v} \over c} \times \bm{H} )$, and  using the relaxatin time approximation for
 the collision term, $- { 1 \over \tau} ( f - f_{\rm eq})$,
with $f_{\rm eq}$ being the distribution function in thermal equilibrium, 
one can obtain the steady state solution to this Boltzmann equation as
\begin{equation}
f(\bm{p})= f_{\rm eq}(\bm{p}) + g( \bm{p}),
\end{equation}
with 
\begin{equation}
g( \bm{p}) = - \tau e { {\partial f_{\rm eq}} \over {\partial \varepsilon}}
\bm{v} \cdot 
\left( \bm{E} - { { e \tau } \over { mc} } \bm{H} \times \bm{E} \right) ,
  \label{eq:g}
\end{equation}
where order $H^2$ terms have been neglected and 
the magnetic field $H_z$ is assumed to be small, {\it i.e.}
$\omega_c  \tau \ll 1$.
The current density $\bm{J} = - e\int { {d^3 \bm{p}} \over {(2 \pi)^3} }
 { {\bm{p} } \over m} f(\bm{p})$ is obtained from Eq. (\ref{eq:g})
to order $O(E)$ and $O(EH)$ as
\begin{equation}
\bm{J} = \sigma \bm{E} + \sigma_H \bm{H} \times \bm{E},
\end{equation}
where 
\begin{equation}
\sigma= - { 1 \over 3} \int {{d^3 \bm{p}} \over {(2 \pi)^3} }
v^2 e^2 \tau { {\partial f_{\rm eq} } \over {\partial \varepsilon}}
\label{eq:sigma}
\end{equation}
is the conductivity, while
\begin{equation}
\sigma_H =  { 1 \over 3} \int {{d^3 \bm{p}} \over {(2 \pi)^3} }
v^2 e^2 \tau { {\partial f_{\rm eq} } \over {\partial \varepsilon}}
{{ e \tau} \over { mc}}
\label{eq:sigmah}
\end{equation}
is the ordinary Hall conductivity.

Now let us fix the direction of the electric and magnetic fields as 
$\bm{E} = E_y \bm{e}_y$ and $\bm{H} = H_z \bm{e}_z$. 
Then the Hall current is along the $x$-direction and we write it as
\begin{equation}
J_x = \sigma_{H} E_y,
\end{equation}
with the Hall conductivity $\sigma_{H}= n e^3 \tau^2H_z/m^2 c$
with $n$ being the electron density.
The Hall coefficient $R_H$ is defined as the ratio $V_H/J_xH_z$, and from 
Eqs.~(\ref{eq:sigma}) and (\ref{eq:sigmah}), one obtains
\begin{equation}
R_H = - { 1 \over { n e c} }.
\label{eq:Hall}
\end{equation}
This result is useful to determine the electron density $n$ experimentally, 
since it does not contain the relaxation time $\tau$. 
Note here that for fixed electron density $n$, Eq.~(\ref{eq:Hall}) 
means $\sigma_H \propto H \sigma_{xx}^2$ and that 
$\rho_H \propto H$, independent of the relaxation time $\tau$ (or equivalently $\sigma_{xx}$).  
In other words, the ratio $|\sigma_{H}|/\sigma_{xx} =  \omega_c \tau \ll 1$. This relation will  
be compared with the AHE case below.

What happens in the opposite limit
$\omega_c \tau \gg 1$?  In this limit, the cyclotron motion is completed many times 
within the lifetime $\tau$. This implies 
the closed cyclotron motion is repeated many times before being disrupted by scattering. 
When treated quantum mechanically the periodic classical motion leads to kinetic energy 
quantization and Landau level formation, which leads in
turn to the celebrated quantum Hall effect. 
As is well known, in the 2D electron gas realized in
semiconductor heterostructures, Landau-level quantization 
leads to the celebrated quantum Hall effect~\cite{Prange:1987_a}.
If we consider the free electrons,
{\it i.e.} completely neglecting the potential 
(both periodic and random) and the interaction among electrons, 
one can  show that $\sigma_{H}$ is given by 
${ {e^2} \over h } \nu$ where $\nu$ is the filling factor of the Landau levels.  In combination with electron 
localization, gaps at integer filling factors lead to quantization of $\sigma_H$.
 For electrons in a two-dimensional cyrstal, the Hall quantization is still quantized, 
 a property which can be traced to the topological properties of Bloch state wavefunctions discussed below.~\cite{Prange:1987_a}.

\begin{figure}
\incl[width=0.7\columnwidth]{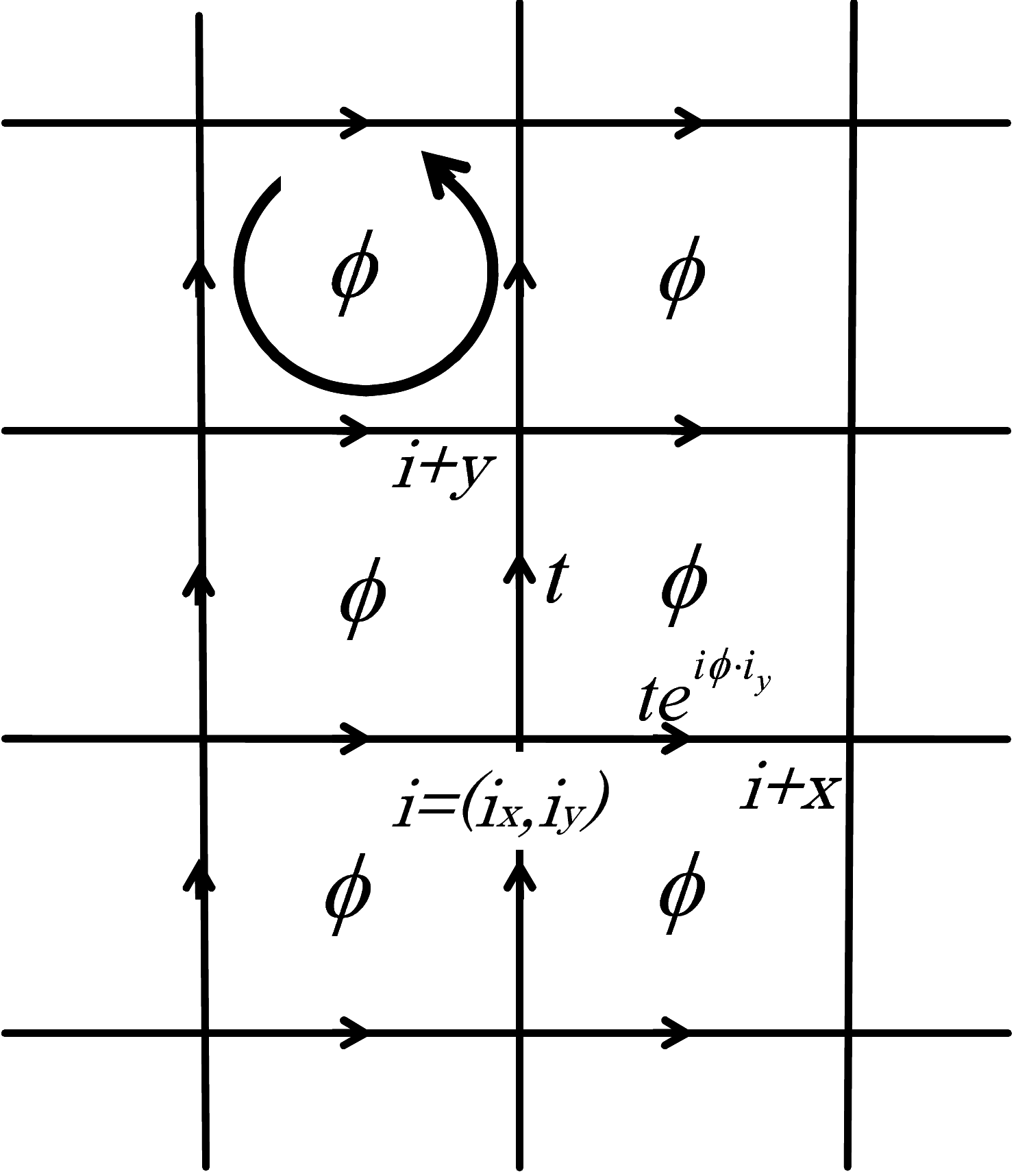}
\caption{\label{fig:Lattice}
Tight-binding model on a square lattice under the magnetic flux $\phi$ for 
each plaquette. We choose the gauge where the phase factor of the transfer 
integral along $x$-direction 
is given by $e^{-i \phi i_y}$ with $i_y$ being the $y$-component of 
the lattice point position $i=(i_x,i_y)$.
}
\end{figure}

Now let us consider the magnetic field effect for electrons under  the influence of
the periodic potential.  
For simplicity we study the tight-binding model 
 \begin{equation}
 H = \sum_{ij}t_{ij}c^\dagger_i c_j.
\end{equation}
The magnetic field adds the Peierls phase factor to the transfer integral
$t_{ij}$ between sites $i$ and $j$, {\it viz.} 
 \begin{equation}
 t_{ij} \to t_{ij} \exp [ia_{ij}],
\end{equation}
 with
\begin{equation}
a_{ij} =  {{ie} \over {\hbar c}} 
 \int_i^j d \bm{r} \cdot \bm{A}(\bm{r}).
 \end{equation}
 Note that the phase factor is periodic with the period $2 \pi$ 
with respect to its exponent.
 Since the gauge transformation
 \begin{equation}
 c_i \to c_i \exp[ i \theta_i]
 \end{equation}
 together with the redefinition 
 \begin{equation}
 a_{ij} \to a_{ij} + \theta_i - \theta_j
 \end{equation}
 keeps the Hamiltonian invariant, 
 $a_{ij}$ itself is not a physical quantity.
 Instead, the flux $\phi = a_{ij} + a_{jk} + a_{kl} + a_{li}$
 per square plaquette is the key quantity in the 
problem (see Fig.\ref{fig:Lattice} )

 For example, one may choose a gauge in which $t_{ij}$ along the
 directions $\pm\bf\hat{y}$ is a real number $t$, while
 $t_{i i+x} = t \exp[- i \phi i_y ]$ 
 to produce a 
uniform flux distribution $\phi$ in each square plaquette. Here $i_y$ is the $y$-component of the lattice point 
position $i=(i_x,i_y)$).
The problem is that the $i_y$-dependence breaks the 
periodicity of the Hamiltonian along the $y$-axis, 
which invalidates the Bloch theorem.
This corresponds to the fact that the vector potential 
$\bm{A}(\bm{r}) = (-H y,0)$ for the uniform magnetic field
$H$ is $y$-dependent, so that momentum component $p_y$
is no longer conserved.
However, when $\phi/(2\pi)$ is a rational number 
$ n/m $, where $n$ and $m$ are integers that do not share a
common factor, one can enlarge the unit cell by 
$m$ times along the $y$-direction to recover the 
periodicity, because 
$\exp[- i \phi (i_y+m) ]= \exp[- i \phi i_y ]$.
Therefore, there appear $m$ sub-bands in the 1st-Brillouin zone,
each of which is characterized by a Chern number 
related to the Hall response as will be described in 
section \ref{sec:top}~\cite{Thouless:1982_a}.
The message here is that an external commensurate magnetic field
leads to a multiband structure with an enlarged unit cell, leading to 
the quantized Hall response when the chemical potential is within the
gap between the sub-bands.   
When $\phi/(2 \pi)$ is an irrational number, on the other hand,
one cannot define the Bloch wavefunction and the electronic 
structure is described by the Hofstadter butterfly~\cite{Hofstadter:1976_a} .

Since flux $2 \pi$ is equivalent to zero flux, 
a commensurate magnetic field can be
equivalent to a magnetic field distribution whose average flux is zero. 
Namely, the total flux penetrating the $m$-plaquettes is
$ 2 \pi n$, which is equivalent to $zero$. 
Interestingly, spatially non-uniform flux distributions can also lead to quantum Hall effects.  This possibility was first considered
by Haldane~\cite{Haldane:1988_a},
who studied a tight-binding model on the honeycomb
lattice. He introduced complex transfer integrals
between the next-nearest neighbor sites in addition to 
the real one between the nearest-neighbor sites.
The resultant Hamiltonian has  translational 
symmetry with respect to the lattice vectors, 
and the Bloch wavefunction can be defined as
the function of the crystal momentum $\bm{k}$ in the 
first Brillouin zone.   
The honeycomb lattice has two sites in the unit cell,
and the tight-binding model produces two bands, separated
by the band gap. The wavefunction of each band is characterized
by the Berry phase curvature, which acts like
a ``magnetic field'' in $\bm{k}$-space, as will be discussed in 
sections \ref{sec:top} and \ref{sec:theory:Boltzmann}. 

The quantum Hall effect results when
the Fermi energy is within this band gap.
Intuitively, this can be interpreted as follows.
Even though the total flux penetrating the 
unit cell is zero, there are loops in the 
unit cells enclosing a nonzero flux.
Each band picks up the flux distribution 
along the loops with different weight and
contributes to the Hall response. 
The sum of the Hall responses from all 
the bands, however, is zero as expected.
Therefore, the ``polarization'' of 
Hall responses between bands in  
a multiband system is a
general and fundamental mechanism of
the Hall response, which is distinct
from the classical picture of the Lorentz force.     

There are several ways to realize a flux distribution within a unit cell in momentum space.
One is the relativistic spin-orbit interaction given by
\begin{equation}
H_{SOI} = { {\hbar e} \over { 2 m^2 c^2}} (\bm{s} \times \nabla V )
\cdot \bm{p},
\label{eq:SOI}
\end{equation}
where $V$ is the potential, $\bm{s}$ the spin, and
$\bm{p}$ the momentum of the electron. This Hamiltonian can
be written as 
\begin{equation}
H_{SOI} = \bm{A}_{SOI} \cdot \bm{p},
\label{eq:SOA}
\end{equation}
with $\bm{A}_{SOI} = { \frac{\hbar e}{ 2 m^2 c^2}} (\bm{s} \times \nabla V )$
acting as the effective vector potential.
Therefore, in the magnetically ordered state, {\it i.e.}, when
$\bm{s}$ is ordered and can be regarded as a $c$-number,
$\bm{A}_{SOI}$ plays a role similar to the 
vector potential of an external magnetic field.
Note that the Bloch theorem is valid even in the presence
of the spin-orbit interaction, since it preserves the 
translational symmetry of the lattice.
However, the unit cell may contain more than two atoms 
and each atom may have multiple orbitals. Hence, the 
situation is very similar to that described above for
the commensurate magnetic field or the Haldane model.

\begin{figure}
\incl[width=6.5cm]{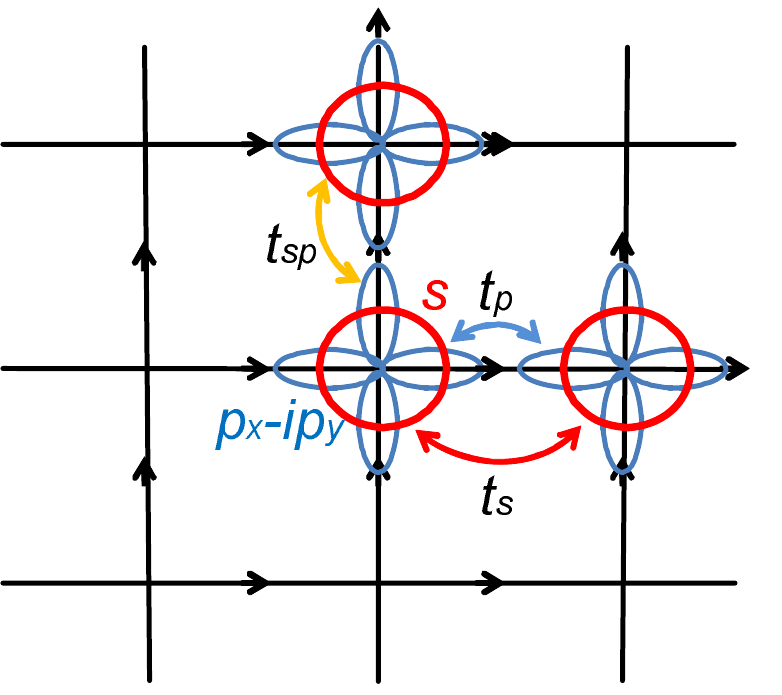}
\caption{\label{fig:sp}
Tight-binding model on a square lattice for a 3 band model made from 
$s$ and $p_x - ip_y$ orbitals with polarized spins. 
The transfer integrals between $s$ and $p_x- i p_y$ orbitals become
complex along the $y$-direction in a way that is equivalent to effective magnetic flux
}
\end{figure}

The following simple model is instructive.  Let us consider 
the tight-binding model on a square lattice given by 
\begin{eqnarray}
H &=& 
- \sum_{i, \sigma, a=x,y}t_s  s^\dagger_{i, \sigma} s_{i + a, \sigma} + h.c.
\nonumber \\
&+& \sum_{i, \sigma, a=x,y}t_p  
p^\dagger_{i,a, \sigma} s_{i + a, a,\sigma} + h.c.
\nonumber \\
&+& \sum_{i, \sigma, a=x,y}t_{sp}  
s^\dagger_{i \sigma} p_{i + a, a, \sigma} + h.c.
\nonumber \\
&+& \lambda \sum_{i,\sigma} \sigma 
( p^\dagger_{i,x, \sigma} -i \sigma p^\dagger_{i,y, \sigma})
 ( p_{i,x, \sigma} + i \sigma p_{i,y, \sigma}).\nonumber\\
\label{eq:spmodel}
\end{eqnarray}
On each site, we put the 3 orbitals $s$, $p_x$ and $p_y$,
associated with the corresponding creation and annihilation 
operators. The first three terms represent the transfer of 
electrons between the neighboring sites as shown by Fig.~\ref{fig:sp}.
The signs in front of $t$'s are determined by the relative sign
of the two orbitals connected by the transfer integrals, and all
$t$'s are assumed to be positive.   
The last term is a simplified SOI 
in which the $z$-component of the spin moment is coupled to that
of the orbital moment.
This 6-band model can be reduced to a 2-band model in the 
ferromagnetic state when only the $\sigma=+1$ component is
retained. By the spin-orbit interaction, 
the $p$-orbitals are split into 
\begin{equation}
| p \pm\rangle =  { 1 \over {\sqrt{2}} } ( |p_x\rangle \pm i |p_y\rangle )
\end{equation}
at each site, with the energy separation 
$2 \lambda$. Therefore, when only the lower
energy state $|p-\rangle$ and the s-orbital $|s\rangle$
are considered, the tight-binding Hamiltonian becomes
$
H = \sum_{\bm{k}} \psi^\dagger(\bm{k}) h(\bm{k}) \psi(\bm{k})
$ 
with $\psi(\bm{k})=[ s_{\bm{k}, \sigma=1}, p_{\bm{k},-,\sigma=1} ]^T$
and 
\begin{equation}
h(\bm{k})
=\left[ \begin{array}{cc}
\varepsilon_s - 2t_s(\cos k_x + \cos k_y ) & \sqrt{2}t_{sp} ( i \sin k_x + \sin k_y )\\
\sqrt{2}t_{sp} (- i \sin k_x + \sin k_y )& \varepsilon_p +t_p(\cos k_x + \cos k_y )
\end{array}\right].\label{eq:Hamiltonian}
\end{equation}
Note  that the complex orbital $|p-\rangle$ is responsible for  the complex 
off-diagonal matrix elements of $h(\bm{k})$.
This produces the Hall response as one can easily see from the 
formula Eq.~(\ref{eq:TKNN}) and Eq.~(\ref{eq:b})
in section \ref{mechanisms-intro}. When the Fermi energy is within the band gap,
the Chern number for each band is $\pm 1$
when $\varepsilon_s - 4 t_s < \varepsilon_p + 2t_p $, and they are zero 
otherwise. This can be understood by considering the effective Hamiltonian
near $\bm{k} = \bm{0}$, which is given by
\begin{equation}
h(\bm{k}) = {\bar \varepsilon} + m \sigma_z + \sqrt{2} t_{sp}(k_y \sigma_x -k_x \sigma_y ),
\label{eq:cont}
\end{equation}
with 
${\bar \varepsilon}= [(\varepsilon_s - 4 t_s)+(\varepsilon_p + 2t_p)]/2$
and 
$m = (\varepsilon_s - 4 t_s)-(\varepsilon_p + 2t_p)$,
which is essentially the Dirac Hamiltonian in (2+1)D.
One can calculate the Berry curvature $\bm{b}_\pm(\bm{k})= \mp b_z(\bm{k}_\perp) \bm{e}_z$
defined in Eq.~(\ref{eq:b}) for the upper(+) and lower ($-$) bands, respectively as
\begin{equation}
b_z(\bm{k}_\perp) = { m \over { 2 [\bm{k}_\perp^2 + m^2]^{3/2}} }.
\label{eq:mono}
\end{equation}
The integral over the two dimensional wavevector 
$\bm{k}_\perp = (k_x,k_y)$ leads to 
Hall conductance $\sigma_H = \frac12 {\rm sgn}(m) e^2/h$
 in this continuum model 
when only the lower band is fully occupied~\cite{Jackiw:1984_a}.
The distribution Eq.~(\ref{eq:mono}) indicates that the
Berry curvature is enhanced when the gap $m$ is small
near this avoided band crossing.  
Note that the continuum model Eq.~(\ref{eq:cont}) 
cannot describe the value of the Hall conductance in the 
original tight-binding model since the information 
is not retained over all the first Brillouin zone.
Actually, it has been proved that the Hall conductance is 
an integer multiple of $e^2/h$ due to the single-valudeness of the 
Bloch wavefunction within the first Brillouin zone. 
However, the {\it change} of the Hall conductance
between positive and negative $m$ can be described 
correctly.

Onoda and Nagaosa~\cite{MOnoda:2002_a} demonstrated that a similar scenario emerges 
also for a 6-band tight binding model of $t_{2g}$-orbitals 
with SOI. They showed that
the Chern number of each band can be nonzero 
leading to a nonzero Hall response arising
from the spontaneous magnetization.
This suggests that the anomalous Hall effect 
can be of topological origin, a 
reinterpretation of the intrinsic 
contribution found by Karplus-Luttinger~\cite{Karplus:1954_a}.
Another mechanism which can produce a flux is 
a non-coplanar spin structure with an
 associated spin chirality as has been
discussed in section \ref{sec:exp:oxcide}.

The considerations explained here have totally neglected the 
finite lifetime $\tau$ due to the impurity 
and/or phonon scattering, a generalization of the parameter 
$\omega_c \tau$ which appeared in the case of the 
external magnetic field. A full understanding of disorder effects usually requires the full power of the detailed 
microscopic theories reviewed in Sec.~\ref{sec:linear_transport}.
The ideas explained in this section though
are sufficient to understand why the intrinsic Berry phase contribution to the Hall effect can be so important.  The SOI
produces an effective
vector potential and a ``magnetic flux distribution'' 
within the unit cell. The strength of 
this flux density is usually much larger than 
the typical magnetic field strength available 
in the laboratory. In particular, the energy scale
``$\hbar \omega_c$'' related to this 
flux is that of the spin-orbit interaction for 
Eq.~(\ref{eq:SOA}), and hence of the order of 30 meV in 
3d transition metal elements. This corresponds to a magnetic
field of the order of 300 T. Therefore, one can expect that the 
intrinsic AHE is easier to observe compared with the quantum Hall effect.

	\subsection{Topological interpretation of the intrinsic mechanism: relation between Fermi sea and Fermi surface properties}\label{sec:top}

The discovery and subsequent investigation of the quantum Hall effect 
led to numerous important conceptual advances~\cite{Prange:1987_a}.  In particular,
the utility of topological considerations in understanding electronic transport in solids was first appreciated. 
 In this section, we provide an introduction to
the application of topological notions such as the Berry phase 
and the Chern number to the AHE problem, and explain the long-unsuspected 
connections between these considerations and KL theory.  A more elementary treatment 
is given in Ref.~\onlinecite{Ong:2005_a}.

The topological expression given in Eq.~(\ref{eq:TKNN}) was
first obtained by Thouless, Kohmoto, Nightingale and Nijs (TKNN)~\cite{Thouless:1982_a} for Bloch electrons in a 2D insulating crystal lattice immersed in a strong magnetic field $H$. These authors showed that 
each band is characterized by a topological integer 
called the Chern number 
\begin{equation}
  C_n\equiv -\int\frac{dk_x dk_y}{(2\pi)^2}b_n^z(\bm{k}).
  \label{eq:Chern}
\end{equation}
According to Eq. (\ref{eq:TKNN}), the Chern number $C_n$ 
gives the quantized Hall conductance of an 
ideal 2D insulator, {\it viz.} $\sigma_{xy}=e^2C_n/h$. 
As mentioned in prior sections, these topological arguments were subsequently 
applied to semiclassical transport theory~\cite{Sundaram:1999_a} and the AHE problem in itinerant ferromagnets~\cite{MOnoda:2002_a,Jungwirth:2002_a}, 
and the equivalence to 
the KL expression~\cite{Karplus:1954_a} for $\sigma_H$ was established.
(Note however, that in multiband systems, induced interband coherence among states with the same crystal momentum is a possibility.)

Although the intrinsic Berry-phase effect involves the entire  Fermi sea, as is clear from 
Eq.~(\ref{eq:TKNN}), it is usually believed that transport properties of Fermi liquids at low temperatures relative
to the Fermi energy should be dependent only on  Fermi-sruface properties. This is a simple consequence of the observation that
at these temperatures only states near the Fermi surface can be excited to produce non-equillibirium transport.

The apparent contradiction between conventional Fermi liquid theory ideas and the Berry phase theory of the AHE  was resolved  
by Haldane~\cite{Haldane:2004_a}, 
showed that the Berry phase 
contribution to the AHE could be viewed
in an alternate way.
Because of its topological nature, the intrinsic AHE, which is most naturally expressed as an integral over
 the occupied Fermi sea, can be rewritten as an integral over the Fermi surface.
Let us start with the topological properties of 
the Berry phase. 
The Berry-phase curvature $\bm{b}_{n\bm{k}}$ is gauge-invariant and divergence-free 
except at quantized monopole and anti-monopole sources with the quantum $\pm2\pi$, {\it i.e.},
\begin{equation}
\bm{\nabla}_k\cdot\bm{b}_{n\bm{k}}=\sum_iq_{ni}\delta^3(\bm{k}-\bm{k}_{ni}), \ \ \ q_{ni}=\pm2\pi.
\label{eq:monopole}
\end{equation}
These monopoles and anti-monopoles appear at isolated $\bm{k}$ points in three dimensions. This is  
because in complex Hermitian eigenvalue problems,  accidental degeneracy can occur by 
tuning the three parameters of $\bm{k}$.

To gain further insight on the topological nature of 3D systems, it is useful to rewrite Eq.~(\ref{eq:TKNN}) as
\begin{equation}
\sigma_{ij}=\epsilon_{ij\ell}\frac{e^2}{h}\frac{K^\ell}{2\pi},
\label{eq:TKNN:3D}
\end{equation}
where
\begin{eqnarray}
\bm{K}&=&\sum_n\bm{K}_n,
\label{eq:TKNN:K}\\
\bm{K}_n&=&-\frac{1}{2\pi}\int_{\mathrm{F.B.Z.}}\hspace*{-18pt}d^3\bm{k}\,
f(\varepsilon_{n\bm{k}})\ \bm{b}_{n\bm{k}}.
\label{eq:TKNN:K_n}
\end{eqnarray}
We note that if the band dispersion $\varepsilon_{n\bm{k}}$ does not cross $\varepsilon_F$, $K_n^\ell$ is quantized 
in integer multiples ($C_{na}$) of a primitive 
reciprocal lattice vector $\bm{G}_a$ at $T$ = 0.  We have
\begin{equation}
\bm{K}_n = \sum_a C_{na}\bm{G}_a,
\end{equation}
where the index $a$ runs over the three independent primitive reciprocal lattice vectors~\cite{Kohmoto:1985_a}.

We next discuss the non-quantized part of the anomalous Hall conductivity. In a real material with multiple Fermi surface (FS) sheets indexed by $\alpha$, we can describe each sheet by 
$\bm{k}_F^{(\alpha)}(\bm{s})$, where $\bm{s}=(s^1,s^2)$ is a parameterization of the surface. It is convenient to redefine the Berry-phase connection and curvature as
\begin{eqnarray}
\tilde{a}^i(\bm{s}) &=& \bm{a}(\bm{k}(\bm{s}))\cdot\partial_{s^i}\bm{k}(\bm{s}),
\label{eq:Berry:ta}\\
\tilde{b}(\bm{s}) &=& \epsilon_{ij}\partial_{s^i}\tilde{a}^j(\bm{s}),
\label{eq:Berry:tb}
\end{eqnarray}
with $\epsilon_{ij}$ the rank-2 antisymmetric tensor.

By integrating Eq.~(\ref{eq:TKNN:K_n}) by parts, eliminating the 
integration over the Brillouin zone boundary for each band, and dropping the band 
indices, we may write (at $T=0$) the vector $\bm{K}$ in Eq.~(\ref{eq:TKNN:K}) as
\begin{eqnarray}
\bm{K}&=&\sum_a C_a\bm{G}_a+\sum_{\alpha}\bm{K}_\alpha,
\label{eq:TKNN:K-K_alpha}\\
\bm{K}_{\alpha}&=&\frac{1}{2\pi}\int_{S_\alpha}\hspace*{-7pt}ds^1\wedge ds^2\ 
\tilde{b}(\bm{s})\bm{k}_F^{(\alpha)}(\bm{s})
\nonumber\\
&&+\frac{1}{4\pi}\sum_i\bm{G}_{\alpha i}\int_{\partial S_\alpha^i}\hspace*{-10pt}ds
\cdot\tilde{a}(\bm{s}).
\label{eq:K_alpha}
\end{eqnarray}
Here, $C_a$ is the sum of $C_{na}$ over the fully occupied bands, $\alpha$ 
labels sheets of the Fermi surface $S_\alpha$, and $\delta S_\alpha^i$ is 
the intersection of the Fermi surface $S_\alpha$ with the Brillouin zone 
boundary $i$ where $\bm{k}_F^{(\alpha)}(\bm{s})$ jumps by $\bm{G}_{\alpha i}$.
Note that the quantity
\begin{equation}
\frac{1}{2\pi}\int_{S_\alpha}\hspace*{-7pt}ds^1\wedge ds^2\ \tilde{b}(\bm{s})
\end{equation}
gives an integer Chern number. Hence gauge invariance requires that 
\begin{equation}
\sum_\alpha\frac{1}{2\pi}\int_{S_\alpha}\hspace*{-7pt}ds^1\wedge ds^2\ \tilde{b}(\bm{s})=0.
\end{equation}
The second term in Eq.~(\ref{eq:K_alpha}) guarantees that $\bm{K}_\alpha$ is unchanged 
by any continuous deformation of the Brillouin zone into another 
primitive reciprocal cell.

	\subsection{Early theoretical studies of the AHE}\label{sec:theory:early}

\subsubsection{Karplus-Luttinger theory and the intrinsic mechanism}
The pioneering work of
Karplus and Luttinger (KL) \cite{Karplus:1954_a} was
the first theory of the AHE fully based on Bloch states
$\psi_{n\bm{k}}$.
As a matter of course, their calculations uncovered the important role played by the mere existence of bands and the associated overlap of Bloch states.
The KL theory neglects all lattice disorder, so that the Hall effect it predicts is based on an
intrinsic mechanism. 
 
In a ferromagnet, the orbital motion of the itinerant electrons couples to spin ordering via the SOI.
Hence all theories of the AHE invoke the SOI term, 
which, as we have mentioned, is described by the Hamiltonian given by Eq. \ref{eq:SOI}, 
where 
$V(\bm{r})$ is the lattice potential. 
The SOI term preserves lattice translation symmetry and
we can therefore define spinors which 
satisfy Bloch's theorem.  When SO interactions are included the Bloch Hamiltonian acts in a direct product of 
orbital and spin-space.
The total Hamiltonian $H_T$ can be separated into contributions as follows:
\begin{equation}
H_T = H_0 + H_{SOI} + H_E
\end{equation}
where $H_0$ is the Hamiltonian in the absence of SOI
and $H_E$ the perturbation due to the applied electric field $\bm{E}$.
In simple models the consequences of 
magnetic order can be represented by replacing the spin operator in Eq.(3.19)  by
the ordered magnetic moment $\bm{M}_s$ as
$\bm{s} \to { {\hbar} \over 2} { {\bm{M}_s } \over {M_0} }$,
where $M_0$ is the magnitude of the saturated moment.
The matrix element of $H_E = - e E_b x_b$ can be written as
\begin{equation}
\langle n \bm{k} | H_E | n' \bm{k}' \rangle
= -e E_b 
\left( i \delta_{n, n'} { {\partial} \over {\partial k_b} }
\delta_{\bm{k}, \bm{k}'}
+ i\delta_{\bm{k}, \bm{k}'}J_b^{n n'}(\bm{k}) \right),
\label{eq:xmat}
\end{equation}
where the ``overlap'' integral
\begin{equation}
J_b^{n n'}(\bm{k})= \int_{\Omega} d \bm{r}\; u^*_{n \bm{k}} (\bm{r})
{ {\partial} \over {\partial k_b} } u_{n' \bm{k}} (\bm{r})
\label{eq:Jb}
\end{equation}
are regular functions of $\bm{k}$.  In hindsight,
$J_b^{nn'}(\bm{k})$ may be recognized as the Berry-phase connection $ \bm{a}_n(\bm{k})$ discussed in
Sec.~\ref{int-intro}.

We next divide $H_E$ into  $H_E^r + H_E^a$,
where $H_E^a$ ($H_E^r$) corresponds to the 
first (second) term on the right hand side of Eq.~(\ref{eq:xmat}).
Absorbing $H_E^r$ into the 
unperturbed Hamiltonian, {\it i.e.}, 
$H_p = H_0 + H_{SOI} + H_E^r$, KL treated the remaining term $H_E^a$ as a
perturbation.  Accordingly, the density matrix $\rho$ was written as
\begin{equation}
\rho = \rho_0( H_p) + \rho_1
\end{equation}
where $\rho_0(H_p)$ is the finite-temperature equilibrium density 
matrix and $\rho_1$ is the correction.
KL assumed that $\rho_1$ gives only the ordinary conductivity,
whereas the AHE arises solely from $\rho_0(H_p)$.
Evaluating the average velocity $\bar{v}_a$ as
$ \bar{v}_a = {\rm Tr}[ \rho_0 v_a ]$, they found that
\begin{equation}
\bar{v}_a= - i e E_b \sum_{n, \bm{k}} 
\rho_0'(E^p_{n \bm{k}}) J_a^{n n}(\bm{k}),
\end{equation}
where $\rho_0'$ is the derivative with respect to the energy.  
As it is clear here, this current is the dissipationless
current in thermal equilibrium under the influence of the 
external electric field, {\it i.e.}, $H_E^r$. The AHE contribution arises from the interband coherence induced by an electronic field, and not from the more complicated rearrangements of states within the partially occupied bands.

A second assumption of KL is that, in 3$d$ metals, 
the SOI energy $H_{SO}\ll \varepsilon_F$
(the Fermi energy) and $W$ (the bandwidth), so that it suffices to consider $H_{SO}$ to leading order.
Using Eq.~(\ref{eq:SOI}), the AHE response
is then proportional to $|\bm{M}_s|$, 
consistent with the empirical relationship Eq.~(\ref{eq:Pugh}).
More explicitly, 
to first-order in
$H_{SOI}$, KL obtained 
\begin{equation}
\bar{\bm{v}} = - {e \over {m \Delta^2}} 
\sum_{\bm{k},n} \rho_0 (\varepsilon_{n \bm{k}}) 
[ \bm{E} \cdot \bm{v}_{n \bm{k}} ] \bf{F}_{ n  \bm{k}},
\end{equation}
where $\varepsilon_{n \bm{k}}$ is the energy 
of the Bloch state for $H_0$, $\Delta$ is the averaged value of 
interband energy separation, and $\bf{F}_{ n  \bm{k}}$ is the 
force $i\langle n \bm{k} | [ H_{SOI}, \bm{p}] | n \bm{k}\rangle$.
This gives the anomalous Hall coefficient
\begin{equation}
R_s \cong { {2 e^2 H_{SO}} \over {m \Delta^2}}
\delta \left\langle  { m \over {m^*}} \right\rangle \rho^2,
\end{equation}
where $| \bm{F} | \cong (e/c) H_{SO} v$, $m^*$ is the effective mass,
 $\delta$ is the number of incompletely filled d-orbitals, and $\rho$ is the resistivity.

Note that the $\rho^2$-dependence implies
that the off-diagonal Hall conductivity $\sigma_H$ 
is independent of the transport 
lifetime $\tau$, {\it i.e.}, it is well-defined even in the absence of disorder, in striking contrast with the diagonal conductivity $\sigma$. 
The implication that $\rho_H = \sigma_H\rho^2$ 
varies as the square of $\rho$ was immediately subjected
to extensive experimental tests, as described in Sec. \ref{sec:brief}.

An important finding in the KL theory is 
that inter-band matrix elements of the current operator 
contribute significantly to the transport currents.
This contrasts with conventional Boltzmann transport 
theory, where the current arises solely from 
the group velocity $\bm{v}_{n,\bm{k}} = \partial \varepsilon_{ n, \bm{k}} / \partial \bm{k}$.  

In the Bloch basis 
\begin{equation}
\langle  \bm{r} | \psi_{n \bm{k}} \rangle = e^{i \bm{k} \cdot \bm{r} }
\langle  \bm{r} | u_{n \bm{k}} \rangle,
\end{equation}
the $N$-orbital Hamiltonian (for a given $\bm{k}$) 
may be decomposed into 
the $2N \times 2N$ matrix $h(\bm{k})$, {\it viz. }
\begin{equation}
h(\bm{k}) = \sum_{n,m} \; \langle n \bm{k}|h(\bm{k}) | m\bm{k}\rangle \; a^\dagger_n(\bm{k}) a_m(\bm{k}).
\end{equation}
The corresponding current operator for a given $\kk$ is 
\begin{equation}
J_\mu(t,\kk) = \sum_{n,m}\; \langle n \bm{k}|{{\partial h(\bm{k})} \over 
{\partial \hbar k_\mu}} | m\bm{k}\rangle \; a^\dagger_n(\bm{k}) a_m(\bm{k}).
\end{equation}
According to Hellman-Feynman's theorem
\begin{equation}
\langle n\bm{k}| { {\partial h(\bm{k})} \over { \partial k_\mu}} 
| n\bm{k}\rangle 
= { {\partial \varepsilon_n(\bm{k})} \over {\partial \bm{k}}},
\end{equation}
the diagonal contribution to the current 
is (with $n=m$) 
\begin{equation}
J^{\rm intra}_\mu(t) = -e \sum_{n\bm{k}}v_{g}(\bm{k}) N_n(\bm{k}),
\label{eq:Jbol}
\end{equation}
where $N_n(\bm{k})=a^\dagger_n(\bm{k}) a_n(\bm{k})$ is the occupation number in the 
state $|n \bm{k}\rangle$. Eq.~(\ref{eq:Jbol}) corresponds to the expression in 
Boltzmann transport theory mentioned before.
The inter-band matrix element $\langle n \bm{k}|{{\partial h(\bm{k})} \over 
{\partial k_\mu}} | m\bm{k}\rangle$ with $n \ne m$ corresponds to 
inter-band transitions. As shown by KL,
the inter-band matrix elements have profound consequences
for the Hall current, as has become apparent from the Berry-phase approach. 

\subsubsection{Extrinsic mechanisms}\label{extrinsic}
\paragraph{Skew scattering} \label{skew-detail}
\indent

In a series of reports, Smit~\cite{Smit:1955_a,Smit:1958_a} 
mounted a serious challenge to the basic findings of KL.
In the linear-response transport regime, the steady-state current balances
the acceleration of electrons by $\bm{E}$ 
against momentum relaxation by scattering from impurities and/or
phonons.  Smit pointed out that this balancing was entirely absent from the KL theory, and purported to show that the KL term vanishes exactly.  
His reasoning wasa that the anomalous velocity
central to KL's theory is proportional to the 
acceleration $\dot{\bm{k}}$ which must vanish on average
at steady state because the force from $\bm{E}$ cancels
that from the impurity potential.  

More significantly, Smit proposed the skew-scattering mechanism (Fig.~\ref{mechanisms}) as the 
source of the AHE~\cite{Smit:1955_a,Smit:1958_a}. As discussed in Sec.~\ref{skew-intro},
in the presence of SOI, the matrix element 
of the impurity scattering potential reads
\begin{equation}
\langle  \bm{k}' s| V | \bm{k}, s\rangle = 
\tilde{V}_{\bm{k}, \bm{k}'} \left( \delta_{s, s'}
+ { {i \hbar^2} \over {4 m^2 c^2} } 
\left( 
\langle  s'| \bm{\sigma} |s\rangle \times \bm{k}') \cdot \bm{k}
\right) \right).
\end{equation} 
Microscopic detailed balance would require that the transition probability 
$W_{n \to m}$ between states $n$ and $m$ is identical to that
proceeding in the opposite direction ($W_{m \to n}$). 
It holds, for example, in the Fermi's golden-rule approximation
\begin{equation}
W_{n \to m}= { { 2 \pi} \over {\hbar}} |\langle n |V| m\rangle|^2 
\delta(E_n - E_m),
\end{equation}
where $V$ is the perturbation inducing the transition.
However, microscopic detailed balance is not generic.  In calculations of
the Hall conductivity, which involve the second Born 
approximation (third order in $V$), detailed balance already  fails.  
In a simple N=1 model, skew scattering can be represented by an asymmetric part of the
transition probability
\begin{equation}
W^A_{\bm{k} \bm{k}'} = - \tau_A^{-1} \bm{k} \times \bm{k}'
\cdot \bm{M}_s.
\end{equation}
When the asymmetric scattering processes is included (dubbed skew scattering),
the scattering probability $W(\bm{k}\rightarrow \bm{k}')$ is 
distinct from $W(\bm{k}'\rightarrow \bm{k})$.

Physically, scattering of a carrier from an impurity introduces a momentum 
perpendicular to both the incident momentum $\bm{k}$ 
and the magnetization $\bm{M}$. This leads to a transverse current 
proportional to the longitudinal current driven by $\bm{E}$.
Consequently, the Hall conductivity $\sigma_H$ and 
the conductivity $\sigma$ are both proportional to the transport lifetime $\tau$. 
Equivalently, $\rho_H = \sigma_H\rho^2$ is proportional to the
resistivity $\rho$.  

\begin{figure}
\incl[width=7cm]{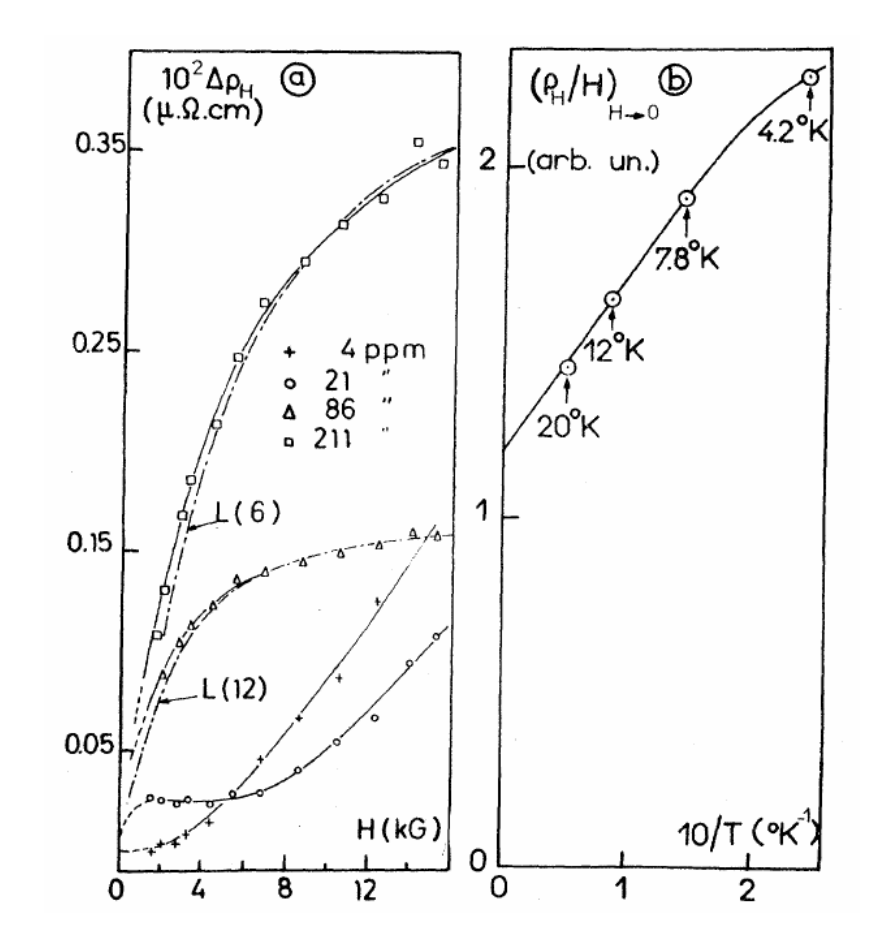}
\caption{\label{fig:Imp}
The dependences of the Hall resistivity in 
AuFe on  magnetic field (Panel a) and temperature
(Panel b). [After Ref. \onlinecite{Hurd:1972_a}.]
}
\end{figure}

As mentioned in Sec. \ref{sec:intro} and in Sec.~\ref{sec:experiments}, 
this prediction -- qualitatively distinct from 
that in the KL theory -- is consistent with experiments, especially on \emph{dilute} Kondo systems.
These are systems which are realized by dissolving magnetic impurities 
(Fe, Mn or Cr) in the non-magnetic hosts Au or Cu. (At 
higher concentrations, these systems become spin glasses.)
Although these systems do not exhibit magnetic ordering
even at $T$ as low as 0.1 K, their Hall profiles 
$\rho_H$ vs. $H$ display an anomalous component derived from 
polarization of the magnetic local moments (Fig.~\ref{fig:Imp} a).

Empirically, the Hall coefficient $R_H$ has the form 
\begin{equation}
R_H = R_0 + A/T
\label{eq:RHskew}
\end{equation}
where $R_0$ is the nominally $T$-independent OHE coefficient and $A$ is a constant (see Fig.~\ref{fig:Imp} b). 
Identifying the second term with the 
paramagnetic magnetization $M = \chi H$, where 
$\chi(T) \propto 1/T$ is the Curie susceptibility of 
the local moments, we see that Eq.~(\ref{eq:RHskew})
is of the Pugh form Eq.~(\ref{eq:Pugh}).  

Many groups have explored the case in
which $\rho$ and $\rho_H$ can be
tuned over a large range by changing the magnetic-impurity concentration $c_i$. Hall experiments on these systems in the 
period 1970-1985 by and large
confirmed the skew-scattering prediction $\rho_H\propto \rho$. 
This led to the conclusion -- often repeated in 
reviews -- that the KL theory had been ``experimentally disproved''.  
As mentioned in Sec. \ref{sec:intro}, this invalid conclusion
ignores the singular role of TRI-breaking in ferromagnets.
While Eq.~(\ref{eq:RHskew}) indeed describes skew scattering, the
dilute Kondo system respects TRI in zero $H$.  
This  essential qualitative differences between ferromagnets and systems 
with easily aligned local moments implies essential differences in the physics of their Hall effects.

\paragraph{Kondo theory}\label{sec:Kondo}
\indent

Kondo has proposed a finite-temperature skew-scattering model
in which spin waves of local moments at finite $T$ 
lead to asymmetric scattering~\cite{Kondo:1962_a}.
In the KL theory, the moments of the ferromagnetic state 
are itinerant: the electrons carrying the transport
current also produce the magnetization. Kondo has considered the
opposite limit in which nonmagnetic $s$ electrons scatter
from spin-wave excitations of the ordered $d$-band local moments 
(with the interaction term $J \bm{S}_n\cdot \bm{s}$). 
Retaining terms linear in the spin-orbit coupling $\lambda$
and cubic in $J$, Kondo derived an 
AHE current that arises from transition probabilities containing
the skew-scattering term $\sim{\bm{k}\times \bm{k}'}$.
The $T$ dependence of $\rho_H$ matches well (especially near $T_C$)
the $\rho_H$ vs. $T$ profiles measured in Ni~\cite{Jan:1952_a,Jan:1952_b,Lavine:1961_a} 
and Fe~\cite{Jan:1952_a} (Fig.~\ref{fig:Kondo}).  

\bfig[h]
\incl[width=7cm]{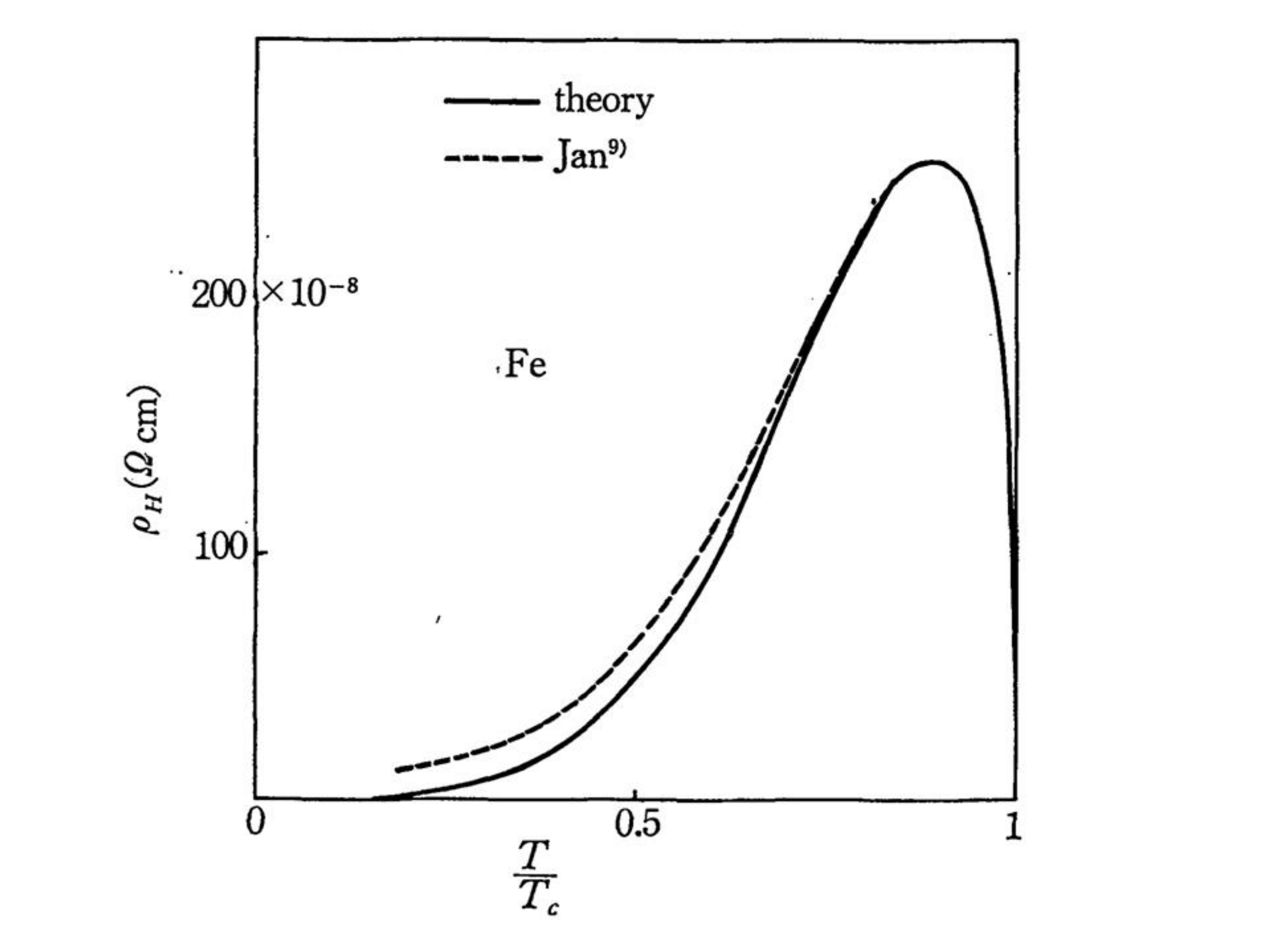}
\caption{\label{fig:Kondo}
Comparison of the curve of $\rho_H$ vs. $T$ measured in 
Fe by Jan with Kondo's calculation [From Ref. \onlinecite{Kondo:1962_a}.]  
}
\efig

Kondo has noted that a problem 
with his model is that it predicts that $\rho_H$ should vanish when the $d$ orbital angular momentum is quenched as in Gd (whereas $\rho_H$ is observed to be large).  
A second problem is that
the overall scale for $\rho_H$ (fixed by the exchange
energies $F_0$, $F_1$ and $F_2$) is too small 
by a factor of 100 compared with experiment.  
Kondo's model with the $s$-$d$ spin-spin interaction replaced by a
$d$(spin)-$s$(orbital) interaction has also been applied 
to antiferromagnets~\cite{Maranzana:1967_a}.

\paragraph{Resonant skew scattering}\label{sec:res_skew}
\indent

\begin{figure}
\incl[width=0.6\columnwidth]{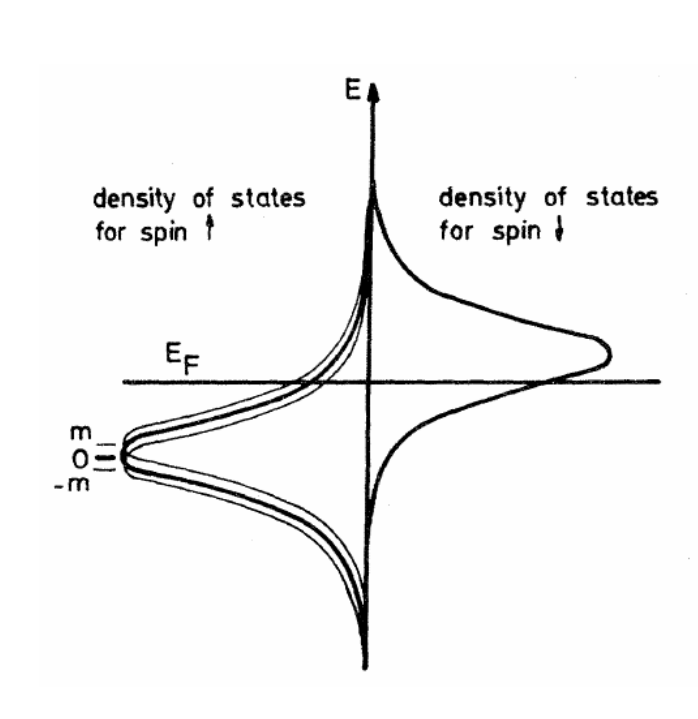}
\caption{
Sketch of the broadened virtual bound states 
of a 3$d$ magnetic impurity dissolved in a 
non-magnetic metallic host.
The SOI lifts the degeneracy of the $d$ levels indexed
by $m$, the magnetic quantum number. [From Ref. \onlinecite{Fert:1972_a}.]
}
\label{fig:Resonance}
\end{figure}

Resonant skew scattering arises from scattering of carriers from virtual bound states in magnetic ions dissolved in a metallic host.  Examples 
are $X$M, where $X$ = Cu, Ag and Au and M = Mn, Cr and Fe. The prototypical model is a $3d$ magnetic ion embedded in a broad $s$ band, as described by the Anderson model \cite{Hewson:1993_a}. As shown in Fig.~\ref{fig:Resonance}, a spin-up $s$ electron transiently occupies the spin-up bound state which lies
slightly below $\varepsilon_F$.  However, a second (spin-down) $s$ electron cannot be captured because the on-site repulsion energy $U$ raises its energy above $\varepsilon_F$. As seen in Fig.~\ref{fig:Resonance}, the SOI causes the energies $E_{d\sigma}^m$ of the $d$ orbitals (labelled by $m$) to be individually resolved. Here, $\sigma=\pm$ is the spin of the bound electron. The applied $\bm{H}$ merely
serves to align spin $\sigma=+$ at each impurity.

The scattering of an incident wave $\mathrm{e}^{i\bm{k\cdot r}}$ is expressed by the phase shifts $\delta_{ls}^{m}(E)$ in the partial-wave expansion of the scattered wave.  The
phase shift is given by
\begin{equation}
\cot\delta_{ls}^m(E) = \frac{(E^m_{d\sigma}-E)}{\Delta},
\label{eq:cot}
\end{equation}
where $\Delta$ is the half-width of each orbital.

In the absence of SOI, $\delta_{ls}^m(E)$ is independent of $m$ and there is no Hall current.  The splitting caused by SOI results in a larger density of states at $\varepsilon_F$ for the orbital $m$ compared with $-m$
in the case where the up spin electron is more than half-filled as shown in Fig.~\ref{fig:Resonance}.  Because the phase shift is sensitive to occupancy of the impurity state
(Friedel sum-rule), we have $\delta_{ls}^{m}\ne\delta_{ls}^{-m}$. This leads to a right-left asymmetry in the scattering, and a large Hall current ensues.  Physically, a conduction electron incident with positive 
$z$-component of angular momentum $m$ hybridizes more strongly with the virtual bound state than one with negative $-m$.  This results in more electrons being scattered to the left than the right.
When the up spin density of states are filled less than half, {\it i.e.}, $E_{d+}^0 >0$, the direction becomes the opposite, leading a sign change of the AHE.

Explicitly, the splitting of $E_{d \sigma}^m$ is given by 
\begin{equation}
E_{d \pm}^m = E_{d \pm}^{0} \pm {\frac12} m \lambda_{\pm},
\label{eq:Ed}
\end{equation}
where $\lambda_{\sigma}$ is the SOI energy for spin $\sigma$. Using Eq.~(\ref{eq:Ed}) in (\ref{eq:cot}), we
have, to order $(\lambda_{\sigma}/\Delta)^2$, 
\begin{equation}
 \delta_{2 \sigma}^m  =  \delta_{2 \sigma}^{0} 
\sigma { {\lambda_{\sigma} m} \over { 2 \Delta} } \sin^2 
\delta_{2 \sigma}^{0} + 
{ {\lambda_{\sigma}^2 m^2} \over { 4 \Delta^2} }
\sin^3 \delta_{2 \sigma}^{0} \cos \delta_{2 \sigma}^{0}.
\end{equation}
Inserting the phase shifts into the Boltzmann transport equation, we find that
the Hall angle $\gamma_{\sigma}\simeq \rho_{xy}^ {\sigma}/\rho_{xx}^{\sigma}$ 
for spin $\sigma$ is, to leading order in 
$\lambda_{\sigma}/\Delta$, 
\begin{equation}
\gamma_{\sigma} = \sigma { 3 \over 5} { {\lambda_{\sigma}} \over \Delta}
\sin ( 2 \delta_{2 \sigma}^{0} - \delta_1) \sin \delta_1,
\end{equation}
where $\delta_1$ is the phase shift of the $p$-wave channel which 
is assumed to be independent of $m$ and $s$~\cite{Fert:1972_a,Fert:1981_a}.
Thermal averaging over $\sigma$, we can obtain
an estimate of the observed Hall angle 
$\gamma$.  Its sign 
is given by the position of the energy level 
relative to $\varepsilon_F$.
With the rough estimate 
$\lambda_{\sigma}/ \Delta \simeq 0.1$,
and $\sin \delta_1 \sim 0.1$, we have 
$|\gamma|\simeq 10^{-2}$. 
Without the resonant 
scattering enhancement, the typical value of 
$\gamma$ is $\sim 10^{-3}$.

\begin{figure}
\incl[width=9cm]{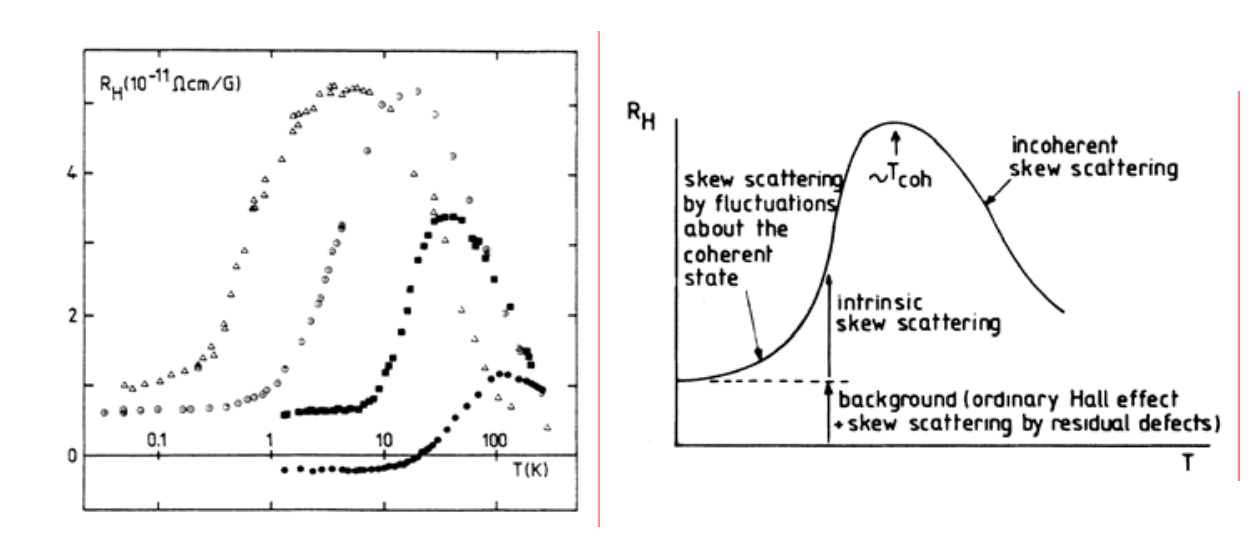}
\caption{\label{fig:heavy}
The Hall coefficient $R_H$ in heavy 
electron systems.   
Left panel: The $T$ dependence 
of $R_H$ in CeAl$_3$ (open triangle),
UPt$_3$ (solid square), UAl$_2$ (solid circle), and
a single crystal of CeRu$_2$Si$_2$ (circle dot symbol).
The field $\bm{H}$ is along the $c$-axis.
Right panel: Interpretation of the $R_H$-$T$ curve 
based on skew scattering from local moments. After Ref. \onlinecite{Hadzicleroux:1986_a}
and \onlinecite{Lapierre:1987_a}. [From Ref. \onlinecite{Fert:1987_a}.]
}
\end{figure}

Heavy-electron systems are characterized by a very large
resistivity $\rho$ above a coherence temperature $T_{coh}$
caused by scattering of carriers from strong 
fluctuations of the local moments formed by $f$-electrons at each lattice site.
Below $T_{coh}$, the local-moment fluctuations decrease rapidly with incipient band-formation involving the
$f$-electrons.  At low $T$, the electrons form a Fermi liquid with a greatly enhanced effective mass.  As shown in Fig.~\ref{fig:heavy}, the Hall coefficient $R_H$ increases to a broad maximum at $T_{coh}$ before decreasing sharply to a small value 
in the low-$T$ coherent-band regime. 
Resonant skew scattering has also been
applied to account for the strong $T$ dependence of the 
Hall coefficient in CeCu$_2$Si$_2$,
UBe$_{13}$, and UPt$_3$~\cite{Fert:1981_a,Coleman:1985_a,Fert:1987_a}.

\paragraph{Side-jump}\label{sj-detail}
\indent

An extrinsic mechanism distinct from skew scattering is 
side-jump~\cite{Berger:1970_a} (Fig.~\ref{mechanisms}).
Berger considered the scattering of a Gaussian wavepacket from
a spherical potential well of radius $R$ given by
\begin{eqnarray}
V(r) &=& \frac{\hbar^2}{2m}( k^2- k_1^2)  \quad (r<R) \nonumber \\
V(r) &=& 0   \quad\quad\quad\quad\quad\quad (r>R),
\end{eqnarray}
in the presence of the SOI term
$
H_{SO} = (1/2 m^2 c^2) (r^{-1} \partial V/ \partial r) S_z L_z,
$
where $S_z$ ($L_z$) is the $z$-component of the spin (orbital) angular momentum. 
For a wavepacket incident with wavevector $\bm{k}$, Berger found that the wavepacket suffers 
a displacement $\Delta y$ transverse to $\bm{k}$ given by
\begin{equation}
\Delta y = { 1\over 6} k \lambda_c^2,
\end{equation}
with $\lambda_c = \hbar/mc$ the Compton wavelength.
For $k \cong k_F\cong 10^{10} {\rm m}^{-1}$ (in typical metals),
$\Delta y\cong 3\times 10^{-16}$ m is far too small to be observed.  

In solids, however, the effective SOI is enhanced by band-structure effects by the 
factor~\cite{Fivaz:1969_a}  
\begin{equation}
{ { 2 m^2 c^2} \over { m^* \hbar}} \tau_q \cong 3.4 \times 10^4,
\label{eq:enhance}
\end{equation}
with $\tau_q = (m^*/ 3 \hbar^2) \sum_{m \ne n} 
(\chi \bar{\xi}  \rho^2 /\Delta E_{nm} ) 
|\langle m | \bm{q} \times \bm{p} | n\rangle|^2$.
Here, $\Delta E_{nm} \cong 0.5$ eV is the gap 
between adjacent $d$-bands, $\chi \cong$ 0.3 is the overlap 
integral, $\rho\cong 2.5 \times 10^{-10}$ m is the nearest-neighbor distance, and $\bar{\xi} = - (\hbar^2/2 m^2 c^2) 
\langle (r^{-1} \partial V/\partial r)\rangle \cong$ 0.1 eV
is the atomic SOI energy.  The factor in Eq. (\ref{eq:enhance})
is essentially the ratio of the
electron rest mass energy $mc^2$ to the energy gap $\Delta E_{nm}$.  With this enhancement, the transverse displacement 
is $\Delta y \simeq 0.8 \times 10^{-11}$ m, which 
renders the contribution relevant to the AHE.

However, because the side-jump contribution to $\sigma_H$ 
is independent of $\tau$, it is experimentally difficult
to distinguish from the KL mechanism. We return to this issue in Sec. \ref{sec:theory:Boltzmann}.

\subsubsection{Kohn-Luttinger theory formalism}

\paragraph{Luttinger theory}\label{sec:luttinger}
\indent

Partly motivated by the objections raised by Smit, Luttinger~\cite{Luttinger:1958_b} 
revisited the AHE problem using the Kohn-Luttinger formalism of 
transport theory~\cite{Kohn:1957_a,Luttinger:1958_a}. 
Employing a systematic expansion in terms of 
the impurity potential $\bar{\varphi}$, 
he solved the transport 
equation for the density matrix and listed several 
contributions to the AHE current, including the intrinsic KL term, the skew-scattering term, 
and other contributions. 
They found that to zero-th order in $\bar{\varphi}$, the average
velocity is obtained as 
the sum of the 3 terms $v_\beta^{(11)}, u_\beta$ and $v_\beta^{(b)}$ defined by~\cite{Luttinger:1958_b}
\begin{equation}
v_\beta^{(11)} = -ie E_\alpha 
\left({ { \partial J_\beta^\ell } \over { \partial k_\alpha}  }
-{ { \partial J_\alpha^\ell } \over { \partial k_\beta}  }
\right)_{\bm{k}=\bm{0}}
\left( {{\varepsilon_F} \over { 3 n_i \bar{\varphi} } }
\right),
\end{equation}
\begin{equation}
u_\beta = -ie E_\alpha 
\left({ { \partial J_\beta^\ell } \over { \partial k_\alpha}  }
-{ { \partial J_\alpha^\ell } \over {\partial k_\beta}  }
\right)_{\bm{k}=\bm{0}},
\end{equation}
and
\begin{equation}
v_\beta^{(b)} = ie E_\alpha
\left[ 
\left({ { \partial J_\beta^\ell } \over { \partial k_\alpha}  }
-{ { \partial J_\alpha^\ell } \over { \partial k_\beta}  }
\right) C
- {\varepsilon_F} T_{\alpha \beta}^{(0)}
\right],
\end{equation}
with $n_i$ the impurity concentration.
(For the definition of $T_{\alpha \beta}^{(0)}$,
see Eq.~(4.27) of Ref. \onlinecite{Luttinger:1958_b}).

The first term $v_{\beta}^{(11)}$ is the skew-scattering 
contribution, while the second 
$u_{\beta}$ is the velocity obtained by KL \cite{Karplus:1954_a}.  
The third term $v_\beta^{(b)}$
is another term of the same order as $u_{\beta}$.

The issues raised by comparing intrinsic vs. extrinsic AHE mechanisms involve several 
fundamental issues
in the theory of transport and quantum systems away from equilibrium.
Following Smit~\cite{Smit:1955_a,Smit:1958_a} and Luttinger~\cite{Luttinger:1958_b}, 
we consider the wavefunction expanded in terms of
Bloch waves $\psi_{\ell}$'s, {\it viz.}
\begin{equation}
\psi(t) = \sum_{\ell} a_{\ell}(t) \psi_{\ell},
\end{equation}
where $\ell=(n, \bm{k})$ stands for the band index $n$ 
and wavevector $\bm{k}$.
The corresponding expectation value of the position 
operator ${\bar x}_\beta$ is given by
\begin{equation}
{\bar x}_\beta = i\sum_{\ell} { {\partial a_{\ell} } \over 
{\partial k_{\beta}} } 
a_{\ell}^* + i\sum_{\ell,\ell'}
a_{\ell'}a_{\ell}^* J_{\beta}^{\ell, \ell'}
\end{equation}
where $J_{\beta}^{\ell, \ell'}$ was defined in Eq.~(\ref{eq:Jb}).
Taking the quantum-statistical average using the density 
matrix $(\rho_T)_{\ell', \ell}=\langle a_{\ell'}a_{\ell}^*\rangle$,  
the expectation value can be written as 
\begin{equation}
\langle \bar{x}_\beta\rangle =i 
\sum_{\ell}{{\partial (\rho_T)_{\ell, \ell} }
\over {\partial k_{\beta} } } |_{\bm{k} = \bm{k}'}+
i\sum_{\ell,\ell'}
(\rho_T)_{\ell' \ell}  J_{\beta}^{\ell, \ell'}.
\label{eq:xbeta}
\end{equation}
It may be seen that the second term in Eq.~(\ref{eq:xbeta}) does not contribute
to the current because it is a regular function of 
$\bm{k}$, and the expectation value of its time-derivative 
is zero. 
Finally, one obtains the expression 
\begin{equation}
\langle \bar{v}_\beta\rangle =i 
\sum_{\ell}{{\partial ({\dot \rho}_T)_{\ell, \ell} }
\over {\partial k_{\beta} } } |_{\bm{k} = \bm{k}'}
\label{eq:vbeta}
\end{equation}
by taking the time-derivative of Eq.(\ref{eq:xbeta}).

Smit~\cite{Smit:1955_a,Smit:1958_a} assumed 
$a_\ell(t) =| a_\ell | e^{- i\varepsilon_{n \bm{k}} t}$
and obtained the diagonal part as 
\begin{equation}
i ( {\dot \rho}_T)_{ \ell , \ell} = { {\partial \varepsilon_{n \bm{k}}} 
\over {\partial k_\beta} } \langle|a_\ell|^2 \rangle,
\label{eq:diag}
\end{equation}
while the off-diagonal part averages to zero because of the 
oscillatory factor $e^{-i(\varepsilon_\ell - \varepsilon_{\ell'})t}$.
Inserting Eq.~(\ref{eq:diag}) in Eq.~(\ref{eq:vbeta})
and using $(\rho_T)_{\ell \ell} = \langle|a_\ell|^2\rangle,$ 
one obtains the usual expression for the velocity, {\it i.e.},
\begin{equation}
\langle \bar{v}_\beta\rangle = \sum_{\ell} (\rho_T)_{\ell, \ell} 
 { {\partial \varepsilon_{n \bm{k}}} \over {\partial k_\beta} } 
\end{equation}
which involves the group velocity, but not the anomalous velocity.  
A subtlety in the expression Eq.~(\ref{eq:vbeta}) is clarified
by writing $\rho= \rho_0 + f e^{st}$ where 
$e^{st}$ is the adiabatic factor, and
$\dot{\rho}_T = s f e^{st}$. This leads to the expression 
\begin{equation}
\langle {\bar v}_\beta\rangle =i s
\sum_{\ell}{ {\partial (f)_{n \bm{k}, n \bm{k}'}}
\over {\partial k_{\beta}} }|_{\bm{k} = \bm{k}'},
\end{equation}
which approaches a finite value in the limit $s \to 0$.
This means that ${ {\partial (f)_{n \bm{k}, n \bm{k}'}}
\over {\partial k_{\beta}} }|_{\bm{k} = \bm{k}'} \sim 1/s$,
{\it i.e.}, a singular function of $s$. This is in sharp contrast 
to $f$ itself, which is a regular function as $s \to 0$,
which KL estimated. They considered the current or velocity 
operator instead of the 
position operator, the latter of which is unbounded and even 
ill-defined for periodic boundary conditions.

Luttinger~\cite{Luttinger:1958_b} argued that the ratio 
$v_{\beta}^{(11)}/u_\beta$ equals $\varepsilon_F/(3 n_i \bar{\varphi})$,
which may become less than 1 for large impurity concentration $n_i$. 
However, if we assume $ \bar{\varphi} $ is comparable to 
$\varepsilon_F$, the expansion parameter is $\varepsilon_F \tau/\hbar$, {\it i.e.},
$v_{\beta}^{(11)}/u_\beta = \varepsilon_F \tau/\hbar$. 

The ``metallicity'' parameter $\varepsilon_F \tau/\hbar$
plays a key role in modern quantum-transport 
theory, especially in the  weak localization 
and interaction theory~\cite{Lee:1985_a}.  Metallic 
conduction corresponds to $\varepsilon_F \tau/\hbar \gg 1$.
More generally, if one assumes that the 
anomalous Hall conductivity $\sigma_H$ is first order in 
the spin-orbit energy $\Delta$, it can be written 
in a scaling form as
\begin{equation}
\sigma_H = { { e^2 } \over {h a} } \cdot  { \Delta \over {\varepsilon_F}}
f \left({{\hbar} \over {\varepsilon_F \tau} } \right),
\end{equation}
where the scaling function can be expanded as
\begin{equation}
f(x) = \sum_{n=-1}^{\infty} c_n x^n.
\end{equation}
Here, $c_n$'s are constants of the order of unity. 
The leading order term $c_{-1} x^{-1}$ corresponds to the 
skew-scattering contribution $\propto\tau$, 
while the second constant term $c_0$
is the contribution found by KL. 
If this expansion is valid, the intrinsic contribution by KL is always smaller than the 
skew scattering contribution in the metallic region with $\hbar/\varepsilon_F \tau \ll 1$.
One should note however that the right hand side of this inequality should really have a 
smaller value to account for the weakness of the skew scattering amplitude. 
When both are comparable, {\it i.e.},  $\hbar/\varepsilon_F \tau \sim 1$,
one needs to worry about the localization effect and the 
system is nearly insulating, with a conduction that is
of the hopping type. This issue is discussed in more detail in Secs. \ref{sec:tm}, \ref{sec:exp:loc}, and \ref{sec:theory:Rashba}.  

\paragraph{Adams-Blount formalism}
\indent

Adams and Blount~\cite{Adams:1959_a} expressed the KL theory in a way that 
anticipates the modern Berry-phase treatment
by introducing the concept of ``field-modified energy bands" and 
the ``intracell" coordinate.
They considered the diagonal part of the coordinate matrix in the 
band $n$ as
\begin{equation}
x_{c}^\mu = 
i \hbar { {\partial} \over {\partial p_\mu } } 
+ X^{nn}_\mu(\bm{p}),
\end{equation}
which is analogous to Eq.~(\ref{eq:xmat}) obtained by KL.
The first term is the Wannier coordinate identifying the
lattice site,
while the second term -- the ``intracell'' coordinate -- 
locates the wavepacket centroid \emph{inside} a unit cell.
Significantly, the \emph{intracell} nature of $X^{nn}_\mu$
implies that it involves virtual interband transitions.
Although the motion of the wavepacket is confined to the
conduction band, its position inside a unit cell involves
virtual occupation of higher bands whose effects 
appear as a geometric phase~\cite{Ong:2005_a}. 
They found that the curl $\nabla_{\bm{k}}\times \bm{X}^{nn}$ acted like an effective magnetic field
that lives in $\bm{ k}$ space.

The application of an electric field $\bm{E}$ 
leads to an anomalous (Luttinger) velocity, which gives a 
Hall current that is manifestly dissipationless.
This is seen by evaluating the commutation relationship
of $x_c^\mu$ and $x_c^\nu$.  We have
\begin{eqnarray}
[ x_c^\mu, x_c^\nu ] &=& i \hbar
\left[
{ { \partial X^{nn}_\nu } \over { \partial p_\mu}  }
- { { \partial X^{nn}_\mu } \over { \partial p_\nu}  }\right]
\nonumber \\
&=& 
i \hbar \varepsilon_{\mu \nu \lambda} 
\bm{B}_n(\bm{p})_\lambda,
\end{eqnarray}
where we have defined the field 
$ \bm{B}_n(\bm{p}) = \nabla_p \times \bm{X}^{nn}(\bm{p})$.
This is analogous to the commutation relationship
among the components of $\bm{\pi} = \bm{p} + (e /c) \bm{A}$
in the presence of the vector potential,{\it i.e.},
$[ \pi_\mu, \pi_\nu ] = i \hbar \varepsilon_{\mu \nu \lambda} 
[ \nabla_r \times A(\bm{r}) ]_\lambda =
i \hbar B_{\lambda} (\bm{r})$.
The anomalous velocity arises from the 
fictitious ``magnetic field" $\bm{B}_n(\bm{p})$ (which lives
in momentum space), and non-commutation of
the gauge-covariant coordinates $x_c^\mu$'s. 
This insight anticipated the modern idea of Berry phase curvature
(see Secs. \ref{sec:top} and \ref{sec:theory:Boltzmann}
for details).

Taking the commutator between $x_c^\mu$ and the
Hamiltonian $H_{nn} = E_n(\bm{p}) - F^\nu x_c^\nu$,
we obtain
\begin{equation}
\bm{v}_{nn} = -i [ \bm{x}_c, H_{nn}] 
= { {\partial E_n(\bm{p})} \over {\partial \bm{p}} }
-  \bm{F} \times \bm{B}_n(\bm{p}). 
\end{equation}
The second term on the right hand side is called the 
anomalous velocity.  Adams and Blount reproduced the 
results of Luttinger~\cite{Luttinger:1958_b} and demonstrated it for the 
simple case of a uniform field
$\bm{B}_n(\bm{p}) = \bm{D}$. 
As recognized by Smit~\cite{Smit:1955_a,Smit:1958_a}, 
the currents associated with the anomalous velocities driven by $\bm{E}$ and by
the impurity potential mutually cancel in steady state. However, $\bm{D}$ introduces corrections to the ``driving term" and the 
``scattering term" in the transport equation.  As a consequence, the
current $\bm{J}$ is 
\begin{eqnarray}
\bm{J} &=& n e^2 \sum_k 
\left( - {2 \over 3} E
{ { d f^{(0)} } \over {d E}} \right)
\nonumber \\
&\times&
\left[ { {\bm{F} \tau} \over m } - 
{{\bm{F} \times \bm{D} } \over {\hbar}} 
+ \left( { { \tau^2 k^2 } \over {3 m \tau_A} } \right)
 ( \bm{F} \times \bm{D} )
\right].
\label{eq:JAdam}
\end{eqnarray}
Note that the current arises entirely from the average
of the ``normal current".   However,
the second term in Eq.~(\ref{eq:JAdam}) is similar to the 
anomalous velocity and is consistent with
the conclusions of KL~\cite{Karplus:1954_a} and Luttinger~\cite{Luttinger:1958_b}.
The consensus now is that the KL contribution exists.  
However, in the clean limit $\tau \to \infty$, the 
leading contribution to the AHE conductivity comes from 
the skew-scattering term.
A more complete discussion of the 
semiclassical treatment is in Sec. \ref{sec:theory:Boltzmann}. 

\paragraph{Nozieres-Lewiner theory}
\indent

Adopting the premises of Luttinger's theory~\cite{Luttinger:1958_b},
Nozieres and Lewiner~\cite{Nozieres:1973_a} investigated a 
simplified model comprised of one conduction and one 
valence band to derive all the possible contributions to the AHE.
The Fermi level $\varepsilon_F$ was assumed to lie near 
the bottom of the conduction band. 
Integrating over the valence band states, Nozieres and 
Lewiner derived an effective Hamiltonian for a state 
$\bm{k}$ in the conduction band.  The derived position operator is
$\bm{r}_{\rm eff} = \bm{r} + \bm{\rho}$, where the new term $\bm{\rho}$ 
which involves the SOI parameter $\lambda$ is
given by 
\begin{equation}
\bm{\rho} = - \lambda \bm{k} \times \bm{S}.
\end{equation}
This polarization or effective shift modifies the 
transport equation to produce several contributions to $\sigma_H$. 
In their simple model, the anomalous Hall current $\bm{J}_{AHE}$
(besides that from the 
skew scattering) is 
\begin{equation}
\bm{J}_{AHE} = 2 N e^2 \lambda \bm{E} \times \bar{\bm{S}}, 
\end{equation}
where $N$ is the carrier concentration, 
and $\bar{\bm{S}}$ is the averaged spin polarization.
The sign is opposite to that of the intrinsic term 
$
\bm{J}_{intrinsic} = -2 N e^2 \lambda \bm{E} \times \bar{\bm{S}} 
$
obtained from the SOI in an ideal lattice.  They identified all the terms as arising from the spin-orbit correction to the scattering potential,
{\it i.e.}, in their theory there is no contribution from the intrinsic mechanism.
However, one should be cautious about this statement because
these cancellations occur among the various contributions and such cancellations can be traced back
to the fact that the Berry curvature is independent of $\kk$.
In particular, it is often the case that the intrinsic contribution survives even 
when $\mu$ lies inside an energy gap, {\it i.e.}, $N=0$ (the side-jump contribution is zero in this case). 
This issue lies at the heart of the discussion of the topological aspect of the 
AHE. 

\section{Linear Transport Theories of the AHE}\label{sec:linear_transport}

In this section we describe the three linear response theories now used to describe the AHE. All three theories are formally equivalent in the
$\varepsilon_F \tau \gg 1$ limit.
The three theories make nearly identical predictions; the small differences between the revised semiclassical
 theory and the two formalism of the quantum theory are thoroughly understood at least for several different toy model systems.
The three different approaches have relative advantages and disadvantages.  Considering all three provides a more 
nuanced picture of AHE physics.
 Their relative
correspondence has been shown analytically in several simple models \cite{Sinitsyn:2008_a,Sinitsyn:2007_a}. 
The  generalized semiclassical Boltzmann transport theory which takes the 
 Berry phase into account is reviewed in Sec. \ref{sec:theory:Boltzmann};
this theory has the advantage of greater physical transparency,
but it lacks the systematic character of the microscopic quantum-mechanical theories whose machinery 
deals automatically with the problems of inter-band coherence. 
Another limitation of the semiclassical Boltzmann approach is that the
``quantum correction'' to the conductivities, {\it i.e.}, the higher order terms in 
$\hbar/\varepsilon_F \tau$, which lead to the Anderson localization and other interesting phenomena~\cite{Lee:1985_a}
cannot be treated systematically.

The microscopic quantum-mechanical linear response theories based on the Kubo formalism
 and the Keldysh (non-equiliribrum Green's function) formalism are reviewed in Sec. \ref{sec:theory:Kubo}
 and Sec.~\ref{sec:theory:quantum}, respectively. 
 These  formulations of   transport theory are organized differently but are essentially 
  equivalent in the linear  regime.
In the Kubo formalism, which is formulated in terms of the equilibrium Green's functions,
the intrinsic contribution is more readily calculated,
especially when combined with first-principles electronic structure theory in applications to complex materials.
The Keldysh formalism can 
more easily account for finite lifetime 
quantum scattering effects and take account of the broadened quasiparticle 
spectral features due to the self-energy, while at the same time maintaining a structure more similar to that of 
semiclassical transport theory ~\cite{SOnoda:2006_b}.
AWe contrast these techniques by comparing their applications to a common model, the ferromagnetic Rashba model
in two-dimensions (2D) which has been studied intensively in recent years (Sec.\ref{sec:theory:Rashba}). 

	\subsection{Semiclassical Boltzmann approach}\label{sec:theory:Boltzmann}

As should be clear from the complex phenomenology of the
AHE in various materials classes discussed in Sec.~\ref{sec:experiments}, 
it is not easy to establish a {\em one-size-fits-all} theory for this phenomenom.
From a microscopic point of view the AHE is a formidable beast.  In subsequent sections
we outline systematic theories of the AHE in metallic systems
which employ Keldysh and Kubo linear response theory formalisms and are organized around an 
expansion in disorder strength, characterized by the 
dimensionless quantity $\hbar/\varepsilon_F\tau$.  A small value for this 
parameter may be taken as a definition of the {\em good metal}. 
We start with semiclassical theory, however, because of its 
greater physical transparency.  In this section
we outline the modern version \cite{Sinitsyn:2008_a,Sinitsyn:2007_a} 
of the semiclassical transport theory of the AHE.  This theory augments standard semiclassical transport theory by 
accounting for coherent band-mixing by the external electric field (which leads to the anomalous velocity contribution) and by a random disorder potential (which leads to side jump).
For simple models in which a comparison is possible, the two theories ({\it i.e.} (i) the Boltzmann theory with all side-jump effects (formulated now in a proper gauge-invariant way)
and the anomalous velocity contribution included and (ii) the metallic limit of the more systematic Keldysh and Kubo formalism treatments)
give identical results for the AHE. 
This section provides a more compact version of the 
material presented in the excellent review by Sinitsyn \cite{Sinitsyn:2008_a},
to which we refer readers interested in further detail. Below we take $\hbar=1$ to simplify notation.

In semiclassical transport theory one
retreats from a microscopic description in terms of delocalized Bloch states to 
a formulation of transport in terms of the dynamics of wavepackets of Bloch states with a well
defined band momentum and position ($n,\kk_c,\rr_c$) and scattering between Bloch states due to 
disorder.  The dynamics of the wavepackets between collisions can be treated by an effective 
Lagrangian formalism. The wavepacket distribution function is assummed to obey a classical
Boltzmann equation which in a spatially uniform system takes the form \cite{Sundaram:1999_a}:
\begin{equation}
\frac{\partial f_{l}}{\partial t} +\hbar \dot{\kk}_c \cdot \frac{\partial f_{l}}{\partial{\kk_c}}=-\sum_{l'}(\omega_{l',l} f_l-\omega_{l,l'} f_{l'}).
\label{boltz_eq}
\end{equation}
Here the label $l$ is a composition of band and momenta $(n,\kk_c)$ labels 
and $\Omega_{l',l}$, the disorder averaged scattering rate between wavepackets defined by states $l$ and $l'$, 
is to be evaluated fully quantum mechanically. 
The semiclassical description is useful in clarifying the physical meaning and origin of the different mechanism 
contributing to the AHE.  However, as explained in this review, contributions to the AHE which are 
important in a relative sense often arise from inter-band coherence effects which are 
neglected in conventional transport theory.
The traditional Boltzmann equation therefore requires elaboration in order to 
achieve a successful description of the AHE.

There is a substantial literature \cite{Smit:1955_a,Berger:1970_a,Jungwirth:2002_a}
on the application of Boltzmann equation concepts to AHE theory (see Sec. \ref{sec:theory:early}). 
However stress was often placed only on one of the several possible mechanisms, creating a lot of confusion. 
A cohesive picture has been lacking until recently \cite{Sinitsyn:2005_a,Sinitsyn:2006_a,Sinitsyn:2008_a}. 
In particular, a key problem with some prior theory was that it incorrectly  ascribed physical meaning to 
gauge dependent quantities.
In order to build the correct gauge-invariant semiclassical theory of AHE we must take the following steps  \cite{Sinitsyn:2008_a}: 
\begin{itemize}
\item[i)]
obtain the equations of motion for a wavepacket constructed from spin-orbit coupled Bloch electrons, 
\item[ii)]
derive the effect of scattering of a wave-packet from a smooth impurity, yielding
the correct gauge-invariant expression for the corresponding side-jump, 
\item[iii)]
use the equations of motion and the scattering rates in Eq.~(\ref{boltz_eq}) and  
solve for the non-equillibirum distribution function, carefully
accounting for the points at which modifications are required to account for side-jump, 
\item[iv)]
utilize the  non-equilibrium distribution function to calculate the dc anomalous Hall currents,
again accounting for the contribution of side-jump to the macroscopic current.
\end{itemize}

The validity of this approach is partially established in the following sections by direct comparison
with fully microscopic calculations for simple model systems in which we are able to identify 
each semiclassically defined mechanisms
with a specific part of the microscopic calculations.

\subsubsection{Equation of motion of Bloch states wave-packets}
We begin by defining a wavepacket centered at position $\rr_c$ with average momentum $\kk_c$:
\begin{equation}
\Psi_{\kk_c,\rr_c}(\rr,t)=\frac{1}{\sqrt{V}}\sum_\kk w_{\kk_c,\rr_c}(\kk)e^{i\kk\cdot(\rr-\rr_c)}u_{n\kk}(\rr).
\end{equation}
A key aspect of this wavepacket is that the complex function, sharply peaked around $\kk_c$,  
must have a very specific
phase factor in order to have the wavepacket centered around $\rr_c$. This can be shown to be \cite{Sundaram:1999_a,Marder:1999_a}:
\begin{equation}
w_{\kk_c,\rr_c}(\kk)=|w_{\kk_c,\rr_c}(\kk)| \exp[i(\kk-\kk_c)\cdot \bm{a}_n],
\end{equation}
where $\bm{a}_n\equiv \langle u_{n\kk}|i\partial_\kk u_{n\kk}\rangle$ is the Berry's connection of the Bloch state (see Sec. \ref{mechanisms-intro}).
We can generate dynamics for the wavepacket parameters  $\kk_c$ and $\rr_c$
by constructing a semiclassical Lagrangian from the quantum wavefunctions:
\begin{eqnarray}
{ \cal L}&=&\langle \Psi_{\kk_c,\rr_c}|i\frac{\partial}{\partial t} -H_0+eV|\Psi_{\kk_c,\rr_c}\rangle \nonumber\\
&=&\hbar \kk_c\cdot \dot{\rr}_c+\hbar\dot{\kk}_c\cdot \bm{ a}_n(\kk_c)-E(\kk_c)+eV(\rr_c).
\end{eqnarray}
All the terms in the above Lagrangian are common to conventional semiclassical theory except 
for the second term, which is a geometric term in phase space depending only on the path of the trajectory in this space. This term is
 the origin of the momentum-space Berry phase \cite{Berry:1984_a} effects in anomalous transport in the semiclassical formalism.  
The corresponding Euler-Lagrange equations of motion are:
 \begin{eqnarray}
 \hbar \dot{\kk}_c &=& -e\bm{E}\\
 \dot{\rr}_c&=&\frac{\partial E_n(\kk_c)}{\partial \kk_c}-\hbar \dot{\kk}_c\times \bm{b}_n(\kk_c)
 \end{eqnarray}
 where $\bm{b}_n(\kk_c)=\bm{\nabla}\times \bm{a}_n$ is the Berry's curvature of the Bloch state.
 Compared to the usual dynamic equations for wave-packets formed by free electrons, a new term emerges
 due to the non-zero Berry's curvature of the Bloch states. This term, which 
 is already linear in electric field $\bm{E}$, is of the Hall type and as such will give rise in the linear transport regime to a Hall current contribution from the
 entire Fermi sea, {\it i.e.}, $j_{Hall}^{int}=-e^2 \bm{E}\times  \frac{1}{V}\sum_\kk f_0(E_{n\kk}) \bm{b}_n(\kk)$.

\subsubsection{Scattering and the side-jump}

From the theory of elastic scattering we know that 
the transition rate $\omega_{l,l'}$ in Eq.~(\ref{boltz_eq}) is given by the  
$T$-matrix element of the disorder potential:
\begin{equation}
\omega_{l'l}\equiv 2\pi |T_{l'l}|^2 \delta(\epsilon_{l'}-\epsilon_{l}).
\label{WT}
\end{equation}
The scattering $T$-matrix is defined by $T_{l'l}=\langle l'| \hat{V}| \psi_{l} \rangle$,
where $\hat{V}$ is the impurity potential operator and 
$| \psi_{l} \rangle$ is the eigenstate of the full Hamiltonian 
$\hat{H}=\hat{H}_0+\hat{V}$ that satisfies the Lippman-Schwinger equation
$| \psi_{l} \rangle = |l\rangle  +({\epsilon_{l}-\hat{H}_0+i\eta})^{-1}{\hat{V}} | \psi_{l} \rangle$. 
$| \psi_{l} \rangle$ is the state which evolves adiabatically from $|l\rangle$ when the disorder potential is turned on slowly.
For weak disorder one can approximate the scattering state $| \psi_{l} \rangle$ 
by a truncated series in powers of $V_{ll'}=\langle l|\hat{V} |l'\rangle$: 
\begin{equation}
| \psi_{l} \rangle \approx |l\rangle +\sum_{l''} \frac{V_{l''l}} {\epsilon_{l}-\epsilon_{l''}+i\eta} | l''\rangle + \ldots
\label{ser1}
\end{equation}
Using this expression in the above definition of the T-matrix and substituting it into 
Eq.~(\ref{WT}), one can expand the scattering rate in powers of the disorder strength
 \begin{equation}
 \omega_{ll'}=\omega^{(2)}_{ll'}+\omega^{(3)}_ {ll'}+\omega^{(4)}_ {ll'}\cdots,
 \label{om1}
 \end{equation}
where $\omega^{(2)}_{ll'}=2\pi \langle |V_{ll'}|^2 \rangle_{dis} \delta (\epsilon_{l} -\epsilon_{l'}),$
\begin{equation}
\omega^{(3)}_ {ll'}=2\pi \left ( \sum_{l''} \frac{\langle V_{ll'}
 V_{l'l''} V_{l''l}\rangle_{dis}}
{\epsilon_{l} -\epsilon_{l''}-i \eta} +c.c. \right) \delta (\epsilon_{l} -\epsilon_{l'}),
\label{om3}
\end{equation}
and so on.  

We can always decompose the scattering rate into components that are symmetric and anti-symmetric in the state indices:
$\omega_{l'l}^{(s/a)} \equiv (\omega_{ll'} \pm \omega_{l'l})/2$.
In conventional Boltzmann theory the AHE is due solely to the anti-symmetric contribution to the scattering rate~\cite{Smit:1955_a}.

The physics of this contribution to the AHE is quite similar to that of the 
longitudinal conductivity.  In particular, the Hall conductivity it leads to is proportional to the Bloch state lifetime $\tau$. 
Since $\omega^{(2)}_{l'l}$ is symmetric, the leading contribution to $\omega_{l'l}^{(a)}$ 
appears at order $V^{3}$.  
Partly for this reason the skew scattering AHE conductivity contributions is always much smaller than the longitudinal conductivity.
(It is this property which motivates the identification below of additional transport mechanism which contribute to the 
AHE and can be analyzed in semiclassical terms.) 
The symmetric part of $\omega^{(3)}_{ll'}$ is not essential since it only renormalizes the second order result 
for $\omega^{(2)}_{l'l}$ and 
the antisymmetric is given by 
\begin{equation}
\begin{array}{l}
\omega^{(3a)}_ {ll'}= -(2\pi)^2  \sum \limits_{l''} \delta (\epsilon_{l}
 -\epsilon_{l''}) {\rm Im} \langle V_{ll'} V_{l'l''} V_{l''l}\rangle_{dis} 
 \delta (\epsilon_{l} -\epsilon_{l'}).
\end{array}
\label{omasym1}
\end{equation}
This term is proportional to the density of scatterers, $n_i$.  
Skew scattering has usually been associated directly with $\omega^{(3a)}$, neglecting in particular the  
higher order term $\omega^{(4a)}_ {ll'}$ which is proportional to $n_i^2$ and 
should not be disregarded because it gives a contribution to the 
AHE which is of the same order as the side-jump contribution considered  below. 
This is a common mistake in the semiclassical analyses of the anomalous Hall effect \cite{Sinitsyn:2008_a}.

Now we come to the interesting side-jump story.  Because the skew scattering conductivity is small, we have to 
include effects which are absent in conventional Boltzmann transport theory.  Remarkably it is possible to 
provide a successful analysis of one of the main additional effects, the side-jump correction, by means 
of a careful semiclassical analysis.  As we have mentioned previously {\em side jump} refers to the microscopic   
displacement $\delta \rr_{l,l'}$ experienced by a wave-packets formed from a spin-orbit coupled Bloch states, when scattering
 from state $l$ to state $l'$  under the influence of a disorder 
potential.  In the presence of an external electric field side jump leads to an energy shift  
$\Delta U_{l,l'}=-e\bm{ E}\cdot \delta \rr_{l,l'}$. Since we are assuming only elastic scattering, an upward 
shift in potential energy requires a downward shift in band energy and {\em vice-versa}.
We therefore need to adjust Eq. (\ref{boltz_eq}) by adding  
$\sum_{l'} \omega^{s}_{l,l'} \frac{\partial f_0(\epsilon_l)}{\epsilon_l}e \bm{ E}\cdot \delta \rr_{l,l'}$
to the r.h.s.  

But what about the side-jump itself? 
An expression for the side-jump $\delta \rr_{l,l'}$ associated with 
a particular transition can be derived by integrating 
$\dot{\rr}_c$ through a transition \cite{Sinitsyn:2008_a}.  We can write 
\begin{eqnarray}
\label{rdot}
\dot{\rr}_c&=&\frac{d}{dt}\langle \Psi_{\kk_c,\rr_c}(\rr,t)| \rr |\Psi_{\kk_c,\rr_c}(\rr,t) \rangle=
\frac{d}{dt}
\left\{ \int \frac{d\rr}{V} \int d\kk\right. \nonumber \\&& \left.   \int d\kk' w(\kk)w^*(\kk')e^{-i\kk'\cdot\rr} \left(\rr e^{i\kk \cdot \rr}\right) 
u_{n\kk'}^*(\rr) \right.  \nonumber \\&& \left. u_{n\kk}(\rr)e^{i(E(\kk')-E(\kk)) t/\hbar}\right \}
=\frac{dE_n(\kk_c)}{d\kk_c}+\\&&
\frac{d}{dt}\left\{ \int_{cell} d\rr \int d\kk w^*(\kk)u_{n\kk}(\rr)\left(i\frac{\partial}{\partial_\kk} w(\kk)u_{n\kk}(\rr) \right)   \right \} \nonumber
\end{eqnarray}
This expression is equivalent to the equations of motion derived within the Lagrangian formalism;
this form has the advantage of making it apparent that in scattering from state  $l$  to a state $l'$ 
a shift in the center of mass coordinate will accompany the velocity deflection.  From Eq.~(\ref{rdot}) it appears that  
the scattering shift will go approximately as:
\begin{equation}
\delta \rr_{l',l}\approx \langle u_{l'}| i\frac{\partial}{\partial {\bm{ k}'}} u_{l'} \rangle - 
 \langle u_{l}| i\frac{\partial}{\partial \bm{ k}} u_{l} \rangle.
\end{equation}
This quantity has usually been associated with the side-jump, although it is gauge-dependent
and therefore arbitrary in value.  The correct expression for the side jump is similar to this one, at least for the  
smooth impurity potentials situation, but was derived only recently by Sinitsyn {\em et al.} \cite{Sinitsyn:2005_a,Sinitsyn:2006_a}:
\begin{equation}
 \delta \bm{ r}_{l'l} = \langle u_{l'}| i\frac{\partial}{\partial {\bm{ k}'}} u_{l'} \rangle - 
 \langle u_{l}| i\frac{\partial}{\partial {\kk}} u_{l} \rangle
 - \hat{\bm{ D}}_{{\kk',\kk}} {\rm arg}[\langle u_{l'}|u_{l}\rangle],
\label{delr4b}
\end{equation}
where ${\rm arg}[a]$ is the phase of the complex number $a$ and
$
\hat{\bm{ D}}_{{\kk',\kk}}= \frac{\partial}{\partial {\kk'}} + \frac{\partial}{\partial {\kk}}
$.
The last term is  essential and makes the expression for the resulting side-jump gauge invariant.
Note that the side-jump is independent of the details of the impurity potential or of the scattering 
process.  As this discussion shows, the side jump contribution to motion during a scattering event is analogous to the 
anomalous velocity contribution to wave-packet evolution between collisions, with the role of the disorder potential 
in the former case taken over in the latter case by the external electric field.

\subsubsection{Kinetic equation for the semiclasscial Boltzmann distribution}
Equations (\ref{WT}) and (\ref{delr4b}) contain the quantum mechanical information necessary to write down a semiclasscial Boltzmann
equation that takes into account both the change of momentum and the coordinate shift during scattering 
in the presence of a driving electric field $\bm{E}$. Keeping only terms up to linear order in the electric field the Boltzmann
equation reads \cite{Sinitsyn:2006_a}:
\begin{equation}
\begin{array}{l}
  \frac{\partial f_l}{\partial t} +e\bm{E}\cdot{\bm{ v}}_{0l} \frac{\partial f_{0} (\epsilon_l) }{\partial {\epsilon_l}}
  = - \sum \limits_{l'}
   \omega_{ll'} [ f_{l}-
   f_{l'}-\frac{\partial f_{0} (\epsilon_l) }{\partial {\epsilon_l}}e\bm{ E} \cdot \delta {\rr}_{l'l}],\\
\end{array} 
\label{beint}
 \end{equation}
 where ${\bm v}_{0l}$ is the usual group velocity 
$
{\bm{ v}}_{0l} = \partial \epsilon_l/\partial {\kk}.
$
Note that for elastic scattering we do not need to take account of the Pauli blocking which yields 
factors like $f_{l} (1-f_{l'})$ on the r.h.s. of Eq. (\ref{beint}), and that the collision terms are 
linear in $f_{l}$ as a consequence. 
(For further discussion of this point see Appendix B in Ref. \onlinecite{Luttinger:1955_a}.)
This Boltzmann equation has the standard form except for the coordinate shift contribution to the collision 
integral explained above.
Because of the side-jump effect, the collision term
does not vanish when the occupation probabilities $f_l$ are 
replaced by their thermal equilibrium values when an external electric field 
is present:
$f_{0}(\epsilon_l)-f_{0}(\epsilon_l-e\bm{ E} \cdot \delta {\rr}_{ll'}) \approx -\frac{\partial f_{0} 
(\epsilon_l) }{\partial {\epsilon_l}}e\bm{ E} \cdot \delta {\rr}_{l'l} \ne 0$. 
Note that the term containing $\omega_{l,l'}f_l$
should be written as $\omega^{(s)}_{l,l'}f_l-\omega^{(a)}_{l,l'} f_l$.  In making this simplification 
we are imagining a typical simple model in which 
the scattering rate depends only on the angle between $\kk$ and $\kk'$.  In that case, $\sum_{l'}\omega^{(a)}_{l,l'}=0$ 
and we can ignore a complication which is primarily notational.

The next step in the Boltzmann theory is to solve for the non-equilibrium distribution function $f_{l}$ to leading order
in the external electric field.  We linearize by writing $f_{l}$ as  
the sum of the equilibrium distribution $f_{0}(\epsilon_l)$ and non-equilibrium corrections: 
\begin{equation}
f_{l}=f_{0}(\epsilon_l)+g_{l}+g_l^{adist},
\label{ffgg}
\end{equation}
where we split the non-equilibrium contribution into two terms $g_l$ and $g^{\rm adist}$ in order 
to capture the skew scattering effect.  $g_l$ and $g^{\rm adist}$ 
solve independent self-consistent time-independent equations \cite{Sinitsyn:2006_a}:
\begin{equation}
 e\bm{E}\cdot \bm{ v}_{0l} \frac{\partial f_{0} (\epsilon_l) }{\partial {\epsilon_l}}
  = -\sum_{l'} \omega_{ll'} (g_{l}-g_{l'} )
\label{bbeint3}
\end{equation}
and
\begin{equation}
\sum_{l'}\omega_{ll'}  \left( g^{adist}_l - g^{adist}_{l'} -\frac{\partial f_0(\epsilon_l)}{\partial \epsilon_l} e\bm{E} \cdot  \delta \bm{ r}_{l'l} \right) =0.
\label{ganl}
\end{equation}
In Eq.~(\ref{delr4b}) we have noted that $ \delta \bm{r}_{l'l} =  \delta \bm{r}_{ll'}$. 
To solve Eq.~(\ref{bbeint3}) we further decompose $g_l=g_{l}^s+g_l^{a1}+g_l^{a2}$ so that 
\begin{equation}
\sum_{l'} \omega_{ll'}^{(3a)} (g_{l}^{s}-g_{l'}^s )+\sum_{l'} \omega_{ll'}^{(2)} (g_{l}^{a1}-g_{l'}^{a1} )=0,
\label{bbeint3a}
\end{equation}
\begin{equation}
\sum_{l'} \omega_{ll'}^{(4a)} (g_{l}^{s}-g_{l'}^s )+\sum_{l'} \omega_{ll'}^{(2)} (g_{l}^{a2}-g_{l'}^{a2} )=0.
\label{bbeint4a}
\end{equation}
Here $g_l^s$ is the usual diagonal non-equilibrium distributions function which can be shown to be proportional
to $n_i^{-1}$.  From Eq.~(\ref{omasym1}) and  $\omega_{ll'}^{(3a)} \sim n_i$, it
follows that $g_l^{3a} \sim n_i^{-1}$. Finally, from $ \omega_{ll'}^{(4a)} \sim n_i^2$ and Eq.~(\ref{bbeint4a}), it follows that $g_l^{4a} \sim n_i^0$;
illustrating the dangers of ignoring the $\omega_{ll'}^{(4a)}$ contribution to $\omega_{l,l'}$.
One can also show from Eq.~(\ref{ganl}) that $g^{adist}_l\sim n_i^0$.

\subsubsection{Anomalous velocities,  anomalous Hall currents, and anomalous Hall mechanisms}

We are now at the final stage where we use the non-equilibrium distribution function derived from
Eqs.~(\ref{bbeint3})-(\ref{bbeint4a}) to compute the anomalous Hall current.
To do so we need first to account for all contributions to the velocity of semiclassical particles
that are consistent with this generalized semiclassical Boltzmann analysis. 
In addition to the band state group velocity $\bm{ v}_{0l}=\partial \epsilon_l /\partial {\kk}$,
we must also take into account the velocity contribution due to the accumulations of 
coordinate shifts after many scattering events, another way in which the side-jump effect
enters the theory, and the velocity contribution from coherent 
band mixing by the electric field (the anomalous velocity effect) \cite{{Nozieres:1973_a},{Sinitsyn:2006_a},{Sinitsyn:2008_a}}:
\begin{equation}
{\bm{v}}_l=\frac{\partial \epsilon_l }{\partial {\kk}} + {\bm{b}}_l  \times e \bm{E} + \sum_{l'} \omega_{l'l} \delta {\rr}_{l'l}.
\label{tovel}
\end{equation}
Combining Eqs.~(\ref{ffgg}) and (\ref{tovel}) we obtain the total current 
\begin{eqnarray}
\bm{ j} &=& e \sum_l f_l \bm{ v}_l=e \sum_l
(f_{0}(\epsilon_l)+g_{l}^s+g_l^{a1}+g_l^{a2}+g_l^{adist}) \nonumber \\ 
&&\times(\frac{\partial \epsilon_l }{\partial {\kk}} + {\vec{\Omega}}_l  \times e \bm{ E} + \sum_{l'} \omega_{l'l} \delta {\rr}_{l'l})
\label{jtotal}
\end{eqnarray}
This gives {\it five} non-zero contributions to the AHE up to linear order in $\bm{ E}$:
\begin{equation}
\sigma_{xy}^{total}=\sigma_{xy}^{int}+\sigma_{xy}^{adist}+\sigma_{xy}^{sj}+\sigma_{xy}^{sk1}
+\sigma_{xy}^{sk2-sj}.
\label{jttt}
\end{equation}
The first term is the  intrinsic contribution which should be by now familiar to the reader:
\begin{equation}
\sigma_{xy}^{int} = -e^2 \sum_l f_0(\epsilon_l) b_{z,l}.
\label{jtotal_i}
\end{equation}
Next are the effects due to coordinate shifts during scattering events (for $\bm{ E}$ along the y-axis):
\begin{equation}
\sigma_{xy}^{adist} = e \sum_l( g_l^{adist}/E_y) (v_{0l})_x
\label{jtotal_adist}
\end{equation}
follows from the distribution function correction due to side jumps  
while
\begin{equation}
\sigma_{xy}^{sj} =  e\sum_l (g_l/E_y) \sum_{l'} \omega_{l'l} (\delta {\rr}_{l'l})_x
\label{jtotal_sj}
\end{equation}
is the current due to the side-jump velocity, {\em i.e.}, due to the accumulation of coordinate shifts after many scattering
events. Since coordinate shifts are responsible both for $\sigma_{xy}^{adist}$ and for $\sigma_{xy}^{sj}$, 
there is, unsurprisingly, an intimate relationship between those two contributions.
In most of the literature, $\sigma_{xy}^{adist}$ is 
usually considered to be part of the side-jump contribution, {\it i.e.}, $\sigma_{xy}^{adist}+\sigma_{xy}^{sj}\rightarrow \sigma_{xy}^{sj}$.
We distinguish between the two because they are physically distinct and appear as separate contributions 
in the microscopic formulation of the AHE theory. 

Finally, $\sigma_{xy}^{sk1}$ and $\sigma_{xy}^{sk2-sj}$ are contributions arising from the asymmetric part of the collision integral
 \cite{Sinitsyn:2008_a}:
\begin{equation}
\sigma_{yx}^{sk1} = -e \sum_l( g_l^{a1}/E_x) (v_{0l})_y \sim n_i^{-1},
\label{jt1}
\end{equation}
\begin{equation}
\sigma_{yx}^{sk2-sj} = -e \sum_l( g_l^{a2}/E_x) (v_{0l})_y \sim n_i^{0}.
\label{jt2}
\end{equation}
According to the old definition of skew scattering both
could be viewed as skew scattering contributions because they originate from the asymmetric 
part of the collision term \cite{Smit:1955_a}. 
However, if instead we {\it define} skew-scattering as the contribution proportional to $n_i^{-1}$, {\it i.e.} linear in $\tau$, as in Sec.~\ref{mechanisms-intro},
it is only the  first  contribution, Eq.~(\ref{jt1}), which is  the skew scattering \cite{Luttinger:1958_b,Leroux-Hugon:1972_a}. 
The second contribution, Eq.~(\ref{jt2}),
was generally discarded in prior semiclassical theories, although it is parametrically of the same size as the side-jump conductivity. 
Explicit quantitative estimates of 
$\sigma_{yx}^{sk2-sj}$ so far exist only for the massive 2D Dirac band \cite{Sinitsyn:2007_a}. 
In parsing this AHE contribution we will incorporate it within the family of side-jump effects due to the fact that
it proportional to $n_i^0$, {\it i.e.} independent of $\sigma_{xx}$. However, it is important to note that this "side-jump" contribution
has no phyiscal link to the side-step experienced by a semiclasscial quasiparticle upon scattering. 
An alternative terminology for this contribution is  {\em intrinsic skew scattering} to distinguish its physical origin 
from side-jump deflections~\cite{Sinitsyn:2008_a}, but, to avoid further confusion, we simply include it as a contribution to the Hall conductivity 
which is of the order of $\sigma_{xx}^0$ and originates from scattering. 

As we will  see below, when connecting  the microscopic formalism to the semiclasscial one, $\sigma_{xy}^{int}$, can be directly identified
with the single bubble (Kubo formalism) contribution to the conductivity, $\sigma_{xy}^{adist}+\sigma_{xy}^{sj}+
+\sigma_{xy}^{sk2}$ constitute the usually termed ladder-diagram vertex corrections to the conductivity due to scattering and therefore it is natural
to group them together although their physical origins are distinct. 
$\sigma_{xy}^{sk1}$ is identified directly with the  three-scattering diagram used in the literature.
This comparison has been made specifically for two simple models, the massive 2D Dirac band  \cite{Sinitsyn:2007_a} and the
2D Rashba with exchange model \cite{Borunda:2007_a}.

	\subsection{Kubo formalism}
	\subsubsection{Kubo   technique for the AHE}\label{sec:theory:Kubo}

The Kubo formalism relates the conductivity to the quilibrium current-current correlation function~\cite{Kubo:1957_a}. It provides a
 fully quantum mechanical formally exact expression for the conductivity in linear response theory \cite{Mahan:1990_a}.
 We do not review the formal machinery for this approach here since can be found in many textbooks. Instead,
 emphasize the key issues in studying the AHE within this formalism and how it relates to the semiclassical formalism
 described in the previous section.
 
For the purpose of studying the AHE  it is best to reformulate the current-current Kubo formula for the conductivity
in the form of the  Bastin formula (see appendix A in \cite{Crepieux:2001_a})
which can be manipulated into the more familiar form for the conductivity of the
Kubo-Streda formula for the $T=0$ Hall conductivity
$\sigma_{xy}=\sigma_{xy}^{I(a)}+\sigma_{xy}^{I(b)}+\sigma_{xy}^{II}$ where:
\begin{equation}
\sigma_{xy}^{I(a)}=\frac{e^2}{2\pi V} {\rm Tr} \langle \hat{v}_x G^R(\epsilon_F) \hat{v}_y G^A(\epsilon_F)\rangle_c,
\label{sigmaIa}
\end{equation}
\begin{equation}
\sigma_{xy}^{I(b)}=-\frac{e^2}{4\pi V} {\rm Tr} \langle \hat{v}_x G^R(\epsilon_F) \hat{v}_y 
G^R(\epsilon_F)+\hat{v}_i G^A(\epsilon_F) \hat{v}_j G^A(\epsilon_F)\rangle_c,
\label{sigmaIb}
\end{equation}
\begin{eqnarray}
\sigma_{xy}^{II}&=&
\frac{e^2}{4\pi V} \int_{-\infty}^{+\infty} d\epsilon f(\epsilon) 
{\rm Tr}[v_x G^R(\epsilon) v_y \frac{G^R(\epsilon)}{d\epsilon} 
\nonumber\\&&
-v_x \frac{G^R(\epsilon)}{d\epsilon}v_yG^R(\epsilon)+c.c.].
\label{sigmaII}
\end{eqnarray}
Here the subscript $c$ indicates a disorder configuration average. 
The last contribution, $\sigma^{II}_{xy}$, was first derived by Streda in the context of studying the 
quantum Hall effect \cite{Streda:1982_a}. 
In these equations
$G^{R/A}(\epsilon_F)=(\epsilon_F-H\pm i\delta)^{-1}$ are the retarded and advanced Green's functions evaluated
at the Fermi energy of the total Hamiltonian. 

Looking more closely $\sigma_{xy}^{II}$ we notice that every term depends on  products of retarded Green's functions only
or products of advanced Green's functions only. It can be shown
that only the disorder free part of $\sigma_{xy}^{II}$ is important in the weak disorder limit, {\em i.e.} ,
this contribution is zeroth order in the parameter $1/k_F l_{sc}$. The only effect of disorder
on this contribution (for metals) is to broaden the Green's functions (see below) through the introduction of a finite
lifetime \cite{Sinitsyn:2007_a}.
 It can therefore be shown by a similar argument that in general 
$\sigma_{xy}^{Ib}$, is of order $1/k_F l_{sc}$ and can be neglected in the weak scattering limit \cite{Mahan:1990_a}.
Thus, important disorder effects beyond simple quasiparticle lifetime broadening
are contained only in $\sigma_{xy}^{Ia}$.  For these reasons, it is standard within the Kubo formalism to 
neglect $\sigma_{xy}^{Ib}$ and evaluate the $\sigma_{xy}^{II}$ contribution with a simple lifetime broadening approximation to the
Green's function.

\begin{figure}
\begin{center}
\includegraphics[width=\columnwidth]{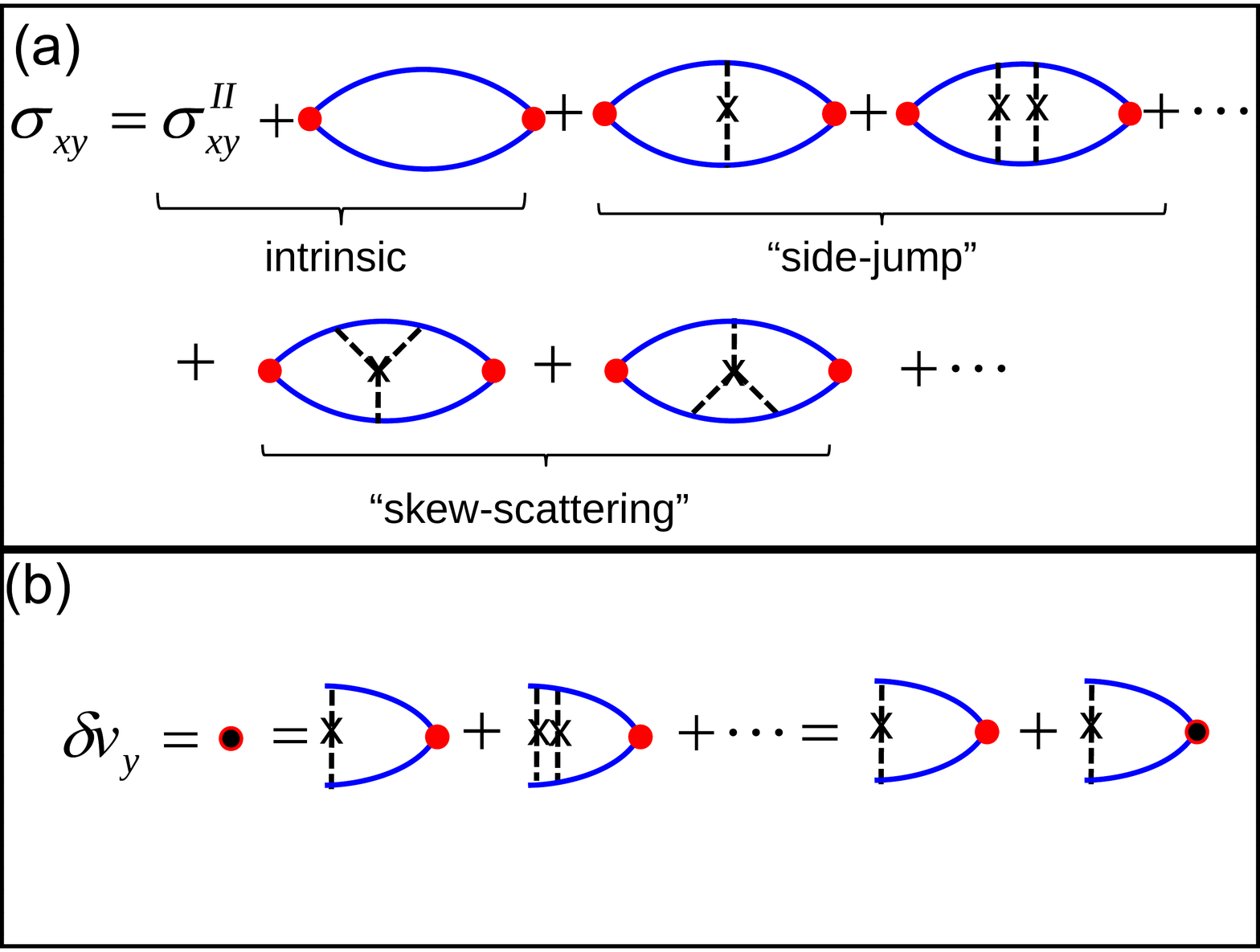}
\end{center}
\caption{Graphical representation of the Kubo formalism application to the AHE. The solid blue lines are the disorder averaged 
Green's function, $\bar{G}$, the red circles the bare velocity vertex $v_\alpha=\partial\hat{H_0}/{\partial \hbar k_\alpha}$, and the
dashed lines with crosses represent disorder scattering ($n_iV_0^2$ for the delta-scatter model).  (b) $\delta v_y$
is the velocity vertex renormalized by vertex corrections.}
\label{fig:Kubo}
\end{figure}

Within this formalism the effect of disorder on the disorder-configuration averaged Green's function is 
captured by the use of the T-matrix, defined by the integral equation $T=W+W G_0 T$, where $W=\sum_i V_0\delta(r-r_i)$
is a delta-scatterers potential
and $G_0$ are the Green's function of the pure lattice. 
From this one obtains 
\begin{equation}
\bar{G}=G_0+G_0 T G_0=G_0+ G_0 \Sigma \bar{G}.
\label{G}
\end{equation}
  Upon disorder averaging we obtain 
\begin{equation}
\Sigma=\langle W\rangle_c+\langle W G_0 W\rangle_c
+\langle W G_0 W G_0 W\rangle_c +...
\end{equation}
To linear order in  the impurity concentration, $n_i$, this translates to
\begin{equation}
\Sigma(z,\kk)= n_i V_{\kk,\kk}+\frac{n_i}{V}\sum_\kk V_{\kk,\kk'} G_0(\kk',z) V_{\kk',\kk}+\cdots,
\end{equation}
with $V_{\kk,\kk'}=V(\kk-\kk')$ being the Fourier transform of the single impurity
potential, which in the case of delta scatterers is simply $V_0$ (see Fig.~\ref{fig:Tdiagrams} for a graphical reprentation).
Note that  $\bar{G}$ and $G_0$ are diagonal in momentum but, due to the presence of spin-orbit coupling,  non-diagonal in 
spin-index in the Pauli spin-basis. Hence, the blue lines depicted in Fig.~\ref{fig:Kubo} represent $\bar{G}$ and are 
in general matrices in band levels.
 
One effect of disorder on the anomalous Hall conductivity is taken into account by inserting the 
disorder averaged Green's function, $\bar{G}^{R/A}$, directly into the expressions for  $
\sigma_{xy}^{Ia}$ and $\sigma_{xy}^{II}$, Eq.~(\ref{sigmaIa}) and Eq.~(\ref{sigmaII}).
This step captures the intrinsic contribution to the AHE and the effect of disorder on it, which is generally weak 
in metallic systems.  This contribution is separately identified in Fig.~\ref{fig:Kubo} a.

The so-called ladder diagram vertex corrections, 
also separately identified in Fig.~\ref{fig:Kubo}, contribute to the AHE at the same order in
$1/k_Fl$ as the intrinsic contribution. It is useful to define a ladder-diagram 
corrected velocty vertex $\tilde{v}_\alpha(\epsilon_F)\equiv v_\alpha+\delta\tilde{v}_\alpha(\epsilon_F)$, where 
\begin{equation}
\delta\tilde{v}_\alpha(\epsilon_F)=\frac{n_i V_0^2}{V}\sum_\kk \bar{G}^R(\epsilon_F) (v_\alpha+\delta\tilde{v}_\alpha(\epsilon_F)) \bar{G}^A(\epsilon_F),
\label{vertex_corr}
\end{equation}
as depicted in Fig.~\ref{fig:Kubo} b. Note again that $\tilde{v}_\alpha(\epsilon_F)$ and $ v_\alpha=\partial\hat{H_0}/{\partial \hbar k_\alpha}$ are matrices in the 
spin-orbit coupled band basis.
The skew scattering contributions are obtained by evaluating, without doing an infinite partial sum as in the case of the
ladder diagrams, third order processes in the disorder scattering shown in Fig.~\ref{fig:Kubo}.

As  may seem obvious from the above machinery, calculating the intrinsic contribution is not very difficult, while calculating the full effects of the disorder
in a systematic way (beyond calculating a few diagrams)
is  challenging for any disorder model beyond the simple delta-scattering model. 

Next we illustrate the full use of this formalism for the simplest nontrivial model, massive 2D Dirac fermions, with the goals of illustrating the complexities 
present in each contribution to the AHE {\it and}  the equivalence of quantum and semiclassical approaches. This model is of course not directly
 linked to any real material reviewed
in Sec.~\ref{sec:experiments} and its main merit is the possibility of obtaining full simple analytical expressions for each of the contributions.
A more realistic model of 2D fermions with Rashba spin-orbit coupling will be discussed in Sec. \ref{sec:theory:quantum}. 
The ferromagnetic Rashba model has been used to 
 propose a minimal model of AHE for materials in which band crossing near the Fermi surface dominate the AHE physics \cite{SOnoda:2008_a}.

The massive 2D Dirac fermion model is specified by:
\begin{equation}
\hat{H}_0=v(k_x \sigma_x +k_y \sigma_y) +\Delta \sigma_z +V_{dis},
\label{dh1}
\end{equation}
where $V_{dis}=\sum_i V_0 \delta ({\bf r}-{\bf R}_i)$, $\sigma_x$ and $\sigma_y$ are Pauli matrices and the impurity free spectrum is $\epsilon_{{\bf k}}^{\pm}=\pm \sqrt{\Delta^2+(v k)^2}$
where $k=|{\bf k}|$  and the labels $\pm$ distinguish bands with positive and negative energies. We ignore in this simple model spin-orbit coupled disorder contributions which can 
be directly incorporated through similar calculations as in Crepieux \etal  ~\cite{Crepieux:2001_a}.
Within this model the disordered averaged Green's function is
\begin{equation}
\begin{array}{l}
\bar{G}^R=\frac{1}{1/G_0^R - \Sigma^{R}}
=\frac{\epsilon_F+i\Gamma +v(k_x \sigma_x +k_y \sigma_y)+(\Delta - i \Gamma_1) \sigma_z}
{(\epsilon_F-\epsilon^+ +i\Gamma^+)(\epsilon_F -\epsilon^- +i\Gamma^-)}
\end{array},
\label{gf2}
\end{equation}
where $\Gamma=\pi n_iV_0^2/(4 v^2)$, $\Gamma_1 = \Gamma \cos(\theta)$, $\gamma^\pm =\Gamma_0 \pm\Gamma_1 \cos(\theta)$,
and $\cos\theta=\Delta/\sqrt{(vk)^2+\Delta^2}$. Note that within this disorder model $\tau\propto 1/n_i$. Using the result in Eq.~(\ref{gf2}) one can calculate  the ladder diagram correction
to the bare velocity vertex given by Eq.~(\ref{vertex_corr}):
\begin{equation}
\tilde{v}_y=8v\Gamma \cos\theta\frac{(1+\cos^2\theta)}{\lambda(1+3\cos^2\theta)^2}\sigma_x
+\left(v+v\frac{\sin^2\theta}{(1+3\cos^2\theta)}\right)\sigma_y,
\end{equation}
where $\theta$ is evaluated at the Fermi energy.
The details of the calculation of this vertex correction is described in Appendix A of Ref. \onlinecite{Sinitsyn:2007_a}.
Incorporating this result in $\sigma_{xy}^{Ia}$ we obtain the intrinsic and side-jump contributions to the conductivity for
$\epsilon_F> \Delta$ 
\begin{equation}
\begin{array}{l}
\sigma_{xy}^{int}=\frac{e^2}{2\pi \hbar V}\sum_\kk {\rm Tr}\left[ v\sigma_x G {v}\sigma_yG \right]
=-\frac{e^2 \cos\theta}{4\hbar \pi}
\end{array},
\label{eq_sint}
\end{equation}
\begin{equation}
\begin{array}{l}
\sigma_{xy}^{sj}=\frac{e^2}{2\pi \hbar V}\sum_\kk {\rm Tr}\left[ v\sigma_x G \delta \tilde{v}_yG \right] \\
\\
=-\frac{e^2 \cos\theta}{4\pi\hbar}\left(\frac{3\sin^2\theta}{(1+3\cos^2\theta)} + \frac{4\sin^2\theta}{(1+3\cos^2\theta)^2}\right)
\end{array}.
\label{eq_ssj}
\end{equation}
The direct calculation of the skew-scattering diagrams of Fig.~\ref{fig:Kubo} is $\sigma_{xy}^{sk}=
 -  \frac{ e^2  }{2\pi \hbar nV_0 } \frac{\ (vk_F)^4 \Delta}{  (4\Delta^2+(vk_F)^2)^2}$
and the final total result is given by
\begin{eqnarray}
\sigma_{xy}&=-\frac{e^2\Delta }{4\pi \hbar \sqrt{(vk_F)^2+\Delta^2}} [1+
\frac{4(vk_F)^2}{4\Delta^2+(vk_F)^2}\nonumber\\&+
\frac{3(vk_F)^4}{(4\Delta^2+(vk_F)^2)^2} ]  -  \frac{ e^2 }{2\pi \hbar nV_0 } \frac{\ (vk_F)^4 \Delta}{  (4\Delta^2+(vk_F)^2)^2}.
\label{sxy2}
\end{eqnarray}
\begin{widetext}

\begin{figure}[h]
\includegraphics[width=16 cm]{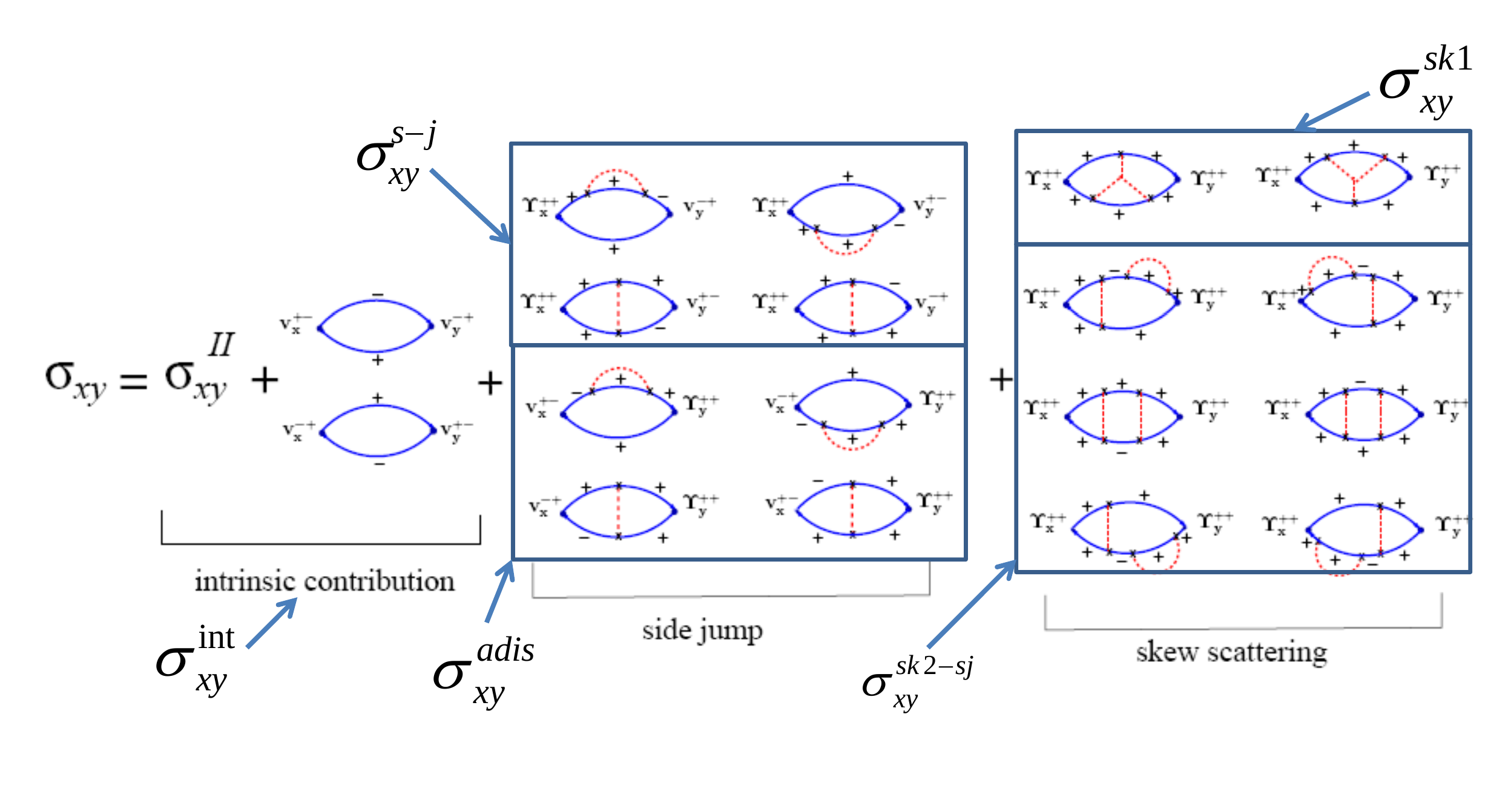}
\caption{Graphical representation of the AHE conductivity in the chiral (band eigenstate) basis.
The two bands of the two-dimensional Dirac model are labeled "$\pm$". The subset of diagrams
that correspond to specific terms in the semiclassical Boltzmann formalism
are indicated.}
\label{fig:Kubo_semiclassical}
\end{figure}

\end{widetext}
\subsubsection{Relation between the Kubo and the semiclassical formalisms}\label{kubo_semi_compar}
When comparing the semiclassical formalism to the Kubo formalism one has to keep in mind that in the semiclassical formalism
the natural basis is the one that diagonalize the spin-orbit coupled Hamiltonian. In the case of the 2D massive Dirac model this is sometimes called chiral basis in the 
literature. On the other hand in the application of the Kubo formalism it is simplest to compute the different Green's functions and vertex 
corrections in the Pauli basis and take the trace at the end of the calculation. In the case of the above model
one can apply the formalism of Sec. \ref{sec:theory:Boltzmann} and obtain the following results for the five distinct 
contributions: $\sigma_{xy}^{int}$,  $\sigma_{xy}^{adist}$, $\sigma_{xy}^{sj}$,
$\sigma_{xy}^{sk2-sj}$ and $\sigma_{xy}^{sk1}$. Below we quote the results for the complicated 
semiclassical calculation of each term (see Sec. IV of Sinitsyn \etal.~\cite{Sinitsyn:2007_a} for details):
\begin{equation}
\sigma_{xy}^{intr}=-\frac{e^2\Delta } {4\pi \hbar \sqrt{\Delta^2+(vk_F)^2}};
\label{sxyintr1}
\end{equation}
\begin{equation}
\sigma_{xy}^{sj}=\sigma_{xy}^{adist}=-\frac{e^2\Delta k_F^2}{2\pi \hbar \sqrt{k_F^2 +\Delta^2} (k_F^2+4\Delta^2)};
\label{vsj4}
\end{equation}
\begin{equation}
\sigma_{xy}^{sk-sj}=-\frac{e^2 3\Delta (vk_F)^4}{4\pi \hbar \sqrt{(vk_F)^2+\Delta^2} [4\Delta^2+(vk_F)^2]^2}; 
\label{sk-sj}
\end{equation}
\begin{equation}
\sigma_{xy}^{sk}=- \frac{ e^2 }{2\pi \hbar n_i V_0 } \frac{\ (vk_F)^4 \Delta}{  (4\Delta^2+(vk_F)^2)^2}. 
\label{skew}
\end{equation}
The correspondance with the Kubo formalism results can be seen after a few algebra steps. The contributions $\sigma_{xy}^{int}$
and $\sigma_{xy}^{sk1}$ are equal in both cases (Eq.~(\ref{eq_sint}) and Eq.~(\ref{sxy2})). As expected, the intrinsic contribution, $\sigma_{xy}^{intr}$, is independent
of disorder in the weak scattering limit and the skew scattering contribution is inversely proportional to the density of scatterers.
However, recall that in Sec. \ref{mechanisms-intro} we have defined the side-jump contribution as the disorder contributions of zeroth order in $n_i$, 
{\it i.e.}, $\tau^0$, as opposed 
to being directly linked to a side-step in the scattering process in the semiclassical theory. Hence, it is the sum of the three physically distinct processes
$\sigma_{xy}^{adist}+\sigma_{xy}^{sj}+\sigma_{xy}^{sk2-sj}$ which can be shown to be identical to Eq. (\ref{eq_ssj}) after some algebraic manipulation. 
Therefore, the old notion of associating the skew scattering directly with the asymmetric part of the collision integral  and the 
side-jump with the side-step scattering alone leads to contradictions with their usual association with respect to the dependence on $\tau$ (or equivalenty $1/n_i$).
We also note, that unlike what happens in simple models where the Berry's curvature is a constant in momentum space, {\it e.g.}, the standard model for electrons in a 
3D semiconductor conduction band~\cite{Nozieres:1973_a}, 
the dependence of the intrinsic and side-jump contributions are quite different with respect to 
parameters such as Fermi energy, exchange splitting, etc.

	\subsection{Keldysh formalism}\label{sec:theory:quantum}

 \label{sec:theory:Keldysh}

Keldysh has developed a Green's function formalism applicable even to the nonequilibrium quantum states,
for which the diagram techniques based on Wick's theorem can be used~\cite{Keldysh:1965_a,Baym:1961_a,Kadanoff:1962_a,Rammer:1986_a,Mahan:1990_a}. 
Unlike with the thermal (Matsubara) Green's functions, the Keldysh Green's functions are defined for any quantum state.
The price for this flexibility is that one needs to introduce the path-ordered product for the contour from
$t=- \infty \to t=\infty$ and back again from $t=\infty \to - \infty$.
Correspondingly, four kinds of the Green's functions $G^R$, $G^A$, $G^<$ and $G^>$ need to be considered,
although only three are independent~\cite{Keldysh:1965_a,Baym:1961_a,Kadanoff:1962_a,Rammer:1986_a,Mahan:1990_a}.
Therefore, the diagram technique and the Dyson equation for the Green's function have a matrix form.

In  linear response theory, one can use the usual thermal Green's function and Kubo formalism.
Since approximations are 
normally required in treating disordered systems, it is important to make them in 
a way which at least satisfies gauge invariance.  In both formalisms this is an 
important theoretical requirement which requires some care.
Roughly speaking, in the Keldysh formalism, 
$G^R$ and $G^A$ describe the single particle states, while 
$G^<$ represents the non-equilibrium particle occupation distribution and  contains vertex corrections.
Therefore, the self-energy and vertex corrections can be treated in a unified way by solving the
matrix Dyson equation. This facilitates the analysis of some models especially when  multiple bands
are involved.  

Another and more essential advantage of Keldysh formalism over the semiclassical formalism 
is that one can go beyond a finite order perturbative treatment of  impurity scattering strength by 
solving a self-consistent equation, as will be discussed in the next subsection.  In essence, we are assigning a finite spectral width to the semiclassical
wave packet to account for an important consequence of  quantum scattering effects. 
In the Keldysh formalism, the semiclassical limit corresponds to ignoring the history of 
scattering particles by keeping the two 
time labels in the Greens functions identical.

We restrict ourselves below to the steady and uniform solution. 
For more generic cases of electromagnetic fields, see Ref.~\onlinecite{Sugimoto:2007_a}.
Let $x=(t,\bm{x})$ be the time-space coordinate.  
Green's functions depend on two space-time points $x_1$ and $x_2$, and the matrix Dyson equation
for the translationally invariant system reads~\cite{Rammer:1986_a}:
\begin{eqnarray}
  (\varepsilon-\hat{H}(\bm{p})-\underline{\hat{\Sigma}}(\varepsilon,\bm{p}))\otimes\underline{\hat{G}}(\varepsilon,\bm{p})&=&1,
  \nonumber\\
  \underline{\hat{G}}(\varepsilon,\bm{p})\otimes(\varepsilon-\hat{H}(\bm{p})-\underline{\hat{\Sigma}}(\varepsilon,\bm{p}))&=&1,
  \label{eq:keldysh:Dyson:pi}
\end{eqnarray}
where we have changed the set of variables $(x_1;x_2)$ to the center-of-mass and the relative coordinates, 
and then proceeded to the Wigner representation $(X;p)$ by means of the Fourier transformation of the relative coordinate;
\begin{equation}
(X,x)\equiv\left(\frac{x_1+x_2}{2},x_1-x_2\right) \to \int\!dt\!\int\!d\bm{x}\,e^{i(\varepsilon t-\bm{p}\cdot\bm{x})/\hbar}\cdots,
\end{equation}
with $p=(\varepsilon,\bm{p})$.
In this Dyson equation, the product $\otimes$ iis reserved for matrix products in band indices, like those that also appear in the Kubo formalism.

In the presence of the external electromagnetic field $A_\mu$,  we must
 introduce the mechanical or kinetic energy-momentum variable 
\begin{equation}
  \pi_\mu(X;p) =p_\mu+eA_\mu(X).
  \label{eq:keldysh:pi:X;p}
\end{equation}
replacing $p$ as the argument of the Green's function, as shown by Onoda \etal~\cite{SOnoda:2006_a}.
In this representation, $\hat{G}^<(X;\pi)/2\pi i$ is  the quantum-mechanical generalization of the semi-classical distribution function.
When an external electric field  $\bm{E}$ is present, the equation of motion, or equivalently, the Dyson equation, retains
the same form as Eq.~(\ref{eq:keldysh:Dyson:pi}) when the product $\otimes$ is replaced by the 
so-called Moyal product~\cite{Moyal:1949_a,SOnoda:2006_a} given by 
\begin{equation}
  \otimes = \exp\left[\frac{i\hbar (-e)}{2}\bm{E}\cdot
(\overleftarrow{\partial}_\varepsilon\overrightarrow{\bm{\nabla}}_p-\overleftarrow{\bm{\nabla}}_p\overrightarrow{\partial}_\varepsilon)\right].
  \label{eq:keldysh:Moyal:pi}
\end{equation}
Henceforth, $\overrightarrow{\partial}$ and $\overleftarrow{\partial}$ denote the derivatives operating on the 
right-hand and left-hand sides, respectively, and the symbol $p=(\varepsilon,\bm{p})$ is used to represent the mechanical energy-momentum $\pi$.
In this formalism, only  gauge invariant quantities appear. For example, the electric field $\bm{E}$ appears instead of
the vector potential $\bm{A}$. 

Expanding Eq.~(\ref{eq:keldysh:Moyal:pi}) in $\bm{E}$ and inserting the result in Eq.~(\ref{eq:keldysh:Dyson:pi}), 
one obtains the Dyson equation to linear order in $\bm{E}$, corresponding to  linear response theory.
The linear order terms $ \hat{G}^{\alpha}_{\bm{E}}$ and $\hat{\Sigma}^{\alpha}_{\bm{E}}$ in $\bm{E}$ are 
decomposed into two parts as
\begin{eqnarray}
  \hat{G}^<_{\bm{E}}&=&
  \hat{G}^<_{\bm{E},I}\partial_\varepsilon f(\varepsilon)
  +\left(\hat{G}^A_{\bm{E}}-\hat{G}^R_{\bm{E}}\right)f(\varepsilon),
  \ \ \ \ \ 
  \label{eq:keldysh:G^<:E}\\
  \hat{\Sigma}^<_{\bm{E}}&=&\hat{\Sigma}^<_{\bm{E},I}\partial_\varepsilon f(\varepsilon)
  +\left(\hat{\Sigma}^A_{\bm{E}}-\hat{\Sigma}^R_{\bm{E}}\right)f(\varepsilon).
  \label{eq:keldysh:Sigma^<:E}
\end{eqnarray}
Here, $f(\varepsilon)$ represents the Fermi distribution function.
In these decompositions, the first term on the r.h.s. corresponds to 
the nonequilibrium deviation of the distribution function due to the electric field 
$\bm{E}$. The second term, on the other hand, represents the change in  quantum 
mechanical wavefunctions due to $\bm{E}$, and arises due to the multiband effect~\cite{Haug:1996_a} 
through the noncommutative nature of the matrices.

\begin{figure}[htb]
  \begin{center}
    \includegraphics[width=7.0cm]{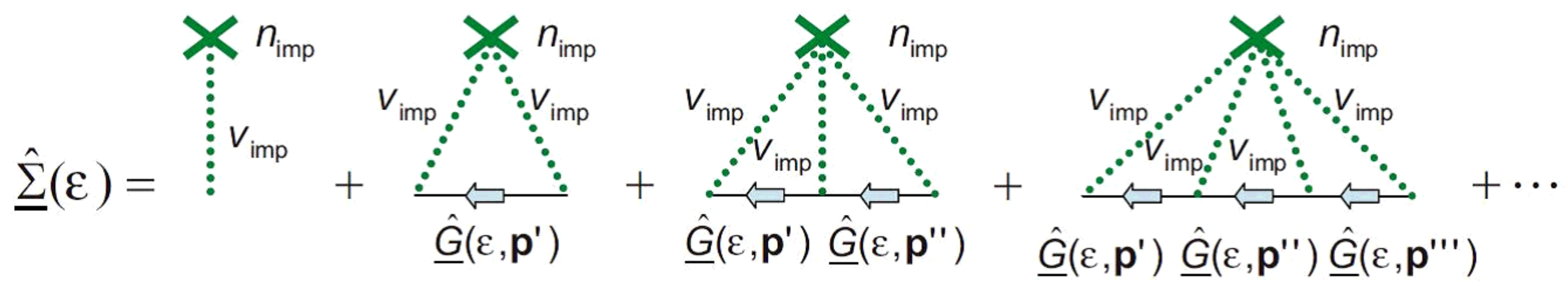}
  \end{center}
  \caption{\label{fig:Tdiagrams} Diagrammatic representation of the self-energy in the self-consistent $T$-matrix 
approximation in the Keldysh space, which is composed of the infinite series of multiple Born scattering amplitudes. [From Ref.~\onlinecite{SOnoda:2008_a}.]}
\end{figure}

The corresponding separation of conductivity contributions is: 
$\sigma_{ij}=\sigma_{ij}^I +\sigma_{ij}^{II}$ with 
\begin{eqnarray}
  \sigma_{ij}^I \!&\!=\!&\!e^2\hbar^2\!\int\!\frac{d^{d+1}p}{(2\pi\hbar)^{d+1}i}\!
  \text{Tr}\left[\hat{v}_i(\bm{p})\hat{G}^<_{E_j,I}(p)\right]
  \partial_\varepsilon f(\varepsilon),
  \label{eq:keldysh:sigma^I}\\
  \sigma_{ij}^{II} \!&\!=\!&\!e^2\hbar^2
  \!\int\!\frac{d^{d+1}p}{(2\pi\hbar)^{d+1}i}{\rm Tr}
  \left[\hat{v}_i(\bm{p}) \left(\hat{G}^A_{E_j}(p)-\hat{G}^R_{E_j}(p)\right)\right] f(\varepsilon).
  \nonumber\\
  \label{eq:keldysh:sigma^II}
\end{eqnarray}
This is in the same spirit as the Str\u{e}da version \cite{Streda:1982_a} of the Kubo-Bastin formula~\cite{Kubo:1957_a,Bastin:1971_a}.
The advantage here is that we can use the diagrammatic technique to connect the self-energy and the Green's function.
For  dilute impurities, one can take the series of diagrams shown in Fig.~\ref{fig:Tdiagrams} corresponding to 
the $T$-matrix approximation.
In this approximation, the self-consistent integral equation for the self-energy and Green's function can be
solved and the solution used to evaluate the first and second terms in Eq.~(\ref{eq:keldysh:G^<:E}) and 
Eq.~(\ref{eq:keldysh:Sigma^<:E}).

In general, Eqs.~(\ref{eq:keldysh:sigma^I}) and (\ref{eq:keldysh:sigma^II}), together with the self-consistent equations
for $G^{R}$,$G^{A}$, and $G^<$ \cite{SOnoda:2008_a} defines a systematic diagrammatic method for calculating $\sigma_{ij}$ in the Str\u{e}da decomposition~\cite{Streda:1982_a} of the Kubo-Bastin formula~\cite{Kubo:1957_a,Bastin:1971_a}.

    \subsection{Two-dimensional ferromagnetic Rashba model -- a minimal model} \label{sec:theory:Rashba}

A very useful model to study  fundamental aspects of the AHE is the ferromagnetic
two-dimensional (2D) Rashba model~\cite{Bychkov:1984_a}:
\begin{equation}
\hat{H}(\bm{p})_\mathrm{tot}=\frac{\bm{p}^2}{2m}-\lambda\bm{p}\times\hat{\bm{\sigma}}\cdot\bm{e}^z-\Delta_0\hat{\sigma}^z+\hat{V}(\bm{x}).
\label{eq:H_tot}
\end{equation}
Here $m$ is the  electron mass,  $\lambda$ is the Rashba spin-orbit interaction strength,  $\Delta_0$ is the mean field exchange splitting,
$\hat{\bm{\sigma}}=(\hat{\sigma}_x,\hat{\sigma}_y,\hat{\sigma}_z)$   and $\hat{\sigma}^0$ are the Pauli and identity matrices, 
$\bm{e}^z$ is  the unit vector  in the $z$ direction, and $\hat{V}(\bm{x})={V}_{0}\sum_{i}\delta(\bm{r}-\bm{r}_{i})$ is a $\delta$-scatterer impurity potential with impurity density $n_i$.
Quantum transport properties of this simple but non-trivial model
have been intensively studied in order to understand  fundamental properties of the AHE in itinerant metallic ferromagnets.
The metallic Rashba is a simple, but its AHE has both intrinsic and extrinsic contributions
and both minority and majority spin Fermi surfaces.  It therefore captures most of the features that are important in real materials with a minimum of 
complicating detail.  The model has therefore received a lot of attention~\cite{Dugaev:2005_a,Inoue:2006_a,Borunda:2007_a,Nunner:2007_a,Kato:2007_a,SOnoda:2006_a,SOnoda:2008_a,Kovalev:2008_a,Kovalev:2009_a}.

\begin{figure}
\begin{center}
\includegraphics[width=7.0cm]{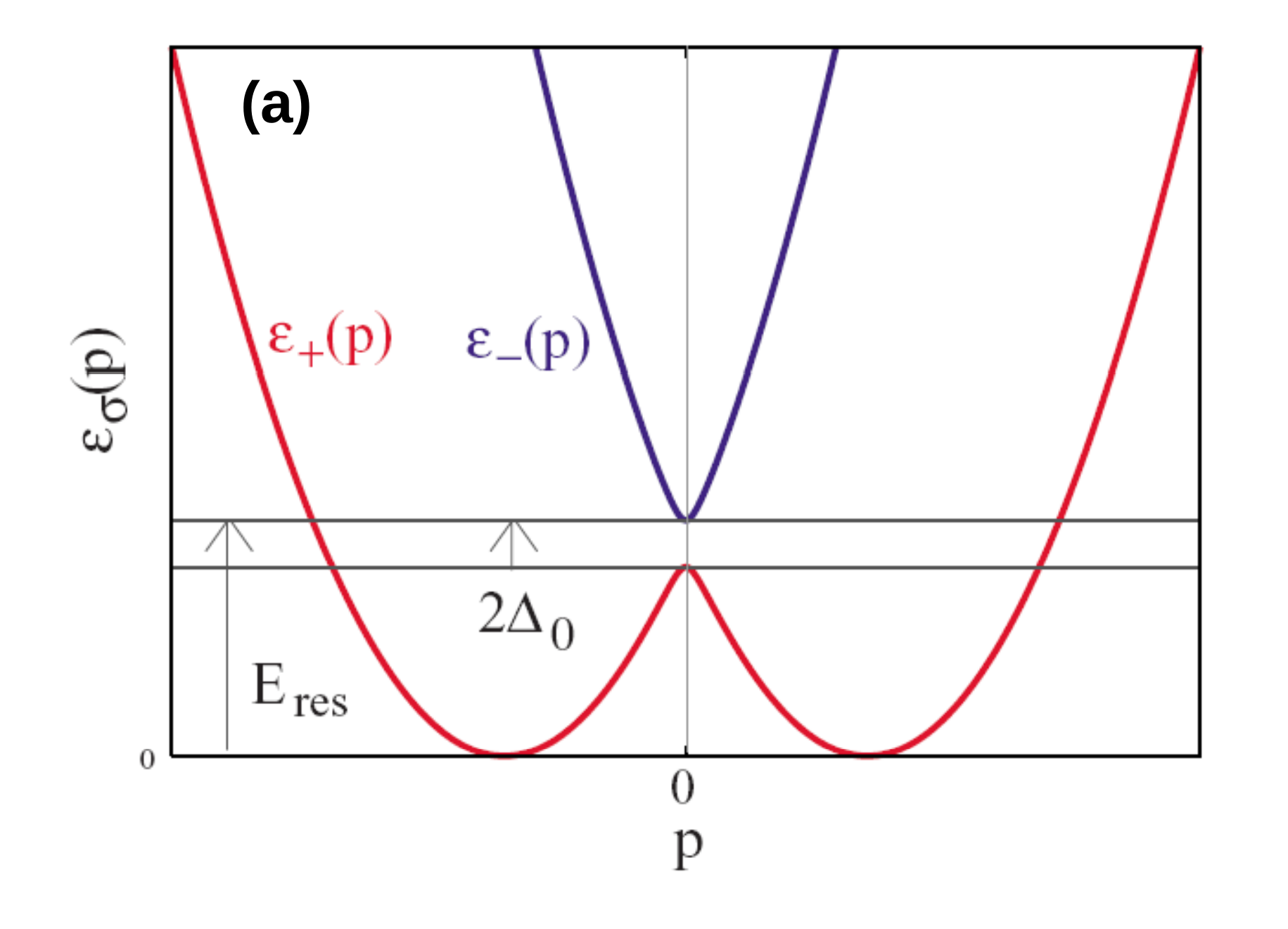}
\includegraphics[width=7.0cm]{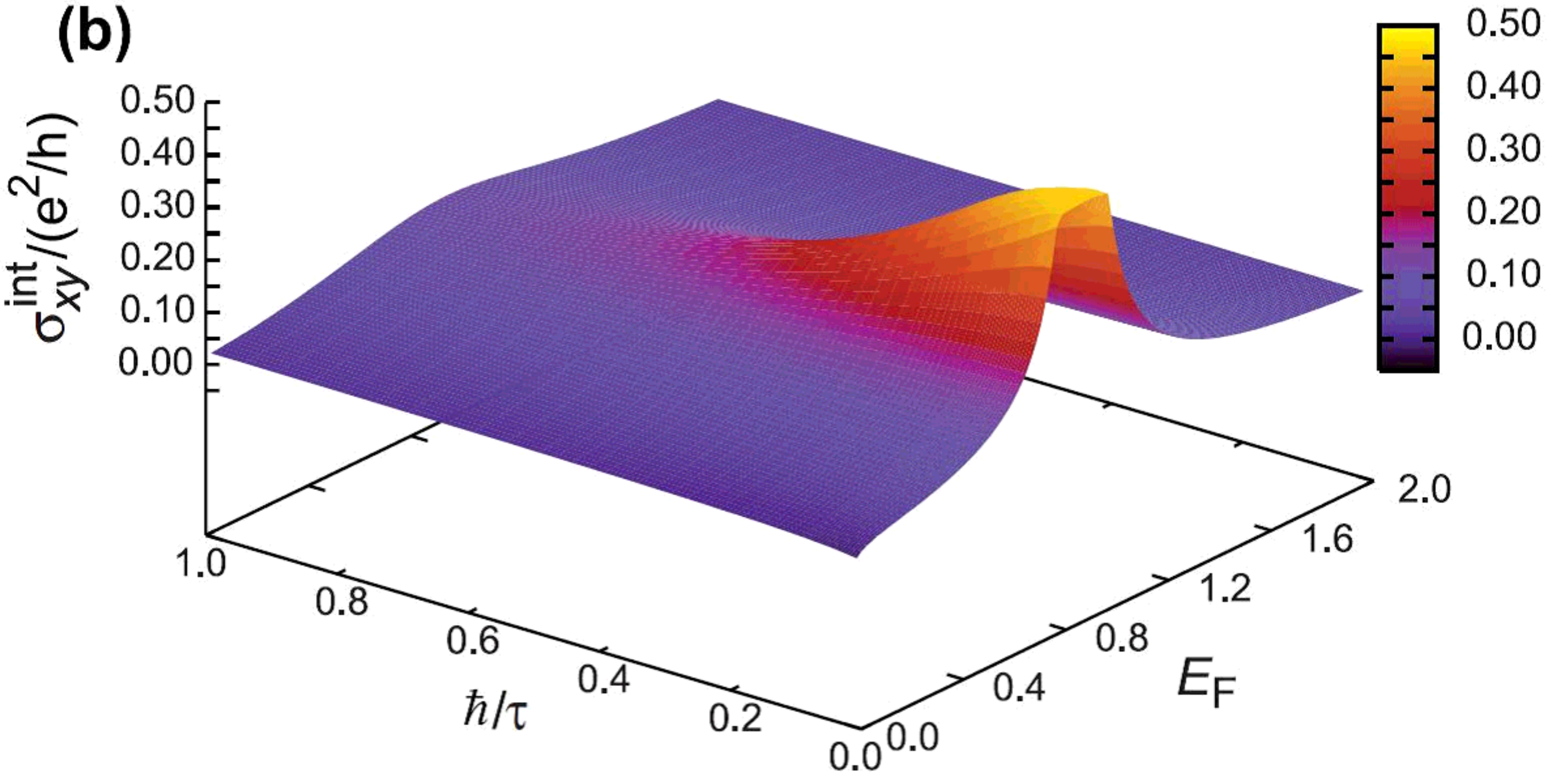}
\end{center}
\caption{(a) Band dispersion of the ferromagnetic 2D Rashba model in the clean limit. (b) The intrinsic anomalous Hall conductivity of the Hamiltonian Eq.~(\ref{eq:H_tot}) as functions of the Fermi level $\varepsilon_F$ measured from the bottom of the majority band and $\hbar/\tau\equiv mn_{\mathrm{imp}}V_0^2/\hbar^2$, which is the Born scattering amplitude for $\lambda=\Delta_0=0$. 
This figure is for the parameter set 
$\Delta_0=0.1$, $2m\lambda=3.59$, and $2mV_0/\hbar^2=0.6$, with energy unit has been taken as $\varepsilon_{0,-}$. [From Ref.~\onlinecite{SOnoda:2006_a}.]}
\label{fig:SOnoda_PRLFig2b}
\end{figure}

The bare Hamiltonian has the band dispersion:
\begin{equation}
\varepsilon_\sigma(\bm{p})=\frac{\bm{p}^2}{2m}-\sigma\Delta_{\bm{p}},
\ \ \ \ \Delta_{\bm{p}}=\sqrt{\lambda^2\bm{p}^2+\Delta_0^2},
\end{equation}
illustrated in Fig.~\ref{fig:SOnoda_PRLFig2b} a, and  Berry-phase curvature
\begin{equation}
b^z_\sigma(\bm{p})=\hbar^2\left[\bm{\nabla}_p\times(i\langle\bm{p},\sigma|\bm{\nabla}_p|\bm{p},\sigma\rangle)\right]^z=\frac{\lambda^2\hbar^2\Delta_0\sigma}{2\Delta_{\bm{p}}^3},
\end{equation}
where $\sigma = \pm$ labels the two eigenstates $|\bm{p},\sigma\rangle$ at momentum $\bm{p}$.

With this we can then obtain the intrisic contribution to the AHE by
integrating over  occupied states at zero temperature~\cite{Culcer:2003_a,Dugaev:2005_a}:
\begin{equation}
\sigma_{xy}^{\mathrm{AH-int}}=\frac{e^2}{2h}\sum_{\sigma}\sigma\left[1-\frac{\Delta_0}{\Delta_{p_\sigma}}\right]\theta
(\mu-\varepsilon_\sigma(\bm{p}_\sigma)),
\label{Rashba_sigma_int}
\end{equation}
where $p_\pm$ denotes the Fermi momentum for the band $\sigma=\pm$. 

An important feature of $\sigma_{xy}^{{AH-int}}$ is its 
enhancement in the interval of $\varepsilon_{0,+} < \mu < \varepsilon_{0,-}$, where it approaches a maximum value close to $e^2/2h$, 
no matter how small $\Delta_0$ is (provided that it is larger than $\hbar/\tau$). 
Near $\bm{p}=0$ the Berry curvatures of the two bands are large and opposite in sign. 
The large Berry curvatures translate into large intrinsic Hall conductivities 
only when the chemical potential lies between the local maximum of one band and the local minimum of the other.
This enhancement of the intrinsic AHE near avoided band crossings
is  illustrated in Fig.~\ref{fig:SOnoda_PRLFig2b} b, where it is seen to survive moderate disorder broadening of several times $\Delta_0$. This peaked feature arises  from the topological nature 
of $\sigma_{xy}^{{AH-int}}$. As a consequence it is important to note that the result is non-perturbative in SOI; 
only a perturbative expansion on $\hbar/\varepsilon_F\tau$ is justified.

Cucler \etal.~\cite{Culcer:2003_a}  were the first to study this model, obtaining Eq.~(\ref{Rashba_sigma_int}). 
They were was followed by Dugaev {\it et al.}~\cite{Dugaev:2005_a} where the intrinsic contribution was calculated
within the Kubo formalism. Although these studies found  a non-zero $\sigma_{xy}^{{AH-int}}$, they did not calculate 
all the contributions arising from disorder (some aspects of the disorder treatment in \cite{Dugaev:2005_a} where corrected
by these authors in \cite{Sinitsyn:2007_a}). The intrinsic AHE $\sigma_{xy}^{{AH-int}}$ comes form 
both $\sigma_{xy}^I$ and $\sigma_{xy}^{II}$. The first part $\sigma_{xy}^I$ contains the intra-band contribution $\sigma_{xy}^{I(a)}$ which is 
sensitive to the impurity scattering vertex correction.

The calculation of $\sigma_{xy}^{I(a)}$ incorporating the effects of disorder using the Kubo formalism, {\it i.e.}, 
incorporating the ladder vertex corrections ("side-jump")  and the leading $O(V_0^3)$ skew-scattering contributions (Sec.~\ref{sec:theory:Kubo}),
yields a vanishing $\sigma_{xy}^{AH}$ for the case where $\varepsilon_F$ is above the gap at $\bm{p}=\bm{0}$ ({\it i.e.} both subbands are occupied), irrespective of the strength of the spin-independent scattering amplitude~\cite{Inoue:2006_a,Borunda:2007_a,Nunner:2008_a}. 
On the other hand, when only the majority band $\sigma=+$ is occupied, $\sigma_{xy}^{I(a)}$ is given by the 
skew-scattering contribution~\cite{Borunda:2007_a},
\begin{equation}
\sigma_{xy}^{I(a)} \approx -\frac{e^2}{h}\frac{1}{n_{\mathrm{imp}}u_{\mathrm{imp}}}\frac{\lambda^2p_+^4D_+(\mu)\Delta_0\Delta_{p_+}}{(3\Delta_0^2+\Delta_{p_+}^2)^2},
\end{equation}
in the leading-order in $(1/n_{\mathrm{imp}})$. 

Some properties of the ferromagnetic Rashba model appear unphyiscal in the limit  $\tau\rightarrow\infty$~\cite{SOnoda:2008_a}: $\sigma_{xy}$  vanishes discontinuously as the chemical potential $\mu$ crosses the edge $\varepsilon_{0,-}$ of the minority band which leads to a diverging  anomalous Nernst effect 
 at $\mu=\varepsilon_{0,-}$, irrespective of the scattering strength, if one assumes the Mott relation  to  be valid for anomalous transport~\cite{Smrcka:1977_a}.
However, this unphysical property does not really hold, in fact,
$\sigma_{xy}$ does not vanish even when both subbands are occupied, as shown by including all  higher-order Born scattering amplitudes as
it is done automatically in the numerical Keldysh approach~\cite{SOnoda:2006_b}. In particular, the skew-scattering contribution arises from the odd-order Born scattering ({\it i.e.}, even order in the impurity potential) beyond the conventional level of approximation, $O(V_0^3)$, that gives rise to the normal skew scattering contributions~\cite{Kovalev:2008_a}. 
This yields the unconventional behavior $\sigma_{xy}^{AH-skew}\propto1/n_{\mathrm{imp}}$ independent of $V_0$~\cite{Kovalev:2008_a}. The possible appearance 
of the AHE in the case where both subbands are occupied was also suggested in the numerical diagonalization calculation of the Kubo formula~\cite{Kato:2007_a}. The 
influence of spin-dependent impurities has also been analyzed~\cite{Nunner:2008_a}.

\begin{figure}
\begin{center}
\includegraphics[width=7.0cm]{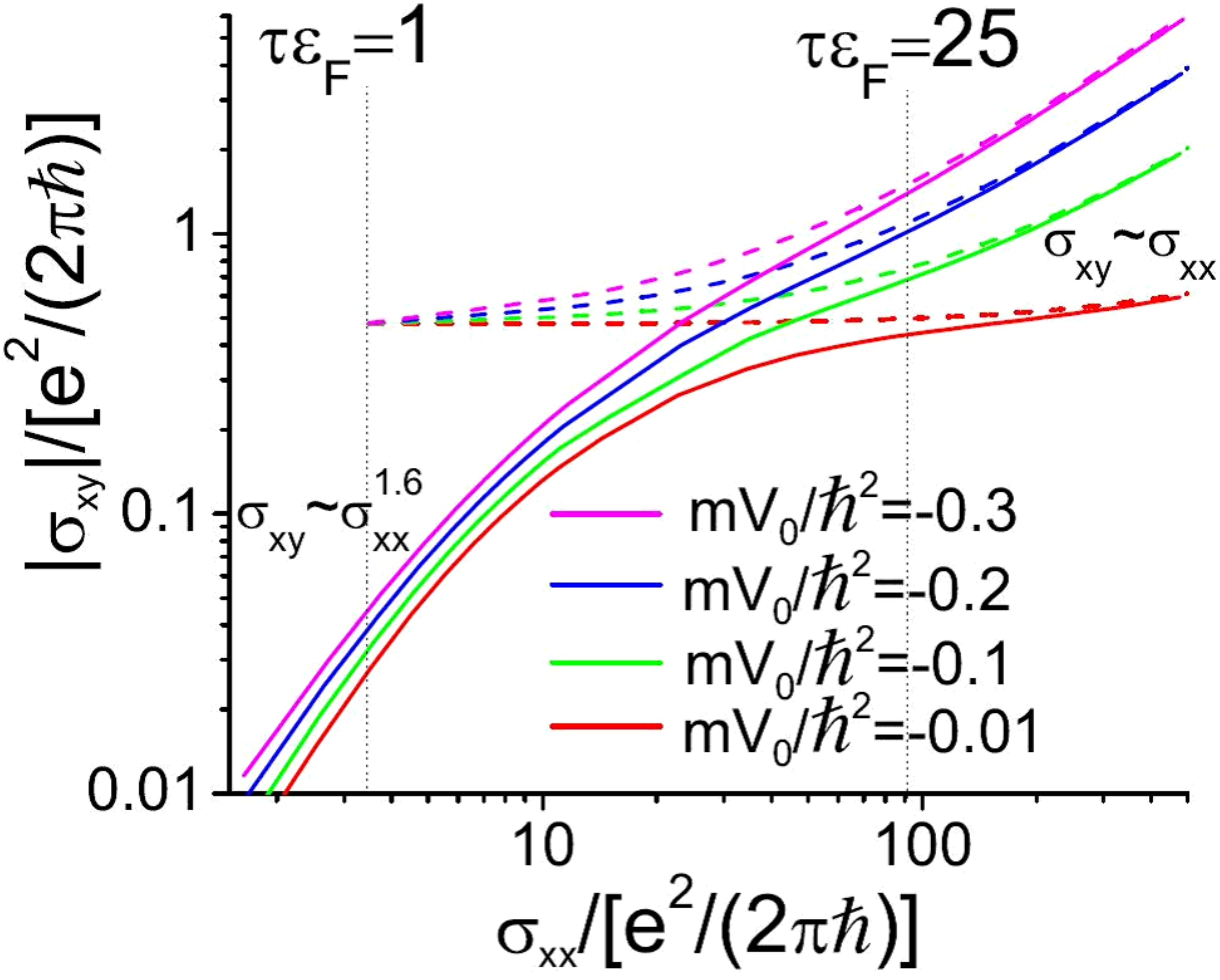}
\end{center}
\caption{\label{fig:Kovalev_Fig9}Total anomalous Hall conductivity {\em vs.} $\sigma_{xx}$ for the Hamiltonian Eq.~(\ref{eq:H_tot}) obtained in the self-consistent $T$-matrix approximation to the Keldysh approach~\cite{SOnoda:2006_a,SOnoda:2008_a,Kovalev:2009_a}.
Curves are for a variety of disorder strengths.
The same parameter values  have been taken as in Fig.~\ref{fig:SOnoda_PRLFig2b} with the chemical potential being located at the center of the two subbands. 
The dashed curves represent the corresponding semiclassical results. [After Ref.~\onlinecite{Kovalev:2009_a}.]}
\end{figure}

A numerical calculation of $\sigma_{xy}^{AH}$ based on the Keldysh formalism using the self-consistent $T$-matrix approximation,
shown in Fig.~\ref{fig:Kovalev_Fig9}, suggests three distinct regimes for the AHE as a function of $\sigma_{xx}$ at 
 low temperatures~\cite{SOnoda:2006_b,SOnoda:2008_a}. In particular, it shows a crossover from the predominant 
skew-scattering region in the clean limit ($\sigma_{xy}\propto\sigma_{xx}$) to an intrinsic-dominated metallic 
region ($\sigma_{xy}\sim\mathrm{constant}$). In this simple model no well defined plateau is observed. 
These results also suggests another crossover to a regime, referred to as the  {\it incoherent regime} by~\cite{SOnoda:2008_a}, 
where $\sigma_{xy}$ decays with the disorder following the 
scaling relation of $\sigma_{xy}\propto\sigma_{xx}^n$ with $n\approx1.6$. 
This scaling arises in the calculation due to the influence of finite-lifetime disorder broadening on  $\sigma_{xy}^{AH-int}$,
while the skew-scattering contribution is quickly dimished by disorder as expected.
Kovalev \etal.~\cite{Kovalev:2009_a} revisited the Keldysh calculations for this model, studying them 
 numerically and analytically. In particular, their study extended the calculations to include
 the dependence of the skew-scattering contribution 
on the chemical potential $\mu$ both for $\mu<\varepsilon_{0,-}$
and for $\mu>\varepsilon_{0,-}$ (as shown in Fig.~\ref{fig:Kovalev_Fig9}). 
These authors demonstrated that changing the sign of the impurity potential changes the sign of the skew-scattering contribution. 
The data collapse illustrated in Fig.~\ref{fig:Kovalev_Fig9} then fails, especially near the intrinsic-extrinsic crossover.  Data collapse in $\sigma^{xy}$ 
{\em vs.} $\sigma_{xx}$ plots  is therefore not a general property of 2D Rashba models and should not be expected in 
real materials.  There is, however, a sufficient tendency in this direction to motivate analyzing experiments by plotting data in this way.

{\it Minimal model--} The above results have the following implications for the generic nature of the AHE~\cite{SOnoda:2006_a,SOnoda:2008_a}.
The 2D ferromagnetic Rashba model can be viewed as a minimal model that takes into account both the ``parity anomaly''~\cite{Jackiw:1984_a} 
associated with an avoided-crossing of dispersing bands, as well as  impurity scattering in a system with two Fermi surface sheets. 
Consider then a general 3D ferromagnet.
When the SOI is neglected, majority and minority spin
Fermi surfaces will intersect along lines in 3D.  For a particular projection $k_x$ of  Bloch momentum along the magnetization direction, 
the Fermi surfaces will touch at points.  SOI will 
generically lift the band degeneracy at these points.
At the  $\kk$-point where the energy gap is a minimum the contribution to $\sigma_{xy}^{AH-int}$ will be a maximum
and therefore the region around this point should account for most of $\sigma_{xy}^{AH}$,
as the first-principles calculations seem to indicate \cite{Fang:2003_a,Yao:2004_a,Wang:2006_a,Yao:2007_a,Wang:2007_a}. 
We emphasize that the Berry phase contributions from the two bands are nearly opposite, so that a large contribution from this region of k-space accrues only if the Fermi level  lies between the split bands.
Expanding the Hamiltonian at this particular $\kk$-point, one can then hope to obtain an effective Hamiltonian of the
form of the ferromagnetic Rashba model. Note  that the gap $\Delta_0$ in this effective model Hamiltonian, Eq.~(\ref{eq:H_tot}),
comes physically from the SOI splitting at this particular $\kk$-point, while the "Rashba SOI" $\lambda$ 
is proportional to the  Fermi velocity near this crossing point. 
In 3D, the anomalous Hall conductivity is then given by the 2D contribution integrated over $p_z$
near this minimum gap region, and remains of  order of  $e^2/ha$ ($a$ the lattice constant) if no accidental cancellation occurs.  

We point out that a key assumption made in the above reasoning
is that the effective Hamiltonian obtained in this expansion 
is sufficiently similar to the ferromagnetic Rashba model.  We note that in the Rashba model it is $\Delta_0$ and not the SOI which opens the gap, 
so this is already one key difference.   The SO interactions in any effective Hamiltonian of this type should in general contain at least Rashba-like and 
Dresselhaus-like  contributions.  
Further studies examining the crossing points more closely near these minimum gap regions will
shed further light on the relationship between the AHE in real materials and the AHE in simple models for which detailed perturbative studies are feasible. 

The  minimal model outlined above suggests the presence of  three regimes: (i) the superclean regime dominated by the skew-scattering contribution over the intrinsic one, (ii) the intrinsic metallic regime where $\sigma_{xy}$ becomes more or less insensitive to the scattering 
strength and $\sigma_{xx}$, and (iii) the dirty regime with $k_F\ell=2\varepsilon_F\tau/\hbar\lesssim1$ exhibiting a sublinear dependence of $\rho_{xy}\propto\rho_{xx}^{2-n}$, or equivalently $\sigma_{xy}\propto\sigma_{xx}^n$ with $n\approx1.6$. It is important to note that
this minimal model is based on elastic scattering and cannot explain the scaling observed in the localized hopping conduction regime as $\sigma_{xx}$
is tuned by changing $T$. Nevertheless, 
if we multiply the 2D anomalous Hall conductivity by $a^{-1}$ with the lattice constant $a\sim 5\ \AA$ for comparison to the experimental results on three-dimensional bulk samples, then an enhanced $\sigma_{xy}^{AH}$ of the order of the quantized value $e^2/h$ in the intrinsic regime should be interpreted as $\sim e^2/ha\sim10^3\ \Omega^{-1}\ \mathrm{cm}^{-1}$. This value compares well with the empirically observed cross over seen in the
 experimental findings on Fe and Co, as discussed in Sec.~\ref{sec:tm}.

\section{Conclusions: Future problems and perspectives on AHE}\label{sec:summary}

In this concluding section, we summarize what has been achieved by 
the recent studies of the AHE and what is not yet understood, pointing 
out possible directions for future research on this fascinating phenomenon.
To keep this section brief, we exclude the historical summary presented in Sec.~\ref{sec:brief}
and Sec.~\ref{sec:theory:early} which outline the early debate on the origin of the AHE.
We avoid repeating all thel points highlighted already in
Sec.~\ref{sec:brief} and focus on the most salient ones. Citations are kept to a minimum as well
and we refer mostly to the sections in which the material was presented.

\paragraph{Recent developments}
\indent

The renewed interest in the AHE, which has lead to a richer and  more cohesive understanding 
of the problem, began in 1998 and was fueled by 
other connected developments in solid state physics. These were: i) the development of geometrical and topological concepts 
useful in understanding electronic properties such as quantum phase interference and the quantum Hall effect.~\cite{Lee:1985_a,Prange:1987_a}, 
ii) the demostration of the 
close relation between the Hall conductance and the 
toplogical Chern number revealed by the TKNN formula~\cite{Thouless:1982_a}, iii)
the development of accurate first-principles
band structure calculation which account realistically for SOI, and iv) the association of
 the Berry-phase concept~\cite{Berry:1984_a} 
with the noncoplanar spin configuration  proposed in the
context of the resonating valence bond (RVB) theory of 
cuprate high temperature superconductors~\cite{Lee:2006_a}.

\noindent {\it Intrinsic AHE--} The concept of an intrinsic AHE,  debated for a long time,
was brought back to the forefront of the AHE problem because of studies which successfully connected 
the topological properties of the quantum states of matter and the transport
 Hall response of a system. In Sec.~\ref{int-intro} we have defined $\sigma_{xy}^{AH-int}$
{\it both} experimentally  and theoretically. From the latter, it is rather straightforward to write $\sigma_{xy}^{AH-int}$
in terms of the Berry curvature
in the $\bm{k}$-space, from which the topological
nature of the intrinsic AHE can be easily  recognized immediately. 
The topological non-perturbative quality of $\sigma_{xy}^{AH-int}$
is highlighted by the finding that  for simple models with
spontaneous magnetization and SOI,  bands can have nonzero Chern numbers 
even without an external magnetic field present.
This means that  expansion with respect 
to  SOI strength is sometimes dangerous since it lifts the degeneracy between the up- and down-spin
bands, leading to  avoided band crossings which can invalidate such expansion. 

Even though the interpretations of the AHE in real systems
are still subtle and complicated, the view that $\sigma_{xy}^{AH-int}$ can be the dominant 
contribution to $\sigma_{xy}^{AH}$ in certain regimes has been strengthened by recent comparisons
of experiment and theory. 
The intrinsic AHE can be calculated 
from  first-principles calculations or, in the case of semiconductors, using  
$\bm{k} \cdot \bm{p}$ theory. These calculations have been compared to recent
experimental measurements for several materials such as
Sr$_{1-x}$Ca$_x$RuO$_3$ (section \ref{sec:exp:oxcide}), 
Fe (section  \ref{sec:tm}),   CuCr$_2$Se$_{4-x}$Br$_x$
(section \ref{sec:exp:other}), and dilute magnetic semiconductors
(section \ref{sec:exp:semi}).
The calculations and experiments show semi-quantitative agreement. 
More importantly however,  violations of the empirical relation
$\sigma_H \propto M$ have been established both theoretically and
experimentally. This suggests that the intrinsic contribution    
has some relevance to the observed AHE. On the other hand, these
studies do not always provide a compelling explanation for dominance of the intrinsic mechanism 

\noindent {\it Fully consistent metallic linear response theories of the AHE --}
Important progress has been achieved in
AHE theory.  The semiclassical theory, appropriately modified to account for interband coherence effects, has been shown
to be consistent with fully microscopic theories based on Kubo and Keldysh formalisms. 
 All three theories have been shown to be equivalent in the
$\varepsilon_F \tau \gg 1$ limit, with each having their advantages and disadvantages (Sec.~\ref{sec:linear_transport}). 
Much of the debate and confusion  in early AHE literature originated from discrepancies and farraginous results from
earlier inconsistent application of these linear response theories.

A semiclassical treatment based on 
the Boltzmann transport equation, but taking into account the 
Berry curvature and inter-band coherence effects, has been formulated 
(Sec. \ref{sec:theory:Boltzmann}). The physical picture for 
each process of AHE is now understood reasonably well in the case of elastic impurity scattering.

More rigorous treatments taking into account the 
multi-band nature of the Green's functions in terms of Kubo and 
Keldysh formalism have been developed fully (section \ref{sec:theory:Keldysh}). 
These have been applied to a particular model, {\it i.e.}, the ferromagnetic Rashba model,
with a static impurity potential which produces elastic scattering.  The ferromagnetic Rashba model has an 
avoided crossing which has been identified as a key player in the AHE of any material.
 These calculations have shown
 a region of  disorder strength over which the 
anomalous Hall conductivity stays more or less 
constant as a function of $\sigma_{xx}$, 
corresponding to the intrinsic-dominated regime. 
The emergence of this regime has been linked to the 
topological nature of the intrinsic contribution, analogous 
to the topologically protexted quantized Hall effect. 
 
\noindent {\it Emergence of three empirical AHE regimes---} 
Based on the large collections of  experimental results and
indications from some theoretical calculations, it is now becoming clear that 
there are at least three different regimes for the behavior of AHE as a function of $\sigma_{xx}$: 
(i) ($\sigma_{xx}> 10^6$ $(\Omega{\rm cm})^{-1}$) A high conductivity regime in which  $\sigma_{xy}^{AH}\sim \sigma_{xx}$,
skew scattering dominates $\sigma_{xy}^{AH}$, and the anomalous Hall angle $\sigma_H/\sigma_{xx}$ 
is  constant. In this regime however 
 the normal Hall conductivity from the Lorentz force, proportional to $\sigma_{xx}^2 H$, is large 
 even for the small magnetic field $H$ used to align  ferromagnetic domains and separating $\sigma_{xy}^{AH}$
 and  $\sigma_{xy}^{NH}$ is therefore challenging. 
(ii) ($10^4$ $(\Omega{\rm cm})^{-1} < \sigma_{xx} <10^6$ $(\Omega{\rm cm})^{-1}$)
An intrinsic or scattering-independent regime  in which 
$\sigma_{xy}^{AH}$ is roughly independent of $\sigma_{xx}$.
In this  intermediate metallic region,  where the comparison between the experiments and 
band structure calculations have been discussed, the intrinsic mechanism is {\it assumed} to be
dominant as mentioned above. The dominance of the intrinsic mechanism over side-jump is hinted at in some model calculations, 
but there is no firm understanding of the limits of this simplifying assumption.
(iii) ($\sigma_{xx} <10^4$ $(\Omega{\rm cm})^{-1}$) 
A bad-metal regime in which $\sigma_{xy}^{AH}$ decreases with decreasing $\sigma_{xx}$ at a rate faster than linear.
In this strong disorder region, a scaling  $\sigma_H \propto \sigma_{xx}^{n}$ with $1.6<n<1.7$  
has been reported experimentally for a variety of materials discussed in Sec.\ref{sec:experiments}. 
This scaling is primarily observed in insulating materials exhibiting variable range hopping transport
and where $\sigma_{xx}$ is tuned by varying $T$. The origin of this scaling is not yet understood
and is a major challenge for  AHE theory in the future.
For metallic ultrathin thin films exhibiting this approximate scaling, it is natural that $\sigma_H$ is suppressed by the strong disorder
(excluding weak localization corrections).  Simple considerations from the Kubo formula where the 
energy denominator includes a $(\hbar/\tau)^{2}$ is that $\sigma_H \propto \tau^{-2}$ when this broadening
is larger than the energy splitting between bands due to the SOI. Since in this large broadening regime
$\sigma_{xx}$ is usually no longer linear in $\tau$, an upper limit of $\beta=2$ for the scaling relation
$\sigma_H \propto \sigma_{xx}^{\beta}$ is expected. The numerical Keldysh studies of the ferromagnetic 2D Rashba model
 indicates that this power is $\beta\sim1.6$, close to what is observed in the limited dirty-metallic range considered in the experiments.
 It is a surprising feature that this scaling seems to hold for both  the metallic and 
insulating samples. 

\paragraph{Future challenges and perspectives}
\indent

In the classical Boltzmann transport theory, the 
resistivity or conductivity at the lowest temperature
is simply related to the strength of the disorder.
However,  quantum interference of the scattered
waves gives rise to a quantum correction to the
conductivity and eventually leads to 
the Anderson localization depending on the dimensionality.
At finite temperature, inelastic scattering by 
electron-electon  and/or electron-phonon interactions
give additional contributions to the resistivity, while 
suppressing  localization effects through a reduction
of the phase coherence length. In addition one needs to 
consider quantum correction due to the 
electron-electon interaction in the  presence of the disorder. 
These issues, revealed in the 80's, must be considered to scrutinize the microscopic 
mechanism of $\sigma_{xx}$ or $\rho_{xx}$ before studying the AHE. This means that 
it seems unlikely that only $\sigma_{xx}(T)$ characterizes the AHE at each temperature.
We can define the Boltzmann transport $\sigma_{xx}^B$
only when the residual resistivity is well defined 
at low temperature before the weak localization effect 
sets in.  Therefore, we need to understand first the microscopic origin of the resistivity. 
Separating the resistivity into  elastic and inelastic contributions via Mathhiessen's rule is the first step in this direction.

To advance understanding in this important issue, one needs to 
develop the theoretical understanding for the 
effect of  inelastic scattering on AHE at finite temperature.
This issue has been partly treated in the hopping theory 
of the AHE described in section \ref{sec:exp:oxcide} where 
 phonon assisted hopping was assumed.
However, the effects of  inelastic scattering on 
the intrinsic and extrinsic mechanisms are not clear at the moment. 
Especially,  spin fluctuation at finite $T$ 
remains the most essential and difficult problem in the theory 
of magnetism, and usually the mean field approximation 
breaks down there. The approximate treatment in terms of 
the temperature dependent exchange splitting, {\it e.g.}, for SrRuO$_3$ (Sec. \ref{sec:exp:oxcide}),
needs to be reexamined by more elaborated method such as the 
dynamical mean field theory, taking into account the 
quantum/thermal fluctuation of the ferromagnetic moments.
These type of studies of the AHE may shed some light on 
the nature of the spin fluctuation in ferromagnets.
Also the interplay between localization and the AHE  should be pursued further in 
the intermediate and strong disorder regimes.
These are all vital issues for quantum 
transport phenomena in solids in general, as well as for AHE specifically.

\begin{figure}
\incl[width=\columnwidth]{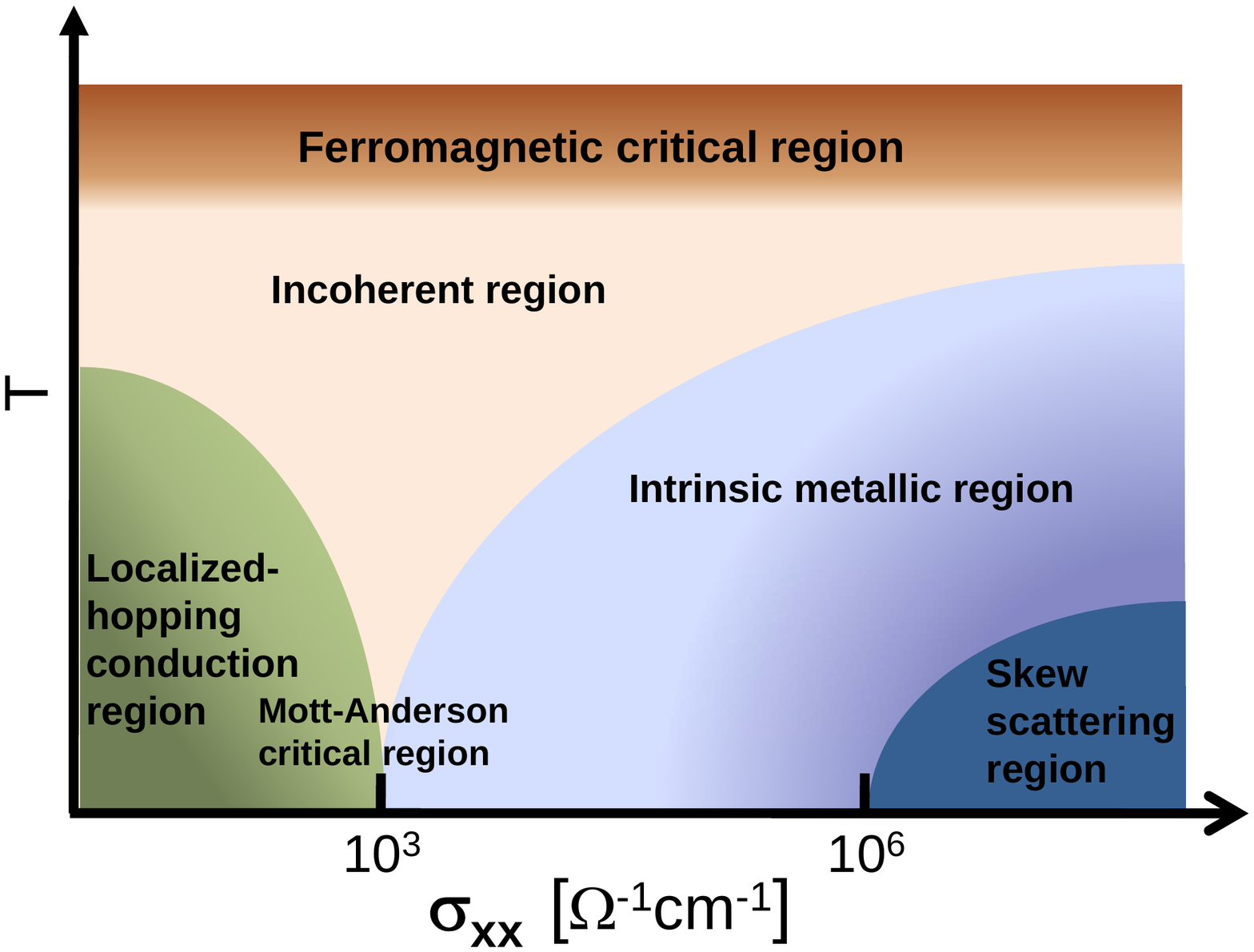}
\caption{\label{fig:PhaseD}
A speculative and schematic phase diagram for 
the anomalous Hall effect in the plane of the 
diagonal conductivity $\sigma_{xx}$ and the 
temperature $T$. 
}
\end{figure}

Admitting that more work needs to be done,
in Fig. \ref{fig:PhaseD} we propose a speculative 
and schematic crossover diagram in the 
plane of diagonal conductivity $\sigma_{xx}$ of the 
Boltzmann transport theory (corresponding
to the disorder strength) and the temperature $T$.
Note that a real system should move
along the $y$-axis as  temperature is changed,
although the observed $\sigma_{xx}$ changes with T.
This phase diagram reflects the empirical fact 
that  inelastic scattering
kills off the extrinsic skew scattering contribution more
effectively,  leaving the intrinsic  and side-jump contributions as 
dominant  at finite temperature. 
We want to help stress that the aim of this figure is to 
promote further studies of the AHE
and to identify the location of each region/system of 
interest. Of course, the generality of this 
diagram is not guaranteed and it is possible that
the crossover boundaries and even the topology of
the phase diagram might depend on the strength of 
the spin-orbit interaction and other details of the 
system.

There still remain many other issues to be studied in the future.
First-principles band structure calculations for 
AHE are still limited to a few number of materials,
and should be extended to many other ferromagnets.
Especially, the heavy fermion systems are an
important class of materials to be studied in detail.
Concerning  this point, a more economical  numerical technique is 
now available~\cite{Marzari:1997_a}.
This method employs the maximally localized Wannier 
functions, which can give the best tight-binding model parameters in an energy 
window of several tens of eV's near the Fermi level. The algorithm for the 
calculation of $\sigma_{xy}$ in the Wannier interpolation scheme has also 
been developed~\cite{Wang:2006_a}. Application of these newly developed 
methods to a large class of materials should be a high priority in the future.

AHE in the dynamical regime is a related interesting problem.
The magneto-optical effects such as the Kerr/Faraday rotation have been  
the standard experimental methods to detect the ferromagnetism. 
These techniques usually focus, however, on the high energy region 
such as the visible light. In this case, an atomic or local picture 
is usually sufficient to interpret the data, and the spectra are not directly 
connected to the d.c. AHE. Recent studies have revealed that the
small energy scale comparable to the spin-orbit interaction is relevant
to AHE which is typically $\sim 10$ meV for 3d transition metal 
and $\sim 100$ meV in DMSs \cite{Sinova:2003_a}. 
This means that the dynamical response, {\it i.e.}, $\sigma_{xy}(\omega)$, in
the THz and infrared region will provide important information on the AHE.
     
A major challenge for experiments is to find examples of a quantized anomalous Hall effect.
There are two candidates at present: (i) a ferromagnetic insulator with a 
band gap~\cite{Liu:2008_a}, and (ii) a disorder induced Anderson insulator with a
quantized Hall conductance~\cite{MOnoda:2003_a}. 
Although theoretically expected,
it is an important issue to establish experimentally
that the quantized Hall effect can be realized even without an external magentic field.   
Such a finding would be the ultimate achievement in identifying an intrinsic AHE.
The dissipationless nature of the anomalous Hall current will manifest itself
in this quantized AHE; engineering systems using  quantum wells 
or  field effect transistors is a promising direction to realize this novel effect.

There are many promising directions for extensions of ideas developed through studies of the AHE.
For example, one can consider
several kinds of ``current'' instead of the charge current.
An example is the thermal/heat current, which can be also induced in a similar fashion 
by the anomalous velocity. The thermoelectric effect has been discussed briefly in 
Sec. \ref{sec:exp:semi}, where combining all the measured thermoelectric transport coefficients
helped settle the issue of the scaling relation $\sigma_{xy}^{AH}\sim\sigma_{xx}^2$ 
in metallic DMSs~\cite{Pu:2008_a}. Recent studies in Fe alloys doped with Si and Co discussed in Sec.~\ref{sec:tm} 
followed a similar strategy~\cite{Shiomi:2009_a}. 
From the temperature dependence of the Lorentz number, they identified the crossover
between the intrinsic and extrinsic mechanisms. 
Further studies of thermal transport will shed some
light on the essence of AHE from a different side. 

Spin current is also a quantity of recent great interest.
A direct generalization of AHE to the spin current is the spin Hall 
effect~\cite{Murakami:2003_a,Sinova:2004_a,Kato:2004_a,Wunderlich:2005_a}, which can be regarded as
the two copies of AHE for up and down spins with the opposite
sign of $\sigma_{xy}$. In this effect, a spin current is 
produced perpendicular to the charge current. 
An interesting recent development in spin Hall effect is
that the quantum spin Hall effect and topological 
insulators have been theoretically predicted and experimentally
confirmed. We did not include this exciting new 
and still developing topic in this review article. Interested readers are 
referred to the original papers and references therein~\cite{Kane:2005_a,Bernevig:2006_a,Koenig:2007_a}.

\section*{Acknowledgements}
We thank the following people for collaboration and useful discussion. A. Asamitsu, P. Bruno, D. Culcer, V. K. Dugaev, Z. Fang, J. Inoue, T. Jungwirth, M. Kawasaki, S. Murakami, Q. Niu,  T. S. Nunner, K. Ohgushi, M. Onoda, Y. Onose, J. Shi, R. Shindou, N. A. Sinitsyn, N. Sugimoto,   Y. Tokura,
and J. Wunderlich.
The work was supported 
by Grant-in-Aids (No. 15104006, No. 16076205, No. 17105002, No. 19048015, N0. 19048008) 
and NAREGI Nanoscience Project from the Ministry of Education, Culture, Sports, Science, and Technology of Japan.
S. O. acknowledges partial support from Grants-in-Aid for Scientific Research under Grant No. 19840053 from the Japan Society of the Promotion of Science, and under Grant No. 20029006 and No. 20046016 from the MEXT of Japan.
J. S. acknoledges partial support from  ONR under Grant No. ONR-N000140610122, by NSF
under Grant No. DMR-0547875, by SWAN-NRI, by the
Cottrell Scholar grant of the Research Corporation. 
N. P. O. acknowledges partial support from a MRSEC grant
from the U. S. National Science Foundation (DMR-0819860).
AHM acknowledges support from the Welch Foundation and from the DOE.

\bibliographystyle{apsrmp}

\end{document}